\newif\ifPDFLaTeX
\expandafter\ifx\csname pdfoutput\endcsname\relax\else\PDFLaTeXtrue\fi

\ifPDFLaTeX
  \documentclass[12pt,a4paper,notitlepage,chapters]{flex}
  \usepackage{type1cm}
  \usepackage[pdftex,colorlinks]{hyperref}
  \usepackage[pdftex]{graphicx,feynmp,emp}
  \usepackage{thophys,mcite}
  \usepackage{array,amsmath,amssymb,amscd,acronym,units}
  \empaddtoTeX{\usepackage{amsmath,amssymb,units}}
  \empaddtoprelude{input graph;}
  \empaddtoprelude{input boxes;}
  \empaddtoTeX{\usepackage{type1cm}}
\else
  \documentclass[12pt,a4paper,notitlepage,chapters]{flex}
  \usepackage{thophys,mcite}
  \usepackage{graphicx,array,amsmath,amssymb,amscd,feynmp,emp,acronym,units}
  \empaddtoTeX{\usepackage{amsmath,amssymb,units}}
  \empaddtoprelude{input graph;}
  \empaddtoprelude{input boxes;}
  \empaddtoTeX{\usepackage{euler,beton}}           
\fi
\allowdisplaybreaks

\makeatletter
\makeatletter
  \def\fps@figure{t}
  \def\fps@table{b}
  \long\def\@makecaption#1#2{%
    \vskip\abovecaptionskip
    \sbox\@tempboxa{#1: \textit{#2}}%
    \ifdim\wd\@tempboxa>\hsize
      #1: \textit{#2}\par
    \else
      \global\@minipagefalse
      \hb@xt@\hsize{\hfil\box\@tempboxa\hfil}%
    \fi
    \vskip\belowcaptionskip}
\makeatother
\def\Circe/{\texttt{circe}}
\def\Kirke/{\texttt{circe}}
\def\AutosEpha/{$\alpha\upsilon\tau o\varsigma$ $\varepsilon\varphi\alpha$}
\makeatletter
\newenvironment{cutequote@}[1]%
 {\def\cutequoteref{#1}%
  \begin{raggedleft}%
    \samepage%
    \leftskip.3\textwidth plus.6\textwidth%
    \bgroup\itshape}%
 {\\\egroup[\cutequoteref]\\
  \end{raggedleft}
  \vspace*{\baselineskip}
  \@afterheading}
\newenvironment{cutequote}[1]%
 {\begin{cutequote@}{\textsc{#1}}}
 {\end{cutequote@}}
\newenvironment{cutequote*}[2]%
 {\begin{cutequote@}{\textsc{#1:} \textit{#2}}}
 {\end{cutequote@}}
\makeatother
\newenvironment{epochs}[1]%
 {\begin{list}{}%
   {\setlength{\leftmargin}{3em}%
    \setlength{\rightmargin}{3em}%
    \setlength{\itemindent}{1em}%
    \setlength{\listparindent}{0pt}%
    \settowidth{\labelwidth}{\textbf{#1:}}%
    }}%
 {\end{list}}
\newtheorem{theorem}{Theorem}

\newcommand{\program}[1]{\texttt{#1}}
\DeclareMathOperator{\Vol}{Vol}
\DeclareMathOperator{\atan}{atan}
\DeclareMathOperator{\tr}{tr}
\DeclareMathOperator{\Aut}{\mathbf{Aut}}
\newcommand{\greenfunction}{Green's function}
\renewcommand{\Box}{\partial^2}
\makeatletter
  \@ifundefined{frontmatter}%
    {\def\frontmatter{\pagenumbering{roman}}%
     \def\mainmatter{\cleardoublepage\pagenumbering{arabic}}}
    {}
  \@ifundefined{abstract}%
    {\newenvironment{abstract}%
      {\newpage\section*{Abstract}\begin{quote}}%
      {\end{quote}\newpage}}
    {}
\makeatother

{\def\$#1: #2 ${#2}}

\begin{document}
\bibliographystyle{prsty-mod}

\title{Electroweak~Gauge~Bosons at Future~Electron-Positron~Colliders}

\author{%
  Thorsten Ohl\\
  \hfil\\
  Darmstadt University of Technology\\
  Schlo\ss gartenstr.~9 \\
  D-64289 Darmstadt, Germany\\
  \hfil\\
  \texttt{<ohl@hep.tu-darmstadt.de>}}

\date{\hfil}

\bgroup  
\makeatletter
\def\@oddhead{%
  \parbox{\textwidth}{%
    \begin{flushright}
      LC-REV-1999-005\\
      IKDA~99/11\\
      June 1999
    \end{flushright}}}
\makeatother
\renewcommand{\thispagestyle}[1]{\relax}
\maketitle
\frontmatter
\begin{abstract}
  The interactions of electroweak gauge bosons are severely constrained
  by the symmetries inferred from low energy observables.  The
  exploration of the dynamics of electroweak symmetry breaking
  requires therefore both an experimental sensitivity and a
  theoretical precision that is better than the natural scale of
  deviations from minimal extrapolations of the low energy effective
  theory.  Some techniques for obtaining improved theoretical
  predictions for precision tests at a future $e^+e^-$~linear collider
  are discussed.
\end{abstract}
\clearpage
\egroup

\chapter*{Preface}

Several excellent, up-to-date reviews of the wide range of physics
that can be probed at a future $e^+e^-$~\ac{LC} are
available~\cite{%
  Zerwas:1996:LC_review,
  Murayama/Peskin:1997:LC_review,
  Accomando/etal:1998:TESLAReview}.
The present review discusses some physical and technical aspects of
electroweak gauge bosons, that will be important at a~\ac{LC}.
The importance of gauge invariance even in a conservative \ac{EFT}
approach, which does not pretend to be valid to arbitrarily short
distances, is emphasized.  The implementation of gauge invariance in
perturbation theory is discussed.  Techniques for improving the
convergence and speed of high dimensional phase space integrations are
described.

\section*{Acknowledgments}

I thank Panagiotis Manakos, Hans Dahmen, and Thomas Mannel for their
support over the years.  Howard Georgi deserves the credit for the
most inspirational year of physics so far (Howard provided an open
door, physics insight, and inspiration, while DFG provided the
financial support).

I thank Harald Anlauf, Edward Boos, Wolfgang Kilian, and Alex Nippe
for enjoyable and productive collaborations that have contributed to
some of the results presented in the following pages.


I thank the organizers of the Maria Laach school for the invitation to
climb on the Effective Field Theory soapbox, as reproduced in part in
appendix~\ref{sec:RG}.  The students make this a fun place to lecture
and the hospitality of the congregation of Maria Laach makes this
school unique.

I thank Deutsche Forschungsgemeinschaft (DFG) and Bundesministerium
for Bildung, Wissenschaft, Forschung und Technologie (BMBF) for
financial support.


\tableofcontents
\mainmatter
\begin{empfile}
\begin{fmffile}{\jobname 1}
\fmfset{arrow_ang}{10}
\fmfset{curly_len}{2mm}
\fmfset{wiggly_len}{3mm}
\fmfcmd{%
  style_def isomorphism expr p =
    cdraw (subpath (0, 1 - arrow_len/pixlen(p,10)) of p);
    cfill (harrow (p, 1))
  enddef;
  style_def morphism expr p =
    draw_dots (subpath (0, 1 - arrow_len/pixlen(p,10)) of p);
    cfill (harrow (p, 1))
  enddef;}
\def\fmfcd(#1,#2){%
  \begin{minipage}{#1\unitlength}%
    \vspace*{.5\baselineskip}%
    \begin{fmfgraph*}(#1,#2)%
    \fmfset{arrow_len}{3mm}%
    \fmfset{arrow_ang}{10}%
    \fmfstraight}
\def\endfmfcd{%
    \end{fmfgraph*}%
    \vspace*{.5\baselineskip}%
  \end{minipage}}
\newcommand{\fmfcdmorphism}[4]{%
  \fmf{#1,label.side=#2,label.dist=3pt,label={\small $#4$}}{#3}}
\newcommand{\fmfcdisomorph}[3][left]{%
  \fmfcdmorphism{isomorphism}{#1}{#2}{#3}}
\newcommand{\fmfcdmorph}[3][left]{%
  \fmfcdmorphism{morphism}{#1}{#2}{#3}}
\newcommand{\fmfcdeq}[1]{\fmf{double}{#1}}
\def\fmfcdsetaux[#1]#2{%
  \fmfv{decor.shape=circle,decor.size=18pt,foreground=white,
        label.dist=0,label=$#1$}{#2}}
\makeatletter
  \def\fmfcdset{\@dblarg{\fmfcdsetaux}}
\makeatother

\chapter{Introduction}
\label{sec:intro}

\section{The Art and Craft of Elementary~Particle~Physics}
\label{sec:craft}

\begin{cutequote*}{Bertrand Russell}{A~History of Western Philosophy}
  To teach how to live without certainty, and yet without being
  paralyzed by hesitation, is perhaps the chief thing that philosophy,
  in our age, can still do for those who study it.
\end{cutequote*}

In a highly idealized scenario, \ac{TEPP} proceeds by condensing the results
of experimental observations up to some energy scale into a model, which is
designed to remain consistent at higher energies.  In a second
step, predictions for experiments at these higher energy scales are derived.
Finally, comparisons with subsequent experimental results are used to
refine the model, if necessary, and the process starts anew.

Therefore, we can identify two complementary tasks: \emph{model
building} on one hand and \emph{calculation of observables} on the
other.
Below, we will focus our attention in the area of model building to
the systematic procedure of~\ac{EFT} and its r\^ole in the modern
interpretation of the \ac{SM} with simple extensions.  The methods for
the calculation of observables will be discussed in greater detail.
Such calculations are naturally divided into two separate parts:
\emph{calculation of amplitudes and differential cross sections} and
\emph{phase space integration}.  Some methods for both parts
are described with a noticeable bias towards the
author's research interests.

\section{Success and Failure of the Standard Model}
\label{sec:sm?}

\begin{cutequote}{Friedrich Nietzsche}
  Tout comprendre,\\
  c'est tout mepriser.
\end{cutequote}

\begin{empcmds}
vardef describe_@# (expr x, y, xl, yl, s, d) text w = 
  Garw_:=1;
  Gaddto_ doublepath Gtcvi_ (xl, yl) {d} ... {d} Gtcvi_ (x, y)
    withpen currentpen Gwithlist_ _op_ w;
  glabel.@# (s, xl, yl) w;
enddef;
vardef describel (expr x, y, xl, yl, s) text w = 
  describe_.lft (x, y, xl, yl, s, right) w
enddef;
vardef describer (expr x, y, xl, yl, s) text w = 
  describe_.rt (x, y, xl, yl, s, left) w
enddef;
\end{empcmds}

\begin{empcmds}
vardef lepton (expr roots, start, finish, name) = 
  save physics; path physics;
  augment.physics (start, roots);
  augment.physics (finish, roots);
  pickup pencircle scaled 2pt;
  gdraw physics;
  pickup pencircle scaled 0.5pt;
  describer (finish + 1, roots, finish + 3, roots, name);
  pickup pencircle scaled 1pt;
enddef;
\end{empcmds}

\begin{empcmds}
vardef hadron (expr proots, roots, start, finish, name) = 
  save physics; path physics;
  augment.physics (start, proots);
  augment.physics (finish, proots);
  augment.physics (finish, roots);
  augment.physics (start, roots);
  gfill (physics -- cycle) withcolor 0.7white;
  pickup pencircle scaled 0.5pt;
  describel (start - 1, roots, start - 3, roots, name);
  pickup pencircle scaled 1pt;
enddef;
\end{empcmds}

In the decade after the commissioning of~LEP1, many aspects of
the~\acl{SM}~\cite{Glashow:1961:SM,*Weinberg:1967:SM,*Salam:1968:SM}
of elementary particle physics have been tested 
successfully to a precision previously
unimaginable~\cite{Bardin/etal:1997:precision,LEPEWWG:1999:global}.
Indeed it is not without irony that one of the main sources of
theoretical uncertainty derives from the insufficient precision of the
old data on hadron production in $e^+e^-$-collisions at a
few~$\unit{GeV}$, despite recent theoretical
progress in the interpretation of this
data~\cite{Eidelman/Jegerlehner:1995:alpha,*Davier/Hoecker:1998:alpha,
*Kuehn/Steinhauser:1998:alpha}.
\begin{figure}
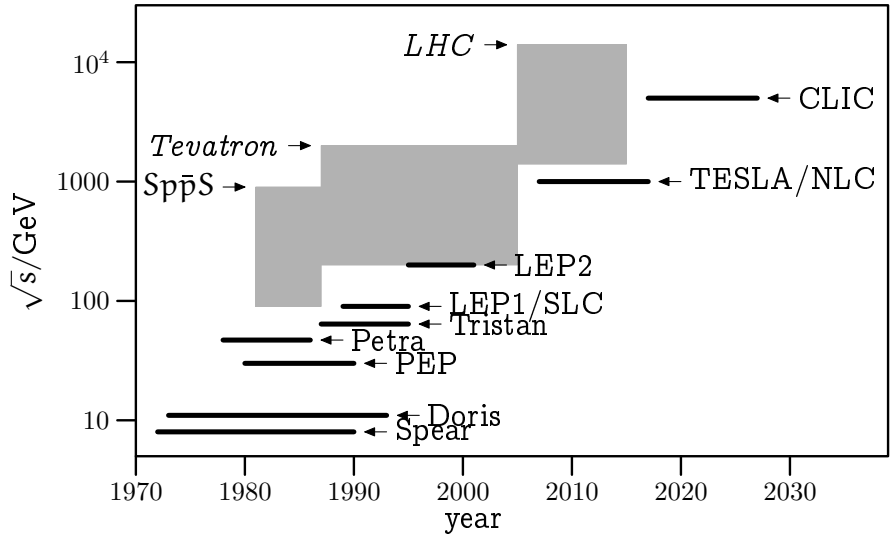

  \begin{center}
    \begin{empgraph}(100,60)
      pickup pencircle scaled 1pt;
      setcoords (linear, log);
      setrange (1970, "5", 2039, "3e4");
      glabel.bot (btex year etex, OUT);
      glabel.lft (btex $\sqrt{s}/\unit{GeV}$ etex rotated 90, OUT);
      hadron (   "90",  "900", 1981, 1987, btex $Sp\bar pS$ etex);
      hadron (  "200",  "2e3", 1987, 2005, btex \textit{Tevatron} etex);
      hadron ("1.4e3", "14e3", 2005, 2015, btex \textit{LHC} etex);
      lepton (    "8",         1972, 1990, btex Spear etex);
      lepton (   "11",         1973, 1993, btex Doris etex);
      lepton (   "30",         1980, 1990, btex PEP etex);
      lepton (   "47",         1978, 1986, btex Petra etex);
      lepton (   "64",         1987, 1995, btex Tristan etex);
      lepton (   "90",         1989, 1995, btex LEP1/SLC etex);
      lepton (  "200",         1995, 2001, btex LEP2 etex);
      lepton (  "1e3",         2007, 2017, btex TESLA/NLC etex);
      lepton (  "5e3",         2017, 2027, btex CLIC etex);
    \end{empgraph}
  \end{center}
  \caption{\label{fig:reaches}%
    Physics reaches of past, present and future colliders.  Hadron
    colliders can not utilize the full beam energy in partonic
    collisions since the momentum is shared by quarks and gluons.  As
    a rule of thumb, the lower border of the grey
    rectangles ($\sqrt{s'}=\sqrt{s}/10$) corresponds to the physics
    reach of hadron colliders. (Data taken
    from~\protect\cite{PDG:1996,PDG:1998}.)}
\end{figure}

The important qualitative features~\cite{Glashow/etal:1970:GIM} of
the~\ac{SM} were established from low energy experiments at a time
before any of the colliders shown in figure~\ref{fig:reaches} were
commissioned.
The symmetry structure of the~\ac{SM} as
a~$\textrm{SU}(2)_L\times\textrm{U}(1)_Y$ gauge theory that is
spontaneously broken to the~$\textrm{U}(1)$ of~\ac{QED} has been firmly
established by the stunning success of the~\ac{SM} at~LEP1.
Nevertheless, direct information regarding the underlying dynamics of
the~\ac{EWSB} has yet to be found.

\begin{figure}
  \begin{center}
    \includegraphics[width=.6\textwidth]{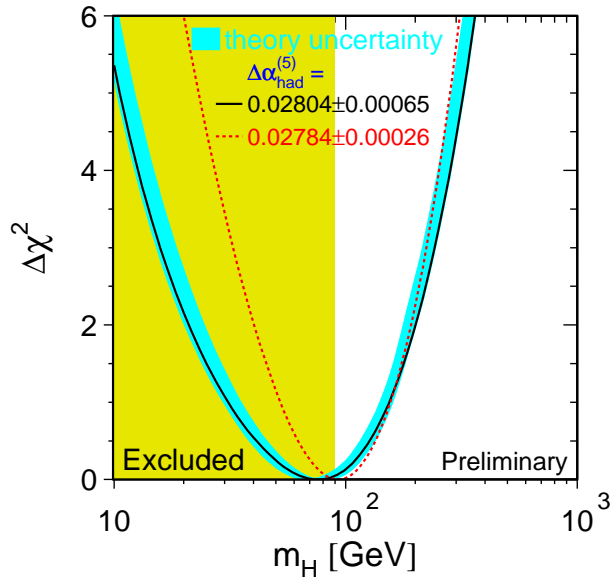}
  \end{center}
  \caption{\label{fig:blueband}%
    The famous ``blue band'' derived from fits of electroweak precision
    data to~\protect\acs{SM}
    predictions~\protect\cite{LEPEWWG:1999:global}, that
    suggests a very light~\protect\acs{SM} Higgs particle, with a mass
    close to the present exclusion limit.  The area excluded by direct
    searches is shaded in light grey on the left.  The parabolas show
    the distance in likelihood from the best fit as a function of the
    Higgs mass for two different determinations of the running
    electromagnetic coupling~$\alpha_{\text{QED}}(M_Z)$.}
\end{figure}
The existing data is compatible with the minimal~\ac{SM} with a light
elementary Higgs particle~\cite{LEPEWWG:1999:global}.  Global fits of
the~\ac{SM} predictions to the~LEP1 precision observables
(including~$m_{\text{top}}$ from~Tevatron and~$M_W$
from~LEP2) suggest a Higgs mass close to the present
exclusion limits (see figure~\ref{fig:blueband}).  However, the Higgs
mass enters the precision observables only logarithmically and the
limits are therefore weak.

Despite the phenomenological hints in its favor, it must be mentioned
that the minimal~\ac{SM} with an elementary Higgs particle is
unattractive from a theoretical point of view.  There is a
quadratically divergent contribution to the Higgs mass renormalization
\begin{equation}
  \delta m_{\text{Higgs}}^2 =
  \parbox{21\unitlength}{%
    \begin{fmfgraph}(20,5)
       \fmfstraight
       \fmfleft{l,dl}
       \fmfright{r,dr}
       \fmf{plain}{l,v,v,r}
       \fmfdot{v}
    \end{fmfgraph}}
   \propto \frac{g}{16\pi^2} \Lambda^2
\end{equation}
that requires an unrealistic fine-tuning of the mass to one part
in~$10^{38}$ in order to keep the Higgs light and the \ac{EWSB} scale
close to~$\unit[1]{TeV}$ (see section~\ref{sec:relevant} for more
details).

There are two ways around this problem and both feature fermions
prominently.  A supersymmetry can relate the scalar mass to a fermion
mass.  Since a fermion mass is protected from renormalization by
chiral symmetry, it can be naturally light and the scalar mass can
naturally remain on the order of the supersymmetry breaking scale.  On
the other hand, Goldstone bosons from the spontaneous breaking of a
continuous symmetry are protected by Goldstone's theorem and can be
naturally light on the order of the symmetry breaking scale as well.
Here the chiral symmetry of fermions is required to protect the
particles that take part in the interaction, which is responsible for the
symmetry breaking, from a large mass renormalization.
The latter models are disfavored by current precision data.

A necessary, though not sufficient, condition for the unification of
gauge interactions in a single simple group at a~\ac{GUT} scale is that the
running couplings (see sections~\ref{sec:anodim}
and~\ref{sec:integrating-out} for details) have a common value at one
scale.  According to the data from the pre-LEP1 era, the~\ac{SM} has
this property.  However, LEP1 precision data revealed that the
coupling do not meet, unless supersymmetric partners are added to the
particle spectrum~\cite{Amaldi/deBoer/Fuerstenau:1991:Unification}.

Even if the naturalness problem is solved by the~\ac{MSSM} and gauge
coupling unification is achieved, models with an elementary Higgs
particle remain
unsatisfactory, because they not only provide no insight into a theory
of flavor~\cite{Georgi:1993:sad}, they also screen any theory of flavor
so effectively that experiments at the~\ac{GUT} scale would be
required to shed any light on the open questions of the physics of
flavor, in particular the ratios of fermion masses and the fermion
mixing angles.

A strongly interacting weak sector that generates~\ac{EWSB}
dynamically has a greater potential of explaining the physics of
flavor.  In such a theory it is conceivable that everything
can be calculated from (renormalized) Clebsh-Gordan coefficients in a
tower of spontaneous symmetry breakings.
Such a reduction from the continuous manifold of free parameters to a
discrete set of symmetry breaking patterns allowed by group theory
would increase the predictive power of~\ac{TEPP} immensely.

Today we have no working theory of flavor that could explain the
ratios of fermion masses and the values of the mixing angles.  We do
not know if Nature was kind enough to hide the answer at the
$\unit{TeV}$-scale.  Of course, she is under no obligation to do so
and could as well have used supersymmetry to put the answers out of
our experimental reach somewhere between the~\ac{GUT} and Planck
scales.  The remarkable success of the \ac{MSSM} with no intermediate
mass scales and gauge coupling
unification~\cite{Amaldi/deBoer/Fuerstenau:1991:Unification} might 
indeed have been the first direct evidence of such a grand desert.

\section{The Mission of the Linear Collider}
\label{sec:mission}

\begin{figure}
  \begin{center}
    \includegraphics[width=.6\textwidth]{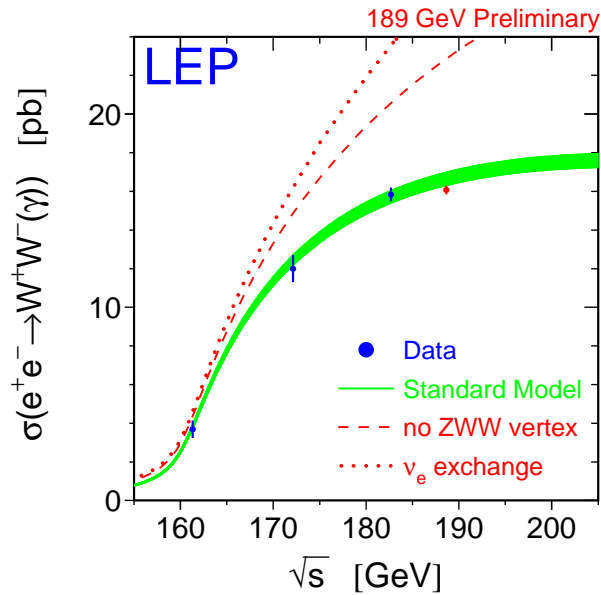}
  \end{center}
  \caption{\label{fig:wwthreshold}%
    $W^+W^-$-production cross
    section~\protect\cite{LEPEWWG:1999:global} close to
    the threshold is in spectacular agreement with the \protect\acs{SM}
    prediction. Already at $\unit[183]{GeV}$, the need for a
    $W^+W^-Z^0$ vertex was clearly visible.}
\end{figure}
\begin{figure}
  \begin{center}
    \includegraphics[width=.3\textwidth]{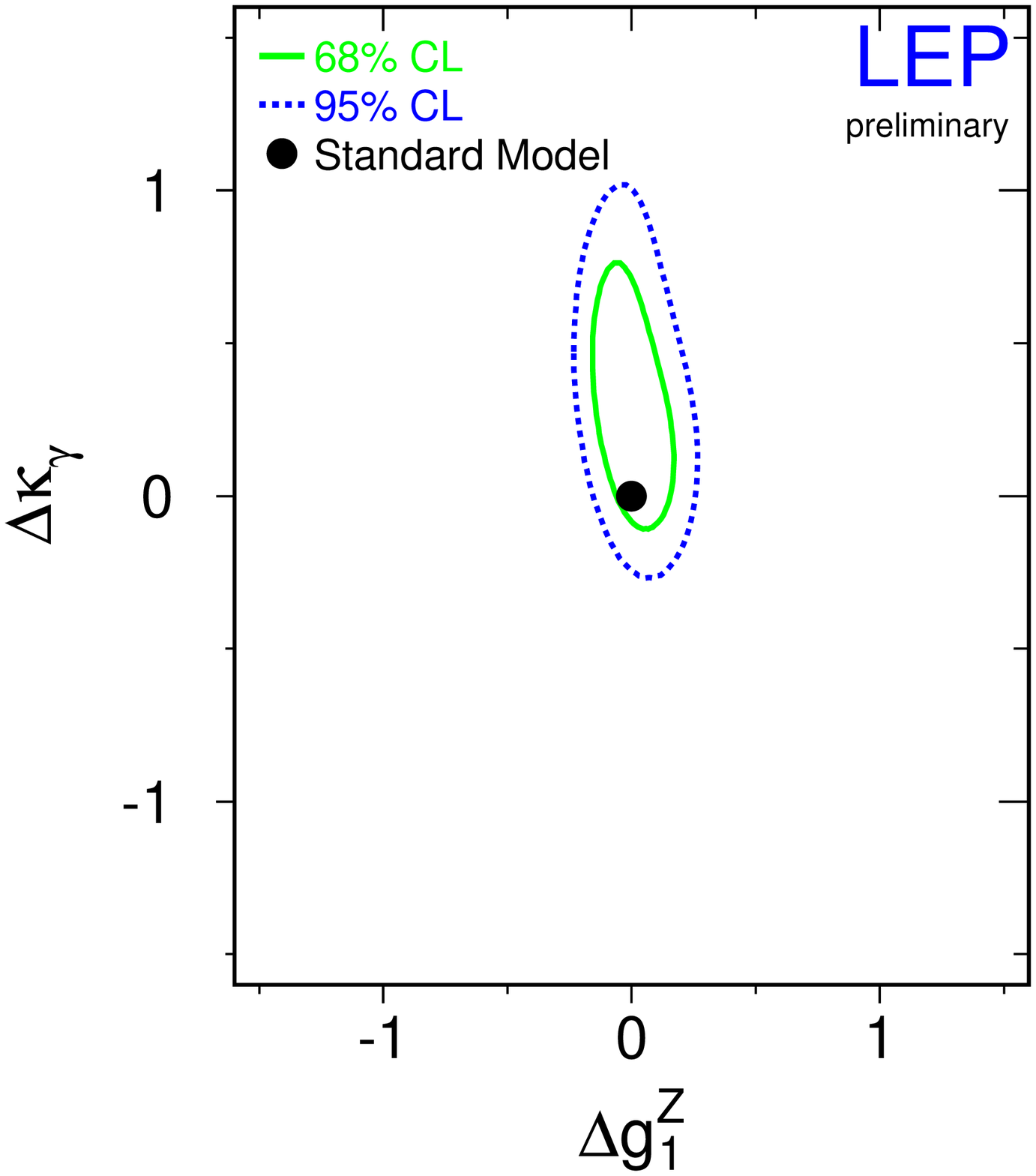}
    \includegraphics[width=.3\textwidth]{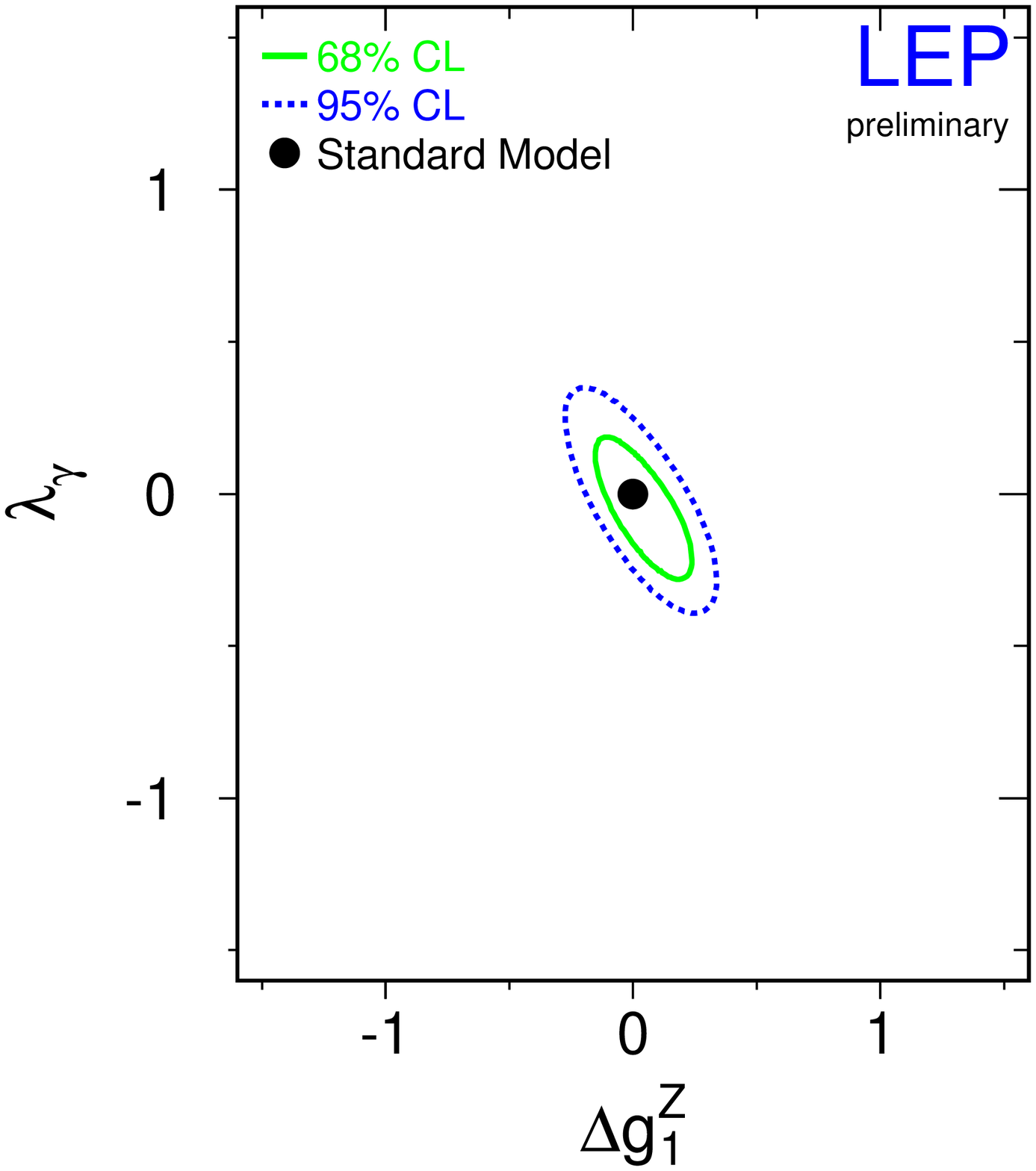}
    \includegraphics[width=.3\textwidth]{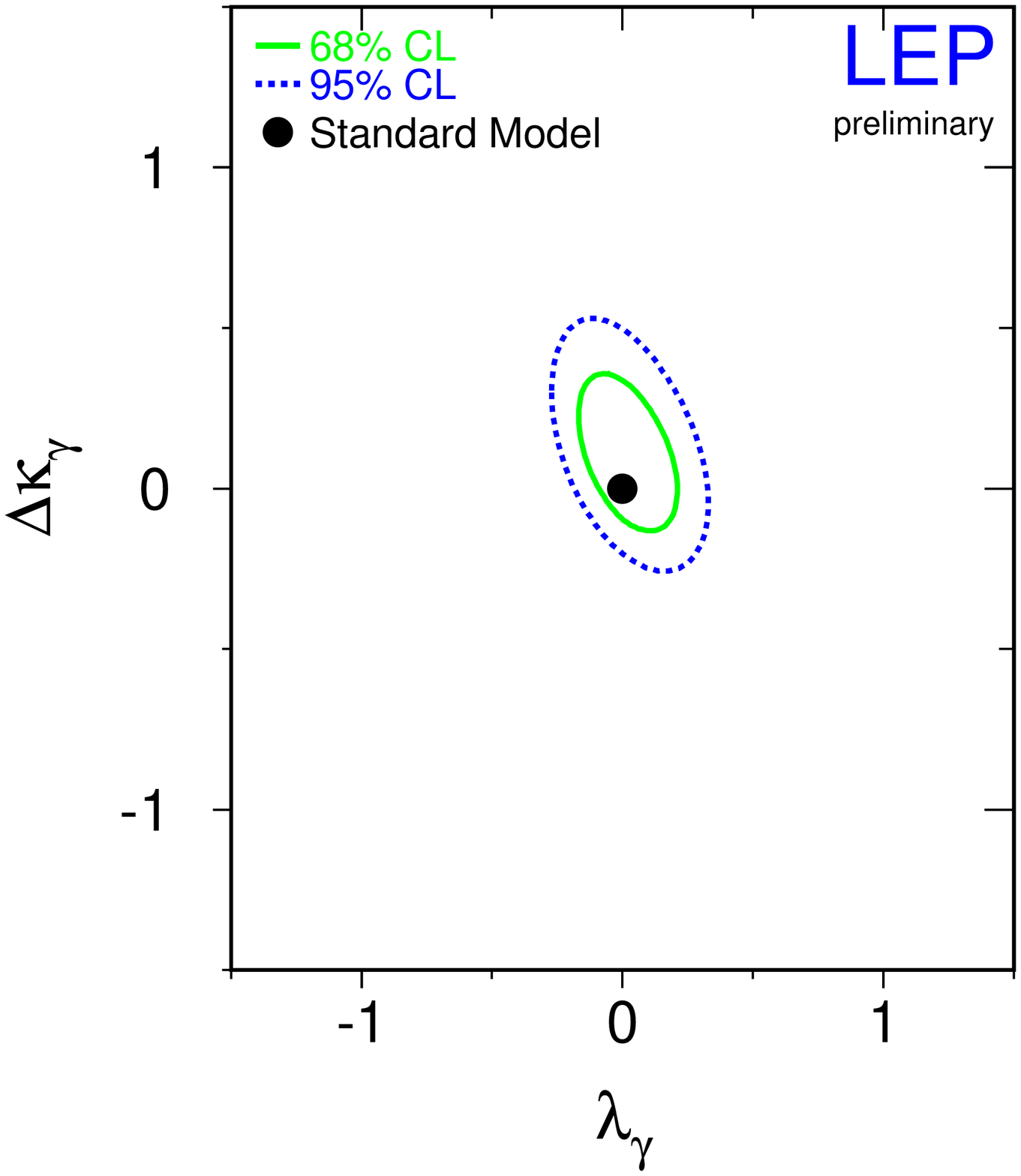}
  \end{center}
  \caption{\label{fig:tgc-vancouver98}%
    Combined LEP2 and Tevatron
    limits on \protect\acl{TGC} at the time of the ICHEP98 conference
    in Vancouver, B.C. (July 1998)~\protect\cite{LEPEWWG:1998:TGC}.
    The~\protect\acs{SM} values are favored, but the $90\%$ CL
    boundaries are still~$O(1)$.  See figure~\ref{fig:tgc-moriond99}
    for a dramatic improvement (note the different scale!).}
\end{figure}
\begin{figure}
  \begin{center}
    \includegraphics[width=.3\textwidth]{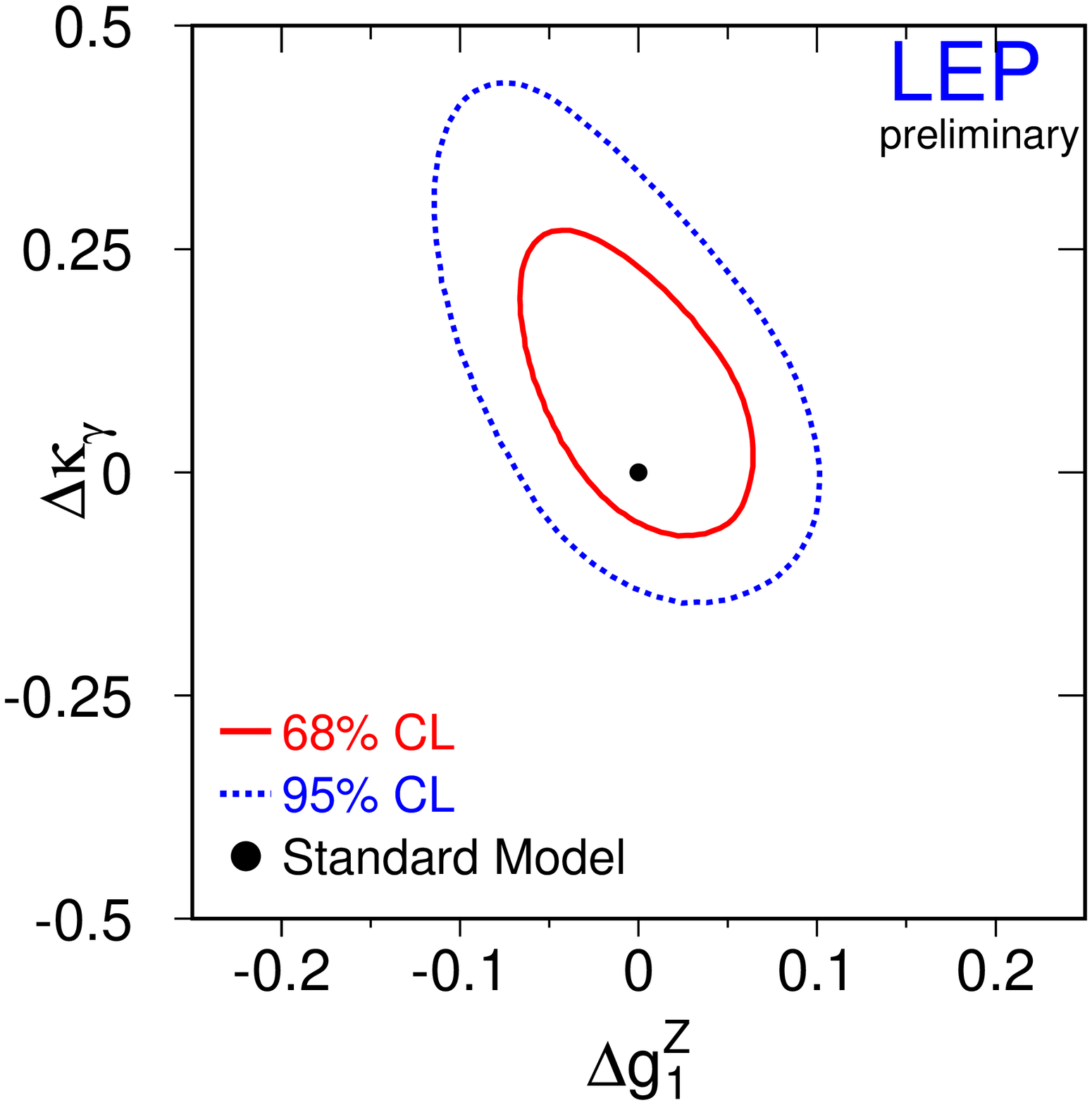}
    \includegraphics[width=.3\textwidth]{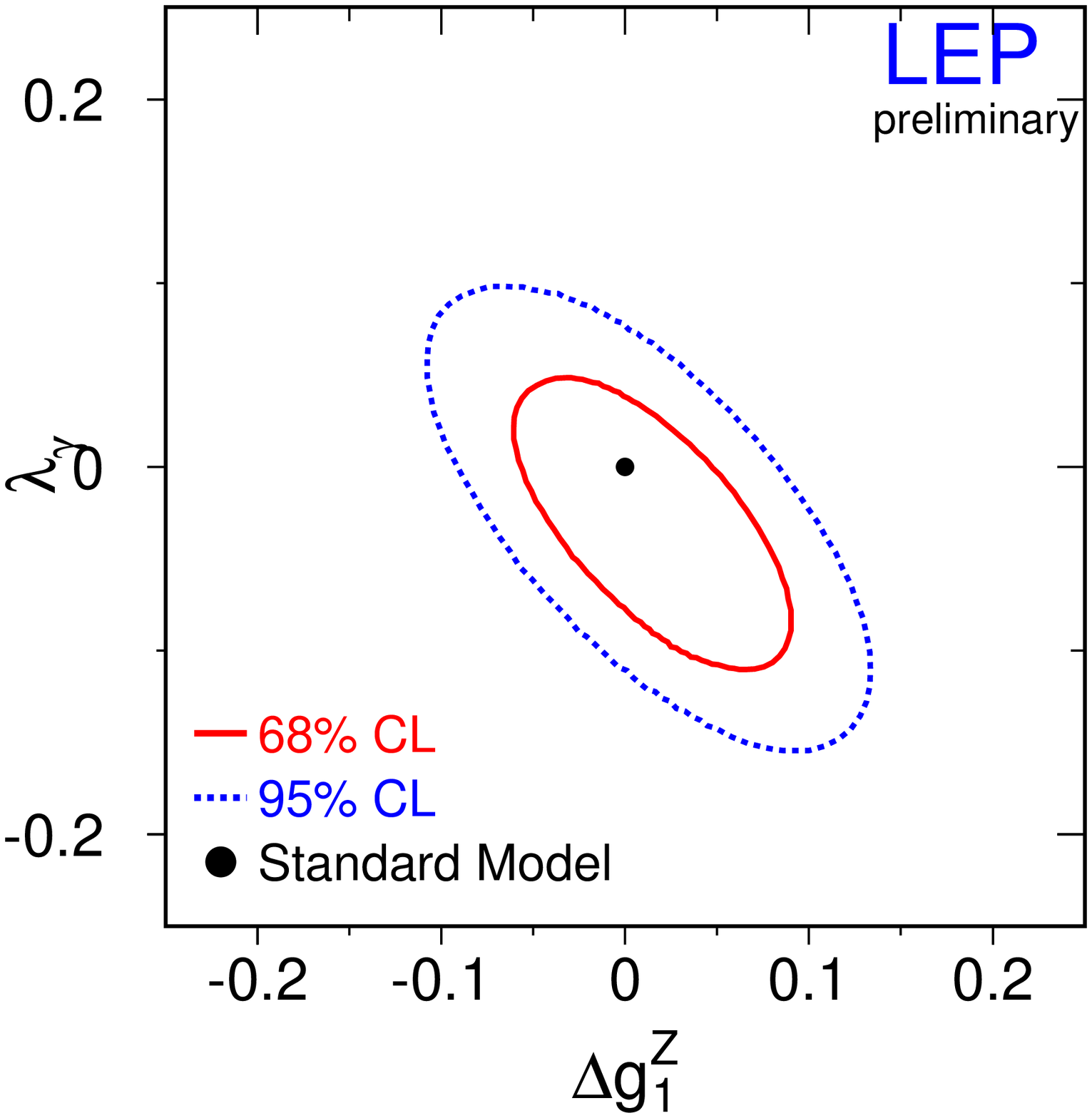}
    \includegraphics[width=.3\textwidth]{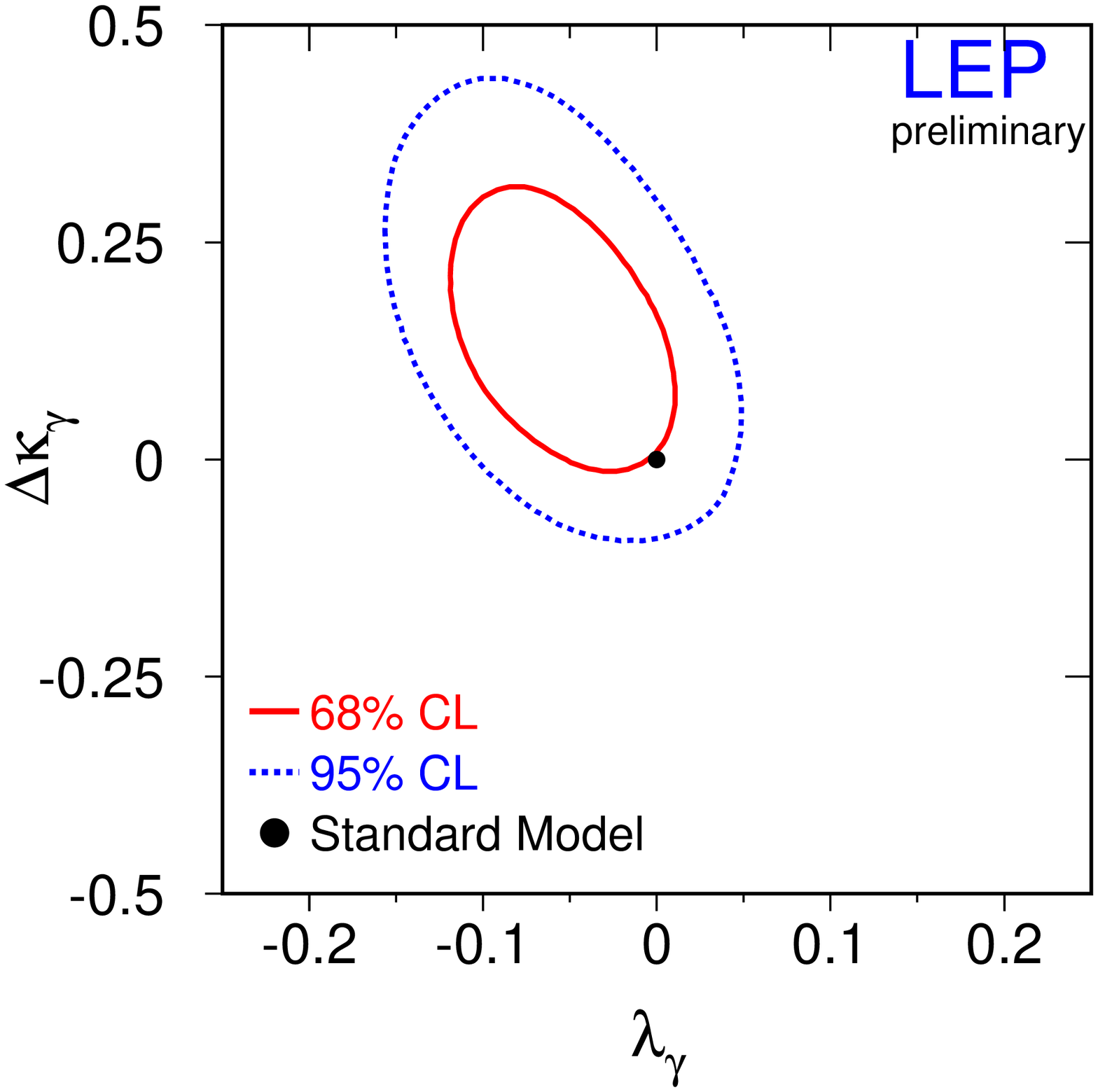}
  \end{center}
  \caption{\label{fig:tgc-moriond99}%
    Combined LEP2 and Tevatron
    limits on \acl{TGC} at the time of the MORIOND 1999 conference
    (March 1999)~\protect\cite{LEPEWWG:1999:global}. The
    $\unit[189]{GeV}$ data
    gives a dramatic improvement over figure~\ref{fig:tgc-vancouver98}
    (note the different scale!) and vanishing~\protect\acs{TGC} are now
    definitely excluded.}
\end{figure}
{}From the previous section, we can identify the main missions for a
future \ac{LC}.  Assuming that neither LEP2, TEV33, or LHC have done
so, the~\ac{LC} has to try to \emph{discover} the Higgs particle and
supersymmetric partners of the particles in the~\ac{SM}.  If a Higgs
particle has already been found at the time of commissioning,
the~\ac{LC} will be in a unique position to determine its nature by measuring
its properties, such as branching ratios, that are difficult to access
at a hadron collider, in detail.

Complementary to this discovery mission, the~\ac{LC} has to perform
\emph{precision} measurements of the properties of the top quark and
of electroweak gauge bosons.  Whatever the mechanism of mass
generation for the fermions will turn out to be, the top quark couples
most strongly of all known fermions to the symmetry breaking sector.
In fact, in the minimal \ac{SM}, the Yukawa coupling of the top quark
to the Higgs particle
is of the order one.  Thus the couplings of the top quark have to be
measured as precisely as possible, because they offer the best chance
of learning something about the interaction of the symmetry breaking
sector with the fermionic sector.

Finally, the~\ac{LC} is in a unique position for precision
measurements of the properties of the electroweak gauge bosons.
LEP2 has firmly established the trilinear $W^+W^-Z^0$
interaction of gauge bosons.  The shape of the $W^+W^-$-production
cross section at the threshold as shown in
figure~\ref{fig:wwthreshold} is as convincing as direct fits
of the coupling parameters in figure~\ref{fig:tgc-moriond99}.

Due to limited statistics and energy, LEP2 will not be able
to measure the $W^+W^-Z^0$ couplings at a level where deviations from
the minimal~\ac{SM} can reasonably be
expected~\cite{Arzt/etal:1995:anomalous,Ellison/Wudka:1998:TGC}
(see chapter~\ref{sec:sm/eft}). 
In a strongly interacting model, with no observed light resonances,
deviations from the couplings
predicted by low energy symmetries are on the order of (see
sections~\ref{sec:msm} and~\ref{sec:NDA})
\begin{equation}
\label{eq:1/16pi2}
  \frac{1}{16\pi^2} \approx 6.3 \cdot 10^{-3}
\end{equation}
and in a weakly interacting model, this factor is further reduced by
coupling constants.  Unless~\ac{QFT} breaks down\footnote{Recently,
the idea of solving the hierarchy problem by introducing large extra
dimensions at the
$\unit{TeV}$-scale~\cite{Arkani-Hamed/etal:1998:extradimensions,
*Arkani-Hamed/etal:1999:extradimensions} has attracted a lot of
attention.  Such a scenario allows many spectacular signatures that
avoid the bounds from \ac{RG}~arguments in
four-dimensional~\ac{QFT}~\cite{Hall/Kolda:1999:extradimensions} and
will not be discussed here.} just above the~LEP2
energy, the LEP2 experiments are not able to establish any deviation
from the~\ac{SM}~\cite{Gounaris/etal:1996:LEP2-TGC}. The
LEP2~measurements are nevertheless important, but we should be
disappointed by the ``failure'' to see any deviation from the~\ac{SM}.
The improvements in the determination of the mass of the
$W^\pm$-bosons will be more significant, because $M_W$ is not
constrained by a low energy symmetry.  An improved measurement
probes the custodial symmetry, if combined with precision calculations
and measurements of~$M_Z$ and~$\sin^2\theta_w$.

The situation will change qualitatively with the commissioning of
the~\ac{LC}.  The~\ac{LC} will be able to detect deviations at the
level of~(\ref{eq:1/16pi2}) and will be the only machine to do
so~\cite{%
  Zerwas:1996:LC_review,
  Murayama/Peskin:1997:LC_review,
  Accomando/etal:1998:TESLAReview}.
Consequently, precision measurements of the properties of electroweak
gauge bosons are the final piece of a \emph{no-lose theorem} for a
high luminosity~($\mathcal{L}\gtrsim\unit[10^{34}]{cm^{-2}sec^{-1}}$),
high energy~($\sqrt{s}\approx\unit[1]{TeV}$) \ac{LC}:
\begin{quote}
  \itshape Either an elementary Higgs particle and possible
  supersymmetry is found at the~\ac{LC} as a signal of a weakly
  interacting symmetry breaking sector or the couplings of electroweak
  gauge bosons will deviate from the minimal~\ac{SM} predictions at a
  measurable level in order to accommodate a strongly interacting
  symmetry breaking sector.
\end{quote}
In order to cover the second part of this \emph{no-lose theorem}, the
craft of precision electroweak gauge boson physics has to be mastered.

\section{Roadmap}
\label{sec:roadmap}

\begin{cutequote*}{William Shakespeare}{The~Winter's~Tale, III.3}
  {\normalfont\textbf{Antigonus:}}
     Thou~art~perfect~then, our~ship~hath~touch'd~upon\\
     The~deserts~of~Bohemia?\\
  {\normalfont\textbf{Mariner:}} Ay,~my~lord,~\ldots
\end{cutequote*}

Chapter~\ref{sec:sm/eft} sketches the phenomenological bottom-up
approach to the field theoretical description of particle physics
experiments and applies it to the~\ac{SM} of electroweak interactions.
Most of this material can be found as part of systematic expositions
in modern textbooks~\cite{Weinberg:QFTv1:Text,Weinberg:QFTv2:Text,
Peskin/Schroeder:QFT:Text}, but is included anyway in a condensed and
slightly provocative form in order to put the other chapters in perspective.
Chapter~\ref{sec:sm/eft} attempts to give a
concise description that separates the features of the~\ac{SM} of
particle physics that are firmly established by phenomenology and the
general principles of~\ac{QFT}, on one side, from model dependent
assumptions that have not been tested so far, on the other side.

Chapter~\ref{sec:sm/eft} in particular assumes familiarity with
the~\acf{RG} for~\ac{QFT} in the Wilsonian
approach~\cite{Wilson:1971,*Wilson:1971b}, roughly on the level
of~\cite{Peskin/Schroeder:QFT:Text}.  Therefore, appendix~\ref{sec:RG}
reproduces a short and deliberately elementary introduction to the
Wilson~\ac{RG} for~\ac{QFT}, based on the first half of lectures
presented at the German Autumn School for High Energy Physics, Maria
Laach, September~9--19, 1997.  The physical intuition provided by the
Wilson~\ac{RG} is crucial for understanding why the~\ac{SM} is so
successful at LEP1 and LEP2 and why we can nevertheless expect a lot
of exciting physics at a~\ac{LC}.

Chapters~\ref{sec:me} and~\ref{sec:ps} dive deeper into the more
technical parts of the phenomenologists' toolbox.
Chapter~\ref{sec:me} discusses specific issues arising in the
calculation of matrix elements and differential cross sections for
gauge bosons in~\ac{PT}.  Chapter~\ref{sec:ps} describes methods for
phase space integration and event generation.  These chapters combine
published original contributions with results that appear to be known
only to the practitioners of the field as ``tricks of the trade.''

Finally, chapter~\ref{sec:lc} discusses some aspects of the forthcoming
physics of electroweak gauge bosons at a \ac{LC}.

\chapter{The~Standard~Model as Effective~Quantum~Field~Theory}
\label{sec:sm/eft}

\begin{cutequote*}{Thomas Hardy}{The~Convergence of the Twain}
  And as the smart ship grew\\
  In stature, grace, and hue,\\
  In shadowy silent distance grew the Iceberg too.
\end{cutequote*}
This chapter explores some consequences of the general principles
of~\ac{QFT} for the phenomenology of elementary particles.  This
material is hardly original (see the texts~\cite{Weinberg:QFTv1:Text,
Weinberg:QFTv2:Text,Peskin/Schroeder:QFT:Text,Georgi:1984:BlueBook}),
but is included in a condensed and intentionally provocative form in
order to set the stage for the following chapters. The
purpose of this exercise is to allow a separation of kinematics and
symmetry on one side, from dynamics on the other side.  The result of
the application of this formalism to the~\ac{SM} is Janus-faced:
many predictions are very solid, because they depend on kinematics and
symmetry only, but, by the same token, the independence from dynamical
details reveals how little we still know about the underlying dynamics
of the~\ac{SM}.
This view on the virtues of the~\ac{SM} is purposefully pessimistic.
It demonstrates that, despite the tremendous achievements of the
past, there is still a lot of exciting physics waiting to be discovered
at the high energy frontier.

Perturbative renormalizability has often been presented as the main
guiding principle behind the construction of the~\ac{SM}.  But when we
acknowledge our ignorance about the ``true'' short distance physics
that is out of the reach of the currently operative, as well as the
currently planned, colliders, we can not adopt it as a guiding principle
anymore.  After all, \ac{QFT} might just be cut off at some string
scale.  However, as stressed in section~\ref{sec:relevant}, among all
equivalent theories that describe Nature, there is \emph{always} one
perturbatively renormalizable~\ac{QFT}, that contains only relevant
and marginal interactions at the highest energy scale.  Since
perturbatively renormalizable~\acs{QFT}s are a technically more
convenient starting point, it is often more economical to use them as
the initial condition.  Nevertheless, irrelevant interactions will be
observed at low energies and one has to explain them as the
result of matching conditions at intermediate thresholds as described
in section~\ref{sec:important-irrelevant} in order to show the
consistency of the description.

A comprehensive and systematical modern treatment of all the
theoretical issues is provided by Weinberg's
text~\cite{Weinberg:QFTv1:Text,Weinberg:QFTv2:Text}. A more
pedagogical treatment from a similarly modern perspective is given
in~\cite{Peskin/Schroeder:QFT:Text}, while more on the phenomenology
can be found in~\cite{Donoghue/Golowich/Holstein:1992:text}.

\section{Currents}
\label{sec:currents}

\begin{cutequote}{Concise~Oxford~Dictionary}
  {\normalfont\textbf{current:}}
  in general circulation or use; \ldots
\end{cutequote}

The \emph{asymptotic single particle states} in a scattering experiment are
characterized by
quantum numbers belonging to the Poincar\'e group (energy~$E=P_0$,
momentum~$P_{1,2,3}$, and angular momentum~$L_3,{\vec L}^2$) as well
as by internal quantum numbers corresponding to conserved charges~$Q_n$:
\begin{subequations}
\label{eq:quantum-numbers}
\begin{align}
  P_\mu \ket{p,l,l_3,q_1,\ldots,q_n}
    &= p_\mu \ket{p,l,l_3,q_1,\ldots,q_n} \\
  {\vec L}^2 \ket{p,l,l_3,q_1,\ldots,q_n}
    &= l(l+1) \ket{p,l,l_3,q_1,\ldots,q_n} \\
  L_3 \ket{p,l,l_3,q_1,\ldots,q_n}
    &= l_3 \ket{p,l,l_3,q_1,\ldots,q_n} \\
  Q_n \ket{p,l,l_3,q_1,\ldots,q_n}
    &= q_n \ket{p,l,l_3,q_1,\ldots,q_n} \,.
\end{align}
\end{subequations}
\begin{figure}
  \begin{center}
    \begin{fmfgraph*}(40,25)
      \fmfleftn{i}{2}
      \fmfrightn{o}{4}
      \fmflabel{$\ket{p,l,l_3,q_1,\ldots}$}{i2}
      \fmflabel{$\ket{p',l',l_3',q_1',\ldots}$}{i1}
      \fmflabel{$\ket{p'',l'',l_3'',q_1'',\ldots}$}{o4}
      \fmflabel{$\ket{p''',l''',l_3''',q_1''',\ldots}$}{o3}
      \fmflabel{$\ket{p'''',l'''',l_3'''',q_1'''',\ldots}$}{o2}
      \fmflabel{$\ket{p''''',l''''',l_3''''',q_1''''',\ldots}$}{o1}
      \fmf{plain}{i1,v,o1}
      \fmf{plain}{i2,v,o4}
      \fmffreeze
      \fmf{plain}{o2,v,o3}
      \fmfblob{.6h}{v}
    \end{fmfgraph*}\\[\baselineskip]
  \end{center}
  \caption{\label{fig:Qn}%
    Scattering amplitude.}
\end{figure}
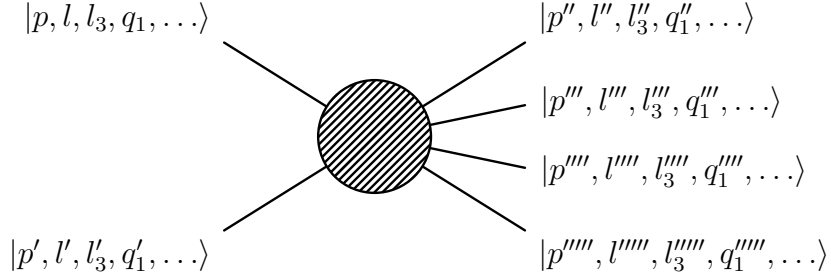
For this to make sense, the~$Q_n$ have to commute with all generators
of the Poincar\'e group\footnote{We do not include supersymmetry
charges among the~$Q_n$.}
\begin{subequations}
\begin{align}
  \left[P_\mu,Q_n\right] &= 0 \\
  \left[L_{\mu\nu},Q_n\right] &= 0
\end{align}
and in particular
\begin{equation}
  \left[H,Q_n\right] = 0\,.
\end{equation}
\end{subequations}
In general, the~$Q_n$ will represent a compact Lie algebra
\begin{equation}
\label{eq:CCR/Qn}
  \left[Q_{n_1},Q_{n_2}\right]
    = \sum_{n_3} if^Q_{n_1n_2n_3} Q^{\vphantom{Q}}_{n_3}
\end{equation}
that factorizes into simple non-abelian and abelian subalgebras.
The quantum numbers~(\ref{eq:quantum-numbers}) define the asymptotic
states in scattering experiments completely, if we employ the cluster
decomposition principle for asymptotic multi particle states.
However, the operators appearing in~(\ref{eq:quantum-numbers}) are not
sufficient for a Lorentz invariant description of the
\emph{interaction} of the states corresponding to tensor products of
asymptotic single particle states.

The creation and annihilation operators in the Fock space built from
the states in~(\ref{eq:quantum-numbers}) are non-local, i.\,e.~their
Fourier transforms neither commute nor anti-commute at space like
distances.  However, the only known way
(see \cite{Weinberg:QFTv1:Text} for a discussion) of
constructing a relativistic (i.\,e.~Lorentz invariant) scattering
matrix uses a \emph{local} interaction, i.\,e.~a Lagrangian density
that is a Lorentz invariant polynomial of \emph{local} operators that
commute or anti-commute at space like distances
\begin{equation}
\label{eq:CCR}
  \left[\phi(x),\phi'(x')\right]_\pm = 0\;\;\; \text{for}\; (x-x')^2 < 0
\end{equation}
and their derivatives.
Therefore, the particle creation and the anti-particle annihilation
operators have to be combined into local \emph{field operators} that
satisfy~(\ref{eq:CCR}).  As a result, it is impossible to have
interactions that conserve particle number and the correct description
of relativistic scattering will involve an infinite number of
degrees of freedom.

The most profound physical consequence of the mathematical result that
there are an infinite number of degrees of freedom is the phenomenon
of~\ac{SSB}.  For a finite number of degrees of freedom, Wigner has
shown long ago that all symmetries of the interaction have to be
symmetries of the ground state and the symmetry has to be represented
linearly.  However, Wigner's theorem breaks down for an infinite
number of degrees of freedom and symmetries can be broken
spontaneously, guaranteeing the existence of massless scalar particles
through Goldstone's theorem.  This allows non-linear
realizations of these symmetries~\cite{Coleman/Wess/Zumino:1969:CCWZ,
*Callan/Coleman/Wess/Zumino:1969:CCWZ}.
The Goldstone bosons are protected from obtaining large masses
through radiative corrections, unlike regular scalars.  On the other
hand, non-linearly realized gauge symmetries provide a consistent
description of massive spin-$1$ particles that does not suffer from
bad high-energy behavior (see section~\ref{sec:bosons} for a detailed
discussion).

The charges~$Q_n$ can be constructed from local
quantities as the integral of a charge density~$j_n^0(x)$ on a space-like
hypersurface~$x_0=t$
\begin{equation}
\label{eq:Q(t)}
  Q_n(t) = \int\limits_{x_0=t}\!\mathrm{d}^3x\,j_n^0(x)\,,
\end{equation}
where~$j_n^0(x)$ is the time component of a conserved current~$j_n^\mu(x)$
\begin{equation}
\label{eq:dj}
  \partial_\mu j_n^\mu(x) = 0\,.
\end{equation}
If the matrix elements of the~$j_n^\mu(x)$ fall off sufficiently
rapidly at large distances, the charge~$Q_n$ is conserved
\begin{equation}
  \frac{\mathrm{d}Q_n}{\mathrm{d}t}(t)
      = \int\limits_{x_0=t}\!\mathrm{d}^3x\,\partial_0j_n^0(x)
      = \int\limits_{x_0=t}\!\mathrm{d}^3x\,\vec\nabla\vec\jmath_n(x)
      = 0 \,.
\end{equation}
These conserved currents are particularly useful in
the phenomenology of interacting field theories,
because they are not renormalized.  For the non abelian
factors of the Lie algebra generated by the charges~$Q_n$, this
is trivial, because any renormalization~$j_\mu\to Zj_\mu$ would induce
a renormalization~$Q\to ZQ$ and violate the commutation
relations~(\ref{eq:CCR/Qn}).  For the abelian factors, we have to
make the 
further assumption that there are operators~$\psi$ that form a non
trivial representation
\begin{equation}
  \left[Q_{n},\psi\right] = q_{n,\psi}\psi\,.
\end{equation}
In this relation, any renormalization of the~$\psi$ will cancel and we
find again a non renormalization theorem for the current.

Matrix elements of the conserved currents in asymptotic states are
constrained by symmetry, independent of dynamical assumptions.  For
example, the matrix element of a current between two scalar single
particle states with identical mass ($p^2={p'}^2$) must have the
form\footnote{The eigenstates are orthogonal for hermitian
charges~$Q$: $\braket{q'|Q|q} \propto \delta_{q,q'}$.}
\begin{equation}
\label{eq:current-ME}
  \braket{p',q'_n|j_{n,\mu}(0)|p,q_n}
     = (p'_\mu + p_\mu) q_n F((p'-p)^2) \delta_{q'_nq_n},\;\;\;
  F(0) = 1
\end{equation}
from Lorentz symmetry and current conservation alone, where
the normalization of the form factor at zero momentum
transfer~$F(0)=1$ is fixed by 
\begin{multline*}
    q_n 2p_0 (2\pi)^3 \delta^3(\vec{p'} - \vec p)
   = q_n\braket{p',q_n|p,q_n}
   = \braket{p',q_n|Q_n|p,q_n} \\
   = \int\limits_{x_0=t}\!\mathrm{d}^3x\,\braket{p',q_n|j_{n,0}(x)|p,q_n}
   = \int\limits_{x_0=t}\!\mathrm{d}^3x\,e^{i\vec{x}(\vec{p}-\vec{p'})}
       \braket{p',q_n|j_{n,0}(0)|p,q_n} \\
   = (2\pi)^3 \delta^3(\vec{p'} - \vec p) \braket{p',q_n|j_{n,0}(0)|p,q_n}\,,
\end{multline*}
i.\,e.\
\begin{equation}
  q_n 2p_0 = \braket{p,q_n|j_{n,0}(0)|p,q_n}\,.
\end{equation}
Similar relations can be derived for matrix elements of particles with
spin, but in general there is a larger number of independent form
factors for which symmetry provides less constraints.

\section{Weak Interactions}
\label{sec:weak}

The low energy manifestations of the interactions described by
the~\acf{SM} are the long range electromagnetic interaction and the
short range weak interaction.  In Nature, we observe weak interactions
in the burning of the solar fuel
\begin{equation}
\label{eq:solar-fuel}
  4p + 2e^- \to {{}^4\!\mathrm{He}} + 2\nu_e + \unit[26.73]{MeV} - E_\nu\,.
\end{equation}
While the released energy is the binding energy of
the ${{}^4\!\mathrm{He}}$~nucleus due to the strong
interaction of protons and neutrons, the process~(\ref{eq:solar-fuel})
would be impossible without the weak interaction, because the strong
interaction conserves isospin, which forbids $p+e^-\to n+\nu_e$
transitions.  The other natural manifestation of weak interactions is
in the nuclear $\beta$-decay, for example
\begin{equation}
  {{}^{137}\!\mathrm{Cs}} \to {{}^{137}\!\mathrm{Ba}} + e^- + \bar\nu_e\,.
\end{equation}
Decades of particle physics experiments with accelerators have found
many more examples of weak interactions that can be observed because
they violate the isospin and parity invariance of strong interactions
(see \cite{PDG:1998} for the most exhaustive and up-to-date list).

\section{Fermi Theory}
\label{sec:fermi}
\begin{cutequote}{Pythagoras' pupils}
  \AutosEpha/
\end{cutequote}

As mentioned in the previous section, there are two defining
characteristics of weak interactions at low energies: parity violation
and isospin non-conservation.  The former is the result of the experimental
observation that all electrons (and muons and taus) produced in
$\beta$-decay are left handed, while all positrons (and anti-muons and
anti-taus) produced in $\beta^+$-decay are right handed\footnote{This
observation summarizes many complicated experiments and
ingenious theoretical interpretations.}.  Therefore, the weak
interaction couples in the leptonic sector to $(V-A)$ charged currents
of the form
\begin{equation*}
  \bar\nu\gamma_\mu(1-\gamma_5)e + \text{h.c.}\,.
\end{equation*}
Using the~$\tau^\pm=(\tau_1\pm i\tau_2)/2$ isospin shift operators,
such charged currents can be combined in the $\pm$-components of a
conserved isospin vector current
\begin{multline}
\label{eq:leptonCC}
  j^\pm_\mu(x) = 
    \left( \bar \nu(x) \;\bar e(x) \right)
      \left[\tau^\pm\otimes\gamma_\mu\frac{1-\gamma_5}{2}\right]
    \begin{pmatrix} \nu(x) \\ e(x) \end{pmatrix} \\
    +
    \left( \bar \nu_\mu(x) \; \bar\mu(x) \right)
      \left[\tau^\pm\otimes\gamma_\mu\frac{1-\gamma_5}{2}\right]
    \begin{pmatrix} \nu_\mu(x) \\ \mu(x) \end{pmatrix} \\
    +
    \left( \bar \nu_\tau(x) \; \bar\tau(x) \right)
      \left[\tau^\pm\otimes\gamma_\mu\frac{1-\gamma_5}{2}\right]
    \begin{pmatrix} \nu_\tau(x) \\ \tau(x) \end{pmatrix}\,,
\end{multline}
which is written more compactly as
\begin{equation}
\label{eq:V-A}
  \vec \jmath_\mu(x) =
    \bar\psi \left[\frac{\vec\tau}{2}\otimes
        \gamma_\mu\frac{1-\gamma_5}{2}\right] \psi\,,
\end{equation}
where the summation over all isospin doublets is implied.  The neutral
component~$j^0_\mu(x)$ of the current~(\ref{eq:V-A}) is masked at low
energies by much stronger electromagnetic interactions of charged
leptons and strong interactions of hadrons and was observed only much
later, after neutrino beams became available.

Making the educated guess that the hadronic part is described by a
similar isospin current, the simplest short range interaction
imaginable couples these currents
\begin{equation}
\label{eq:Fermi}
   L(x) = \frac{g}{\Lambda^2} j^+_\mu(x) j^{-,\mu}(x)
\end{equation}
with~$g$ a suitable coupling constant and~$\Lambda$ a suitable scale.
According to section~\ref{sec:relevant}, the interaction~(\ref{eq:Fermi}) is
irrelevant and expected to be very weak at low energies (compared
to~$\Lambda$).

The interaction~(\ref{eq:Fermi}) has been tested quantitatively
(including radiative corrections,
see e.\,g.~\cite{Donoghue/Golowich/Holstein:1992:text}) in the leptonic
sector, where the matrix elements of the products of isospin currents
can be calculated without further assumptions.  From the comparisons
with experimental results, lepton universality, i.\,e.~that each
doublet of leptons contributes \emph{identically} to the current as
in~(\ref{eq:leptonCC}), is firmly established.

For the hadronic part of the current, no ab-initio calculations are
possible yet, due to our limited understanding of the non-perturbative
dynamics of the strong interactions.  In semi-leptonic processes like
$B^+\to\bar{D^0}\mu^+\nu_\mu$, the matrix element of the interaction
can be factorized in terms of a leptonic and a hadronic current
\begin{equation}
\label{eq:semileptonic}
  \Braket{\mu^+\nu_\mu \bar{D^0}|j^+_\mu(x) j^{-,\mu}(x)|B^+}
    = \Braket{\mu^+\nu_\mu\vphantom{\bar{D^0}}|j^+_\mu(x)|0}
      \Braket{\bar{D^0}|j^{-,\mu}(x)|B^+}\,.
\end{equation}
Model independent symmetry arguments for the conserved vector current
and partially conserved axial current similar to~(\ref{eq:current-ME})
can be used to express~(\ref{eq:semileptonic}) in terms of a single
unknown form factor. For heavy hadrons, the form factor has a
known normalization a zero momentum transfer, up to small
corrections~\cite{Isgur/Wise:1988:HQET_classic,
*Isgur/Wise:1989:HQET_classic,*Eichten/Hill:1990:HQET_classic,
*Georgi:1990:HQET_classic,*Grinstein:1990:HQET_classic}.
This provides non-trivial cross checks for such processes and the
interaction~(\ref{eq:Fermi}) has passed these tests.  Similar
symmetry arguments have less predictive power for light baryons and
vector mesons, because there are more independent form 
factors, but some non-trivial checks can still be performed.

Compared to this, purely hadronic weak processes remain poorly
understood, because the matrix elements can not be factorized in terms
of hadronic currents and ab-initio calculations in terms of quarks are
not yet possible.  However, this reflects more on our still poor
understanding of
non-perturbative \ac{QFT} than on the accuracy of the description of
weak interactions with the Fermi interaction~(\ref{eq:Fermi}).

\section{Flavor Symmetries}
\label{sec:flavor}

\begin{empcmds}
  def connect (suffix from, ff, to, tf) =
    (ff[from.sw,from.se] -- tf[to.nw,to.ne]
     cutbefore bpath.from cutafter bpath.to)
  enddef;
\end{empcmds}
\begin{figure}
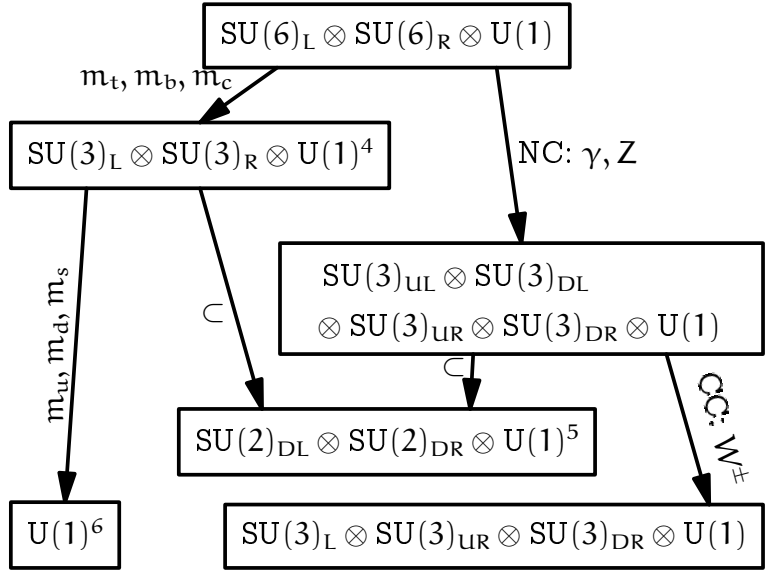

  \begin{center}
    \begin{emp}(100,\ifPDFLaTeX90\else75\fi)
      ahlength := 4mm;
      ahangle := 30;
      pickup pencircle scaled 1.5pt;
      defaultdx := 6pt;
      defaultdy := 6pt;
      boxit.SU[66] (btex $\textrm{SU}(6)_L\otimes
                          \textrm{SU}(6)_R\otimes\textrm{U}(1)$ etex);
      boxit.SU[33] (btex $\textrm{SU}(3)_L\otimes
                          \textrm{SU}(3)_R\otimes\textrm{U}(1)^4$ etex);
      boxit.SU[3333] (btex
        \parbox{60mm}{\vspace*{-1.3\baselineskip}
          \begin{multline*}
            \textrm{SU}(3)_{UL}\otimes\textrm{SU}(3)_{DL}\\
              \otimes\textrm{SU}(3)_{UR}\otimes\textrm{SU}(3)_{DR}
              \otimes\textrm{U}(1)
          \end{multline*}
        \vspace*{-1.7\baselineskip}} etex);  
      boxit.SU[22] (btex $\textrm{SU}(2)_{DL}\otimes
                          \textrm{SU}(2)_{DR}\otimes\textrm{U}(1)^5$ etex);
      boxit.SU[333] (btex $\textrm{SU}(3)_L\otimes
                           \textrm{SU}(3)_{UR}\otimes
                           \textrm{SU}(3)_{DR}\otimes\textrm{U}(1)$ etex);
      boxit.SU[1] (btex $\textrm{U}(1)^6$ etex);
      SU[66]n = (.5w,h);
      SU[3333]e = (w,whatever);
      SU[33]w = (0,whatever);
      SU[22]c = (.5w,whatever);
      SU[1]sw = (0,whatever);
      SU[333]se= (w,0);
      SU[66]s = SU[33]n + (whatever,d);
      SU[33]s = SU[3333]n + (whatever,d);
      SU[3333]s = SU[22]n + (whatever,d);
      SU[22]s = SU[1]n + (whatever,d/2);
      SU[1]s = SU[333]s + (whatever,0);
      drawboxed (SU[66], SU[33], SU[3333], SU[22], SU[333], SU[1]);
      path sb[][];
      sb[66][33] = connect (SU[66], .2, SU[33], .5);
      sb[33][22] = connect (SU[33], .5, SU[22], .2);
      sb[33][1] = connect (SU[33], .2, SU[1], .5);
      sb[66][3333] = connect (SU[66], .8, SU[3333], .5);
      sb[3333][22] = connect (SU[3333], .4, SU[22], .7);
      sb[3333][333] = connect (SU[3333], .8, SU[333], .9);
      drawarrow sb[66][33];
      label.ulft (btex $m_t, m_b, m_c$ etex, point .5 of sb[66][33]);
      drawarrow sb[33][22];
      label.llft (btex $\subset$ etex, point .5 of sb[33][22]);
      drawarrow sb[33][1];
      label.lft (btex $m_u, m_d, m_s$ etex rotated 90, point .5 of sb[33][1]);
      drawarrow sb[66][3333];
      label.rt (btex NC: $\gamma, Z$ etex, point .5 of sb[66][3333]);
      drawarrow sb[3333][22];
      label.ulft (btex $\subset$ etex, point .5 of sb[3333][22]);
      drawarrow sb[3333][333];
      label.rt (btex CC: $W^\pm$ etex rotated -75, point .5 of sb[3333][333]);
    \end{emp}
  \end{center}
  \caption{\label{fig:SUNs}%
    Chiral symmetry breaking patterns of hadrons by weak interactions
    and quark masses.  On the left hand side, the symmetry is broken by
    an as yet unknown mechanism that creates heavy quark masses.  In
    the right hand side, the embedding of the weak interaction gauge
    group in the global chiral symmetry groups results in quite
    different symmetry breaking patterns.}
\end{figure}
What are the hadronic currents that the weak interaction couples to?
A comprehensive study of the transitions among hadrons induced by the
weak interaction currents
\begin{subequations}
\label{eq:currents}
\begin{align}
  j^a_{\mu,\text{meson}} &=
      \sum_{j,k} i \Phi_j^\dagger T^a_{jk}
          \overleftrightarrow{\partial_\mu} \Phi_k \\
  j^a_{\mu,\text{baryon}} &=
      \sum_{j,k} \bar\Psi_j T^a_{jk} \Gamma_\mu \Psi_k
\end{align}
\end{subequations}
together with the hadron masses~$M_j$
\begin{subequations}
\label{eq:masses}
\begin{align}
  L_{0,\text{meson}} &=
    \sum_j \left[
      \frac{1}{2} \left(\partial_\mu\Phi_j\right)^\dagger
                  \left(\partial^\mu\Phi_j\right)
    - \frac{1}{2} M_j^2 \Phi_j^\dagger\Phi_j \right] \\
  L_{0,\text{baryon}} &=
    \sum_j \left[i\bar\Psi_j\fmslash{\partial}\Psi_j
      - M_j \bar\Psi_j\Psi_j \right]
\end{align}
\end{subequations}
reveals intriguing symmetry patterns for the meson fields~$\Phi_j$ and
baryon fields~$\Psi_j$, which can be summarized by the succession of
global (i.\,e.~ungauged) $\textrm{SU}(N)$~flavor symmetry groups in
figure~\ref{fig:SUNs}\footnote{Note that the
$\textrm{SU}(6)_{L,R}$~symmetry groups in figure~\ref{fig:SUNs}
contain some revisionistic historiography, because no hadrons with top
flavor are observed in nature.  Furthermore, the Fermi interaction is
not applicable at energies as high as the top threshold.  Therefore, a
strictly phenomenological analysis would use~$\textrm{SU}(5)_{L,R}$
instead.  Nevertheless, for the purpose of studying the symmetry
properties, it is useful to pretend temporarily that the top is
lighter than it actually is.}.  The strong interactions appear to be
completely flavor blind and hadrons can be classified in
representations of a global
$\textrm{SU}(6)_L\otimes\textrm{SU}(6)_R\otimes\textrm{U}(1)$
symmetry group\footnote{The six different flavors will later
correspond to quarks, coming with left handed and right handed
chiralities. Note that the~$\textrm{SU}(6)$'s must not be confused with
the nonrelativistic $\textrm{SU}(6)$~symmetry containing
the~$\textrm{SU}(3)$ of the light hadrons and the~$\textrm{SU}(2)$ of
spin angular momentum.}. 

\begin{table}
  \caption{\label{tab:SM}%
    The \protect\acf{SM} of elementary particle physics. We have
    color~($\mathbf{C}$) singlets~($\mathbf{1}$) and
    triplets~($\mathbf{3}$), weak isospin~($\mathbf{T}$) singlets and
    doublets~($\mathbf{2}$) and various hypercharge~($Y$) assignments.}
  \begin{center}
    \setlength{\extrarowheight}{1ex}
    \newcommand{\threegen}[4]{%
        $\displaystyle #2$%
      & $\displaystyle #3$%
      & $\displaystyle #4$}
    \newcommand{\threegenx}[4]{%
          $\displaystyle #2$%
      &   $\displaystyle #3$%
      &   $\displaystyle #4$}
    \newcommand{\QN}[4]{
        $(\mathbf{#1},{\mathbf{#2}})_{{#3}}$
      & ${\displaystyle#4}$}
      \begin{tabular}{ccclc}
      \multicolumn{3}{c}{Leptons}
        & \QN{C}{T}{Y}{Q = T_3 + \frac{Y}{2}} \\
          \threegen{leptons}{\nu_{e,R}\,(?)}{\nu_{\mu,R}\,(?)}{\nu_{\tau,R}\,(?)}
        & \QN{1}{1}{0}{0} \\
          \threegenx{leptons}{e_R}{\mu_R}{\tau_R}
        & \QN{1}{1}{-2}{-1} \\
          \threegenx{leptons}%
            {\begin{pmatrix} \nu_{e,L} \\ e_L \end{pmatrix}}%
            {\begin{pmatrix} \nu_{\mu,L} \\ \mu_L \end{pmatrix}}%
            {\begin{pmatrix} [\nu_{\tau,L}] \\ \tau_L \end{pmatrix}}
        & \QN{1}{2}{-1}{\begin{pmatrix} 0 \\ -1 \end{pmatrix}} \\
      \multicolumn{3}{c}{Quarks} \\
          \threegenx{quarks}{u_R}{c_R}{t_R}
        & \QN{3}{1}{4/3}{\frac{2}{3}} \\
          \threegenx{quarks}{d_R}{s_R}{b_R}
        & \QN{3}{1}{-2/3}{-\frac{1}{3}} \\
          \threegenx{quarks}%
            {\begin{pmatrix} u_L \\ d_L \end{pmatrix}}%
            {\begin{pmatrix} c_L \\ s_L \end{pmatrix}}%
            {\begin{pmatrix} t_L \\ b_L \end{pmatrix}}
        & \QN{3}{2}{1/3}{\begin{pmatrix} \frac{2}{3} \\ -\frac{1}{3} \end{pmatrix}} \\
      \multicolumn{3}{c}{Gauge Bosons} \\
      \multicolumn{3}{c}%
        {{$\gamma$\qquad$Z^0$\qquad$W^\pm$\qquad$g$}} \\
      \multicolumn{3}{c}{Higgs} \\
      \multicolumn{3}{c}{$\Phi\,(?)$}
        & \QN{1}{?}{?}{?} \\
    \end{tabular}
  \end{center}
\end{table}
The classification of the hadronic currents and their interactions
according to the symmetries described in figure~\ref{fig:SUNs} is
complicated by the fact that the observed hadrons do not live in the
fundamental representation~$N$ of~$\textrm{SU}(N)$. Instead they live
in the irreducible components of product representations that carry no
overall charge of the color~$\textrm{SU}(3)$, responsible for
\ac{QCD}, the strong interaction in the quark picture.  These
representations of~$\textrm{SU}(N)$
are~$N\otimes\bar N=1\oplus(N^2-1)$ for the mesons
and~$N\otimes N\otimes N$ for the baryons.  However,
the leptons do form a fundamental representation of the symmetries in
figure~\ref{fig:SUNs}, as shown in the upper left part of
table~\ref{tab:SM}. Furthermore, there is independent experimental
evidence, in particular from deep-inelastic lepton hadron scattering,
for quarks as constituents of hadrons, which live in a fundamental
representation (see table~\ref{tab:SM}).  In particular, the
properties of hadrons with heavy flavors are well described by
heavy quarks surrounded by a
$\textrm{SU}(3)_L\otimes\textrm{SU}(3)_R$~flavor symmetric ``cloud''
of light quarks and gluons~\cite{Isgur/Wise:1988:HQET_classic,
*Isgur/Wise:1989:HQET_classic,*Eichten/Hill:1990:HQET_classic,
*Georgi:1990:HQET_classic,*Grinstein:1990:HQET_classic}.   The product
representations for the observed hadron spectra are therefore
understood as the bound states of quarks.

The $\textrm{SU}(6)_L\otimes\textrm{SU}(6)_R$ symmetry is badly broken
by the heavy flavor masses~$m_t$, $m_b$, and~$m_c$, which leaves the
$\textrm{SU}(3)_L\otimes\textrm{SU}(3)_R$ of the eightfold way of the
three light flavors~\cite{GellMann/Neeman:1964:8foldway}. This is
already a useful symmetry that can be used to derive quantitative
results, provided that the breaking of the flavor-$\textrm{SU}(3)$ to
the isospin-$\textrm{SU}(2)$ due to the finite strange quark mass is
taken into account properly, together with the spontaneous chiral
symmetry breaking
$\textrm{SU}(3)_L\otimes\textrm{SU}(3)_R\to\textrm{SU}(3)_{L+R}$
caused by~\ac{QCD}.  The product representations are the
familiar singlet plus octet for the light
mesons~$3\otimes\bar3=1\oplus8$ and the singlet, two octets and a
decuplet for the light
baryons~$3\otimes3\otimes3=1\oplus8\oplus8\oplus10$.

On the other hand, the $\textrm{SU}(6)_L\otimes\textrm{SU}(6)_R$ is
also broken by differing charges for weakly interacting neutral current
interactions (i.\,e.~$Z^0$-exchange and~\ac{QED}) for up-type quarks
and down-type quarks.  The
remaining~$\textrm{SU}(3)_{UL}\otimes\textrm{SU}(3)_{UR}$ corresponds
to the flavors~$u$, $c$, and $t$ of the up-type quarks and 
the~$\textrm{SU}(3)_{DL}\otimes\textrm{SU}(3)_{DR}$ corresponds to the
flavors~$d$, $s$, and $b$ of the down-type quarks.
This symmetry in broken further by the weakly interacting charged
current interactions to a~$\textrm{SU}(3)_L$ for the left handed part
that interacts and a~$\textrm{SU}(3)_{UR}\otimes\textrm{SU}(3)_{DR}$
for the non-interacting right handed parts.  Again, these symmetries
are badly broken by heavy flavor masses, but
the~$\textrm{SU}(3)_L\otimes\textrm{SU}(3)_R$ and
the~$\textrm{SU}(3)_{UL}\otimes\textrm{SU}(3)_{DL}\otimes
\textrm{SU}(3)_{UR}\otimes\textrm{SU}(3)_{DR}$ have a common
subgroup~$\textrm{SU}(2)_{DL}\otimes\textrm{SU}(2)_{DR}$ that is a
good symmetry and has interesting phenomenological
consequences~\cite{Georgi:1992:DDbar,*Ohl/etal:1993:DDbar}.

An additional complication is provided by the fact that
the $\textrm{SU}(3)_{UL}\otimes\textrm{SU}(3)_{UR}$, the
$\textrm{SU}(3)_{DL}\otimes\textrm{SU}(3)_{DR}$, and the isospin
$\textrm{SU}(2)$ of the weak interactions twisted by the CKM mixing
matrix.  Without the intuition provided by quarks, this low energy
symmetry structure of the weak interactions would be very hard to
understand.

The~\ac{SM} provides a complete phenomenological description of the
currents described in this section.  However, of all the symmetry
breakings in figure~\ref{fig:SUNs}, it only explains the two steps on
the right, which are due to neutral currents and charged currents.
The other symmetry breakings, as well as the CKM mixing have to be put
in by hand, introducing most of the free parameters in the~\ac{SM}.

The strong interaction physics that is responsible for the relation
between the fundamental representation and the product representations
is interesting in itself, but not of fundamental importance for the
weak interaction physics under consideration.  The significant observation
of the present section is that the leptonic and semi-leptonic low
energy weak interactions can indeed be described by the Fermi
interaction~(\ref{eq:Fermi}) for observable hadronic and leptonic
currents.  The axial part of the short range neutral current weak
interaction has also been observed more recently in low energy
experiments through their parity violating effects.  Advances in
experimental atomic physics
have made atomic parity violation a viable tool for precision
physics~\cite{Wood/et/al:1997:APV,*Bennett/Wiemann:1999:APV}.

The emerging picture of the weak interactions at low energy is that of
interacting isospin $\textrm{SU}(2)$ currents, where the
charged components act on the left handed fields and the neutral
component couples differently to the two members of the isospin
doublet particles.  The symmetry is incomplete, as long as the charged
current and neutral current pieces of the interaction appear to be
unrelated.  Below in section~\ref{sec:msm}, we will see that the
corresponding gauge bosons are 
indeed related by a custodial $\textrm{SU}(2)_c$
symmetry~\cite{Sikivie/etal:1980:custodial}.

\section{Higher Energy Scattering}
\label{sec:unitarity}

Having established the Fermi interaction~(\ref{eq:Fermi}) for low
energy weak interactions, we must study if and when this description
will break down at higher energies, either phenomenologically or from
theoretical inconsistencies.  From power counting, the interaction is
irrelevant and the \ac{RG} flow in the ultraviolet will eventually
leave the perturbative domain (see section~\ref{sec:rgflow}).  But
instead of studying radiative corrections, we can also study a tree
level process like $e\nu$~scattering.  The differential cross section
is found to be quadratically rising with energy
\begin{equation}
  \parbox{22\unitlength}{%
    \fmfframe(3,4)(3,4){%
      \begin{fmfgraph*}(15,10)
      \fmfleft{e,nue} \fmfright{nue',e'}
      \fmf{fermion}{e,gf,e'}
      \fmf{fermion}{nue,gf,nue'}
      \fmfdot{gf}
      \fmfv{label=$e^-$}{e}
      \fmfv{label=$e^-$}{e'}
      \fmfv{label=$\nu_e$}{nue}
      \fmfv{label=$\nu_e$}{nue'}
    \end{fmfgraph*}}}
  \Longrightarrow
    \frac{\textrm{d}\sigma}{\textrm{d}\Omega} =
    \frac{G_F^2}{4\pi^2}\cdot E_{CM}^2\,.
\end{equation}
On the other hand, we can project the
scattering amplitude on the $S$-wave component and verify the
unitarity of the partial wave
\begin{equation}
  |\mathcal{M}_{J=0}|\le1\,.
\end{equation}
This gives an upper bound on differential cross section
\begin{equation}
   \frac{\textrm{d}\sigma}{\textrm{d}\Omega} \le \frac{1}{E_{CM}^2}\,.
\end{equation}
\begin{empcmds}
  rootgf := 1/0.2928;
  pi := 3.14159;
  emax := 6;
\end{empcmds}
\begin{figure}
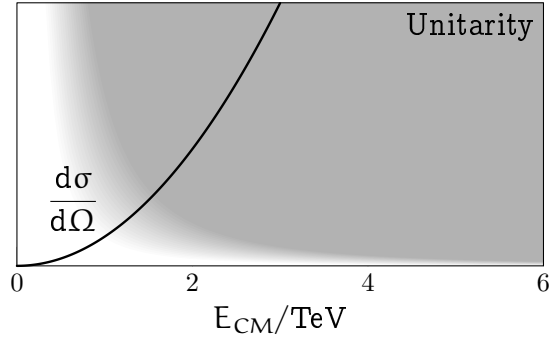

  \begin{center}
    \begin{empgraph}(70,35)
      smax :=  1/2 * (rootgf/(2pi)*emax)**2;
      pickup pencircle scaled 0.5pt;
      setrange (0, 0, emax, smax);
      autogrid (itick.bot,);
      path back, sigma;
      augment.back (0,0);
      augment.back (emax,0);
      augment.back (emax,smax);
      augment.back (0,smax);
      gfill back--cycle withcolor white;
      path unitarity[];
      for t = 0, 1:
        emin := 1 / (sqrt (smax/(1+1.5*(t-.5))));
        for e = emin step 0.05 until (emax+0.1):
          ee := min (e, emax);
          augment.unitarity[t] (ee, (1+1.5*(t-.5))/(ee*ee));
        endfor
      endfor
      for t = 0 step 0.05 until 1.01:
        path utmp;
        utmp = interpath (t, unitarity0, unitarity1);
        augment.utmp (emax, smax);
        gfill utmp--cycle withcolor t[white,.7white];
      endfor
      pickup pencircle scaled 1pt;
      for e = 0 step 0.01 until emax:
        s := (rootgf/(2pi)*e)**2;
        exitif s >= smax;
        augment.sigma (e, s);
      endfor
      gdraw sigma;
      pickup pencircle scaled 0.5pt;
      glabel.ulft (btex $\displaystyle
                         \frac{\textrm{d}\sigma}{\textrm{d}\Omega}$ etex,
                   100);
      glabel.llft (btex Unitarity etex, (emax-0.05,smax-0.05));
      glabel.bot (btex $E_{CM}/\text{TeV}$ etex, OUT);
    \end{empgraph}
  \end{center}
  \caption{\label{fig:fermitrouble}%
    Differential cross section and $S$-wave tree-level unitarity bound
    for~$e^-\nu_e\to e^-\nu_e$.}
\end{figure}
Both the differential cross section and the unitarity bound are shown
in figure~\ref{fig:fermitrouble}, where the unitarity bound is turned
on smoothly, because we have to allow for higher order corrections
close to the bound. In any case, it is obvious from
figure~\ref{fig:fermitrouble}, that the Fermi
interaction~(\ref{eq:Fermi}) has to be replaced before~$\unit[1]{TeV}$.

The solution to this problem is provided by softening the interaction
at high energies, e.\,g.
\begin{equation}
\label{eq:softening}
  \frac{1}{\Lambda^2} \to \frac{-1}{p^2-\Lambda^2}
\end{equation}
for some combination~$p^2$ of kinematical invariants that scales
as~$E_{CM}^2$.
The particular example in~(\ref{eq:softening}) is just one of many
possibilities that reduce to the Fermi interaction at low energy.
However, it corresponds to replacing the local
interaction~(\ref{eq:Fermi}) by the exchange of vector bosons
\begin{equation}
\label{eq:boson-exchange}
  \parbox{27\unitlength}{%
    \fmfframe(3,4)(3,4){%
      \begin{fmfgraph*}(20,15)
        \fmfleft{e,nue} \fmfright{nue',e'}
        \fmf{fermion}{e,cc,nue'}
        \fmf{fermion}{nue,cc',e'}
        \fmf{boson,label=$W^\pm$}{cc,cc'}
        \fmfdot{cc,cc'}
        \fmfv{label=$e^-$}{e}
        \fmfv{label=$e^-$}{e'}
        \fmfv{label=$\nu_e$}{nue}
        \fmfv{label=$\nu_e$}{nue'}
      \end{fmfgraph*}}}
  \Longrightarrow
    \frac{\textrm{d}\sigma}{\textrm{d}\Omega}
      = \frac{g^4}{32\pi^2}
        \frac{1}{E_{CM}^2}\frac{1}{\left(1-\cos\theta
            + \frac{2M_W^2}{E_{CM}^2}\right)^2}
\end{equation}
and can therefore be described again by a local interaction.  Unlike
an arbitrary form factor, a local interaction produces a Lorentz
invariant scattering matrix (see section~\ref{sec:currents}).
The formal Lorentz invariance of an ad-hoc form factor in a non-local
interaction does \emph{not} suffice for Lorentz invariance of the
scattering matrix for all matrix elements if the form factor is taken
into account consistently~\cite{Weinberg:QFTv1:Text}. Space-like
correlations will ruin Lorentz invariance.

Incidentally, the interaction that produces the vector boson exchange
to replace the irrelevant current-current interaction,
\begin{equation}
\label{eq:Aj}
  L(x) = \sum_a A^a_\mu(x) j^{a,\mu}(x),
\end{equation}
is marginal (see \ref{sec:relevant}) and the scale in the
irrelevant interaction is replaced
by the mass of the new particle.  This fits of course nicely within
the~\ac{EFT} paradigm, where a low energy~\ac{EFT} breaks down
at the scale where new particles and interactions have to be introduced.

\section{Intermediate Vector Bosons}
\label{sec:bosons}

The introduction of intermediate vector
bosons~(\ref{eq:boson-exchange}) is not without problems, because
unitarity requires that all particles that are exchanged must also
appear in external states.  Massive vector
bosons have three degrees of freedom and massless vector bosons
(photons) have only two degrees of freedom.  However, a naive counting
of the components of a Lorentz vector finds four degrees of freedom.

One approach to a consistent theory would construct the vector bosons
as narrow spin one resonances of scalars or spin-$1/2$ fermions.  In
such a theory, unitarity could be proven in a Hilbert space of
asymptotic states built from scalars and spin-$1/2$ fermions, avoiding
all problems with unphysical degrees of freedom.  However,
constituents that could form resonances with the correct quantum
numbers are not observed and Occam's razor suggests to attempt a
description in terms of the vectorial degrees of freedom before
inventing new constituents.

The current operators have all the right quantum numbers, but they
have a mass dimension of three and can therefore not be identified
with intermediate vector bosons with a mass dimension of one.
Unfortunately, the very fact that conserved currents are not
renormalized prevents them from acquiring an anomalous dimension and
turning into a vector field with canonical dimension for sufficiently
large coupling\footnote{See however~\cite{Kosower:1993:Vectors} for
an interesting speculation along this line.}.

For non-interacting massive vector bosons, a simple solution is
that the equation of motion
\begin{equation}
  (\Box + M^2) A^\mu - \partial^\mu\partial_\nu A^\nu = 0
\end{equation}
derived from the Proca Lagrangian with an explicit mass term
\begin{equation}
\label{eq:proca}
  L = - \frac{1}{4} F_{\mu\nu} F^{\mu\nu} + \frac{1}{2} M^2 A_\mu A^\mu,
  \;\;\; F_{\mu\nu} = \partial_\mu A_\nu - \partial_\nu A_\mu
\end{equation}
eliminates the unphysical degree of freedom automatically, because of
\begin{equation}
\label{eq:M2dA}
  M^2 \partial_\mu A^\mu = 0\,.
\end{equation}
The operator equation~(\ref{eq:M2dA}) removes precisely the degree of
freedom that would correspond to a negative norm state in a naive
canonical quantization of all four components.

\subsection{Power Counting}
\label{sec:longitudinal}

If the vector boson~$A$ is coupled to a \emph{conserved} current
in~(\ref{eq:Aj}), the operator equation~(\ref{eq:M2dA}) remains
intact.  However, even if the vector boson couples to conserved
currents only, it is not true in general that contributions from the
longitudinal part of the propagator
\begin{equation}
\label{eq:proca-propagator}
  - \frac{i}{p^2-M^2+i\epsilon}
     \left( g_{\mu\nu} - \frac{p_\mu p_\nu}{M^2} \right)
\end{equation}
cancel in physical amplitudes.  Instead, except for exceptional cases,
(\ref{eq:M2dA})~has to be enforced explicitely in intermediate states.
As a consequence, the power counting rules~(\ref{eq:NDA}) used to classify
effective Lagrangians are violated by the longitudinal polarization state.

If the currents in~(\ref{eq:currents}) are conserved and correspond to
an abelian symmetry, i.\,e.
\begin{equation}
\label{eq:abelianNC}
  \left[T^a,T^b\right] = 0\,,
\end{equation}
the longitudinal polarization state decouples nevertheless.  This is
trivial from the equations of motion for on-shell currents, but in
order to show that the contributions from off-shell propagators
cancel, one has to sum over all Feynman diagrams~(see also
section~\ref{sec:groves}).  For example, in the process~$e^+e^-\to
Z^0Z^0$, the diagrams
\begin{equation}
\label{eq:eeZZ}
  M_{\mu\nu}(p_+,p_-;k,q) = 
  \parbox{34\unitlength}{%
    \fmfframe(1,6)(12,6){%
      \begin{fmfgraph*}(20,15)
        \fmfleft{nu,nubar} \fmfright{wp,wm}
        \fmf{fermion}{nu,cc,cc',nubar}
        \fmf{boson}{cc,wp}
        \fmf{boson}{cc',wm}
        \fmfdot{cc,cc'}
        \fmfv{label=$e^-(p_-)$}{nu}
        \fmfv{label=$e^+(p_+)$}{nubar}
        \fmfv{label=$Z^0(q)$}{wp}
        \fmfv{label=$Z^0(k)$}{wm}
      \end{fmfgraph*}}} +
  \parbox{34\unitlength}{%
    \fmfframe(12,6)(1,6){%
      \begin{fmfgraph*}(20,15)
        \fmfleft{nu,nubar} \fmfright{wp,wm}
        \fmf{fermion}{nu,cc,cc',nubar}
        \fmf{phantom}{cc,wp}
        \fmf{phantom}{cc',wm}
        \fmfdot{cc,cc'}
        \fmffreeze
        \fmf{boson}{cc,wm}
        \fmf{boson,rubout}{cc',wp}
        \fmfv{label=$e^-(p_-)$}{nu}
        \fmfv{label=$e^+(p_+)$}{nubar}
        \fmfv{label=$Z^0(q)$}{wp}
        \fmfv{label=$Z^0(k)$}{wm}
      \end{fmfgraph*}}}
\end{equation}
contribute equally to the Ward identity
\begin{equation}
  k^\mu\epsilon^\nu(q) M_{\mu\nu}(p_+,p_-;k,q) = 0\,.
\end{equation}
In general, (\ref{eq:abelianNC}) is satisfied for the neutral current
piece of the weak interaction, because these particular~$T^a$ are
generated from the Pauli matrices~$\mathbf{1}$ and~$\tau_3$ by linear
combination, direct product, and direct sum.

Due to this decoupling, it is possible to replace the Proca
propagator~(\ref{eq:proca-propagator}) for abelian currents by a
St\"uckelberg propagator 
\begin{equation}
\label{eq:stuckelberg-propagator}
  - \frac{i}{p^2-M^2+i\epsilon}
     \left( g_{\mu\nu} -
             (1-\xi)\frac{p_\mu p_\nu}{p^2 - \xi M^2 + i\epsilon} \right)\,,
\end{equation}
which is compatible with the power counting rules~(\ref{eq:NDA}).
Then the choice~$\xi=1$ is particularly
convenient for quick calculations, while the cancellation of~$\xi$
dependent pieces provides useful cross checks in more complicated
calculations. With~(\ref{eq:stuckelberg-propagator}), $\partial A$ can
be eliminated consistently from matrix elements
\begin{equation}
\label{eq:dA}
  \Braket{\text{phys.}|\partial_\mu A^\mu(x)|\text{phys.}'} = 0
\end{equation}
as a \emph{free} field, but it does not necessarily vanish as an
operator in the full indefinite metric Hilbert space.

The situation is more involved for charged current interactions in
which~(\ref{eq:abelianNC}) is violated.  For example, in the
process~$e^+e^-\to W^+W^-$ only one of the diagrams
corresponding to~(\ref{eq:eeZZ}) does not vanish
\begin{equation}
\label{eq:eeWW1}
  M_{\mu\nu}^{(1)}(p_+,p_-;k,q) = 
  \parbox{30\unitlength}{%
    \fmfframe(3,4)(6,4){%
      \begin{fmfgraph*}(20,15)
        \fmfleft{nu,nubar} \fmfright{wm,wp}
        \fmf{fermion}{nu,cc,cc',nubar}
        \fmf{boson}{cc,wm}
        \fmf{boson}{cc',wp}
        \fmfdot{cc,cc'}
        \fmfv{label=$e^-$}{nu}
        \fmfv{label=$e^+$}{nubar}
        \fmfv{label=$W^-$}{wm}
        \fmfv{label=$W^+$}{wp}
      \end{fmfgraph*}}}
\end{equation}
and it does \emph{not} satisfy the Ward identity by itself
\begin{equation}
\label{eq:deeWW1}
  k^\mu\epsilon^\nu(q) M_{\mu\nu}^{(1)}(p_+,p_-;k,q)
     \propto \bar v(p_+) \fmslash{\epsilon} (q) u(p_-) \,.
\end{equation}
Since the~$W^\pm$ are charged, there is a second diagram with photon
exchange
\begin{equation}
\label{eq:eeWW2}
  M_{\mu\nu}^{(2)}(p_+,p_-;k,q) = 
  \parbox{30\unitlength}{%
    \fmfframe(6,4)(3,4){%
      \begin{fmfgraph*}(20,15)
        \fmfleft{nu,nubar} \fmfright{wm,wp}
        \fmf{fermion}{nu,cc,nubar}
        \fmf{boson,label=$\gamma$}{cc,cc'}
        \fmf{boson}{cc',wm}
        \fmf{boson}{cc',wp}
        \fmfdot{cc,cc'}
        \fmffreeze
        \fmfv{label=$e^-$}{nu}
        \fmfv{label=$e^+$}{nubar}
        \fmfv{label=$W^+$}{wp}
        \fmfv{label=$W^-$}{wm}
      \end{fmfgraph*}}}\,,
\end{equation}
but this does not cancel~(\ref{eq:deeWW1}) for minimal
coupling~$\partial_\mu\to\partial_\mu-ieA_\mu$ of the~$W^\pm$.
Therefore, we must \emph{not} use a St\"uckelberg
propagator~(\ref{eq:stuckelberg-propagator}) for massive vector bosons
coupled to charged currents naively.  In an attempt to salvage the power
counting, we can take a hint from unbroken non-abelian gauge theories
(a.\,k.\,a.~Yang-Mills theories) and investigate another diagram
\begin{equation}
\label{eq:eeWW3}
  \parbox{37\unitlength}{%
    \fmfframe(3,4)(3,4){%
      \begin{fmfgraph*}(20,15)
        \fmfleft{nu,nubar} \fmfright{wm,wp}
        \fmf{fermion}{nu,cc,nubar}
        \fmf{boson,label=$Z^0$}{cc,cc'}
        \fmf{boson}{cc',wm}
        \fmf{boson}{cc',wp}
        \fmfdot{cc,cc'}
        \fmfv{label=$e^-$}{nu}
        \fmfv{label=$e^+$}{nubar}
        \fmfv{label=$W^+$}{wp}
        \fmfv{label=$W^-$}{wm}
      \end{fmfgraph*}}}\,,
\end{equation}
arising from a coupling of the~$W^\pm$'s to a third gauge boson,
the~$Z^0$, which couples to the $\tau_3$~component of the current, in
order to close the gauge algebra generated by the~$\tau^\pm$.  Indeed
using the non-abelian gauge vertex and putting the outgoing~$W^\pm$'s
on their mass-shell
\begin{equation}
\label{eq:deeWW3}
  k^\mu\epsilon^\nu(q) M_{\mu\nu}^{(1)}(p_+,p_-;k,q)
     \propto \frac{s-M_W^2}{s-M_Z^2}
       \bar v(p_+) \fmslash{\epsilon} (q) u(p_-) \,.
\end{equation}
Thus~(\ref{eq:deeWW1}) could be canceled with charge assignments that
are compatible with the gauge algebra, as long as~$M_W=M_Z$.
However, the sum of~(\ref{eq:eeWW1}), (\ref{eq:eeWW2})
and~(\ref{eq:eeWW3}) \emph{still} couples to unphysical polarization
states if the external bosons are massive, because it is impossible to
cancel the massless photon pole this way. Moreover, adding the Yang-Mills
interaction to the Proca Lagrangian~(\ref{eq:proca}) results in
equations of motion that violate~(\ref{eq:M2dA}).  Therefore, a
successive plugging of the holes is not likely to succeed and we
should take a step back and reconsider, how unitarity is implemented
in covariantly quantized \emph{massless} Yang-Mills theories.

In addition to the two physical degrees of freedom there are four
unphysical degrees of freedom in the covariant quantization of gauge
theories: the scalar and longitudinal polarization
states~$\partial_\mu A^\mu(x)$
and~$A_L(x)$ of the gauge bosons are accompanied by two Faddeev-Popov
ghosts~$\eta(x)$ and $\bar\eta(x)$~\cite{Faddeev/Popov:1967:ghosts}. The
quartet-mechanism~\cite{Kugo/Ojima:1979:quartet} for
\ac{BRS}~\cite{Becchi/etal:1975:BRS} invariant interactions guarantees that
these unphysical degrees of freedom decouple from the two physical
degrees of freedom.  Only in an abelian gauge theory, as~\ac{QED}, do
the unphysical polarization states and ghost decouple separately.
Thus the physical Hilbert space in a Yang-Mills theory is not characterized
by the simple Gupta-Bleuler condition~(\ref{eq:dA}), but by the
cohomology of the conserved BRS-charge 
\begin{multline}
 Q_{\text{BRS}}(t) = \frac{1}{2} \int_{x_0=t}\!\mathrm{d}^3\vec x
    \tr\Biggl( \pi_0(x) \overleftrightarrow{\partial}_0 \eta(x)
       - ig\pi_0(x)\left[A_0(x),\eta(x)\right] \\
       + \frac{ig}{2}\left[\eta(x),\eta(x)\right]\partial_0\bar\eta(x) \Biggr)\,.
\end{multline}
The cohomology arises because
the BRS-charge $Q_{\text{BRS}}$ projects on zero norm states
\begin{equation}
 \left\|Q_{\text{BRS}}\Ket{\Psi}\right\|^2
    = \Braket{\Psi|Q_{\text{BRS}}^2|\Psi} = 0\,,
\end{equation}
due to
\begin{subequations}
\begin{align}
  Q_{\text{BRS}}^\dagger &= Q_{\text{BRS}} \\
  Q_{\text{BRS}}^2 &=0\,,
\end{align}
\end{subequations}
and such states have to factored out of the physical states
characterized by
\begin{equation}
\label{eq:Q(phys)}
 Q_{\text{BRS}}\Ket{\text{phys.}} = 0\,,
\end{equation}
in order to construct a proper Hilbert space.
For massive vector bosons, some other degree of freedom has to take
the place of the now physical longitudinal polarization
state~$A_L(x)$ in the quartet-mechanism.  This degree of freedom
can be provided by the Goldstone bosons of a spontaneous symmetry
breaking and the quantum numbers will match, if, and only if, the
broken symmetry is the gauge symmetry of the massive vector bosons.

\subsection{Hidden Symmetry}
\label{sec:hidden}

Spontaneously broken gauge theories provide a consistent description
of massive vector bosons, in which the dimensions of the fields are
indeed given by~(\ref{eq:dim(phi)}) and~(\ref{eq:free-propagators}).
There is no argument that they are the \emph{only} consistent
description.  However, such an argument is not necessary, because it
is possible to use a non-linear field redefinition to rewrite any low
energy effective interaction of vector bosons as the unitarity gauge
version of a non linearly realized gauge
theory~\cite{Cornwall/Levin/Tiktopoulos:1974:SM,
Chanowitz/etal:1987:vectorbosons,Burgess/London:1993:effective}.
In an abelian toy example\footnote{In the non-abelian generalization, 
the physics is obscured slightly by the need to parametrize the factor
group manifold~$G/H$.  Nevertheless, it is a straightforward
application of the formalism set forth
in~\protect\cite{Coleman/Wess/Zumino:1969:CCWZ,
*Callan/Coleman/Wess/Zumino:1969:CCWZ} and explicit formulae are given
in~\cite{Chanowitz/etal:1987:vectorbosons,Burgess/London:1993:effective}.}
there are the correspondences
\begin{equation}
\label{eq:gaugeit}
  \begin{pmatrix} \psi(x) \\ V_\mu(x) \end{pmatrix}
    \Longleftrightarrow
  \begin{pmatrix} \chi'(x) \\ \frac{-1}{gf} D_\mu\phi(x) \end{pmatrix}
    =
  \begin{pmatrix}
    e^{-i\phi(x)/f} \chi(x) \\
     A_\mu(x) - \frac{1}{gf} \partial_\mu\phi(x)
  \end{pmatrix}\,.
\end{equation}
$\chi$ is a matter field and~$V_\mu$ is massive vector field,
while~$\phi$ is a Goldstone boson field and~$A_\mu$ a gauge field of a
non-linearly realized~$\textrm{U}(1)$
\begin{equation}
  \begin{pmatrix} \chi(x) \\ \phi(x) \\ A_\mu(x) \end{pmatrix}
    \rightarrow
  \begin{pmatrix}
    e^{i\omega(x)} \chi(x) \\
    \phi(x) + f\omega(x) \\
    A_\mu(x) + \frac{1}{g} \partial_\mu\omega(x)
  \end{pmatrix}\,.
\end{equation}
Without a specification of the subsidiary conditions for the
quantization of the vector fields, it appears that~$(A_\mu,\phi)$ has
one more degree of freedom than~$V_\mu$.  However, $V_\mu$ is
understood to be quantized with~$\partial_\mu V^\mu=0$ enforced
explicitely, while~$(A_\mu,\phi)$ is quantized in an indefinite norm
Hilbert space where~$\phi$ and~$\partial_\mu A^\mu$ conspire with the
Faddeev-Popov ghosts to decouple from physical
amplitudes~\cite{Kugo/Ojima:1979:quartet} and the same number of
degrees of freedom survive.  Since the non-linear
field redefinitions do not change the on-shell scattering matrix
elements~\cite{Haag:1958:HaagRuelle,*Borchers:1960:HaagRuelle,
*Ruelle:1962:HaagRuelle}, the two formulations will yield the same
predictions for observable quantities.

The preceding reasoning does not prove that all vector bosons are
gauge bosons in disguise, but it comes very close, because it shows
that for every theory with massive vector bosons, there is another
theory in which the vector bosons are parametrized gauge bosons of
a non-linearly realized gauge symmetry.  These theories are
\emph{indistinguishable} at energy scales below the symmetry breaking
scale~$\Lambda=4\pi f$ characterizing the non-linear realization.
Above the symmetry breaking scale, the two theories are of course very
different, because the gauge symmetry will then be realized linearly.

In a covariant $R_\xi$-gauge, a special case of~(\ref{eq:Q(phys)})
is the new subsidiary condition
\begin{equation}
\label{eq:dA+phi}
  \Braket{\text{phys.}|\partial_\mu A^\mu(x)
      - \xi m_A \phi(x)|\text{phys.}'} = 0\,.
\end{equation}
that takes the place of~(\ref{eq:dA}).
At energies large compared to the boson mass, the longitudinal
polarization vector~$\epsilon_L^\mu(k)$ can be approximated
by~$k^\mu/m_A$.  Then~(\ref{eq:dA+phi}) turns into the celebrated
\emph{equivalence
theorem}~\cite{Cornwall/Levin/Tiktopoulos:1974:SM,
Chanowitz/Gaillard:1985:equivalence}
and the scattering amplitudes for longitudinal vector bosons can be
calculated more efficiently by calculating the corresponding
amplitudes for Goldstone bosons.

\subsection{Tree Level Unitarity}

A complementary analysis arrives at the same conclusion. A systematic
study of the tree level unitarity of four-point functions and
five-point functions calculated with arbitrary interactions reveals
that \emph{only} spontaneously broken gauge theories result in amplitudes
that satisfy tree level unitarity at high
energy~\cite{Cornwall/Levin/Tiktopoulos:1974:SM} (see
also~\cite{Quigg:1983:textbook} for a less systematical but more
pedagogical account).  As could be expected from the
study of the Ward identities, the production of \emph{longitudinal}
vector bosons
\begin{equation}
  \parbox{27\unitlength}{%
    \fmfframe(3,4)(3,4){%
      \begin{fmfgraph*}(20,15)
        \fmfleft{nu,nubar} \fmfright{wp,wm}
        \fmf{fermion}{nu,cc,cc',nubar}
        \fmf{boson}{cc,wp}
        \fmf{boson}{cc',wm}
        \fmfdot{cc,cc'}
        \fmfv{label=$\nu$}{nu}
        \fmfv{label=$\bar\nu$}{nubar}
        \fmfv{label=$W_L^+$}{wp}
        \fmfv{label=$W_L^-$}{wm}
      \end{fmfgraph*}}}
  \Longrightarrow
    \frac{\textrm{d}\sigma}{\textrm{d}\Omega}
      = \frac{G_F^2}{32\pi^2}\cdot E_{CM}^2 \cdot \sin^2\theta
\end{equation}
is responsible for this result. $P$-wave unitarity
($|\mathcal{M}_{J=1}|\le1$) demands
\begin{equation}
  \frac{\textrm{d}\sigma}{\textrm{d}\Omega} \le \frac{18\sin^2\theta}{E_{CM}^2}
\end{equation}
which is displayed in figure~\ref{fig:nogaugetrouble}. Unitarity is
maintained up to considerably higher scales than in
figure~\ref{fig:fermitrouble},
but will eventually be violated at a few~$\unit{TeV}$.  Continuing
this heuristic ``patching up'' will
unambiguously~\cite{Cornwall/Levin/Tiktopoulos:1974:SM} lead to
a spontaneously broken gauge theory, identical to the one proposed in
sections~\ref{sec:longitudinal} and~\ref{sec:hidden}. 
\begin{figure}
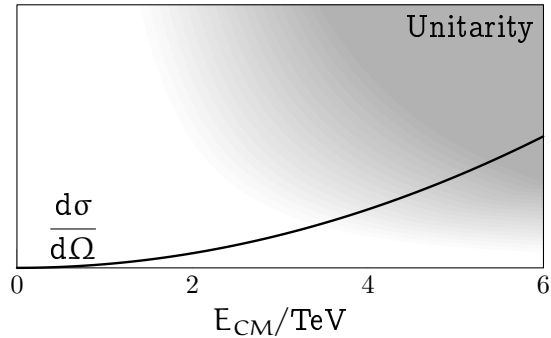

  \begin{center}
    \begin{empgraph}(70,35)
      smax := 2 * ((rootgf/pi*emax)**2)/32;
      emin := sqrt (18/smax);
      pickup pencircle scaled 0.5pt;
      setrange (0, 0, emax, smax);
      autogrid (itick.bot,);
      path back, sigma;
      augment.back (0,0);
      augment.back (emax,0);
      augment.back (emax,smax);
      augment.back (0,smax);
      gfill back--cycle withcolor white;
      path unitarity[];
      for t = 0, 1:
        emin := 1 / (sqrt (smax/18./(1+1.5*(t-.5))));
        for e = emin step 0.05 until (emax+0.1):
          ee := min (e, emax);
          augment.unitarity[t] (ee, 18.*(1+1.5*(t-.5))/(ee*ee));
        endfor
      endfor
      for t = 0 step 0.05 until 1.01:
        path utmp;
        utmp = interpath (t, unitarity0, unitarity1);
        augment.utmp (emax, smax);
        gfill utmp--cycle withcolor t[white,.7white];
      endfor
      pickup pencircle scaled 1pt;
      for e = 0 step 0.01 until emax:
        s := ((rootgf/pi*e)**2)/32;
        exitif s >= smax;
        augment.sigma (e, s);
      endfor
      gdraw sigma;
      pickup pencircle scaled 0.5pt;
      glabel.ulft (btex $\displaystyle
                         \frac{\textrm{d}\sigma}{\textrm{d}\Omega}$ etex,
                   100);
      glabel.llft (btex Unitarity etex, (emax-0.05,smax-0.05));
      glabel.bot (btex $E_{CM}/\text{TeV}$ etex, OUT);
    \end{empgraph}
  \end{center}
  \caption{\label{fig:nogaugetrouble}%
    Tree-level unitarity bound for~$\nu\bar\nu\to W_L^+W_L^-$.}
\end{figure}

\section{The Minimal Standard Model}
\label{sec:msm}

Combining the low energy information on the flavor symmetries with the
observation of the top quark and the description of weak interactions
by gauge bosons of a non-linearly
realized~$\textrm{SU}(2)_L\otimes\textrm{U}(1)_Y/\textrm{U}(1)_Q$
and the unbroken~$\textrm{SU}(3)_C$ of~\ac{QCD}, results in the particles of
the minimal~\ac{SM} lined up with their quantum number assignments in
table~\ref{tab:SM}.  Instead of
realizing~$\textrm{SU}(2)_L\otimes\textrm{U}(1)_Y/\textrm{U}(1)_Q$
non-linearly, it is also possible to introduce an elementary Higgs
particle to break the electroweak symmetry.

Both options provide a consistent theoretical framework for
calculations and are not ruled out by experiment.  However, a linear
realization with a light Higgs near~$\unit[100]{GeV}$ appears to be
favored by current experimental results (see figure~\ref{fig:blueband}).

As discussed at length in section~\ref{sec:bosons}, the low energy
non-abelian charged currents require that among the equivalent
effective theories describing the vector bosons, there is one in
which the bosons are the non-abelian gauge bosons of a broken
symmetry.  Therefore the non-abelian couplings shown in
figure~\ref{fig:tgc-moriond99} are unavoidable and deviations from the
gauge theory predictions have to be consistent with the power
counting rules of~\ac{EFT}, i.\,e.~\ac{NDA}, as described in
section~\ref{sec:NDA}.

\ac{NDA} is based on the hierarchy of two scales, here the Fermi
scale~$v\approx\unit[246]{GeV}$ and $\Lambda_{\text{EWSB}} = 4\pi
v\approx\unit[3.1]{TeV}$, with no intermediate scale.  Therefore its
predictions can be invalidated by the observation of a new physical
threshold between~$v$ and~$\Lambda_{\text{EWSB}}$.  However, such a
threshold would be observed and \ac{NDA} is a valid tool for studying
electroweak gauge bosons on the $\unit{TeV}$ scale, if no new
particles will have been discovered until then.

Unless a new threshold is discovered or something very spectacular
happens that would undermine most our present understanding
of~\ac{QFT}, the deviations from the minimal~\ac{SM} must be on the
order of
\begin{equation}
\label{eq:ya1/16pi2}
  \frac{1}{16\pi^2} \approx 6.3 \cdot 10^{-3}\,.
\end{equation}
{}From the predictions for the sensitivity at LEP2 on gauge boson
couplings~\cite{Gounaris/etal:1996:LEP2-TGC}, we see
that~(\ref{eq:ya1/16pi2}) will not be reached.  Thus LEP2 will
\emph{not} be able to distinguish among models for~\ac{EWSB}, based on
measurements of the gauge boson couplings alone.  The direct search
for a very light Higgs is more promising.
This situation changes will change dramatically at the~\ac{LC}, due to
the increased energy and luminosity (see section~\ref{sec:tgc}).

\section{Precision Tests}
\label{sec:precision}

The observation that low energy physics can be described by an
effective field theory, does \emph{not} mean that precision
experiments can not access at least some aspects of the physics at
higher scales through loop corrections~\cite{Georgi:1991:STU}. If the
low energy~\ac{EFT} has less symmetry than the~\ac{EFT} relevant at a
higher energy scale, the parameters of the low energy~\ac{EFT} can be
tested for the remnants of the high energy symmetry.  The most
celebrated example of this phenomenon at LEP1 are the radiative
corrections to the $\rho$-parameter and the evidence for a custodial
$\textrm{SU}(2)_c$~symmetry~\cite{Sikivie/etal:1980:custodial} of
the~\ac{EWSB} sector (see also section~\ref{sec:decoupling}).

At low energies, in the current-current couplings~(\ref{eq:Fermi}),
the relative strengths of the weak and neutral current couplings is
a free parameter.  After the introduction of intermediate vector
bosons, the two dimensionful parameters are interpreted as the ratio
of dimensionless couplings and vector boson masses. Above the
production threshold of the vector bosons, their masses can be
measured independently from the strength of weak interactions.  From
the observed values, they appear to be related because the
$\rho$-parameter
\begin{equation}
\label{eq:rho}
  \rho = \frac{M_W^2}{\cos^2\theta_w M_Z^2}
\end{equation}
is very close to unity, with the renormalization scheme
dependent\footnote{There are mass dependent renormalization schemes
that actually enforce~$\rho=1$ at one scale.  The non-trivial
observation is then that $\rho$~does not ``run away'' from this
value.} deviations in the~\ac{SM} on the order of~$1\%$.
In~(\ref{eq:rho}), the weak mixing angle~$\theta_w$ is defined
\begin{equation}
\label{eq:theta}
  e = g\sin\theta_w
\end{equation}
by the ratio of the left handed charged current couplings and the
electromagnetic coupling.

Of course, $\rho=1$~could be just a numerical accident, unless it can
be explained by a symmetry.  In the minimal~\ac{SM}, $\rho=1$~is
explained by the fact that all three components of the isospin-triplet
charged gauge boson receive the \emph{same} mass from the coupling to
the~\ac{EWSB} sector, \emph{before} the neutral
component mixes with the hypercharge gauge boson. In the linear
realization of the~\ac{EWSB}, the global
$\textrm{SU}(2)_c\otimes\textrm{SU}(2)_L$~symmetry is equivalent to
the~$\textrm{SO}(4)$ of the four components of the complex Higgs
doublet.  In a Higgs-less realization, the
$\textrm{SU}(2)_c\otimes\textrm{SU}(2)_L\otimes
 \textrm{U}(1)_Y/\textrm{SU}(2)_{C+L}\otimes\textrm{U}(1)_Q$ has to be
realized non-linearly~\cite{Coleman/Wess/Zumino:1969:CCWZ,
*Callan/Coleman/Wess/Zumino:1969:CCWZ}.  Apparently, whatever
the detailed dynamics of~\ac{EWSB} is, it respects the \emph{custodial
symmetry} relating~$W^+$, $W^-$,
and~$W^3=\cos\theta_wZ^0+\sin\theta_wA$.

The most important deviations from $\rho=1$ in the minimal~\ac{SM}
depend on~$m_t$ (see section~\ref{sec:decoupling}).  Therefore, the
joint efforts of~LEP1, LEP2, SLC and Tevatron in the measurement
of~$M_Z$, $M_W$, $\sin\theta_w$, and~$m_t$ have provided strict
constraints on the symmetry of the~\ac{EWSB} sector.  The
$\unit{TeV}$-scale physics probed in this way exceeds the direct
physics reach of the contributing colliders
(see figure~\ref{fig:reaches}) by an order of magnitude.

\chapter{Matrix Elements}
\label{sec:me}

The quest for a theory of flavor demands precise calculations of high
energy scattering processes in the framework of the~\ac{SM} and
its extensions.  At the Tevatron, the~\ac{LHC}, and at the~\ac{LC}, final
states with many detected particles and with tagged flavor will be the
primary handle for testing theories of flavor.

\section{Three Roads from Actions to Matrix Elements}
\label{sec:action->me}

\subsection{Manual Calculation}

The traditional textbook approach to deriving amplitudes from actions
are \emph{manual calculations}, which can be aided by computer algebra
tools.  The time-honored method of calculating the squared amplitudes
directly using trace techniques is no longer practical for today's
multi particle final states and has generally been replaced by
helicity amplitude methods (see e.\,g.~\cite{Kleiss/Stirling:1985:helamp,
Berends/Giele:1987:WeylvanderWaerden,*Dittmaier:1999:WeylvanderWaerden,
Gastmans/Wu:1990:photon}).

Manual calculations have the disadvantage of consuming a lot of
valuable physicist's time, but can provide insights that are hidden
by the other approaches discussed below.  Regarding basic processes,
where extremely fast and accurate predictions are required for
numerical fits, manual calculations can provide efficient formulae
that are still unrivaled.

\subsection{Computer Aided Calculation}

An increasingly popular technique is to
use a well tested \emph{library of basic helicity amplitudes} for the
building blocks of Feynman diagrams and to construct the complete
amplitude directly in the program in the form of function calls.  A
possible disadvantage is that the differential cross section is nowhere
available as a formula, but the value of such a formula is limited
anyway, since they can hardly be printed on a single page anymore.

\subsection{Automatic Calculation}

Fully \emph{automatic calculations} are a further step in the same
direction.  The Feynman rules (or equivalent prescriptions) are no
longer applied manually, but encoded algorithmically.  This method
will become more and more important in the future, but more work is
needed for the automated construction of efficient unweighted event
generators (see section~\ref{sec:unweighted-MC})
and for the automated factorization of collinear singularities in
radiative corrections (see~\cite{Ohl:1997:MCReview}).

\section{Forests and Groves}
\label{sec:groves}

Calculations of cross section with many-particle final states remain
challenging despite all technical advances and it is of crucial
importance to be able to concentrate on the important parts of the
scattering amplitude for the phenomena under consideration.

In gauge theories, however, it is impossible to simply select a few
signal diagrams and to ignore irreducible backgrounds.  The same
subtle cancellations among the diagrams in a gauge invariant subset
that lead to the celebrated good high energy behavior of gauge
theories such as the~\ac{SM} (see section~\ref{sec:bosons}),
come back to haunt us if we
accidentally select a subset of diagrams that is not gauge invariant.
The results from such a calculation have \emph{no} predictive power,
because they depend on unphysical parameters introduced during the
gauge fixing of the Lagrangian.  It must be stressed that not all
diagrams in a gauge invariant subset have the same pole structure and
that a selection based on ``signal'' or ``background'' will not suffice.

The subsets of Feynman diagrams selected for any calculation must
therefore form a \emph{gauge invariant subset}, i.\,e.~together they
must already satisfy the Ward- and Slavnov-Taylor-identities to ensure
the cancellation of contributions from unphysical degrees of freedom.

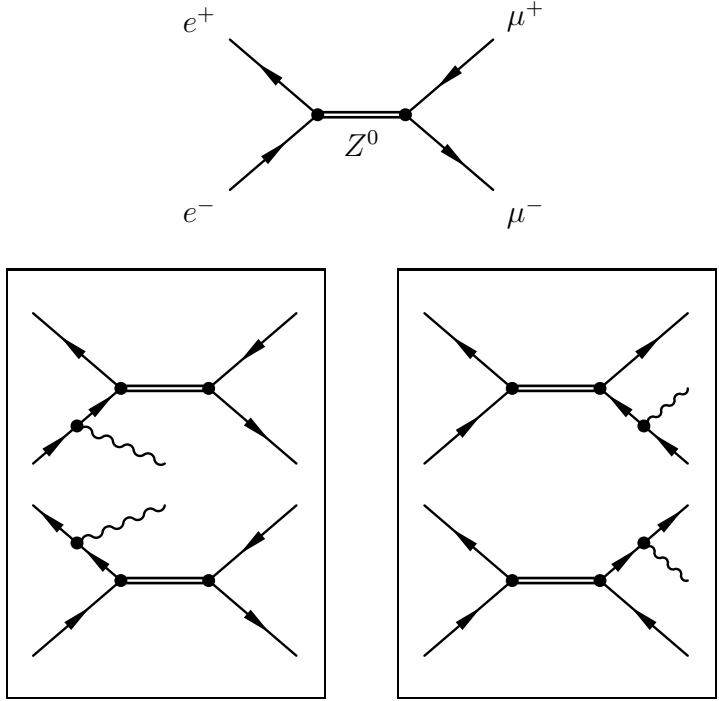
\begin{figure}
  \begin{center}
    \begin{fmfgraph*}(35,20)
      \fmflabel{$e^+$}{ep}
      \fmflabel{$e^-$}{em}
      \fmflabel{$\mu^+$}{ep'}
      \fmflabel{$\mu^-$}{em'}
      \fmfleft{em,ep}
      \fmfright{em',ep'}
      \fmf{fermion}{em,v,ep}
      \fmf{fermion}{ep',v',em'}
      \fmf{double,label=$Z^0$,tension=2}{v,v'}
      \fmfdot{v,v'}
    \end{fmfgraph*}\\
    \hfil\\
    \hfil\\
    \fbox{%
      \parbox{40\unitlength}{%
        \begin{center}
          \begin{fmfgraph}(35,20)
            \fmfleft{em,ep}
            \fmfright{em',ep'}
            \fmfbottom{g'}
            \fmf{phantom}{em,v,ep}
            \fmf{fermion}{ep',v',em'}
            \fmf{double,tension=2}{v,v'}
            \fmffreeze
            \fmf{fermion}{em,g,v,ep}
            \fmf{photon,tension=0}{g,g'}
            \fmfdot{v,v',g}
          \end{fmfgraph}\\
          \hfil\\
          \begin{fmfgraph}(35,20)
            \fmfleft{em,ep}
            \fmfright{em',ep'}
            \fmftop{g'}
            \fmf{phantom}{em,v,ep}
            \fmf{fermion}{ep',v',em'}
            \fmf{double,tension=2}{v,v'}
            \fmffreeze
            \fmf{fermion}{em,v,g,ep}
            \fmf{photon,tension=0}{g,g'}
            \fmfdot{v,v',g}
          \end{fmfgraph}
        \end{center}}}
    \qquad
    \fbox{%
      \parbox{40\unitlength}{%
        \begin{center}
          \begin{fmfgraph}(35,20)
            \fmfleft{em,ep}
            \fmfright{em',g',ep'}
            \fmf{fermion}{em,v,ep}
            \fmf{phantom}{ep',v',em'}
            \fmf{double,tension=2}{v,v'}
            \fmffreeze
            \fmf{fermion}{em',g,v',ep'}
            \fmf{photon,tension=0}{g,g'}
            \fmfdot{v,v',g}
          \end{fmfgraph}\\
          \hfil\\
          \begin{fmfgraph}(35,20)
            \fmfleft{em,ep}
            \fmfright{em',g',ep'}
            \fmf{fermion}{em,v,ep}
            \fmf{phantom}{ep',v',em'}
            \fmf{double,tension=2}{v,v'}
            \fmffreeze
            \fmf{fermion}{em',v',g,ep'}
            \fmf{photon,tension=0}{g,g'}
            \fmfdot{v,v',g}
          \end{fmfgraph}
        \end{center}}}
  \end{center}
  \caption{\label{fig:ee2mmg}%
    The gauge invariant subsets in~$e^+e^-\to\mu^+\mu^-\gamma$ are
    obtained by adding the photon in all possible ways to connected
    lines of charged particles.}
\end{figure}
In abelian gauge theories, such as QED, the classification of gauge
invariant subsets is straightforward and can be summarized by the
requirement of inserting any additional photon into \emph{all}
connected charged propagators, as shown in figure~\ref{fig:ee2mmg}.
This situation is similar for gauge
theories with simple gauge groups, the difference being that the gauge
bosons are carrying charge themselves.  For non-simple gauge groups
like the~\ac{SM}, which even includes mixing, the
classification of gauge invariant subsets is much more involved.

In early calculations, the classification of gauge invariant subsets
in the~\ac{SM} has been performed in an ad-hoc fashion
(see \cite{Bardin/etal:1994:4f-classification,Boos/Ohl:1997:gg4f}),
which was later superseded by the explicit construction of
\emph{groves}~\cite{Boos/Ohl:1999:groves}, the smallest gauge
invariant classes of tree Feynman diagrams in gauge theories.  This
construction is not restricted to gauge theories with simple gauge
groups.  Instead, it is applicable to gauge groups with any number of
factors, which can even be mixed, as in the~\ac{SM}.
Furthermore, it does not require a summation over complete multiplets
and can therefore be used in flavor physics when members of weak
isospin doublets (such as charm or bottom) are detected separately.
The method constructs the smallest gauge invariant subsets and
examples below illustrate that they are indeed smaller than those
derived from looking at the final state
alone~\cite{Bardin/etal:1994:4f-classification,Boos/Ohl:1997:gg4f}.

The method will probably also have applications in loop calculations.
However, the published proofs use properties of tree diagrams and
further research is required in this area.

In unbroken gauge theories, the permutation symmetry of external gauge
quantum numbers can be used to subdivide the scattering amplitude
corresponding to a grove further into gauge invariant sub-amplitudes
(see~\cite{Berends/Giele:1987:color-decomposition,
*Mangano/Parke/Xu:1988:color-decomposition,
*Mangano/Parke:1991:review} and references cited therein).  In this
decomposition, each Feynman diagram contributes to more than one
sub-amplitude.  It is not yet known how to perform a similar
decomposition systematically in the~\ac{SM}, because the
entanglement of gauge couplings and gauge boson masses complicates the
structure of the amplitudes.  The construction of the groves provides
a necessary first step towards the solution of this important problem.

\subsection{Gauge Cancellations}

Even if the general arguments are well known, the explicit
cancellations of unphysical contributions remain amazing in explicit
perturbative calculations in gauge theories.  The intricate nature of
these cancellations complicates the development of systematic
calculational procedures.  Nevertheless, phenomenology needs
numerically well behaved matrix elements and cross sections, which
should be as compact as possible.  The development of an automated
procedure for obtaining compact expressions with explicit gauge
cancellations remains a formidable challenge.  The selection of gauge
invariant subsets of Feynman diagrams described in the remainder of
this section is one necessary step towards this goal.

As discussed in chapter~\ref{sec:sm/eft}, there is
overwhelming phenomenological evidence that the strong and electroweak
interactions are mediated by vector particles.  A naive description of
vector particles would assign to them four degrees of freedom or
polarization states, of which only two (for massless particles) or
three (for massive particles) can be physical.  The only known
theories that can consistently decouple the unphysical degrees of
freedom for interacting fields are gauge theories (for massless
particles) or equivalent to spontaneously broken gauge theories (for
massive particles).

However, perturbative calculations require an explicit breaking of
gauge invariance for technical reasons and the cancellation of
unphysical contributions is not manifest in intermediate stages of
calculations.  The contribution of a particular Feynman diagram to a
scattering amplitude depends in the gauge fixing procedure and has no
physical meaning, in general.

Among the possible gauge fixing schemes, two complementary approaches
can be identified: \emph{unitarity gauges} and \emph{renormalizable
gauges}.  A massive gauge boson in unitarity gauge has a propagator
\begin{equation}
\label{eq:unitarity}
  - \frac{i}{p^2-M^2+i\epsilon}
     \left( g_{\mu\nu} - \frac{p_\mu p_\nu}{M^2} \right)
\end{equation}
and unitarity of the time evolution can be proved straightforwardly,
because only physical
degrees of freedom propagate and the condition~$\partial_\mu V^\mu=0$
is maintained.  Unfortunately, the longitudinal part of the
propagator~(\ref{eq:unitarity}) does not fall off with high energy and
momentum.  This spoils the power counting for renormalization and, even
worse, violates tree-level unitarity at high energies, because the
scattering amplitudes for longitudinal vector bosons do not fall off
rapidly enough to satisfy the optical theorem at high energies.

On the other hand, the family of $R_\xi$-gauge fixings results in
propagators that propagate unphysical degrees of freedom
\begin{equation}
\label{eq:Rxi}
  - \frac{i}{p^2-M^2+i\epsilon}
     \left( g_{\mu\nu} -
             (1-\xi)\frac{p_\mu p_\nu}{p^2 - \xi M^2 + i\epsilon} \right)
\end{equation}
but with good high energy behavior and standard power counting even
for the longitudinal components.  The cancellation of the unphysical
contributions in scattering matrix elements turns out to be a
formidable combinatorial problem~\cite{tHooft:1971:1,
  *tHooft:1971:2,*tHooft/Veltman:1972:Combinatorics}.

The second approach is superior for formal proofs to all orders
(see~\cite{Piguet/Rouet:1981:review,*Piguet:1995:book} and references
therein), because it allows to treat the two complicated problems
independently.  In a first step, the theory is defined using
a~\ac{BPHZ} subtraction
procedure~\cite{Bogoliubov/Parasiuk:1957:BPHZ,*Hepp:1966:BPHZ,*Zimmermann:1969:BPHZ},
which violates gauge invariance and unitarity, but allows to prove
the~\ac{QAP}~\cite{Lowenstein:1971:QAP,*Lam:1972:QAP,*Lam:1973:QAP}.
The~\ac{QAP} states that the classical symmetries of the Lagrangian
density~$L$ are respected by the effective action~$\Gamma$, upto a
\emph{local} operator of dimension five (see
also~\cite{Kilian/Ohl:1994:QAP}).  Subsequently, one can prove by
an inspection of all operators of dimension five, that the
\ac{BRS}~invariance of the effective action can be restored by
\emph{finite} and \emph{local} counterterms of dimension four,
provided that the coefficient of the \ac{ABBJ}~anomaly vanishes.
Finally, the unitarity of the \ac{BRS} invariant theory is proven with
the quartet-mechanism~\cite{Kugo/Ojima:1979:quartet}.

This two-step procedure is transparent because it decouples the
combinatorics of renormalized perturbative power counting from the
combinatorics of gauge invariance and \ac{BRS} invariance.  Once the
\ac{QAP} is established, \ac{BRS} invariance is reduced to a problem
in finite-dimensional cohomology.  On the other hand, controlling the
high energy behavior of the unitarity gauge
propagator~(\ref{eq:unitarity}) and the combinatorics of Ward
identities simultaneously is prohibitively complicated, even in low
orders of the loop expansion.

However, this decoupling of kinematical structure and gauge
structure, which makes the abstract argument transparent, is not
useful in concrete calculations, because the abstract arguments apply
only to the sum of all diagrams and shed little light on how the gauge
cancellations ``work''.  One particular problem is that the Ward
identities relate diagrams with different pole structure.  For
example, in~$q\bar q\to gg$
\begin{equation}
  \parbox{26\unitlength}{%
    \begin{fmfgraph}(25,15)
      \fmfleft{i1,i2}
      \fmfright{o4,o3}
      \fmf{fermion}{i1,v1}
      \fmf{fermion,tension=0.5}{v1,v4}
      \fmfdot{v1,v4}
      \fmf{fermion}{v4,i2}
      \fmf{gluon}{o3,v4}
      \fmf{gluon}{o4,v1}
    \end{fmfgraph}} + 
  \parbox{26\unitlength}{%
    \begin{fmfgraph}(25,15)
      \fmfleft{i1,i2}
      \fmfright{o3,o4}
      \fmf{fermion}{i1,v1}
      \fmf{fermion,tension=0.5}{v1,v4}
      \fmfdot{v1,v4}
      \fmf{fermion}{v4,i2}
      \fmf{phantom}{o4,v4}
      \fmf{phantom}{o3,v1}
      \fmffreeze
      \fmf{gluon}{o4,v1}
      \fmf{gluon,rubout}{o3,v4}
    \end{fmfgraph}} + 
  \parbox{26\unitlength}{%
    \begin{fmfgraph}(25,15)
      \fmfleft{i1,i2}
      \fmfright{o3,o4}
      \fmf{fermion}{i1,v1}
      \fmf{fermion}{v1,i2}
      \fmf{gluon}{v5,v1}
      \fmfdot{v1,v5}
      \fmf{gluon}{o4,v5}
      \fmf{gluon}{o3,v5}
    \end{fmfgraph}}
\end{equation}
the numerator factors cancel parts of denominators and the Ward identity
is satisfied only by the sum and not by individual diagrams.

Therefore, the physical amplitudes are determined by an intricate web
of kinematical structure and gauge structure, that is very hard to
disentangle.  Nevertheless, the identification of partial sums of
Feynman diagrams that are gauge invariant by themselves is of
great practical importance for at least two reasons.  Firstly, it is
more economical to spent the time improving Monte Carlo statistics for
the important pieces of the amplitude than to calculate the complete
amplitude all the time. This requires the identification of the
smallest gauge 
invariant part of the amplitude that contains the important
pieces. Secondly, different parts of the amplitude are characterized
by different scales (e.\,g.~$s$ or~$t$) for radiative corrections.
The different scales can only be used if the parts correspond to
separately gauge invariant pieces.

The practical consequence of these arguments for LEP2
physics is that each individual diagram in the~$e^+e^-\to W^+W^-$
amplitude
\begin{equation}
\label{eq:ee2WW}
  \parbox{26\unitlength}{%
    \begin{fmfgraph}(25,15)
      \fmfleft{i1,i2}
      \fmfright{o3,o4}
      \fmf{fermion}{i1,v1}
      \fmf{fermion}{v1,i2}
      \fmf{photon}{v1,v5}
      \fmfdot{v1,v5}
      \fmf{dbl_plain_arrow}{v5,o4}
      \fmf{dbl_plain_arrow}{o3,v5}
    \end{fmfgraph}} + 
  \parbox{26\unitlength}{%
    \begin{fmfgraph}(25,15)
      \fmfleft{i1,i2}
      \fmfright{o3,o4}
      \fmf{fermion}{i1,v1}
      \fmf{fermion}{v1,i2}
      \fmf{dbl_plain}{v1,v5}
      \fmfdot{v1,v5}
      \fmf{dbl_plain_arrow}{v5,o4}
      \fmf{dbl_plain_arrow}{o3,v5}
    \end{fmfgraph}} + 
  \parbox{26\unitlength}{%
    \begin{fmfgraph}(25,15)
      \fmfleft{i1,i2}
      \fmfright{o3,o4}
      \fmf{fermion}{i1,v1}
      \fmf{fermion}{v1,v4}
      \fmfdot{v1,v4}
      \fmf{fermion}{v4,i2}
      \fmf{dbl_plain_arrow}{v4,o4}
      \fmf{dbl_plain_arrow}{o3,v1}
    \end{fmfgraph}}
\end{equation}
has a bad high energy behavior~\cite{Quigg:1983:textbook} and only the
inclusion of subtle interferences in the sum results in good high
energy behavior\footnote{The good high energy behavior is realized
separately for the~$\textrm{SU}(2)_L$ and~$\textrm{U}(1)_Y$ parts of the
amplitude~\protect\cite{Beenakker/Denner:1994:WW}, but this requires
breaking up diagrams.}.

The on-shell pair production~(\ref{eq:ee2WW}) is unfortunately not the
observed process.  This causes two problems: the finite width
necessitates Dyson-resummation and higher orders violate gauge
invariance~\cite{Argyres/etal:1995:BHF}.  Furthermore, the observed
final state consists of many observed particles with tagged flavor.
Therefore there are many tree-level background diagrams that
contribute to the same process.  For this reason, the calculations are
complicated and it is important to identify the important parts.

\subsection{Flavor Selection Rules}
\label{sec:FlavorSelectionRules}

The interesting flavor physics at the~\ac{LHC} and at the~\ac{LC} is
different from~\ac{QCD} with light quarks because the gauge symmetry
is broken spontaneously and the gauge group is not simple.
Furthermore, the factors are mixed according to
\begin{equation}
  \begin{pmatrix} Z^0 \\ A \end{pmatrix} =
  \begin{pmatrix}
     \cos\theta_w & - \sin\theta_w \\
     \sin\theta_w & \cos\theta_w
  \end{pmatrix}
  \begin{pmatrix} W^3 \\ B \end{pmatrix}
\end{equation}
and the combinatorics are consequently more intricate.  Finally, the
flavor of leptons and heavy quarks is tagged and the flavor
transformations do not commute with the gauge
transformations and no summation over complete gauge multiplets is
possible.  For these reasons, the~\ac{QCD} recursion relations or
string inspired methods~\cite{Berends/Giele:1987:color-decomposition, %
  *Mangano/Parke:1991:review}
are not directly applicable.

One method of identifying gauge invariant subsets is to identify
subsets that contribute to a particular final
state~\cite{Bardin/etal:1994:4f-classification,Boos/Ohl:1997:gg4f}.
For example, there are ten tree diagrams contributing to~$e^+e^-\to
\mu^-\bar\nu_\mu u\bar d$ and therefore the corresponding subset of
the 20~diagrams in~$e^+e^-\to e^-\bar\nu_e u\bar d$ and its complement
must be both gauge invariant by themselves.  The subsets that can be
derived in this way are in general not minimal, but this example shows
that selection rules of flavor symmetries that commute with the gauge
group appear to be a useful tool.

\begin{figure}
  \begin{center}
    \fbox{%
      \parbox{38\unitlength}{%
        \fmfframe(6,5)(6,5){%
          \begin{fmfgraph*}(25,15)
            \fmfleft{ep,em}
            \fmfright{ep',em'}
            \fmflabel{$e^+$}{ep}
            \fmflabel{$e^-$}{em}
            \fmflabel{$e^+$}{ep'}
            \fmflabel{$e^-$}{em'}
            \fmf{photon,tension=0.5,label=$\gamma,,Z^0$}{v,v'}
            \fmf{fermion}{em,v,ep}
            \fmf{fermion}{ep',v',em'}
            \fmfdot{v,v'}
          \end{fmfgraph*}}}} \qquad
    \fbox{%
      \parbox{38\unitlength}{%
        \fmfframe(6,5)(6,5){%
          \begin{fmfgraph*}(25,15)
            \fmfleft{ep,em}
            \fmfright{ep',em'}
            \fmflabel{$e^+$}{ep}
            \fmflabel{$e^-$}{em}
            \fmflabel{$e^+$}{ep'}
            \fmflabel{$e^-$}{em'}
            \fmf{photon,tension=0.5,label=$\gamma,,Z^0$}{v,v'}
            \fmf{fermion}{em,v,em'}
            \fmf{fermion}{ep',v',ep}
            \fmfdot{v,v'}
          \end{fmfgraph*}}}}
  \end{center}
  \caption{\label{fig:bhabha}%
    Gauge invariant subsets of Feynman diagrams in Bhabha scattering.}
\end{figure}
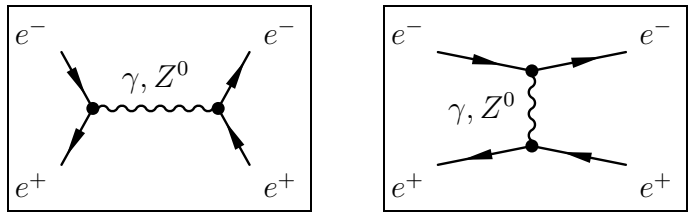
The simplest example of these flavor selection rules is provided by
Bhabha scattering: the $s$-channel and the $t$-channel diagram in
figure~\ref{fig:bhabha} have to be gauge-invariant separately, because
both~$e^+e^-\to \mu^+\mu^-$ and~$e^+\mu^-\to e^+\mu^-$ are physical
processes that must give gauge invariant amplitudes by themselves.
The first process has only the $s$-channel diagram and the second only
the $t$-channel diagram.  As a result, we see that conserved currents
of (real or fictitious) horizontal (i.\,e.~generation) symmetries can
serve as a tool to separate gauge invariant classes of Feynman
diagrams.

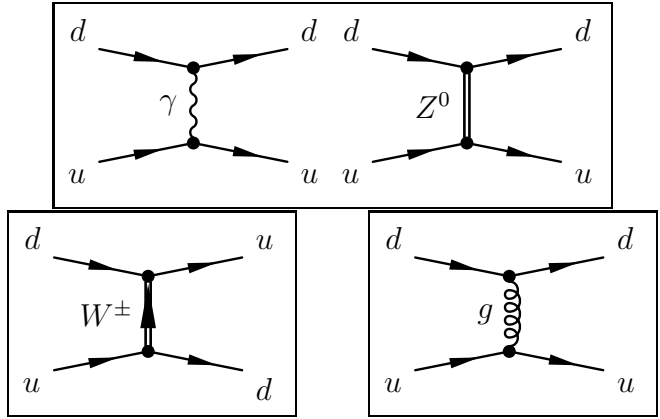
\begin{figure}
  \begin{center}
    \fbox{%
      \parbox{72\unitlength}{%
        \fmfframe(5,5)(5,5){%
          \begin{fmfgraph*}(25,15)
            \fmfleft{u1,d1}
            \fmflabel{$u$}{u1}
            \fmflabel{$d$}{d1}
            \fmfright{u2,d2}
            \fmflabel{$u$}{u2}
            \fmflabel{$d$}{d2}
            \fmf{photon,tension=0.5,label=$\gamma$}{uv,dv}
            \fmf{fermion}{u1,uv,u2}
            \fmf{fermion}{d1,dv,d2}
            \fmfdot{uv,dv}
          \end{fmfgraph*}}
        \fmfframe(5,5)(5,5){%
          \begin{fmfgraph*}(25,15)
            \fmfleft{u1,d1}
            \fmflabel{$u$}{u1}
            \fmflabel{$d$}{d1}
            \fmfright{u2,d2}
            \fmflabel{$u$}{u2}
            \fmflabel{$d$}{d2}
            \fmf{dbl_plain,tension=0.5,label=$Z^0$}{uv,dv}
            \fmf{fermion}{u1,uv,u2}
            \fmf{fermion}{d1,dv,d2}
            \fmfdot{uv,dv}
          \end{fmfgraph*}}}}\\
    \fbox{%
      \parbox{36\unitlength}{%
        \fmfframe(5,5)(5,5){%
          \begin{fmfgraph*}(25,15)
            \fmfleft{u1,d1}
            \fmflabel{$u$}{u1}
            \fmflabel{$d$}{d1}
            \fmfright{d2,u2}
            \fmflabel{$u$}{u2}
            \fmflabel{$d$}{d2}
            \fmf{dbl_plain_arrow,tension=0.5,label=$W^\pm$}{uv,dv}
            \fmf{fermion}{u1,uv,d2}
            \fmf{fermion}{d1,dv,u2}
            \fmfdot{uv,dv}
          \end{fmfgraph*}}}} \qquad
    \fbox{%
      \parbox{36\unitlength}{%
        \fmfframe(5,5)(5,5){%
          \begin{fmfgraph*}(25,15)
            \fmfleft{u1,d1}
            \fmflabel{$u$}{u1}
            \fmflabel{$d$}{d1}
            \fmfright{u2,d2}
            \fmflabel{$u$}{u2}
            \fmflabel{$d$}{d2}
            \fmf{gluon,tension=0.5,label=$g$}{uv,dv}
            \fmf{fermion}{u1,uv,u2}
            \fmf{fermion}{d1,dv,d2}
            \fmfdot{uv,dv}
          \end{fmfgraph*}}}}
  \end{center}
  \caption{\label{fig:ud2ud}%
    Gauge invariant subsets of Feynman diagrams in $ud\to ud$ scattering.}
\end{figure}
A less trivial example is provided by the three separately gauge
invariant sets in $ud\to ud$~scattering shown in
figure~\ref{fig:ud2ud}.  The separate gauge invariance of the gluon
exchange diagram is obvious because the~\ac{QCD} generators commute
with the electroweak generators and the strong coupling can be
switched on and off without violating gauge invariance.  The charged
current diagram is separately gauge-invariant, because we may assume
that the CKM mixing matrix is diagonal and then the charged current
diagonal is absent in $us\to us$~scattering, which is related to
$ud\to ud$ by a horizontal symmetry that commutes with the gauge
group.

These example in $2\to2$ scattering are not useful by themselves,
unless we can handle the combinatorics in the more complicated
realistic applications.

\subsection{Forests}
\label{sec:forests}
In order to develop tools for understanding the combinatorics, we have
to take a step back from the intricacies of gauge theories and study
unflavored scalar $\phi^3$- and $\phi^4$-theory.  We assume that
no~$\phi^5$ or higher vertices are required by phenomenology.  In
principle, the combinatorial methods describe below can be extended
to non-renormalizable theories.  However, the required arguments would
have be more involved, without providing additional insight into the
gauge sector.  Since there are no selection rules, the diagrams~$S_1$,
$S_2$, and~$S_3$ in
\begin{equation}
\label{eq:flips}
  T_4 = \{t_4^{S,1},t_4^{S,2},t_4^{S,3},t_4^{S,4}\} = \left\{
   \parbox{16\unitlength}{%
     \begin{fmfgraph}(15,12)
       \fmfleft{g1,g2}
       \fmfright{g1',g2'}
       \fmf{plain}{g1,v,g2}
       \fmf{plain}{g1',v',g2'}
       \fmf{plain}{v,v'}
       \fmfdot{v,v'}
     \end{fmfgraph}},
   \parbox{16\unitlength}{%
     \begin{fmfgraph}(15,12)
       \fmfleft{g1,g2}
       \fmfright{g1',g2'}
       \fmf{plain}{g1,v1,g1'}
       \fmf{plain}{g2,v2,g2'}
       \fmf{plain}{v1,v2}
       \fmfdot{v1,v2}
     \end{fmfgraph}},
   \parbox{16\unitlength}{%
     \begin{fmfgraph}(15,12)
       \fmfleft{g1,g2}
       \fmfright{g1',g2'}
       \fmf{phantom}{g1,v1,g1'}
       \fmf{phantom}{g2,v2,g2'}
       \fmf{plain}{v1,v2}
       \fmfdot{v1,v2}
       \fmffreeze
       \fmf{plain}{g1,v1,g2'}
       \fmf{plain,rubout}{g2,v2,g1'}
     \end{fmfgraph}},
   \parbox{16\unitlength}{%
     \begin{fmfgraph}(15,12)
       \fmfleft{g1,g2}
       \fmfright{g1',g2'}
       \fmf{plain}{g1,v,g2}
       \fmf{plain}{g1',v,g2'}
       \fmfdot{v,v}
     \end{fmfgraph}}
    \right\}.
\end{equation}
must have the same coupling strength to ensure crossing invariance.
If there are additional symmetries, as in the case of gauge theories,
the coupling of~$t_4^{S,4}$ will also be fixed relative
to~$t_4^{S,1}$, $t_4^{S,2}$, and~$t_4^{S,3}$.

Henceforth, we will call each exchange $t\leftrightarrow t'$ of two members of the
set~$T_4$ of all tree graphs with four external particles an
\emph{elementary flip}.  The elementary flips define a trivial
relation~$t\circ t'$ on~$T_4$, which is true, if and only if~$t$
and~$t'$ are related by a flip.  The relation~$\circ$ is trivial
on~$T_4$ because \emph{all} pairs are related by an elementary flip.

However, the elementary flips in~$T_4$ induce \emph{flips} in~$T_5$
(the set of all tree diagrams with five external particles) if they
are applied to an arbitrary four particle subdiagram
\begin{subequations}
\begin{equation}
   \parbox{19\unitlength}{%
     \begin{fmfgraph}(18,15)
       \fmfleft{g1,g2}
       \fmfright{g1',g2',g3'}
       \fmf{plain}{g1,v,g2}
       \fmf{plain,tension=0.5}{g1',v'}
       \fmf{plain}{v',v''}
       \fmf{dashes,tension=0.5}{v'',g2'}
       \fmf{dashes,tension=0.5}{v'',g3'}
       \fmf{plain}{v,v'}
       \fmfdot{v,v',v''}
     \end{fmfgraph}}
  \Longrightarrow
  \left\{
   \parbox{19\unitlength}{%
     \begin{fmfgraph}(18,15)
       \fmfleft{g1,g2}
       \fmfright{g1',g2',g3'}
       \fmf{plain}{g1,v1}
       \fmf{plain,tension=0.5}{v1,g1'}
       \fmf{plain}{g2,v2,v''}
       \fmf{dashes,tension=0.5}{v'',g2'}
       \fmf{dashes,tension=0.5}{v'',g3'}
       \fmf{plain}{v1,v2}
       \fmfdot{v1,v2,v''}
     \end{fmfgraph}},
   \parbox{19\unitlength}{%
     \begin{fmfgraph}(18,15)
       \fmfleft{g1,g2}
       \fmfright{g1',g2',g3'}
       \fmf{phantom}{g1,v1}
       \fmf{phantom,tension=0.5}{v1,g1'}
       \fmf{phantom}{g2,v2,v''}
       \fmf{phantom,tension=0.5}{v'',g2'}
       \fmf{dashes,tension=0.5}{v'',g3'}
       \fmf{plain}{v1,v2}
       \fmfdot{v1,v2,v''}
       \fmffreeze
       \fmf{dashes}{v'',g2'}
       \fmf{plain}{g1,v1,v''}
       \fmf{plain,rubout}{g2,v2,g1'}
     \end{fmfgraph}},
   \parbox{19\unitlength}{%
     \begin{fmfgraph}(18,15)
       \fmfleft{g1,g2}
       \fmfright{g1',g2',g3'}
       \fmf{plain}{g1,v,g2}
       \fmf{plain,tension=0.5}{g1',v}
       \fmf{plain}{v,v''}
       \fmf{dashes,tension=0.5}{v'',g2'}
       \fmf{dashes,tension=0.5}{v'',g3'}
       \fmfdot{v,v''}
     \end{fmfgraph}}
    \right\}
\end{equation}
Obviously, there is more than one element of~$T_4$ embedded in a
particular $t\in T_5$ and the same diagram is member of other
quartets, e.\,g.
\begin{equation}
   \parbox{19\unitlength}{%
     \begin{fmfgraph}(18,15)
       \fmfleft{g1,g2}
       \fmfright{g1',g2',g3'}
       \fmf{dashes}{g1,v,g2}
       \fmf{plain,tension=0.5}{g1',v'}
       \fmf{plain}{v',v''}
       \fmf{plain,tension=0.5}{v'',g2'}
       \fmf{plain,tension=0.5}{v'',g3'}
       \fmf{plain}{v,v'}
       \fmfdot{v,v',v''}
     \end{fmfgraph}}
  \Longrightarrow
  \left\{
   \parbox{19\unitlength}{%
     \begin{fmfgraph}(18,15)
       \fmfleft{g1,g2}
       \fmfright{g1',g2',g3'}
       \fmf{dashes}{g1,v,g2}
       \fmf{plain}{v,v'}
       \fmf{plain,tension=0.5}{v',g3'}
       \fmf{plain,tension=0.5}{g1',v'',g2'}
       \fmf{plain}{v',v''}
       \fmfdot{v,v',v''}
     \end{fmfgraph}},
   \parbox{19\unitlength}{%
     \begin{fmfgraph}(18,15)
       \fmfleft{g1,g2}
       \fmfright{g1',g2',g3'}
       \fmf{dashes}{g1,v,g2}
       \fmf{phantom}{v,v'}
       \fmf{phantom,tension=0.5}{v',g3'}
       \fmf{phantom,tension=0.5}{g1',v'',g2'}
       \fmf{plain}{v',v''}
       \fmffreeze
       \fmf{plain}{v,v',g2'}
       \fmf{plain,rubout}{g1',v'',g3'}
       \fmfdot{v,v',v''}
     \end{fmfgraph}},
   \parbox{19\unitlength}{%
     \begin{fmfgraph}(18,15)
       \fmfleft{g1,g2}
       \fmfright{g1',g2',g3'}
       \fmf{dashes}{g1,v,g2}
       \fmf{plain}{v,v'}
       \fmf{plain,tension=0.5}{v',g3'}
       \fmf{plain,tension=0.5}{g1',v',g2'}
       \fmfdot{v,v'}
     \end{fmfgraph}}
    \right\}
\end{equation}
\end{subequations}
\begin{figure}
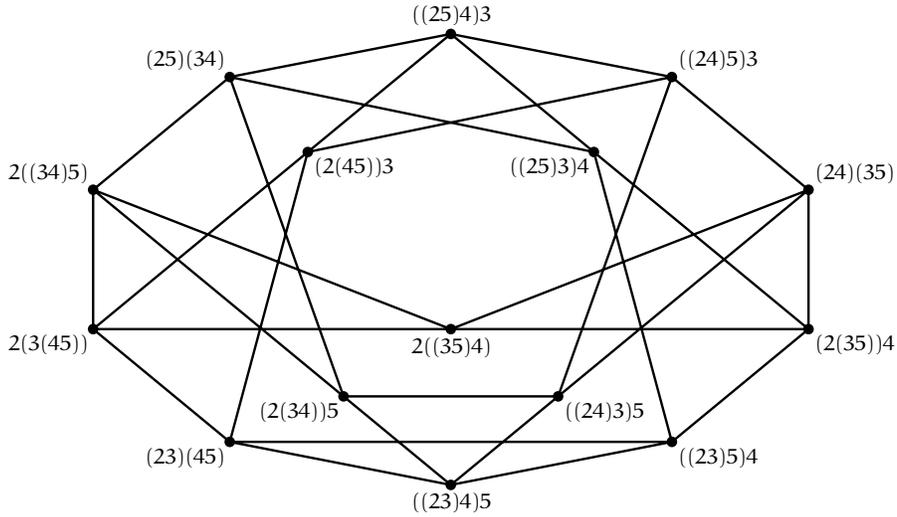

  \begin{center}
    \begin{emp}(100,60)
      pair t_i[], t_o[];
      for i := -10 upto 10:
        t_o[i] := (0,min(w,h)) rotated -36i xscaled (w/h) shifted (w,h)/2;
      endfor
      t_i[0] := .5[t_o[-3],t_o[3]];
      t_i[1] := .3[t_o[5],t_o[2]];
      t_i[-1] := .3[t_o[-5],t_o[-2]];
      t_i[2] := .4[t_o[0],t_o[3]];
      t_i[-2] := .4[t_o[0],t_o[-3]];
      pickup pencircle scaled 1pt;
      draw for i := 0 upto 9: t_o[i]-- endfor
          t_o[0]--t_i[2]--t_o[3]--t_i[0]--t_o[-2]
        --t_i[-1]--t_o[-5]--t_i[1]
        --t_i[-1]--t_o[-1]--t_i[2]--t_o[4]
        --t_o[-4]--t_i[-2]--t_o[1]--t_i[1]
        --t_o[2]--t_i[0]--t_o[-3]--t_i[-2]--cycle;
      pickup pencircle scaled 4pt;
      for i := -4 upto 5: drawdot t_o[i]; endfor
      for i := -2 upto 2: drawdot t_i[i]; endfor
      label.llft (btex $\scriptstyle (23)(45) $ etex, t_o[-4]);
      label.llft (btex $\scriptstyle 2(3(45)) $ etex, t_o[-3]);
      label.ulft (btex $\scriptstyle 2((34)5) $ etex, t_o[-2]);
      label.ulft (btex $\scriptstyle (25)(34) $ etex, t_o[-1]);
      label.top  (btex $\scriptstyle ((25)4)3 $ etex, t_o[0]);
      label.urt  (btex $\scriptstyle ((24)5)3 $ etex, t_o[1]);
      label.urt  (btex $\scriptstyle (24)(35) $ etex, t_o[2]);
      label.lrt  (btex $\scriptstyle (2(35))4 $ etex, t_o[3]);
      label.lrt  (btex $\scriptstyle ((23)5)4 $ etex, t_o[4]);
      label.bot  (btex $\scriptstyle ((23)4)5 $ etex, t_o[5]);
      label.lrt  (btex $\scriptstyle (2(45))3 $ etex, t_i[-2]);
      label llft (btex $\scriptstyle (2(34))5 $ etex, t_i[-1]);
      label.bot  (btex $\scriptstyle 2((35)4) $ etex, t_i[0]);
      label.lrt  (btex $\scriptstyle ((24)3)5 $ etex, t_i[1]);
      label.llft (btex $\scriptstyle ((25)3)4 $ etex, t_i[2]);
      setbounds currentpicture to (0,-5)--(w,-5)--(w,h)--(0,h)--cycle;
    \end{emp}
  \end{center}
  \caption{\label{fig:5phi3}%
    The forest~$F_5$ of the 15 five-point tree diagrams in unflavored
    $\phi^3$-theory. The diagrams are specified by fixing vertex~1 and
    using parentheses to denote the order in which lines are joined at
    vertices (see the footnote on
    page~\protect\pageref{fn:parenthesis} for examples).}
\end{figure}
There are 15~five-point tree diagrams in~$\phi^3$ and inspection shows
that only four other inequivalent diagrams can be reached from any
diagram by flips of a four-point subdiagram.  Thus there must be a
non-trivial mathematical structure on the set~$T_5$ with the
relation~$\circ$.  This structure can be visualized by the graph in
figure~\ref{fig:5phi3}.  Similarly, there are 25~five-point tree
diagrams in combined~$\phi^3$ and~$\phi^4$ and at most six can be
reached from any diagram by a flip.

In the same way, the trivial relation on~$T_4$ has a non-trivial
natural extension to the set~$T_n$ of all $n$-point tree diagrams:
$t\circ t'$ is true if and only if~$t$ and~$t'$ are identical up to a
single flip of a four-point subdiagram
\begin{equation}
   t\circ t' \Longleftrightarrow
     \exists t_4\in T_4,t'_4\in T_4:
       t_4 \circ t'_4 \land t\setminus t_4 = t'\setminus t'_4
\end{equation}
Note that this relation is not transitive and therefore not an
equivalence relation.  Instead, this relation allows us to view the
elements of~$T_n$ as the vertices
of a graph~$F_n$, where the edges of the graph are formed by the pairs
of diagrams related by a single flip
\begin{equation}
\label{eq:forest}
  F_n = \bigl\{(t,t')\in T_n\times T_n \bigl| t\circ t'\bigl\}\,.
\end{equation}
To avoid confusion, we will refer to graph~$F_n$ as \emph{forest} and
to its vertices as Feynman \emph{diagrams}.

The most important property of the forests for our applications is
\begin{theorem}
\label{th:vanilla}
  The unflavored forest~$F_n$ is connected for all~$n$.
\end{theorem}
which is easily proved by mathematical induction on~$n$.  Indeed, all
members of~$T_{n+1}$ can be derived from~$T_n$ by connecting a new
external line to a propagator (forming a new $\phi^3$-vertex) or an existing
$\phi^3$-vertex (transforming it into $\phi^4$-vertex).  Obviously,
two such insertions are related by a finite number
of flips, which proves the induction step.  This theorem shows that it
is possible to construct all Feynman diagrams by visiting the nodes
of~$F_n$ along successive applications of the flips
in~(\ref{eq:flips}).

Already the simplest non-trivial example of a unflavored (``vanilla'')
forest, the 15~tree diagrams with five external particles in
unflavored $\phi^3$-theory, as shown in figure~\ref{fig:5phi3},
displays an intriguing symmetry structure: there are 120 permutations
of the vertices of~$F_5$ that leave this forest invariant.  The
automorphism group~$\Aut(F_5)$ is therefore with 120 elements much
larger than one would
expect.  The classification of vanilla forests is probably an
interesting mathematical problem, but without obvious physical
applications.  However, as we will see below, not all flips are
created equal and only some flips are required by gauge invariance,
leading to a partition of the forest into gauge invariant subsets,
called \emph{groves}~\cite{Boos/Ohl:1999:groves}.

\subsubsection{Vanilla Trees}
\label{sec:vanilla}

\begin{empcmds}
  vardef bb =
    setbounds currentpicture to (0,-12pt)--(0,h)--(w,h)--(w,-12pt)--cycle
  enddef;
\end{empcmds}
\begin{empcmds}
  def tree_setup_one_two = 
    pair n[][];
    n[0][0] := (1/2w,3/3h);
    n[1][0] := (1/2w,2/3h);
    n[2][0] := (1/4w,1/3h);
    n[2][1] := (3/4w,1/3h);
    n[3][0] := (2/4w,0/3h);
    n[3][1] := (4/4w,0/3h);
    pickup pencircle scaled 1pt;
    draw n[0][0]--n[1][0];
    draw n[1][0]--n[2][0];
    draw n[1][0]--n[2][1];
    draw n[2][1]--n[3][0];
    draw n[2][1]--n[3][1];
  enddef;
\end{empcmds}
\begin{empdef}[treeabc](10,12)
  tree_setup_one_two;
  label.bot (btex $\vphantom{abc}a$ etex, n[2][0]);
  label.bot (btex $\vphantom{abc}b$ etex, n[3][0]);
  label.bot (btex $\vphantom{abc}c$ etex, n[3][1]);
  bb;
\end{empdef}
\begin{empdef}[treecab](10,12)
  tree_setup_one_two;
  label.bot (btex $\vphantom{abc}c$ etex, n[2][0]);
  label.bot (btex $\vphantom{abc}a$ etex, n[3][0]);
  label.bot (btex $\vphantom{abc}b$ etex, n[3][1]);
  bb;
\end{empdef}
\begin{empdef}[treebac](10,12)
  tree_setup_one_two;
  label.bot (btex $\vphantom{abc}b$ etex, n[2][0]);
  label.bot (btex $\vphantom{abc}a$ etex, n[3][0]);
  label.bot (btex $\vphantom{abc}c$ etex, n[3][1]);
  bb;
\end{empdef}
\begin{empdef}[treebca](10,12)
  tree_setup_one_two;
  label.bot (btex $\vphantom{abc}b$ etex, n[2][0]);
  label.bot (btex $\vphantom{abc}c$ etex, n[3][0]);
  label.bot (btex $\vphantom{abc}a$ etex, n[3][1]);
  bb;
\end{empdef}
\begin{empcmds}
  def tree_setup_one_two_rev = 
    pair n[][];
    n[0][0] := (1/2w,3/3h);
    n[1][0] := (1/2w,2/3h);
    n[2][0] := (1/4w,1/3h);
    n[2][1] := (3/4w,1/3h);
    n[3][0] := (0/4w,0/3h);
    n[3][1] := (2/4w,0/3h);
    pickup pencircle scaled 1pt;
    draw n[0][0]--n[1][0];
    draw n[1][0]--n[2][0];
    draw n[1][0]--n[2][1];
    draw n[2][0]--n[3][0];
    draw n[2][0]--n[3][1];
  enddef;
\end{empcmds}
\begin{empdef}[treebcar](10,12)
  tree_setup_one_two_rev;
  label.bot (btex $\vphantom{abc}b$ etex, n[3][0]);
  label.bot (btex $\vphantom{abc}c$ etex, n[3][1]);
  label.bot (btex $\vphantom{abc}a$ etex, n[2][1]);
  bb;
\end{empdef}
\begin{empcmds}
  def tree_setup_three = 
    pair n[][];
    n[0][0] := (1/2w,3/3h);
    n[1][0] := (1/2w,2/3h);
    n[2][0] := (0/2w,1/3h);
    n[2][1] := (1/2w,1/3h);
    n[2][2] := (2/2w,1/3h);
    pickup pencircle scaled 1pt;
    draw n[0][0]--n[1][0];
    draw n[1][0]--n[2][0];
    draw n[1][0]--n[2][1];
    draw n[1][0]--n[2][2];
  enddef;
\end{empcmds}
\begin{empdef}[treeabc3](10,12)
  tree_setup_three;
  label.bot (btex $\vphantom{abc}a$ etex, n[2][0]);
  label.bot (btex $\vphantom{abc}b$ etex, n[2][1]);
  label.bot (btex $\vphantom{abc}c$ etex, n[2][2]);
  bb;
\end{empdef}
\begin{empcmds}
  def tree_setup_four = 
    pair n[][];
    n[0][0] := (1/2w,4/4h);
    n[1][0] := (1/2w,3/4h);
    n[2][0] := (1/5w,2/4h);
    n[2][1] := (4/5w,2/4h);
    n[3][0] := (0/3w,1/4h);
    n[3][1] := (1/3w,1/4h);
    n[3][2] := (2/3w,1/4h);
    n[3][3] := (3/3w,1/4h);
    pickup pencircle scaled 1pt;
    draw n[0][0]--n[1][0];
    draw n[1][0]--n[2][0];
    draw n[1][0]--n[2][1];
    draw n[2][0]--n[3][0];
    draw n[2][0]--n[3][1];
    draw n[2][1]--n[3][2];
    draw n[2][1]--n[3][3];
  enddef;
\end{empcmds}
\begin{empdef}[treeabcd4](15,15)
  tree_setup_four;
  label.bot (btex $\vphantom{abcd}a$ etex, n[3][0]);
  label.bot (btex $\vphantom{abcd}b$ etex, n[3][1]);
  label.bot (btex $\vphantom{abcd}c$ etex, n[3][2]);
  label.bot (btex $\vphantom{abcd}d$ etex, n[3][3]);
  bb;
\end{empdef}
\begin{empcmds}
  def tree_setup_one_one_two = 
    pair n[][];
    n[0][0] := (1/2w,4/4h);
    n[1][0] := (1/2w,3/4h);
    n[2][0] := (1/3w,2/4h);
    n[2][1] := (2/3w,2/4h);
    n[3][0] := (3/6w,1/4h);
    n[3][1] := (5/6w,1/4h);
    n[4][0] := (2/3w,0/4h);
    n[4][1] := (3/3w,0/4h);
    pickup pencircle scaled 1pt;
    draw n[0][0]--n[1][0];
    draw n[1][0]--n[2][0];
    draw n[1][0]--n[2][1];
    draw n[2][1]--n[3][0];
    draw n[2][1]--n[3][1];
    draw n[3][1]--n[4][0];
    draw n[3][1]--n[4][1];
  enddef;
\end{empcmds}
\begin{empdef}[treeabcd](15,15)
  tree_setup_one_one_two;
  label.bot (btex $\vphantom{abcd}a$ etex, n[2][0]);
  label.bot (btex $\vphantom{abcd}b$ etex, n[3][0]);
  label.bot (btex $\vphantom{abcd}c$ etex, n[4][0]);
  label.bot (btex $\vphantom{abcd}d$ etex, n[4][1]);
  bb;
\end{empdef}
\begin{empdef}[treecdab](15,15)
  tree_setup_one_one_two;
  label.bot (btex $\vphantom{abcd}c$ etex, n[2][0]);
  label.bot (btex $\vphantom{abcd}d$ etex, n[3][0]);
  label.bot (btex $\vphantom{abcd}a$ etex, n[4][0]);
  label.bot (btex $\vphantom{abcd}b$ etex, n[4][1]);
  bb;
\end{empdef}
\begin{empcmds}
  def tree_setup_one_one_two_rev = 
    pair n[][];
    n[0][0] := (1/2w,4/4h);
    n[1][0] := (1/2w,3/4h);
    n[2][0] := (1/4w,2/4h);
    n[2][1] := (3/4w,2/4h);
    n[3][0] := (2/4w,1/4h);
    n[3][1] := (4/4w,1/4h);
    n[4][0] := (1/3w,0/4h);
    n[4][1] := (2/3w,0/4h);
    pickup pencircle scaled 1pt;
    draw n[0][0]--n[1][0];
    draw n[1][0]--n[2][0];
    draw n[1][0]--n[2][1];
    draw n[2][1]--n[3][0];
    draw n[2][1]--n[3][1];
    draw n[3][0]--n[4][0];
    draw n[3][0]--n[4][1];
  enddef;
\end{empcmds}
\begin{empdef}[treebcdar](15,15)
  tree_setup_one_one_two_rev;
  label.bot (btex $\vphantom{abcd}b$ etex, n[2][0]);
  label.bot (btex $\vphantom{abcd}c$ etex, n[4][0]);
  label.bot (btex $\vphantom{abcd}d$ etex, n[4][1]);
  label.bot (btex $\vphantom{abcd}a$ etex, n[3][1]);
  bb;
\end{empdef}
\begin{empdef}[treedabcr](15,15)
  tree_setup_one_one_two_rev;
  label.bot (btex $\vphantom{abcd}d$ etex, n[2][0]);
  label.bot (btex $\vphantom{abcd}a$ etex, n[4][0]);
  label.bot (btex $\vphantom{abcd}b$ etex, n[4][1]);
  label.bot (btex $\vphantom{abcd}c$ etex, n[3][1]);
  bb;
\end{empdef}
\begin{empcmds}
  def tree_setup_two_two = 
    pair n[][];
    n[0][0] := (1/2w,3/3h);
    n[1][0] := (1/2w,2/3h);
    n[2][0] := (1/6w,1/3h);
    n[2][1] := (3/6w,1/3h);
    n[2][2] := (4/5w,1/3h);
    n[3][0] := (2/3w,0/3h);
    n[3][1] := (3/3w,0/3h);
    pickup pencircle scaled 1pt;
    draw n[0][0]--n[1][0];
    draw n[1][0]--n[2][0];
    draw n[1][0]--n[2][1];
    draw n[1][0]--n[2][2];
    draw n[2][2]--n[3][0];
    draw n[2][2]--n[3][1];
  enddef;
\end{empcmds}
\begin{empdef}[treeabcd22](20,15)
  tree_setup_two_two;
  label.bot (btex $\vphantom{abcd}a$ etex, n[2][0]);
  label.bot (btex $\vphantom{abcd}b$ etex, n[2][1]);
  label.bot (btex $\vphantom{abcd}c$ etex, n[3][0]);
  label.bot (btex $\vphantom{abcd}d$ etex, n[3][1]);
  bb;
\end{empdef}

The enumeration of tree diagrams is simplified considerably if we can
find a \emph{unique} representative for each topological equivalence
class~\cite{Ohl:1999:bocages}.  One such
representation is defined by
picking one external line as the root and drawing the reminder of the
diagram as a mathematical tree.  The edges at each node are sorted
according to the index of the external lines beneath each node.  The
sorting lifts the remaining degeneracy and there remains only one
representative for each topological equivalence class. To be specific,
here is how the members of~$T_4$ are represented
\begin{subequations}
\label{eq:t4}
\begin{align}
  \parbox{25\unitlength}{%
    \fmfframe(2,6)(2,4){%
      \begin{fmfgraph*}(20,10)
        \fmfleft{e1,a}
        \fmflabel{$1$}{e1}
        \fmflabel{$a$}{a}
        \fmfright{c,b}
        \fmflabel{$b$}{b}
        \fmflabel{$c$}{c}
        \fmf{plain,tension=0.5}{v,v'}
        \fmf{plain}{e1,v,a}
        \fmf{plain}{b,v',c}
        \fmfdot{v,v'}
      \end{fmfgraph*}}}
  &\Longleftrightarrow
  \parbox{11\unitlength}{\empuse{treeabc}} \\
  \parbox{25\unitlength}{%
    \fmfframe(2,6)(2,4){%
      \begin{fmfgraph*}(20,10)
        \fmfleft{e1,a}
        \fmflabel{$1$}{e1}
        \fmflabel{$a$}{a}
        \fmfright{c,b}
        \fmflabel{$b$}{b}
        \fmflabel{$c$}{c}
        \fmf{plain,tension=0.5}{v,v'}
        \fmf{plain}{e1,v,c}
        \fmf{plain}{a,v',b}
        \fmfdot{v,v'}
      \end{fmfgraph*}}}
  &\Longleftrightarrow
  \parbox{11\unitlength}{\empuse{treecab}} \\
  \parbox{25\unitlength}{%
    \fmfframe(2,6)(2,4){%
      \begin{fmfgraph*}(20,10)
        \fmfleft{e1,a}
        \fmflabel{$1$}{e1}
        \fmflabel{$a$}{a}
        \fmfright{c,b}
        \fmflabel{$b$}{b}
        \fmflabel{$c$}{c}
        \fmf{plain,tension=0.5}{v,v'}
        \fmf{phantom}{e1,v,c}
        \fmf{phantom}{a,v',b}
        \fmffreeze
        \fmfdot{v,v'}
        \fmf{plain}{e1,v,b}
        \fmf{plain,rubout}{a,v',c}
      \end{fmfgraph*}}}
  &\Longleftrightarrow
  \parbox{11\unitlength}{\empuse{treebac}} \\
  \parbox{25\unitlength}{%
    \fmfframe(2,6)(2,4){%
      \begin{fmfgraph*}(20,10)
        \fmfleft{e1,a}
        \fmflabel{$1$}{e1}
        \fmflabel{$a$}{a}
        \fmfright{c,b}
        \fmflabel{$b$}{b}
        \fmflabel{$c$}{c}
        \fmf{plain}{e1,v,c}
        \fmf{plain}{a,v,b}
        \fmffreeze
        \fmfdot{v,v}
      \end{fmfgraph*}}}
  &\Longleftrightarrow
  \parbox{11\unitlength}{\empuse{treeabc3}}
\end{align}
\end{subequations}
In this representation\footnote{\protect\label{fn:parenthesis}%
Instead of the graphical representation used
in~(\protect\ref{eq:t4}---\protect\ref{eq:t5-match''}), a notation
based on parenthesis has been used in figure~\protect\ref{fig:5phi3}.
In this less intuitive but more concise notation, the four trees in
(\protect\ref{eq:t4}) are denoted by~$a(bc)$, $c(ab)$, $b(ac)$,
and~$abc$, respectively.  The matching in~(\protect\ref{eq:t5-match'})
is written $(ab)(cd)\Rightarrow\{a(b(cd)),b((cd)a),c(d(ab)),d((ab)c)\}$.},
the flips are equivalent to a recursive \emph{tree pattern matching}
\begin{subequations}
\label{eq:t5-match}
\begin{equation}
  \parbox{11\unitlength}{\empuse{treeabc}} \lor
  \parbox{11\unitlength}{\empuse{treebcar}} \lor
  \parbox{11\unitlength}{\empuse{treeabc3}}
  \Rightarrow
  \left\{
    \parbox{11\unitlength}{\empuse{treeabc}},
    \parbox{11\unitlength}{\empuse{treebca}},
    \parbox{11\unitlength}{\empuse{treecab}},
    \parbox{11\unitlength}{\empuse{treeabc3}}
  \right\}
\end{equation}
where each of $a$, $b$, $c$ can be either a complete subtree or a
single leaf.  In these pattern matchings, a subtle special case in
which two eligible vertices meet, has to be taken care of
\begin{equation}
\label{eq:t5-match'}
  \parbox{16\unitlength}{\empuse{treeabcd4}}
  \Rightarrow
  \left\{
    \parbox{16\unitlength}{\empuse{treeabcd}},
    \parbox{16\unitlength}{\empuse{treebcdar}},
    \parbox{16\unitlength}{\empuse{treecdab}},
    \parbox{16\unitlength}{\empuse{treedabcr}}
  \right\}
\end{equation}
\end{subequations}
in order not to miss a branch.  However, cases like
\begin{equation}
\label{eq:t5-match''}
  \parbox{21\unitlength}{\empuse{treeabcd22}}
\end{equation}
need \emph{no} special rules because the root vertex will only be
expanded and never be contracted with another vertex.

This unique representation of tree diagrams provides an efficient
equality predicate and the recursive pattern
matching provides a function calculating the neighbors
\begin{equation}
  \phi(t) = \{t'\in T| t'\circ t\}
\end{equation}
of any tree~$t$ in the forest.  Using these two ingredients, the
construction of the forest is a well defined mathematical problem with
efficient textbook solutions, known as \emph{depth first search} and
\emph{breadth first search}.  Essentially, these solutions consist of
recursively following the edges to the neighbors until all nodes have
been visited, keeping track of the visited nodes to avoid loops.

If the forest is connected (from theorem~\ref{th:vanilla}, the vanilla
forests are known to be connected) and a single tree is known, the
construction of the forest is also an efficient algorithm for
generating all Feynman tree diagrams.

\subsection{Flavored Forests}
\label{sec:flavored}

In physics applications, we have to deal with multiple flavors of
particles.  Therefore we introduce \emph{flavored forests}, where the
admissibility of elementary flips~$t\circ t'$ depends on the four
particles involved through the Feynman rules for the vertices in~$t_4$
and~$t'$.

In order to simplify the combinatorics when counting diagrams for
theories with more than one flavor, we will below treat \emph{all}
external particles as outgoing.  The physical amplitudes are obtained
later by selecting the incoming particles in the end.  Ward
identities, etc.~will be proved for the latter physical amplitudes, of
course.

\subsubsection{Flavored Trees}

\emph{Flavored trees} are derived from vanilla trees by adding a label
to each edge.  Feynman rules are then used at each node to identify
valid flavor combinations.

\begin{empcmds}
  def tree_setup_two = 
    pair n[][];
    n[0][0] := (1/2w,2/3h);
    n[1][0] := (1/2w,1/3h);
    n[2][0] := (1/6w,0/3h);
    n[2][1] := (5/6w,0/3h);
    pickup pencircle scaled 1pt;
    draw n[0][0]--n[1][0];
    draw n[1][0]--n[2][0];
    draw n[1][0]--n[2][1];
  enddef;
\end{empcmds}
\begin{empdef}[treeuubar](10,12)
  tree_setup_two;
  label.bot (btex $\vphantom{\bar u}u$ etex, n[2][0]);
  label.bot (btex $\vphantom{\bar u}\bar u$ etex, n[2][1]);
  bb;
\end{empdef}
\begin{empdef}[treeuubargamma](10,12)
  tree_setup_two;
  label.top (btex $\gamma$ etex, n[0][0]);
  label.bot (btex $\vphantom{\bar u}u$ etex, n[2][0]);
  label.bot (btex $\vphantom{\bar u}\bar u$ etex, n[2][1]);
  bb;
\end{empdef}
\begin{empdef}[treeuubarZ](10,12)
  tree_setup_two;
  label.top (btex $Z^0$ etex, n[0][0]);
  label.bot (btex $\vphantom{\bar u}u$ etex, n[2][0]);
  label.bot (btex $\vphantom{\bar u}\bar u$ etex, n[2][1]);
  bb;
\end{empdef}
\begin{empdef}[treeuubarglue](10,12)
  tree_setup_two;
  label.top (btex $g$ etex, n[0][0]);
  label.bot (btex $\vphantom{\bar u}u$ etex, n[2][0]);
  label.bot (btex $\vphantom{\bar u}\bar u$ etex, n[2][1]);
  bb;
\end{empdef}

All flavored trees can be generated from vanilla trees (which
correspond to ``topologies'') by a recursive tensor product of
elementary alternatives. E.\,g.
\begin{equation}
  \parbox{11\unitlength}{\empuse{treeuubar}}
  \mapsto
  \left\{
    \parbox{11\unitlength}{\empuse{treeuubargamma}},
    \parbox{11\unitlength}{\empuse{treeuubarZ}},
    \parbox{11\unitlength}{\empuse{treeuubarglue}}
  \right\}\,.
\end{equation}
The representation described in section~\ref{sec:vanilla} allows to
use efficient tree homomorphisms for the necessary manipulations.  In
particular, the calculation of the algebraic expressions corresponding
to flavored trees is a tree homomorphism:
\begin{equation}
  \parbox{11\unitlength}{\empuse{treeuubargamma}}
  \mapsto
   \bar u(p_1) \gamma_\mu u(p_2)\,.
\end{equation}
If there were no Feynman rules, the flavored forests would be
connected just like the vanilla forests.  However, the Feynman rules
typically prohibit most of the possible flavor assignments and it
turns out that the remaining vertices form more than one connected
component.

\subsection{Groves}
The construction of the groves is based on the observation that the
flips in gauge theories fall into two different classes: the
\emph{flavor flips} among
\begin{equation}
\label{eq:F}
  \{t_4^{F,1},t_4^{F,2},t_4^{F,3}\} = \left\{
   \parbox{16\unitlength}{%
     \begin{fmfgraph}(15,12)
       \fmfleft{f1,f2}
       \fmfright{f1',f2'}
       \fmf{fermion}{f1,v,f2}
       \fmf{fermion}{f2',v',f1'}
       \fmf{boson,tension=0.5}{v,v'}
       \fmfdot{v,v'}
     \end{fmfgraph}},
   \parbox{16\unitlength}{%
     \begin{fmfgraph}(15,12)
       \fmfleft{f1,f2}
       \fmfright{f1',f2'}
       \fmf{fermion}{f1,v1,f1'}
       \fmf{fermion}{f2',v2,f2}
       \fmf{boson,tension=0.5}{v1,v2}
       \fmfdot{v1,v2}
     \end{fmfgraph}},
   \parbox{16\unitlength}{%
     \begin{fmfgraph}(15,12)
       \fmfleft{f1,f2}
       \fmfright{f1',f2'}
       \fmf{phantom}{f1,v1,f1'}
       \fmf{phantom}{f2',v2,f2}
       \fmf{boson,tension=0.5}{v1,v2}
       \fmfdot{v1,v2}
       \fmffreeze
       \fmf{fermion}{f1,v1,f2'}
       \fmf{fermion,rubout}{f1',v2,f2}
     \end{fmfgraph}}
   \right\}\,,
\end{equation}
which involve four matter fields that carry gauge
charge and possibly additional conserved quantum numbers
and the \emph{gauge flips} among
\begin{subequations}
\begin{equation}
\label{eq:G}
  \{t_4^{G,1},t_4^{G,2},t_4^{G,3},t_4^{G,4}\} = \left\{
   \parbox{16\unitlength}{%
     \begin{fmfgraph}(15,12)
       \fmfleft{g1,g2}
       \fmfright{g1',g2'}
       \fmf{boson}{g1,v,g2}
       \fmf{boson}{g1',v',g2'}
       \fmf{boson,tension=0.5}{v,v'}
       \fmfdot{v,v'}
     \end{fmfgraph}},
   \parbox{16\unitlength}{%
     \begin{fmfgraph}(15,12)
       \fmfleft{g1,g2}
       \fmfright{g1',g2'}
       \fmf{boson}{g1,v1,g1'}
       \fmf{boson}{g2,v2,g2'}
       \fmf{boson,tension=0.5}{v1,v2}
       \fmfdot{v1,v2}
     \end{fmfgraph}},
   \parbox{16\unitlength}{%
     \begin{fmfgraph}(15,12)
       \fmfleft{g1,g2}
       \fmfright{g1',g2'}
       \fmf{phantom}{g1,v1,g1'}
       \fmf{phantom}{g2,v2,g2'}
       \fmf{boson,tension=0.5}{v1,v2}
       \fmfdot{v1,v2}
       \fmffreeze
       \fmf{boson}{g1,v1,g2'}
       \fmf{boson,rubout}{g2,v2,g1'}
     \end{fmfgraph}},
   \parbox{16\unitlength}{%
     \begin{fmfgraph}(15,12)
       \fmfleft{g1,g2}
       \fmfright{g1',g2'}
       \fmf{boson}{g1,v,g1'}
       \fmf{boson}{g2,v,g2'}
       \fmfdot{v}
     \end{fmfgraph}}
    \right\}
\end{equation}
and
\begin{equation}
\label{eq:C}
  \{t_4^{G,5},t_4^{G,6},t_4^{G,7},t_4^{G,8}\} = \left\{
   \parbox{16\unitlength}{%
     \begin{fmfgraph}(15,12)
       \fmfright{g,g'}
       \fmfleft{f,f'}
       \fmf{fermion}{f,v}
       \fmf{fermion,tension=0.5}{v,v'}
       \fmf{fermion}{v',f'}
       \fmf{boson}{v,g}
       \fmf{boson}{v',g'}
       \fmfdot{v,v'}
     \end{fmfgraph}},
   \parbox{16\unitlength}{%
     \begin{fmfgraph}(15,12)
       \fmfright{g,g'}
       \fmfleft{f,f'}
       \fmf{fermion}{f,v}
       \fmf{fermion,tension=0.5}{v,v'}
       \fmf{fermion}{v',f'}
       \fmf{phantom}{v,g}
       \fmf{phantom}{v',g'}
       \fmffreeze
       \fmf{boson}{v',g}
       \fmf{boson,rubout}{v,g'}
       \fmfdot{v,v'}
     \end{fmfgraph}},
   \parbox{16\unitlength}{%
     \begin{fmfgraph}(15,12)
       \fmfright{g,g'}
       \fmfleft{f,f'}
       \fmf{fermion}{f,v,f'}
       \fmf{boson,tension=0.5}{v,v'}
       \fmf{boson}{v',g}
       \fmf{boson}{v',g'}
       \fmfdot{v,v'}
     \end{fmfgraph}},
   \parbox{16\unitlength}{%
     \begin{fmfgraph}(15,12)
       \fmfright{g,g'}
       \fmfleft{f,f'}
       \fmf{fermion}{f,v,f'}
       \fmf{boson}{v,g}
       \fmf{boson}{v,g'}
       \fmfdot{v}
     \end{fmfgraph}}
   \right\}\,
\end{equation}
\end{subequations}
which also involve external gauge bosons
(the diagram $t_4^{G,8}$, is only present for
scalar matter fields that appear in extensions of the~\ac{SM}:
SUSY partners, leptoquarks, etc.).  In gauge theories with more than
one factor, like the~\ac{SM}, the gauge flips are extended in
the obvious way to include all four-point functions with at least one
gauge-boson. Commuting gauge group factors lead to separate sets, of
course.  In spontaneously broken gauge theories, the Higgs and
Goldstone boson fields contribute additional flips, in which they are
treated like gauge bosons. Ghosts can be ignored at tree level.

As we have seen for the examples of Bhabha-scattering and $ud\to ud$
scattering in section~\ref{sec:FlavorSelectionRules}, the flavor
flips~(\ref{eq:F}) are special because they can be switched off
without spoiling gauge invariance by introducing a horizontal symmetry
that commutes with the gauge group.  Such a horizontal symmetry is
similar to the replicated generations in the~\ac{SM}, but if
three generations do not suffice, it can also be introduced
artificially.  This observation suggests to introduce two relations:
\begin{subequations}
\begin{equation}
  t\bullet t'  \Longleftrightarrow
    \text{$t$ and~$t'$ related by a \emph{gauge} flip}
\end{equation}
and
\begin{equation}
  t\circ t' \Longleftrightarrow
    \text{$t$ and~$t'$ related by a \emph{flavor or gauge} flip}.
\end{equation}
\end{subequations}
These two relations define two different graphs with the same set~$T(E)$ of
all Feynman diagrams as vertices:
\begin{subequations}
\begin{equation}
  F(E) = \bigl\{(t,t')\in T(E)\times T(E) \bigl| t\circ t'\bigl\}
\end{equation}
and
\begin{equation}
  G(E) = \bigl\{(t,t')\in T(E)\times T(E) \bigl| t\bullet t'\bigr\}.
\end{equation}
\end{subequations}
As we have seen above in Bhabha-scattering, it is in general not
possible to connect all pairs of diagrams in~$G(E)$ by a series of
gauge flips.  Thus there will be more than one connected component.
For brevity, we will continue to denote the \emph{flavor
forest}~$F(E)$ as the \emph{forest} of the external state~$E$ and we
will denote the connected components~$G_i(E)$ of the \emph{gauge
forest}~$G(E)$ as the \emph{groves} of~$E$.
Since~$t\bullet t'\Rightarrow t\circ t'$, we
have~$\bigcup_i G_i(E)=G(E)\subseteq F(E)$, i.\,e.~the groves are a
\emph{partition} of the gauge forest.
\begin{theorem}
  The forest~$F(E)$ for an external state~$E$
  consisting of gauge and matter fields is connected if the fields
  in~$E$ carry no conserved quantum numbers other than the gauge
  charges. The groves~$G_i(E)$ are the minimal
  gauge invariant classes of Feynman diagrams.
\end{theorem}
By construction, the
theorem is true for the four-point diagrams and we can use induction
on the number of external matter fields and gauge bosons. Since the
matter fields are carrying conserved charges, they can only be added
in pairs.

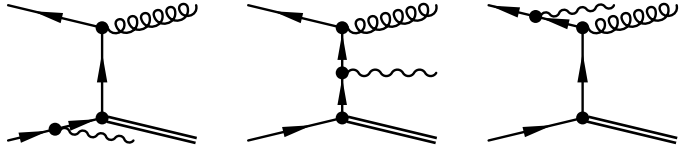
\begin{figure}
  \begin{center}
    \begin{fmfgraph}(25,18)
      \fmfleft{q,qbar}
      \fmfright{Z,G}
      \fmfbottom{d1,d2,g',d3}
      \fmf{phantom}{q,vZ}
      \fmf{phantom,tension=0.5}{vZ,vG}
      \fmf{phantom}{vG,qbar}
      \fmf{double}{vZ,Z}
      \fmf{gluon}{vG,G}
      \fmffreeze
      \fmf{fermion}{q,vg,vZ,vG,qbar}
      \fmf{photon,tension=0}{vg,g'}
      \fmfdot{vZ,vG,vg}
    \end{fmfgraph}
    \quad
    \begin{fmfgraph}(25,18)
      \fmfleft{q,qbar}
      \fmfright{Z,g',G}
      \fmf{phantom}{q,vZ}
      \fmf{phantom,tension=0.5}{vZ,vG}
      \fmf{phantom}{vG,qbar}
      \fmf{double}{vZ,Z}
      \fmf{gluon}{vG,G}
      \fmffreeze
      \fmf{fermion}{q,vZ,vg,vG,qbar}
      \fmf{photon,tension=0}{vg,g'}
      \fmfdot{vZ,vG,vg}
    \end{fmfgraph}
    \quad
    \begin{fmfgraph}(25,18)
      \fmfleft{q,qbar}
      \fmfright{Z,G}
      \fmftop{d1,d2,g',d3}
      \fmf{phantom}{q,vZ}
      \fmf{phantom,tension=0.5}{vZ,vG}
      \fmf{phantom}{vG,qbar}
      \fmf{double}{vZ,Z}
      \fmf{gluon}{vG,G}
      \fmffreeze
      \fmf{fermion}{q,vZ,vG,vg,qbar}
      \fmf{photon,tension=0}{vg,g'}
      \fmfdot{vZ,vG,vg}
    \end{fmfgraph}
  \end{center}
  \caption{\label{fig:NC-flips}%
    The three diagrams in~$q\bar q\to gZ^0\gamma$ are connected by
    successive gauge flips of subdiagrams like~(\protect\ref{eq:C}).
    If the decays of the~$Z^0$ were included, bremsstrahlung from the
    decay products would form a separate grove, because no gauge flip
    will pass the photon across the uncharged~$Z^0$.}
\end{figure}
If we add an additional external gauge boson to a gauge invariant
amplitude, the diagrammatical proof of the Ward and Slavnov-Taylor
identities in gauge theories requires us to sum over all ways to
attach a gauge boson to connected gauge charge carrying components of
the Feynman diagrams.  However, the gauge flips are connecting pairs of 
neighboring insertions and can be iterated along gauge charge carrying
propagators.  Therefore no partition of the forest~$F(E)$ that is
finer than the groves~$G_i(E)$ can preserve gauge invariance.  Two
examples are shown in figures~\ref{fig:NC-flips}
and~\ref{fig:CC-flips}.
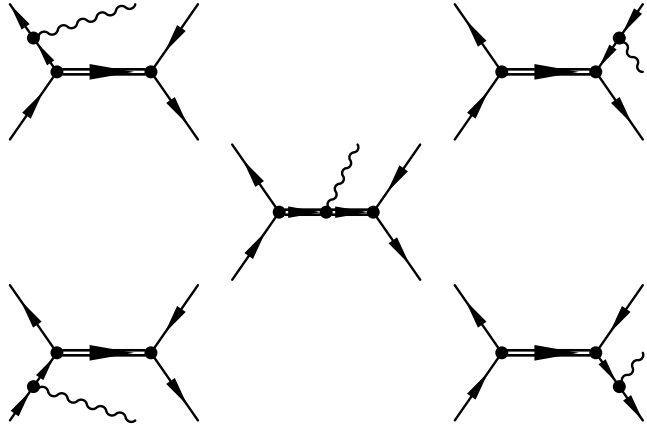
\begin{figure}
  \begin{center}
    \begin{fmfgraph}(25,18)
      \fmfleft{u,dbar}
      \fmfright{c,sbar}
      \fmftop{d1,d2,g,d3}
      \fmf{phantom}{u,v,dbar}
      \fmf{phantom}{v,v'}
      \fmf{phantom}{sbar,v',c}
      \fmffreeze
      \fmf{dbl_plain_arrow}{v,v'}
      \fmf{fermion}{u,v,vg,dbar}
      \fmf{fermion}{sbar,v',c}
      \fmf{photon,tension=0}{vg,g}
      \fmfdot{v,vg,v'}
    \end{fmfgraph}
    \hspace*{30\unitlength}
    \begin{fmfgraph}(25,18)
      \fmfleft{u,dbar}
      \fmfright{c,g,sbar}
      \fmf{phantom}{u,v,dbar}
      \fmf{phantom}{v,v'}
      \fmf{phantom}{sbar,v',c}
      \fmffreeze
      \fmf{dbl_plain_arrow}{v,v'}
      \fmf{fermion}{u,v,dbar}
      \fmf{fermion}{sbar,vg,v',c}
      \fmf{photon,tension=0}{vg,g}
      \fmfdot{v,vg,v'}
    \end{fmfgraph}\\
    \begin{fmfgraph}(25,18)
      \fmfleft{u,dbar}
      \fmfright{c,sbar}
      \fmftop{d1,d2,g,d3}
      \fmf{phantom}{u,v,dbar}
      \fmf{phantom}{v,v'}
      \fmf{phantom}{sbar,v',c}
      \fmffreeze
      \fmf{dbl_plain_arrow}{v,vg,v'}
      \fmf{fermion}{u,v,dbar}
      \fmf{fermion}{sbar,v',c}
      \fmf{photon,tension=0}{vg,g}
      \fmfdot{v,vg,v'}
    \end{fmfgraph}\\
    \begin{fmfgraph}(25,18)
      \fmfleft{u,dbar}
      \fmfright{c,sbar}
      \fmfbottom{d1,d2,g,d3}
      \fmf{phantom}{u,v,dbar}
      \fmf{phantom}{v,v'}
      \fmf{phantom}{sbar,v',c}
      \fmffreeze
      \fmf{dbl_plain_arrow}{v,v'}
      \fmf{fermion}{u,vg,v,dbar}
      \fmf{fermion}{sbar,v',c}
      \fmf{photon,tension=0}{vg,g}
      \fmfdot{v,vg,v'}
    \end{fmfgraph}
    \hspace*{30\unitlength}
    \begin{fmfgraph}(25,18)
      \fmfleft{u,dbar}
      \fmfright{c,g,sbar}
      \fmf{phantom}{u,v,dbar}
      \fmf{phantom}{v,v'}
      \fmf{phantom}{sbar,v',c}
      \fmffreeze
      \fmf{dbl_plain_arrow}{v,v'}
      \fmf{fermion}{u,v,dbar}
      \fmf{fermion}{sbar,v',vg,c}
      \fmf{photon,tension=0}{vg,g}
      \fmfdot{v,vg,v'}
    \end{fmfgraph}
  \end{center}
  \caption{\label{fig:CC-flips}%
    All five diagrams in $u\bar d\to c\bar s\gamma$ are in the same
    grove, because they are connected by gauge flips passing through
    the diagram in the center.  In contrast,
    in~$e^+e^-\to\mu^+\mu^-\gamma$
    (see figure~\protect\ref{fig:ee2mmg}), the center diagram is
    missing and there are two separate groves.}
\end{figure}

If we add an additional pair of matter fields to a gauge invariant
amplitude, we have to consider two separate cases
\begin{equation}
\label{eq:matter-pair}
   \parbox{26\unitlength}{%
     \begin{fmfgraph}(25,15)
       \fmfforce{(.2w,.2h)}{t1}
       \fmfforce{(.8w,.2h)}{t2}
       \fmfblob{.3h}{t1,t2}
       \fmfforce{(.2w,h)}{x1}
       \fmfforce{(.8w,h)}{x2}
       \fmf{fermion}{t2,t1}
     \end{fmfgraph}}\xrightarrow{n_f\to n_f+2}
   \left\{
   \parbox{26\unitlength}{%
     \begin{fmfgraph}(25,15)
       \fmfforce{(.2w,.2h)}{t1}
       \fmfforce{(.8w,.2h)}{t2}
       \fmfblob{.3h}{t1,t2}
       \fmfforce{(.2w,h)}{x1}
       \fmfforce{(.8w,h)}{x2}
       \fmf{fermion}{t2,v1,t1}
       \fmffreeze
       \fmf{fermion}{x1,v2,x2}
       \fmf{boson}{v1,v2}
       \fmfdot{v1,v2}
     \end{fmfgraph}},
   \parbox{26\unitlength}{%
     \begin{fmfgraph}(25,15)
       \fmfforce{(.2w,.2h)}{t1}
       \fmfforce{(.8w,.2h)}{t2}
       \fmfblob{.3h}{t1,t2}
       \fmfforce{(.2w,h)}{x1}
       \fmfforce{(.8w,h)}{x2}
       \fmf{fermion}{t2,v1,x2}
       \fmf{fermion}{x1,v2,t1}
       \fmf{boson}{v1,v2}
       \fmfdot{v1,v2}
     \end{fmfgraph}}
  \right\}.
\end{equation}
If the new flavor
does not already appear among the other matter fields, the only way
to attach the pair is through a gauge boson, as in the left diagram in
the braces in~(\ref{eq:matter-pair}).  If the new flavor is
already present, we can also break up a matter field propagator and
connect the two parts of the diagram with a new gauge propagator.
Since it is always possible to introduce a new flavor, either physical
or fictitious, without breaking gauge invariance, these cases
fork off separately gauge invariant classes every time we add a new
pair of matter fields.  On the other hand, the cases
in~(\ref{eq:matter-pair}) are related by a flavor flip.
Therefore~$F(E)$ remains connected, the~$G_i(E)$ are separately gauge
invariant and the proof is complete.

Earlier
attempts~\cite{Bardin/etal:1994:4f-classification,Boos/Ohl:1997:gg4f}
have used physical final states as a criterion for identifying gauge
invariant subsets.  We have already shown that the groves are minimal
and therefore never form a more coarse partition than the one derived
from a consideration of the final states alone.  Below we shall see
examples where the groves do indeed provide a strictly finer partition.

In a practical application, one calculates the groves for the
interesting combinations of gauge quantum numbers, such as weak
isospin and hypercharge in the~\ac{SM}, using an external state
where all other quantum numbers are equal, since the flavor forest
would otherwise be disconnected.  The combinatorics is simplified by
treating all particles as outgoing.  The physical amplitude is then
obtained by selecting the groves that are compatible with the other
quantum numbers of the process under consideration.  Concrete examples
are considered in the next section.

If only a few flavor combinations are interesting, it can be more
convenient in practical applications to generate all diagrams for the
external states conventionally and to partition the disconnected
flavor forest. The resulting groves will be the same, of course.

\subsection{Application}
\label{sec:application}

\begin{table}
  \caption{\label{tab:6f}%
    The groves for all processes with six external massless fermions
    and bremsstrahlung in the~\protect\acs{SM}
    (without~\protect\acs{QCD}). All particles are considered as
    outgoing.}
  \begin{center}
    \begin{tabular}{lrl}
      external fields ($E$) & \#diagrams & groves \\\hline
      $u \bar u u \bar u u \bar u$
        & $144$ & $18\cdot\mathbf{8}$ \\
      $u \bar u u \bar u u \bar u\gamma$
        & $1008$ & $18\cdot\mathbf{24}+36\cdot\mathbf{16}$ \\
      $u \bar u u \bar u d \bar d$
        & $92$ & $4\cdot\mathbf{11}+6\cdot\mathbf{8}$ \\
      $u \bar u u \bar u d \bar d\gamma$
        & $716$ & $4\cdot\mathbf{95}+6\cdot\mathbf{24}+12\cdot\mathbf{16}$ \\
      $\ell^+ \ell^- u \bar u d \bar d$
        & $35$ & $1\cdot\mathbf{11}+3\cdot\mathbf{8}$ \\
      $\ell^+ \ell^- u \bar u d \bar d\gamma$
        & $262$ & $1\cdot\mathbf{94}+3\cdot\mathbf{24}+6\cdot\mathbf{16}$ \\
      $\ell^+ \nu d \bar u d \bar d$
        & $20$ & $2\cdot\mathbf{10}$\\
      $\ell^+ \nu d \bar u d \bar d\gamma$
        & $152$ & $2\cdot\mathbf{76}$ \\
      $\ell^+ \ell^- \ell^+ \nu d \bar u$
        & $20$ & $2\cdot\mathbf{10}$\\
      $\ell^+ \ell^- \ell^+ \nu d \bar u\gamma$
        & $150$ & $2\cdot\mathbf{75}$ \\
      $\ell^+ \nu \ell^- \bar\nu d \bar d$
        & $19$ & $1\cdot\mathbf{9}+2\cdot\mathbf{4}+1\cdot\mathbf{2}$ \\
      $\ell^+ \nu \ell^- \bar\nu d \bar d\gamma$
        & $107$ & $1\cdot\mathbf{59}+2\cdot\mathbf{12}+2\cdot\mathbf{8}
                  +2\cdot\mathbf{4}$ \\
      $\ell^+ \nu \ell^- \bar\nu \ell^+ \ell^-$
        & $56$ & $4\cdot\mathbf{9}+4\cdot\mathbf{4}+2\cdot\mathbf{2}$ \\
      $\ell^+ \nu \ell^- \bar\nu \ell^+ \ell^-\gamma$
        & $328$ & $4\cdot\mathbf{58}+4\cdot\mathbf{12}+4\cdot\mathbf{8}
                  +4\cdot\mathbf{4}$ \\
      $\ell^+ \nu \ell^- \bar\nu \nu \bar\nu$
        & $36$ & $4\cdot\mathbf{6}+6\cdot\mathbf{2}$ \\
      $\ell^+ \nu \ell^- \bar\nu \nu \bar\nu\gamma$
        & $132$ & $4\cdot\mathbf{26}+2\cdot\mathbf{6}+4\cdot\mathbf{4}$ \\
      $\nu \bar\nu \nu \bar\nu \nu \bar\nu$
        & $36$ & $18\cdot\mathbf{2}$
    \end{tabular}
  \end{center}
\end{table}

In table~\ref{tab:6f}, we list the groves for all processes with six
external massless fermions in the~\ac{SM}, with and without
single photon bremsstrahlung, without~\ac{QCD} and CKM mixing.  We could
also include fermion masses, \ac{QCD}, and CKM mixing within the same
formalism, but the table would have to be much larger, because
additional gluon, Higgs and Goldstone
diagrams appear, depending on whether the fermions are massive or
massless, colored or uncolored. In the table, cases with
identical~$\textrm{SU}(2)_L\otimes \textrm{U}(1)_Y$ quantum numbers
are listed only once and cases with different~$T_3$ and~$Y$ are listed
separately only if the vanishing of the electric charge
removes diagrams from a grove.  E.\,g.~$e^+e^-e^+e^-e^+e^-$ is
identical to~$u\bar u u\bar u u\bar u$ and not listed separately.

The familiar non-minimal gauge invariant classes for $e^+e^-\to
f_1\bar f_2 f_3\bar f_4$ \cite{Bardin/etal:1994:4f-classification} are
included in table~\ref{tab:6f} as special cases.  The LEP2
$WW$-classes CC09, CC10, and~CC11 are immediately obvious.  As a not
quite so obvious example, the
process~$e^+e^-\to\mu^+\mu^-\tau^+\tau^-$ has the same
$\textrm{SU}(2)_L$~quantum numbers as~$u\bar u\to u\bar uu\bar u$.  We
can read off table~\ref{tab:6f} that, in the case of identical pairs,
there are 18~groves, of 8~diagrams each.  If all three pairs are
different, the number of groves has to be divided by~$3!$, because we
are no longer free to connect the three particle-antiparticle pairs
arbitrarily.  Thus there are 24~diagrams contributing to the
process~$e^+e^-\to\mu^+\mu^-\tau^+\tau^-$ and they are organized in
three groves of 8~diagrams each.  Any diagram in a grove can be
reached from the other 7 by exchanging the vertices of the gauge bosons
on one fermion line and be exchanging~$Z^0$ and~$\gamma$.
Since there are no non-abelian 
couplings in this process, the separate gauge invariance of each grove
could also be proven as in QED, by varying the hypercharge of each
particle: $A\propto A_1\cdot q_e^2 q_\mu q_\tau + A_2\cdot q_e q_\mu^2
q_\tau + A_3 \cdot q_e q_\mu q_\tau^2$.  

\begin{figure}
  \begin{center}
    \includegraphics[width=.9\textwidth]{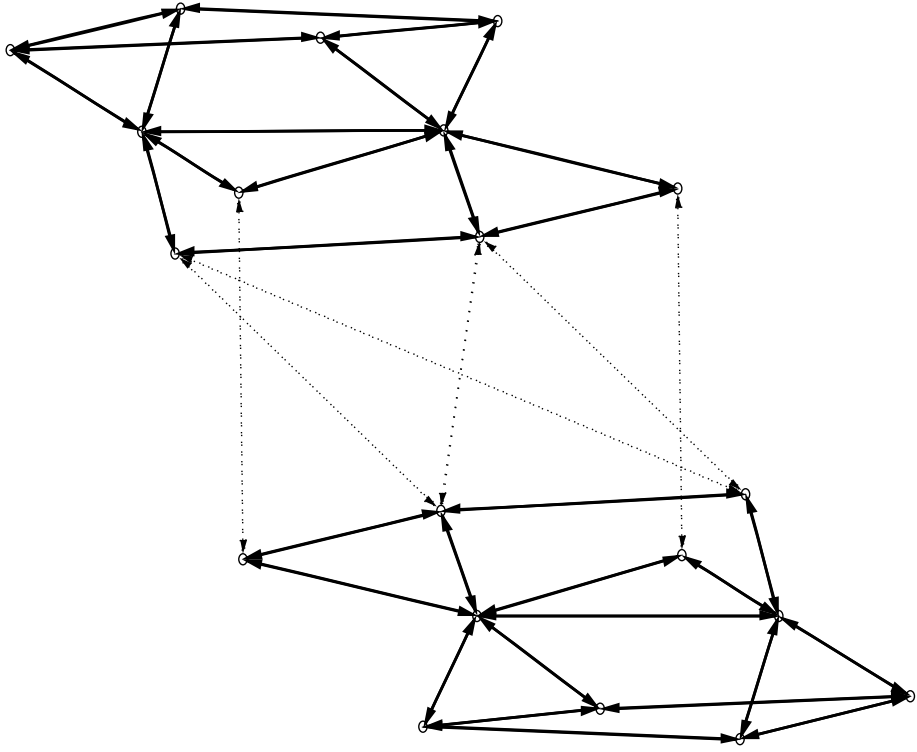}
  \end{center}
  \caption{\label{fig:cc20}%
    The two groves in the CC20 process $e^+e^-\to e^-\bar\nu_e u\bar d$.
    Gauge flips are drawn a solid lines and flavor flips are drawn as
    dotted lines.  These groves can be found in table~\ref{tab:6f}
    as~$\ell^+ \ell^- \ell^+ \nu d \bar u$.}
\end{figure}
As is well known, the diagrams for the CC20 process $e^+e^-\to
e^-\bar\nu_e u\bar d$ fall into two separately gauge invariant
classes, depicted in figure~\ref{fig:cc20} (including processes
related by crossing symmetry).
\begin{figure}
  \begin{center}
    \includegraphics[width=.4\textwidth]{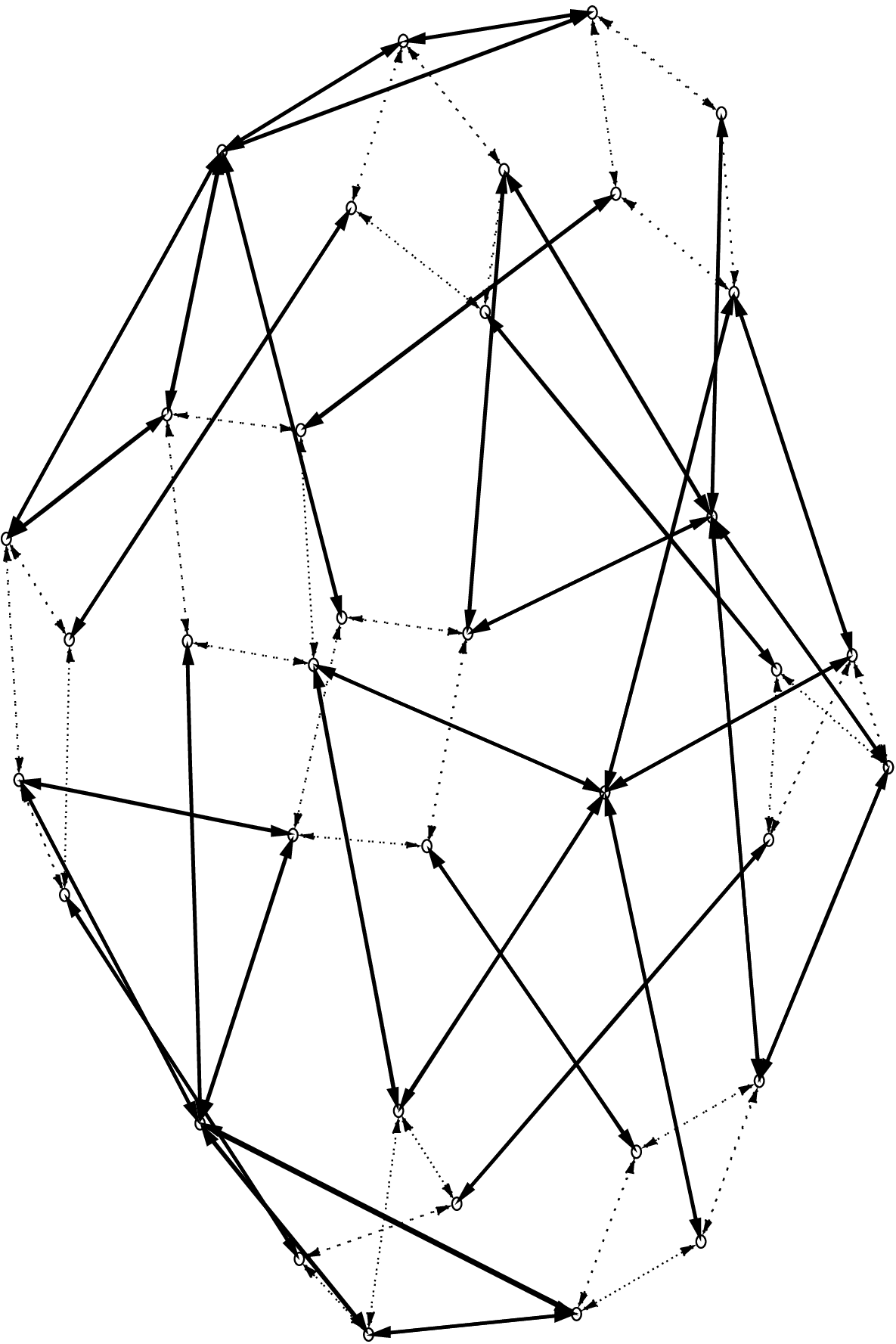}\quad
    \includegraphics[width=.4\textwidth]{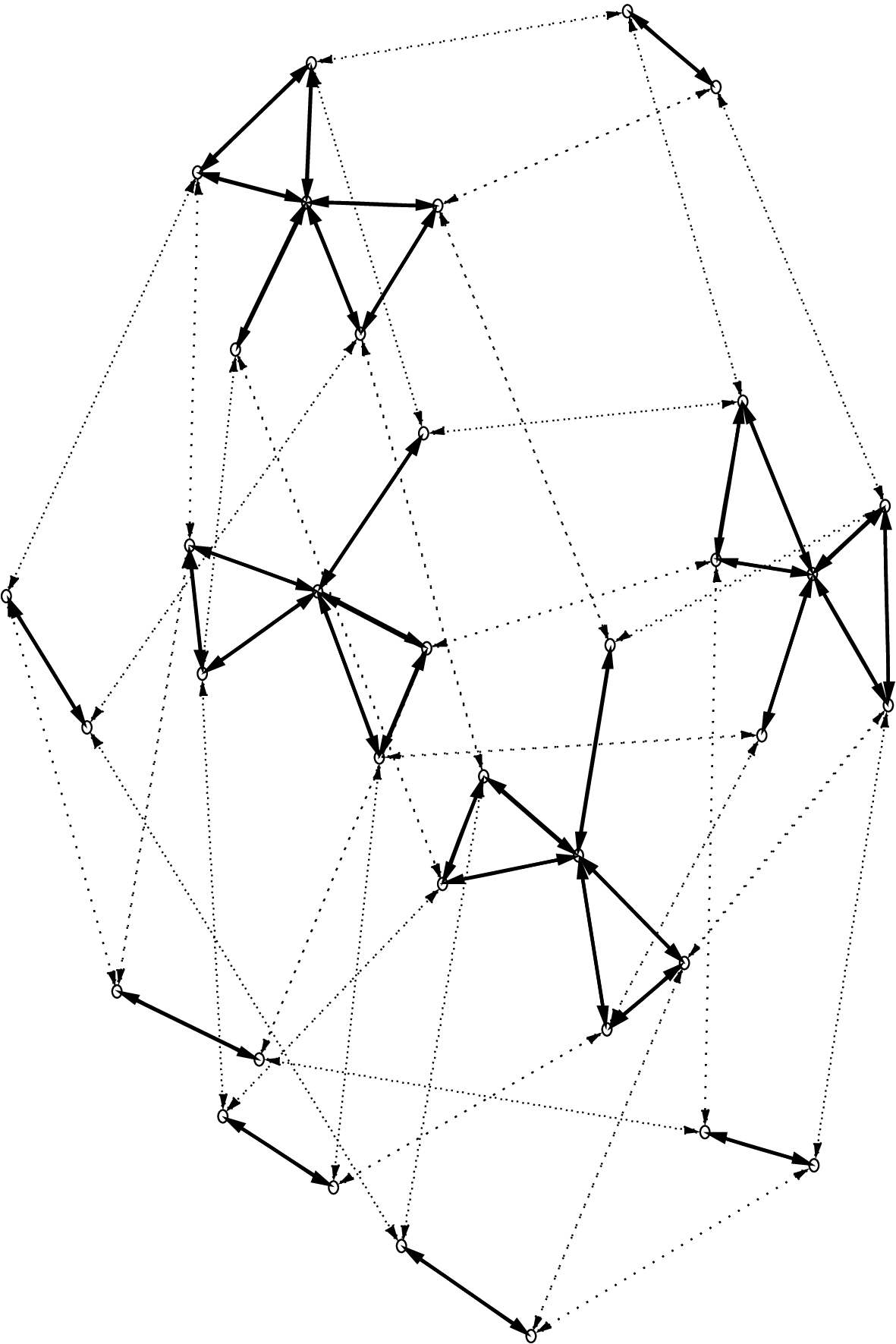}
  \end{center}
  \caption{\label{fig:nc06}%
    Two views of $e^+e^-\to \nu_e\bar\nu_e \nu_e\bar\nu_e$. Gauge
    flips are drawn a solid lines and flavor flips are drawn as dotted
    lines.}
\end{figure}
Figure~\ref{fig:nc06} shows two views of the groves for $e^+e^-\to
\nu_e\bar\nu_e \nu_e\bar\nu_e$ (including processes related by
crossing symmetry).  The left hand side emphasizes the four-fold
permutation symmetry of flavor flips, while the right hand side
emphasizes the groups of six diagrams forming a single grove.
\begin{figure}
  \begin{center}
    \includegraphics[width=.9\textwidth]{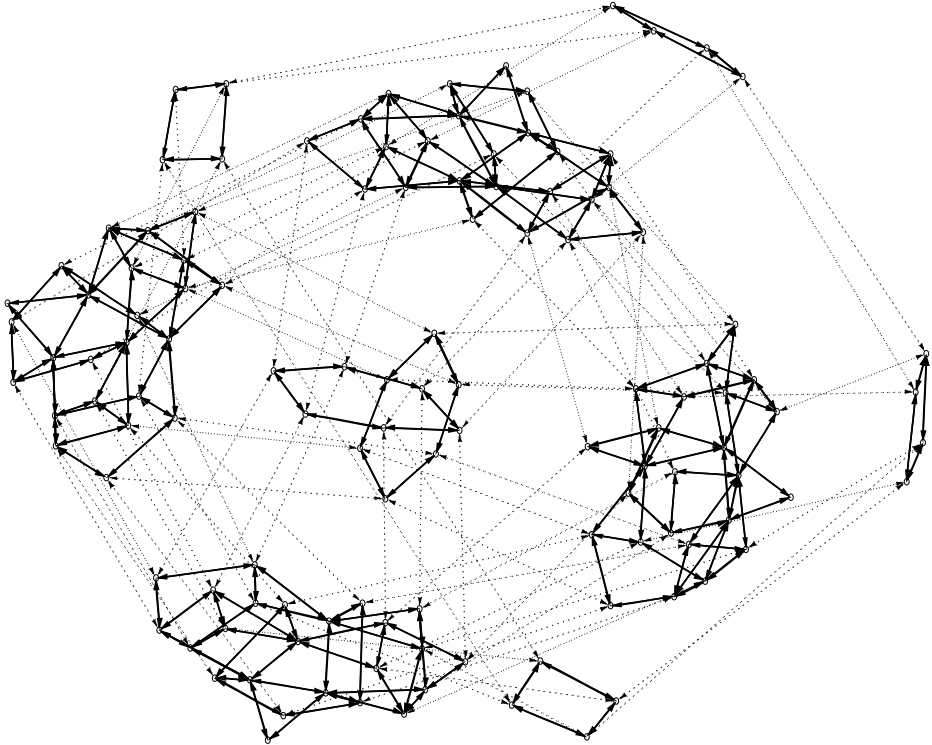}
  \end{center}
  \caption{\label{fig:nc06g}%
    The decomposition $4\cdot\mathbf{26} + 2\cdot\mathbf{6} + 4\cdot\mathbf{4}$
    of~$e^+e^-\to \nu\bar\nu\nu\bar\nu\gamma$. Gauge flips are drawn a
    solid lines and flavor flips are drawn as dotted lines.}
\end{figure}
The process $e^+e^-\to \nu\bar\nu\nu\bar\nu\gamma$ is not yet the most
pressing physics problem, but small enough to produce the nice picture
shown in figure~\ref{fig:nc06g}.  The decomposition $4\cdot\mathbf{26} +
2\cdot\mathbf{6} + 4\cdot\mathbf{4}$ is clearly visible.
\begin{figure}
  \begin{center}
    \includegraphics[width=.9\textwidth]{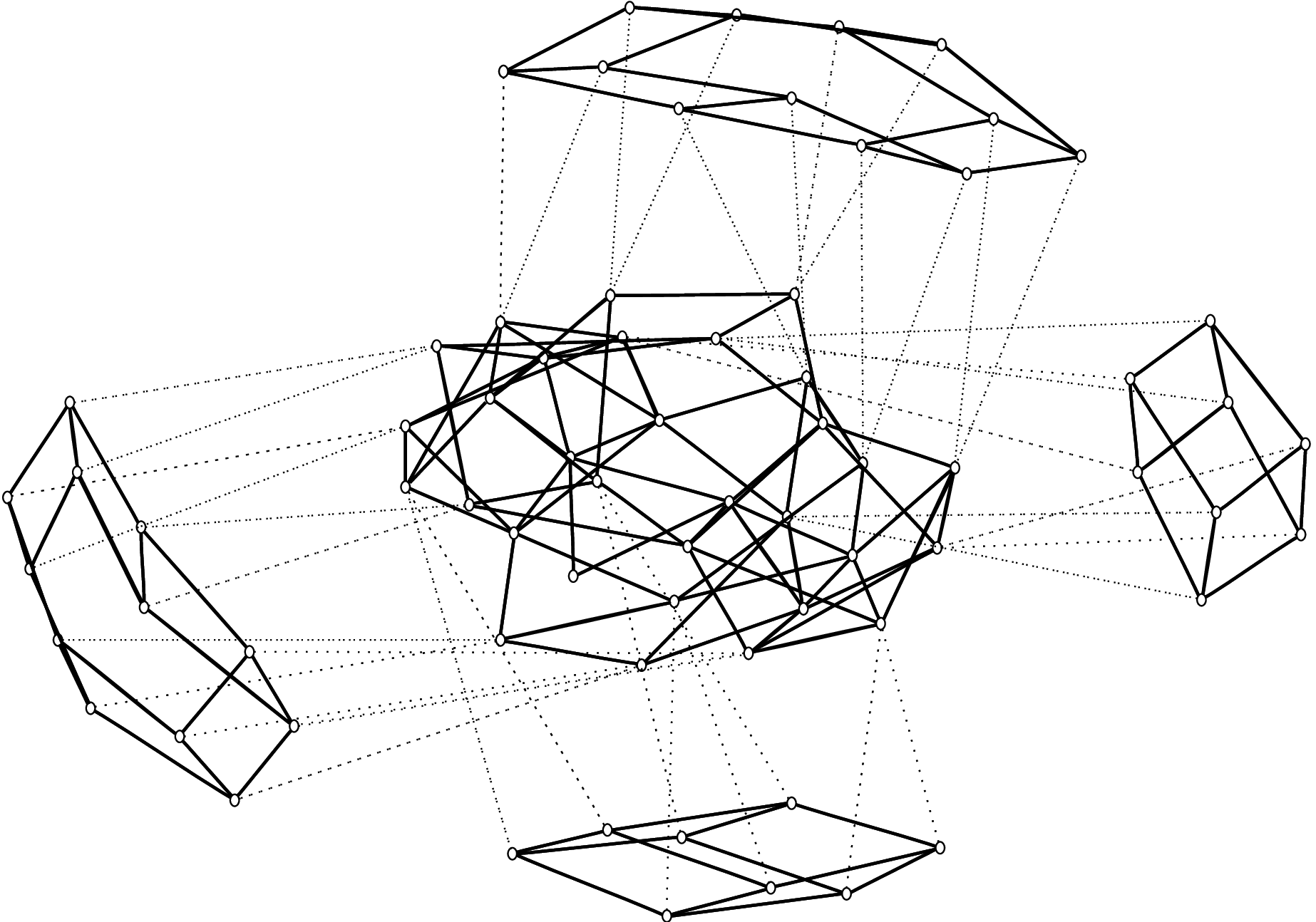}
  \end{center}
  \caption{\label{fig:mix71}%
    The forest of size~71 for the process~$\gamma\gamma\to u\bar d d\bar u$ in
    the~\protect\acs{SM} (without~\protect\acs{QCD}, CKM mixing and masses in
    unitarity gauge) with one grove of size~31, 
    two of size 12 and two of size 8. Solid lines
    represent gauge flips and dotted lines represent flavor flips.}
\end{figure}
Finally, figure~\ref{fig:mix71} displays the forest for the
process~$\gamma\gamma\to u\bar d d\bar u$ in the~\ac{SM}.
The grove in the center consists of the
31~diagrams with charged current interactions (the set CC21
of~\cite{Boos/Ohl:1997:gg4f}).  The four small groves of neutral
current interactions are only connected with the rest of the forest
through the charged current grove.  The automorphism group of this
forest has 128 elements.

The groves can now be used to select the Feynman diagrams to be
calculated by other means.  However, we note that it is also possible
to calculate the amplitude with little additional effort already
during the construction of the groves by keeping track of momenta and
couplings in the diagram flips.

\subsection{Automorphisms}
\label{sec:automorphisms}

The forests and groves that we have studied appear to be very
symmetrical in the neighborhood of any vertex.  However, the global
connection of these neighborhoods is twisted, which makes it all but
impossible to draw the graphs in a way that makes these apparent
symmetries manifest (see figure~\ref{fig:nc06}).

Nevertheless, one can turn to mathematics (in particular to
computational and constructive group
theory~\cite{McKay:1981:nauty-theory,*McKay:1990:nauty}) and
construct the automorphism groups $\Aut(F(E))$ and $\Aut(G_i(E))$ of
the forest~$F(E)$ and the groves~$G_i(E)$, i.\,e.~the group of
permutations of vertices that leave the edges invariant.  These groups
turn out to be larger than one might expect.  For example, the group
of permutations of the 71~vertices of the forest~$F(\gamma\gamma\to
u\bar d d\bar u)$ in figure~\ref{fig:mix71}, that leave the edges
invariant, has 128~elements.  Similarly, the automorphism group of the
forest in figure~\ref{fig:5phi3} has $120=5!$~elements.  The study of
these groups and their relations might enable us to construct gauge
invariant subsets directly.

\subsection{Advanced Combinatorics}
\label{sec:combinatorics}
Even pure gauge forests can provide new insights into the
combinatorics of gauge cancellations among Feynman diagrams.
Consider the process~$gg\to ggg$: there are 25 diagrams (all in the
same grove, of course).  15 of them have three triple gluon vertices
and the remaining 10 diagrams have one triple and one quadruple vertex
each.  The sub-forest of the 15 diagrams with triple vertices is
equivalent to the one shown in figure~\ref{fig:5phi3}.  The 30
endpoints of the gauge flips starting from the 10 diagrams with a
quadruple vertex cover this sub-forest with 15 nodes exactly twice.

Therefore, the gauge cancellations do \emph{not} follow from a simple
recursion of the gauge cancellations in~$gg\to gg$, where the
sub-forest of diagrams with triple couplings is covered exactly once.
Unfortunately, the symmetries in $gg\to gggg$ are much more
complicated.   Nevertheless, the automorphism groups of groves should be
studied in order to understand the gauge cancellations better.

\subsection{Loops}
\label{sec:loops}

The number of loops is invariant under flips, as can be seen from
Euler's formula.  However, some of the proofs above have made explicit
use of properties of tree diagrams.  Therefore further research is
required in this area.

\chapter{Phase Space}
\label{sec:ps}

\begin{cutequote*}{William Shakespeare}{Sonnet XIV}
Not from the stars do I my judgment pluck,\\
And yet methinks I have astronomy
\end{cutequote*}

The calculation of matrix elements using the methods outlined in
chapter~\ref{sec:me} is but the first step towards a comparison of
theoretical predictions with experiment.  In the second step, the
differential cross section obtained from squaring the amplitude has to
be integrated in the region of phase space corresponding to the
detector.  While this step is common to all observables, it
remains highly non-trivial and will be discussed in detail in this
chapter.

In our application, the integration of a function~$f$ with a given
measure~$\mu$ on a manifold~$M$
\begin{equation}
\label{eq:I(f)}
  I(f) = \int_M\!\mathrm{d}\mu(p)\,f(p)
\end{equation}
is complicated by three facts
\begin{enumerate}
  \item the manifold of multi particle phase space and the
    corresponding Lorentz invariant measure
    \begin{equation}
      \label{eq:dmu}
      \mathrm{d}\mu(p)
         = \delta^4(\textstyle\sum_n k_n - P)
           \prod_n \mathrm{d}^4k_n\,\delta(k_n^2-m_n^2)
    \end{equation}
    is non-trivial for three particles in the final state and
    analytically intractable for more than three massive particles,
  \item the function is given by a squared matrix element multiplied
    by kinematical cuts or acceptances
    \begin{equation}
      f(p) = \left| T (k_1,k_2,\ldots) \right|^2
                \cdot C(k_1,k_2,\ldots)
    \end{equation}
    and usually ill-behaved.  Integrable singularities are common in
    theories with massless particles and fluctuations over many orders
    of magnitude are the norm rather than the exception,
  \item in realistic applications, the acceptances and
    efficiencies in~$C(k_1,k_2,\ldots)$ are almost never given as
    analytical functions. Instead, they are themselves the result of
    complicated simulations.
\end{enumerate}

\section{Four Roads from Matrix Elements to Answers}
\label{sec:me2answers}

\subsection{Analytic Integration}
Analytic integration is mentioned only for completeness.  Already for
three or more particles in the final state with halfway realistic
cuts, a fully analytical calculation of integrated cross sections is
impossible.  At least some integrations will have to be performed
numerically.

\subsection{Semianalytic Calculation}
The radiative corrections for LEP1 precision physics on the
$Z^0$-resonance are dominated by soft photons.  The phase space in the
soft limit remains sufficiently simple so that all but a few
integrations can be performed analytically for simple cuts.  Since
programs based on this approach are extremely fast compared to the
more universally applicable, they play an important r\^ole in LEP1
phenomenology.  Due to their speed and precision, they are the
preferred tool for the unfolding of electroweak precision observables
at LEP1.  An additional application is to act as benchmarks for other
programs. The situation at LEP2 and the~\ac{LC} is different, because
more phase space for hard radiation is available and substantial
changes are necessary in these programs.

\subsection{Monte Carlo Integration}

If only~$I(f)$ in~(\ref{eq:I(f)}) has to be calculated, it is possible
to calculate the estimator~$E(f)$, given in~(\ref{eq:E(f)}) below,
directly, using an
appropriate source of pseudo random numbers.  This is known
colloquially as ``weighted event generation,'' because the phase space
points used for sampling are not equally likely, but have a weight
attached to them.  Since $f$~is not required to be bounded for this
approach, the application can focus all attention on minimizing the sampling
error~(\ref{eq:V(f,g)}).  The techniques for achieving this goal are
well developed and the currently most sophisticated are described in
detail in section~\ref{sec:amc}.

\subsection{Monte Carlo Event Generation}
\label{sec:unweighted-MC}

\begin{figure}
  \begin{center}
    \begin{picture}(100,100)
      \put(0,0){\includegraphics[width=100mm]{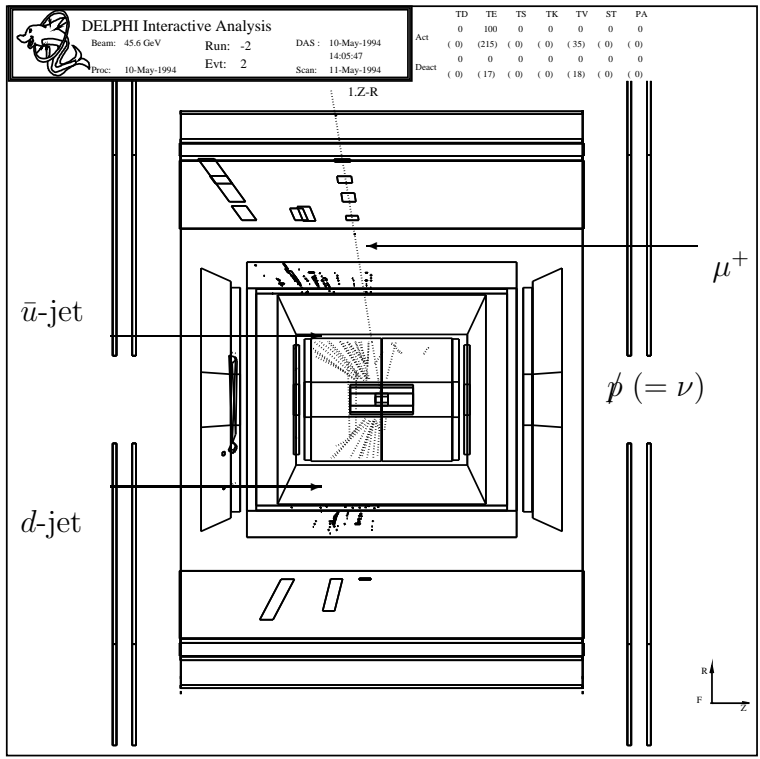}}
      \put(92,68){\vector(-1,0){44}}
      \put(94,64){$\mu^+$}
      \put(80,48){$\fmslash p \; (= \nu)$}
      \put(14,56){\vector(1,0){28}}
      \put(2,58){$\bar u$-jet}
      \put(14,36){\vector(1,0){28}}
      \put(2,30){$d$-jet}
    \end{picture}
  \end{center}
  \caption{\label{fig:wopper}%
    Simulated event $e^+e^- \to d\bar u \mu\bar\nu_\mu$ at $\sqrt s =
    \unit[200]{GeV}$ (``particle identification'' from Monte Carlo
    record).  Event generated by
    \program{WOPPER}~\protect\cite{Anlauf/etal:1994:wopper1.0,
    *Anlauf/etal:1994:wopper1.1,*Anlauf/etal:1996:wopper1.5} and
    passed through DELPHI detector
    simulation~\protect\cite{Phillips:1994:wopper}.}
\end{figure}
\begin{figure}
  \begin{center}
    \includegraphics[width=0.6\textwidth,angle=-90]{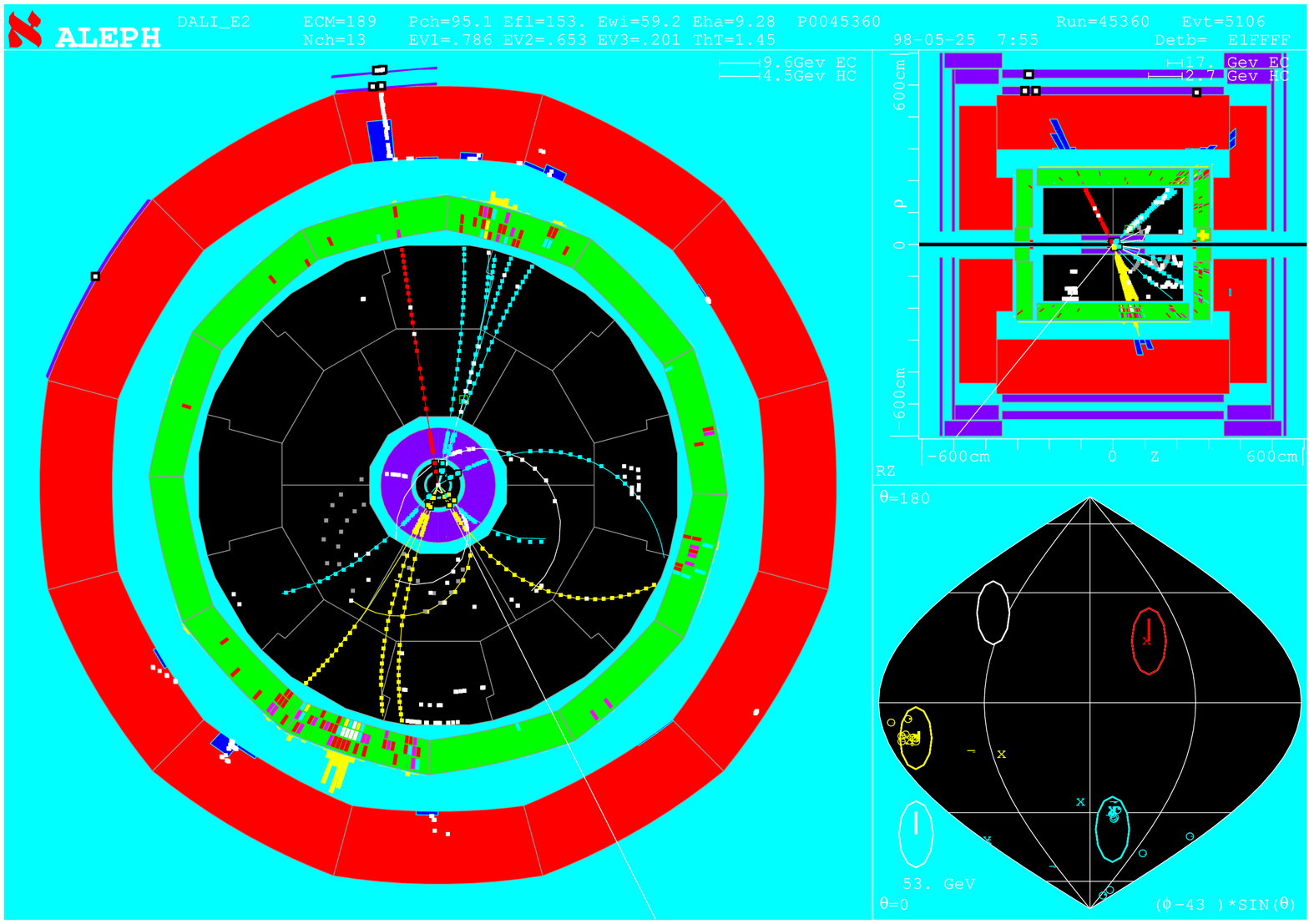}
  \end{center}
  \caption{\label{fig:aleph}%
    Actual LEP2 event at $\sqrt s = \unit[189]{GeV}$:
    $e^+e^-\to q\bar q \mu\bar\nu_\mu$ in the ALEPH detector.}
\end{figure}
In principle, it is possible to use weighted event generators also in
detector simulations similar to the one shown in
figure~\ref{fig:wopper}, as long as
the detector simulation is equipped with the trivial infrastructure
for the bookkeeping of event weights.

However, this approach has the serious drawback that the distribution
of events that are simulated in the detector, can be different from
the realistic distributions.  In the result of the integration, this
is corrected by the weights, but the problem remains that the detector
is exercised mostly in the region of the phase space where few events
are expected in the experiment.

Therefore it is desirable to generated ``unweighted events'',
i.\,e.~events with uniform weight~$1$, which are distributed exactly as
predicted by theory.  For bounded probability distributions, such
samples can be generated by brute force rejection (generate a phase
space point~$x$ and a uniform deviate~$y$; accept~$x$, if $y\le
p(x)/p_{\max}$, else repeat) and if the distribution does not vary too
much, this method can be sufficiently efficient.  Unfortunately, most
distributions in~\ac{HEP} are singular and require sophisticated
mapping techniques to handle the integrable singularities (see
section~\ref{sec:amc}).

\subsection{Why We Can Not Use the Metropolis Algorithm Naively}

At first sight, the Metropolis
algorithm~\cite{Metropolis/etal:1953:Algorithm} appears to be an
attractive alternative to the brute-force rejection algorithm for
unweighted event generation.  The Metropolis algorithm constructs a
Markov process covering the phase space, which can be shown to converge
asymptotically to the desired probability density without having to
reject a single event.  A Markov step is accepted with a probability
that is equal to the ratio of the probability densities at the new and
the old phase space point.  If the move is rejected, the \emph{same} point
must be added to the sample \emph{again}.  The advantage of the procedure is
that probabilities exceeding unity can be treated \emph{as if} they were
unity and probability densities need not be bounded for this agorithm
to work
(see~\cite{Itzykson/Drouffe:1989:textbook,Montvay/Muenster:1994:textbook}
for pedagogical discussions).

If the probability distribution is rather steep, simple numerical
experiments show that the number of rejected moves is high.
Therefore, the number of events that will appear more than once in a
finite sample will be high as well and this causes problems further
down the simulation chain.  Error estimates in detector simulation and
unfolding tools might not be prepared for finite samples with many
duplicate events~\cite{Moenig:1999:parallel}.

In general, the problem is that, unless the step size and the initial
conditions are chosen (and varied!) carefully, errors can be
\emph{under}estimated significantly for steep distributions.
Nevertheless, if the problems with obtaining \emph{robust} error
estimates can be gotten under control, the Metropolis algorithm could also
develop into a powerful part of the calculational tool chest for
event generators.  But until then, a strong \emph{caveat emptor}
regarding results obtained in this way is in order.

\section{Monte Carlo}
\label{sec:mc}

\begin{cutequote}{Concise~Oxford~Dictionary}
  {\normalfont\textbf{simulate:}}
    feign, pretend to have or feel, put on; pretend
    to be, act like, resemble, mimic, take or have an altered form
    suggested by; \ldots
\end{cutequote}
The overview in the preceding section has shown that weighted and
unweighted Monte Carlo methods are the most robust and flexible part
of the phenomenological tool chest for phase space integration.
Therefore they deserve a more detailed discussion. Given
a (pseudo-)random sequence of points in $n$-particle phase space
\begin{equation}
  \frak{p}_g=\{p_1,p_2,\ldots,p_N\}\,,
\end{equation}
where the points are distributed according to
\begin{subequations}
\begin{equation}
  \mathrm{d}\mu_g(p)=g(p)\mathrm{d}\mu(p),
\end{equation}
with
\begin{equation}
\label{eq:g-norm}
  \int_M\!\mathrm{d}\mu_g(p) = 1\;
    \text{and}\; g(p) > 0\; \text{(almost everywhere)}\,,
\end{equation}
\end{subequations}
we get an estimator of
\begin{equation}
  I(f) = \int_M\!\mathrm{d}\mu(p)\,f(p)
       = \int_M\!\mathrm{d}\mu_g(p)\,\frac{f(p)}{g(p)}
\end{equation}
as
\begin{equation}
\label{eq:E(f)}
  E(f) = \left\langle \frac{f}{g} \right\rangle_g
       = \frac{1}{N} \sum_{i=1}^{N} \frac{f(p_i)}{g(p_i)}\,.
\end{equation}
In~(\ref{eq:E(f)}), the volume of~$M$ is encoded in~$g$, because~$g$
is normalized according to~(\ref{eq:g-norm}).  As written, $E(f)$
depends on~$g$ and the particular $\frak{p}_g$, but the dependence
on~$g$ drops out completely in the \emph{ensemble average} over
all~$\frak{p}_g$.  Explicitely spelling out the dependence on~$g$ is
important, because the measure~$\textrm{d}\mu(p)$ (i.\,e.~$\textrm{d}\mu_g(p)$
for~$g(x)=1$) is in general by no means the one that can be generated
most efficiently.  As will be shown shortly, it is also not the most
useful one.  Therefore a discussion should consider a general~$g$ from
the outset.

For large~$\frak{p}_g$, $E(f)$ will converge to~$I(f)$, but when
comparing theoretical predictions with experimental observations, we
have to deal with the more complicated problem of estimating the error
for a finite sample size and minimizing this error.

Besides optimizing~$g$ for this purpose, we could also use a
deterministic~$\frak{p}_g$ derived from \emph{quasi random numbers}.
Under certain conditions, this approach can speed up convergence
dramatically~\cite{Hoogland/James/Kleiss:1997} (from~$1/\sqrt{N}$
to~$1/N$) at the expense of the robustness of the conditions.
Therefore we will not proceed along this line here and just notice
that further research in the area is highly desirable.  Also it is not
clear how to interface deterministic Monte Carlos using quasi random
numbers with detector simulations.

An unbiased estimator for the variance of~$E(f)$ is
given by
\begin{equation}
\label{eq:V(f,g)}
  V(f,g) = \frac{1}{N-1} \left(
         \left\langle \left(\frac{f}{g}\right)^2 \right\rangle_g
           - \left\langle\frac{f}{g}\right\rangle_g^2 \right)\,.
\end{equation}
A direct consequence of~(\ref{eq:V(f,g)}) is $V(f,f)=0$. This
vanishing error is not paradoxical, because~$I(g)$ has to be known
exactly in order to be able to generate~$\frak{p}_g$.
For the following, it is useful to note that~$V(f,g)$ is a biased
estimator for~$W_g(f/g)/(N-1)$, where the \emph{variation}~$W_g(h)$ of
a function~$h$ with respect to the measure~$\mu_g$ is
defined as
\begin{equation}
\label{eq:W(h)}
  W_g(h) = \int_M\!\mathrm{d}\mu_g(p)\,\left( (h(p))^2 - (I(h))^2 \right)
         = \int_M\!\mathrm{d}\mu_g(p)\,\left( h(p) - I(h) \right)^2 \ge 0\,.
\end{equation}
The advantage of~$W_g(f/g)$ over~$V(f,g)$ is that~$W_g(f/g)$ does
\emph{not} depend on the particular~$\frak{p}_g$, which will simplify
the following reasoning considerably.

For Monte Carlo integration, the optimal~$g$ will minimize
\begin{equation}
\label{eq:delta_2}
  \Delta_2(f,g) = W_g(f/g)
\end{equation}
while for event generation, the optimal~$g$ will minimize
\begin{equation}
\label{eq:delta_1}
  \Delta_1(f,g) = \int_M\!\mathrm{d}\mu_g(p)\,
     \left( 1 - \frac{\frac{f(p)}{g(p)}}{\sup_M\left(\frac{f}{g}\right)}
     \right)
        = 1 - \frac{I(f)}{\sup_M\left(\frac{f}{g}\right)}\,,
\end{equation}
in order to maximize the rejection efficiency.  If~$f$ is a wildly
fluctuating function, this optimization of~$g$ is indispensable for
obtaining a useful accuracy.  Typical causes for large fluctuations
are integrable singularities of~$f$ or~$\mu$ inside of~$M$ or
non-integrable singularities very close to~$M$.  Therefore, we will
use the term ``singularity'' for those parts of~$M$ in which there are
large fluctuations in~$f$ or~$\mu$.

If the~$g$ could be chosen from a function space that includes~$f$,
both~$\Delta_1$ and~$\Delta_2$ would find the same optimal~$g=f$,
since~$\Delta_1(f,f)=0$ and~$\Delta_2(f,f)=0$.  However, in practical
applications, we have $g\in G$, with a function space~$G$ that does
\emph{not} include~$f$ (otherwise we would already know both~$I(f)$
and an efficient algorithm to generate phase space points according
to~$\mu_f$). Therefore, which $g\in G$ is closest to~$f$ will in
general depend on whether the metric is~$\Delta_1$ or~$\Delta_2$.

\begin{figure}
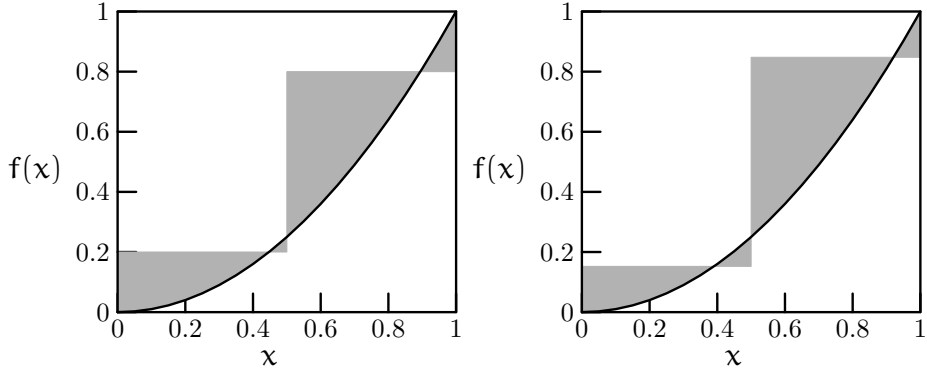

  \begin{center}
    \begin{empgraph}(45,40)
      pickup pencircle scaled 1.0pt;
      setrange (0, 0, 1, 1);
      autogrid (itick.bot,itick.lft);
      numeric a[], s[], t[], u;
      path f, g[], m;
      augment.f (0,0);
      for x = 0 step 0.05 until 1.01:
        augment.f (x, x*x);
      endfor
      a1 = 1 / 5;
      augment.m  (0.5,0.0);  augment.m  (0.5,1.0);
      augment.g1 (0.0,a1);   augment.g1 (0.5,a1);
      augment.g2 (0.5,1-a1); augment.g2 (1.0,1-a1);
      (whatever, u) = m intersectiontimes f;
      (s1, t1) = g1 intersectiontimes f;
      (s2, t2) = g2 intersectiontimes f;
      gfill (subpath (0,t1) of f) -- (subpath (s1,0) of g1) -- cycle
         withcolor .7white;
      gfill (subpath (u,t1) of f) -- (subpath (s1,infinity) of g1) -- cycle
         withcolor .7white;
      gfill (subpath (u,t2) of f) -- (subpath (s2,0) of g2) -- cycle
         withcolor .7white;
      gfill (subpath (infinity,t2) of f) -- (subpath (s2,infinity) of g2) -- cycle
         withcolor .7white;
      gdraw f;
      glabel.lft (btex $f(x)$ etex, OUT);
      glabel.bot (btex $x$ etex, OUT);
    \end{empgraph}
    \begin{empgraph}(45,40)
      pickup pencircle scaled 1.0pt;
      setrange (0, 0, 1, 1);
      autogrid (itick.bot,itick.lft);
      numeric a[], s[], t[], u;
      path f, g[], m;
      augment.f (0,0);
      for x = 0 step 0.05 until 1.01:
        augment.f (x, x*x);
      endfor
      a2 = 1 / (1 + sqrt(31));
      augment.m  (0.5,0.0);  augment.m  (0.5,1.0);
      augment.g1 (0.0,a2);   augment.g1 (0.5,a2);
      augment.g2 (0.5,1-a2); augment.g2 (1.0,1-a2);
      (whatever, u) = m intersectiontimes f;
      (s1, t1) = g1 intersectiontimes f;
      (s2, t2) = g2 intersectiontimes f;
      gfill (subpath (0,t1) of f) -- (subpath (s1,0) of g1) -- cycle
         withcolor .7white;
      gfill (subpath (u,t1) of f) -- (subpath (s1,infinity) of g1) -- cycle
         withcolor .7white;
      gfill (subpath (u,t2) of f) -- (subpath (s2,0) of g2) -- cycle
         withcolor .7white;
      gfill (subpath (infinity,t2) of f) -- (subpath (s2,infinity) of g2) -- cycle
         withcolor .7white;
      gdraw f;
      glabel.lft (btex $f(x)$ etex, OUT);
      glabel.bot (btex $x$ etex, OUT);
    \end{empgraph}
  \end{center}
  \caption{\label{fig:compare-deltas}%
    In the left graph, $\Delta_1(f,g_\alpha)$~is minimized for a
    simple step function, while in the left graph
    $\Delta_2(f,g_\alpha)$~is minimized in the \emph{same} function
    space. $\Delta_2(f,g_\alpha)$~distributes the mismatch more evenly
    to achieve faster convergence for integration, but has lower
    efficiency for event generation.}
\end{figure}
For concreteness, consider the trivial example of~$f(x)=x^2$
on~$M=[0,1]$ with a one parameter family of step functions
\begin{equation}
  g_\alpha(x) = \alpha\cdot\Theta(1/2-x) + (1-\alpha)\cdot\Theta(x-1/2)
\end{equation}
Then
\begin{equation}
  \Delta_2(f,g_\alpha) =
     \frac{1}{160\cdot\alpha} + \frac{155}{160\cdot(1-\alpha)} - \frac{1}{9}
\end{equation}
which is minimized by
\begin{equation}
  \alpha_2 = \frac{1}{1+\sqrt{31}} = 0.15226\ldots\,.
\end{equation}
On the other hand, $\Delta_1(f,g_\alpha)$ is minimized by
\begin{equation}
  \alpha_1 = \frac{1}{5}
\end{equation}
and there is a noticeable difference, even for smooth functions, as
shown in figure~\ref{fig:compare-deltas}.

{}From a technical point of view, $\Delta_2$ is more convenient. For
a given~$g\in G$, $\Delta_2(f,g)$ will exist for many interesting
functions~$f$ for which~$\Delta_1(f,g)$ is useless
because~$\sup_M(f/g)$ diverges.  Furthermore, the estimation
of~$\Delta_2(f,g)$ is statistically robust, while the estimation
of~$\sup_M(f/g)$ required by~$\Delta_1(f,g)$ is much more delicate.
Finally, the optimality equations for~$g$ derived from~$\Delta_2(f,g)$
by functional derivation are linear.

The set~$G$ is not defined precisely, it includes all probability
densities that can be generated ``cheaply.''  This includes all
densities for which the inverse of the anti-derivative can be computed
efficiently, but there are other, less well known,
cases~\cite{Devroye:1986:random_variates}.  Of course, there is no
absolute measure for the ``cheapness'' of generating~$g$.  Instead,
the effort has to be put in context.  As long as the generation of a
phase space point~$p$ does not take longer than the evaluation of the
differential cross section~$f(p)$, the generation can be
considered ``cheap.''  If the variance reduction is substantial, it
can be useful to use even more expensive~$g$'s.

The manual minimization of either~$\Delta_n(f,g)$ for a given~$f$ can be a
tedious exercise, that has to be repeated every time the coupling
parameters or kinematical cuts in~$f$ are changed.  For this reason it
is economical to use adaptive algorithms that minimize~$\Delta_n(f,g)$
automatically, as long as they perform satisfactorily.

\section{Adaptive Monte Carlo}
\label{sec:amc}
\begin{empcmds}
def setup_frame =
  pair ul, ur, ll, lr;
  ypart (ul) = ypart (ur); ypart (ll) = ypart (lr);
  xpart (ul) = xpart (ll); xpart (ur) = xpart (lr);
  numeric weight_width, weight_dist;
  weight_width = 0.1w; weight_dist = 0.05w;
  ll = (.1w,.1w);
  ur = (w-weight_width-weight_dist,h-weight_width-weight_dist);
enddef;
def draw_frame =
  pickup pencircle scaled .7pt;
  draw ll -- lr -- ur -- ul -- cycle;
  label.lrt (btex $0$ etex, ll);
  label.bot (btex $x_1$ etex, .5[ll,lr]);
  label.bot (btex $1$ etex, lr);
  label.ulft (btex $0$ etex, ll);
  label.lft (btex $x_2$ etex, .5[ll,ul]);
  label.lft (btex $1$ etex, ul);
enddef;
\end{empcmds}
Manual optimization of~$g$ is often too time consuming, in particular
if the dependence of the integral on external parameters (in the
integrand and in the boundaries) is to be studied.  Adaptive numerical
approaches are more attractive in these cases.  

In this section, we will use for illustration only integrals over the
unit hypercube
\begin{equation}
  x \in M = \left]0,1\right[^{\otimes n}
\end{equation}
with Cartesian measure
\begin{equation}
  \mathrm{d}\mu(x)
    = \mathrm{d}x_1\wedge\mathrm{d}x_2\wedge\ldots\wedge\mathrm{d}x_n
\end{equation}
and~$\Vol(M)=1$.  The general case follows by covering the
manifold~$M$ with one or more maps and taking into account the
appropriate Jacobians.

\subsection{Multi Channel}
\label{sec:multi-channel}

The basic question is: how can we implement a variable
weight~$g$ in~$\mathrm{d}\mu_g(x)=g(x)\mathrm{d}\mu(x)$ efficiently?
An obvious option is to employ a weighted sum of densities
\begin{equation}
\label{eq:linear-mc}
  g(x) = \sum_{i=1}^n \alpha_i g_i(x)
\end{equation}
with
\begin{equation}
  \int\!\mathrm{d}\mu(x)\,g_i(x) = 1\;
   \text{and}\; \sum_{i=1}^n \alpha_i = 1\;
   \text{and}\; \alpha_i \ge 0
\end{equation}
and to optimize the weights~$\alpha_i$~\cite{Kleiss/Pittau:1994:multichannel}.
This approach has been very successful in Monte Carlo integration
(e.\,g.~\cite{Berends/Kleiss/Pittau:1995:Excalibur,Denner/etal:1999:4fgamma}).
However, in applications with several hundred different
channels~$g_i$~\cite{Denner/etal:1999:4fgamma}, the optimization of
the weights~$\alpha_i$ according to~\cite{Kleiss/Pittau:1994:multichannel} yields
only modest improvements~\cite{Roth/Wackeroth:1999:private}.  It
appears to be more important that a channel is present, than that the
weights are accurately tuned.

In a typical application, the~$g_i$ are constructed manually from
squared signal diagrams with simplified spin correlations.  The
corresponding distributions can be generated easily by successive
integration of the $s$-channel poles and $t$-channel singularities.
The drawback of this approach is that the location and the shape of
the singularities has to be known analytically.  While the location of
the poles is fixed by the particle masses, radiative correction and
interference terms will modify the shape and shift the actual maxima.
A particularly serious problem is that the effects of radiative
corrections depend strongly on kinematical cuts.

\begin{figure}
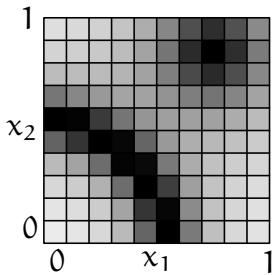

  \begin{center}
    \begin{emp}(40,40)
      setup_frame;
      vardef bwz (expr x, a, b) =
        (b*b / ((xpart(x-a))**2 + (ypart(x-a))**2 + b*b))
      enddef;
      vardef bwr (expr x, r, b) =
        (b*b / (((abs x) - r)**2 + b*b))
      enddef;
      for i = .1 step .1 until 1.01:
        for j = .1 step .1 until 1.01:
          fill (j-0.1)[(i-0.1)[ll,lr],(i-0.1)[ul,ur]]
            -- (j-0.1)[(i-0.0)[ll,lr],(i-0.0)[ul,ur]]
            -- (j-0.0)[(i-0.0)[ll,lr],(i-0.0)[ul,ur]]
            -- (j-0.0)[(i-0.1)[ll,lr],(i-0.1)[ul,ur]]
            -- cycle withcolor (   (1 - bwr ((i,j), 0.6, 0.1)
                                      - bwz ((i,j), (0.8,0.9), 0.2))*white);
        endfor
      endfor
      pickup pencircle scaled .7pt;
      for i = .1 step .1 until 0.99:
        draw i[ll,lr] -- i[ul,ur]; draw i[ll,ul] -- i[lr,ur];
      endfor
      draw_frame;
    \end{emp}
  \end{center}
  \caption{\label{fig:fixed-grid}%
    A fixed grid with variable weights can not adapt to singular
    integrands.}
\end{figure}
A special case of~(\ref{eq:linear-mc}) is given by using the
characteristic functions of fixed bins as~$g_i$, which corresponds to
variable weights for bins in a fixed grid, as depicted in
figure~\ref{fig:fixed-grid}.  This case is not useful for applications
in particle physics, because a grid with a \emph{fixed resolution} can
not adapt to the power law singularities from propagators as are
ubiquitous in particle physics.  Differential cross sections typically
vary over many orders of magnitude and the grid would have to be too
fine to be useful.

\begin{figure}
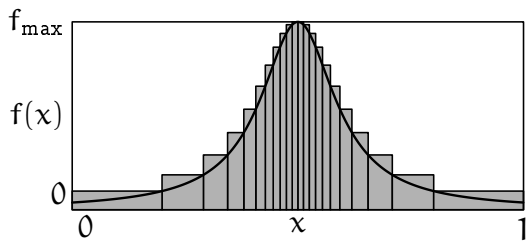

  \begin{center}
    \begin{emp}(80,45)
      setup_frame;
      vardef bwx (expr x, a, b) = b*b / ((x-a)**2 + b*b) enddef;
      vardef tand expr x = (sind x) / (cosd x) enddef;
      vardef tan expr x = (tand (180/3.141598*x)) enddef;
      vardef xy (expr i, j) = j[i[ll,lr],i[ul,ur]] enddef;
      pickup pencircle scaled .7pt;
      numeric ii, oldii, jj, pk, wd;
      path bin;
      pk = 0.5;
      wd = 0.1;
      oldii := pk;
      for i = (pk+.05) step .05 until 1.01:
        ii := wd*(tan((i-pk)/(4wd))) + pk;
        jj := bwx(oldii,pk,wd);
        bin := (xy(ii,0)) -- (xy(ii,jj)) -- (xy(oldii,jj)) -- (xy(oldii,0)) -- cycle;
        fill bin withcolor 0.7white; draw bin;
        oldii := ii;
      endfor
      jj := bwx(oldii,pk,wd);
      bin := (xy(1,0)) -- (xy(1,jj)) -- (xy(oldii,jj)) -- (xy(oldii,0)) -- cycle;
      fill bin withcolor 0.7white; draw bin;
      oldii := pk;
      for i = (pk-.05) step -.05 until -.01:
        ii := wd*(tan((i-pk)/(4wd))) + pk;
        jj := bwx(oldii,pk,wd);
        bin := (xy(ii,0)) -- (xy(ii,jj)) -- (xy(oldii,jj)) -- (xy(oldii,0)) -- cycle;
        fill bin withcolor 0.7white; draw bin;
        oldii := ii;
      endfor
      jj := bwx(oldii,pk,wd);
      bin := (xy(0,0)) -- (xy(0,jj)) -- (xy(oldii,jj)) -- (xy(oldii,0)) -- cycle;
      fill bin withcolor 0.7white; draw bin;
      pickup pencircle scaled 1pt;
      draw (xy(0,bwx (0,pk,wd)))
        for i = .01 step .01 until 1.001:
          .. (xy(i,bwx (i,pk,wd)))
        endfor;
      pickup pencircle scaled .7pt;
      draw ll -- lr -- ur -- ul -- cycle;
      label.lrt (btex $0$ etex, ll);
      label.bot (btex $x$ etex, .5[ll,lr]);
      label.bot (btex $1$ etex, lr);
      label.ulft (btex $0$ etex, ll);
      label.lft (btex $f(x)$ etex, .5[ll,ul]);
      label.lft (btex $f_{\max}$ etex, ul);
    \end{emp}
  \end{center}
  \caption{\label{fig:variable-grid}%
    In one dimension, a variable grid with fixed weights can adapt
    well to singular integrands.}
\end{figure}
In \emph{one} dimension, there is an interesting alternative to
varying the weights of fixed width.  It is possible to vary instead
the size of bins with fixed weight\footnote{Varying the weights as
well would be possible but adds no new flexibility.  It is therefore
better to avoid the added complexity and to keep all weights identical.}
\begin{equation}
\label{eq:VEGAS-onedim}
  g(x) = \frac{1}{N} \sum_{i=1}^N
     \frac{\Theta(x-x_{i-1}) \Theta(x_i-x)}{x_i-x_{i-1}}\,.
\end{equation}
This way it is possible to have bins corresponding to densities that
vary over many orders of magnitude.  It is possible to iteratively
improve the grid, driven by samplings of~$f(x)\textrm{d}\mu_g(x)$.
The improvement can either decrease~$\Delta_1(f,g)$ (\emph{importance
sampling} for event generation) or decrease~$\Delta_2(f,g)$
(\emph{stratified sampling} for high precision integration).  An
optimal~$g$ would solve the functional equation
\begin{equation}
\label{eq:Delta'}
  \frac{\delta}{\delta g} \Delta_{1,2}(f,g) = \Delta'_{1,2}(f,g) = 0.
\end{equation}
While it is straightforward to sample~$\Delta'_{1,2}(f,g)$
simultaneously with~$f/g$, an exact solution of~(\ref{eq:Delta'}) is
not possible, because contributions have to be moved from one bin to
another.  Nevertheless, an approximate solution is
possible under the assumption that the integrand in~(\ref{eq:Delta'})
is piecewise constant~\cite{Lepage:1978:vegas,*Lepage:1980:vegas}.

\begin{figure}
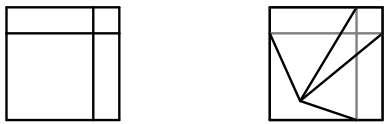

  \begin{center}
    \begin{emp}(60,27)
      pickup pencircle scaled 1pt;
      pair good[][], good'[][], bad[][];
      numeric margin.h, margin.v, size[], dist;
      margin.h = 5mm;
      margin.v = 6mm;
      for i := 0 upto 2:
        for j := 0 upto 2:
          bad[i][j] = good'[i][j] = good[i][j] + (dist,0);
        endfor
      endfor
      for i := 0 upto 1:
        for j := 0 upto 2:
          good[i+1][j] = good[i][j] + (size[i],0);
          good[j][i+1] = good[j][i] + (0,size[i]);
        endfor
      endfor
      size[1] = .3*size[0];
      good[0][0] = (margin.h,margin.v);
      good[0][2] = (margin.h,h-margin.v);
      bad[2][0] = (w-margin.h,margin.v);
      bad[2][2] = (w-margin.h,h-margin.v);
      bad[1][1] := bad[1][1] + (-.5,-.6)*(size[1]+size[0]);
      for i := 0 upto 2:
        for j := 0 upto 1:
          draw good[i][j]--good[i][j+1];
          draw good[j][i]--good[j+1][i];
          draw good'[i][j]--good'[i][j+1] withcolor .5white;
          draw good'[j][i]--good'[j+1][i] withcolor .5white;
          draw bad[i][j]--bad[i][j+1];
          draw bad[j][i]--bad[j+1][i];
        endfor
      endfor
    \end{emp}
  \end{center}
  \caption{\label{fig:non-convex}%
    Quadrangles do not remain convex in general after unconstrained
    shifts of lattice points.}
\end{figure}
Unfortunately, there is no immediate generalization to higher
dimensions, because freely shifting vertices starting from a
hypercubic lattice will in general lead to quadrangles that are no
longer convex, as illustrated in figure~\ref{fig:non-convex}.  The
concave quadrangles are more than a technical inconvenience, because
they lead to numeric instabilities of the grid optimization.

\subsection{Delaunay Triangulation}
Mathematically, the most natural decomposition of the integration
region in higher dimension is into simplices.  In fact, there have
been attempts to use simplical decompositions in adaptive Monte
Carlo~\cite{Manankova/etal:1995:MILX}.  The results of this exercise
have not been impressive so far, in particular for strongly
fluctuating integrands.  The reason for the failure of the naive
approach is that the shape of the simplices is not constrained and
that instabilities will produce ``splinters'' that are numerically
ill-behaved.

The construction of ``well behaved'' simplical decompositions is a
rather non-trivial mathematical problem
(see~\cite{Bern/Eppstein:1992:triangulation} for reviews), that
remains unsolved 
in the general case.  However, there has been progress in two and
three dimensions, driven in part by the needs of the engineering
community for numerically robust finite element codes.  The theory for
two dimensions is rather complete and mathematically very
beautiful~\cite{Knuth:1992:Hulls}.  Therefore it is appropriate to
make a factorized ansatz~\cite{Ohl:????:TRAMP}, in which the ``least
factorizable'' pairs and triplets are handled by simplical
decomposition, using triangles and tetrahedra.

\begin{figure}
  \begin{center}
    \includegraphics[width=0.7\textwidth]{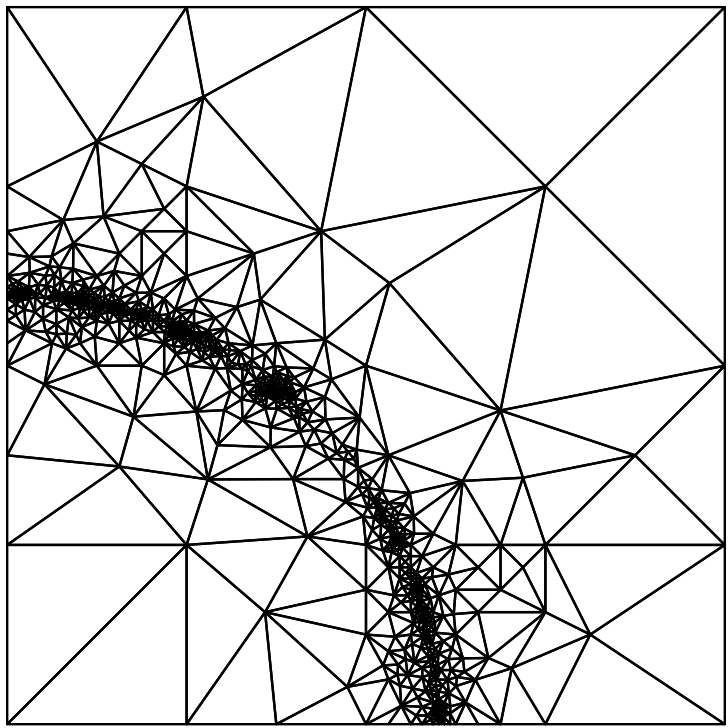}
  \end{center}
  \caption{\label{fig:delaunay}%
    Delaunay triangulation with Steiner points adapted to a
    singularity on a quarter circle.}
\end{figure}
In two dimensions, the most suitable tool are \emph{constrained
Delaunay triangulation}s with \emph{Steiner points}.  For a given set
of points, the Delaunay triangulation is the unique triangulation in
which the center of the circle passing through the corners of each
triangle is inside of that triangle.  This uniqueness helps to
stabilize the optimization of the triangulation, which is no longer
unique, if we allow to add additional points, so-called Steiner
points, to the point set in order to satisfy additional constraints.
Nevertheless, there are well defined and terminating
algorithms~\cite{Ruppert:1995:refinement} that construct constrained
Delaunay triangulations with global and local maximum area constraints,
as well as minimum angle constraints.  The former are what is required
for modeling a function for importance sampling and the latter
stabilize the iterative refinement of the triangulation.

However, there is still much more work needed to put these ideas to
work in particle physics applications beyond a proof of
concept~\cite{Ohl:????:TRAMP}, because the required computational
geometry is at or beyond the state-of-the-art in finite element
analysis for engineering.

\subsection{VEGAS}
\label{sec:vegas}

\subsubsection{Factorization}

The previous section has shown that one-dimensional distributions are
singled out because a mathematically well defined optimization
strategy can be formulated with very little effort.  Therefore it was
tempting to propose a factorized
ansatz~\cite{Lepage:1978:vegas,*Lepage:1980:vegas}
\begin{equation}
\label{eq:VEGAS-ansatz}
  g(x) = g_1(x_1) g_2(x_2) \cdots g_n(x_n)
\end{equation}
and to investigate its properties.  It turns out
that~(\ref{eq:VEGAS-ansatz}) produces satisfactory results with very
little effort and the corresponding
code \program{VEGAS}~\cite{Lepage:1978:vegas,*Lepage:1980:vegas} has
been used as a ``black box'' by countless physics programs.  For two
decades no significant improvements of~\program{VEGAS} have been
proposed.

The factorized ansatz~(\ref{eq:VEGAS-ansatz}) of \program{VEGAS}
guarantees that the resulting grid is hypercubic. Furthermore, the
simple one-dimensional update prescriptions can be used in each
dimension, if the estimators entering the prescriptions are averaged
over the remaining
dimensions.  A final benefit of the factorized ansatz is that the
computational costs grow only linearly with the number of dimensions
and not exponentially.

\begin{figure}
  \begin{center}
    \begin{emp}(35,35)
      setup_frame;
      vardef bw (expr x, a, b) = b*b / ((x-a)**2 + b*b) enddef;
      pickup pencircle scaled .7pt;
      for r = 1 step -0.02  until 0:
        fill fullcircle
          scaled (r*(xpart (lr-ll)))
          shifted .5[.5[ll,lr],.5[ul,ur]]
          withcolor ((1 - bw (r, 0.7, 0.15)) * white);
      endfor
      draw_frame;
    \end{emp}
    \qquad
    \begin{emp}(35,35)
      setup_frame;
      vardef bw (expr x, a, b) = b*b / ((x-a)**2 + b*b) enddef;
      pickup pencircle scaled .7pt;
      path p[], q[];
      p1 := ll--ur; q1 := ll--lr--ur;
      p2 := ur--ll; q2 := ur--ul--ll;
      for r = 1 step -0.02  until 0:
        fill (interpath (r, p1, q1) & interpath (r, p2, q2) & cycle)
          withcolor ((1 - bw (r, 0, 0.25)) * white);
      endfor
      draw_frame;
    \end{emp}
  \end{center}
  \caption{\label{fig:transformation-suffices}%
    Integrands that \protect\program{VEGAS} can handle after a
    coordinate transformation.}
\end{figure}
\begin{figure}
  \begin{center}
    \begin{picture}(0,0)
      \put(5,4){\includegraphics[width=25mm,clip]{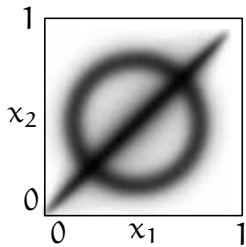}}
    \end{picture}%
    \begin{emp}(35,35)
      setup_frame;
      draw_frame;
    \end{emp}
  \end{center}
  \caption{\label{fig:transformation-does-notsuffice}%
    An integrand that \protect\program{VEGAS} can not handle, even
    after a coordinate transformation.}
\end{figure}
Of course, not all functions of interest in particle physics can be
approximated by a factorized ansatz.  While the functions depicted in
figure~\ref{fig:transformation-suffices} have factorized singularities
after a simple coordinate transformation, their sum in
figure~\ref{fig:transformation-does-notsuffice} can not be sampled
efficiently by~\program{VEGAS}.  Similar example are ubiquitous in
particle physics, because the cross sections are derived from sums of
Feynman diagrams and have singularities in several kinematical
variables which can not be used as coordinates simultaneously.

\begin{figure}
  \begin{center}
    \hfil\\
    \begin{fmfgraph*}(30,20)
      \fmfleft{p1,p2}
      \fmfright{q1,k,q2}
      \fmflabel{$p_1$}{p1}
      \fmflabel{$p_2$}{p2}
      \fmflabel{$q_1$}{q1}
      \fmflabel{$q_2$}{q2}
      \fmflabel{$k$}{k}
      \fmf{fermion}{p1,v,p2}
      \fmf{photon}{v,vq}
      \fmf{fermion,tension=.5}{q1,vq}
      \fmf{fermion,tension=.5,label=$s_2$,label.side=left}{vq,vg}
      \fmf{fermion,tension=.5}{vg,q2}
      \fmffreeze
      \fmf{gluon}{vg,k}
      \fmfdot{v,vq,vg}
    \end{fmfgraph*}
    \qquad\qquad
    \begin{fmfgraph*}(30,20)
      \fmfleft{p1,p2}
      \fmfright{q1,k,q2} 
      \fmflabel{$p_1$}{p1}
      \fmflabel{$p_2$}{p2}
      \fmflabel{$q_1$}{q1}
      \fmflabel{$q_2$}{q2}
      \fmflabel{$k$}{k}
      \fmf{fermion}{p1,v,p2}
      \fmf{photon}{v,vq}
      \fmf{fermion,tension=.5}{q1,vg}
      \fmf{fermion,tension=.5,label=$s_1$,label.side=left}{vg,vq}
      \fmf{fermion,tension=.5}{vq,q2}
      \fmffreeze
      \fmf{gluon}{vg,k}
      \fmfdot{v,vq,vg}
    \end{fmfgraph*}\\
    \hfil
  \end{center}
  \caption{\label{fig:3jets}%
    Three jet production~$e^+e^-\to q\bar qg$.}
\end{figure}
For example, the three jet production~$e^+e^-\to q\bar qg$ in
figure~\ref{fig:3jets} has important singularities in both
variables~$s_{1/2}=(q_{1/2}+k)^2$.  In this case, there are
parametrizations of the phase space that use both~$s_1$ and~$s_2$
simultaneously.  However the corresponding
Jacobian~$|\partial\phi/\partial x|$ contains non-factorizable
singularities in~$s_1$ and~$s_2$ induced by a Gram determinant as
$1/\sqrt{\Delta_4(p_1,p_2,q_1,q_2)}$.  Thus there is no 
parametrizations in which both singularities factorize simultaneously
and \program{VEGAS} can not adjust to both singularities
(see~\cite{Byckling/Kajantie:1973:phasespace}).  A general
solution to this problem will be presented below in
section~\ref{sec:vamp}.

\subsubsection{Stratified vs. Importance Sampling}

The implementation
of~\program{VEGAS}~\cite{Lepage:1978:vegas,*Lepage:1980:vegas} that is
widely used, has three different modes: importance sampling,
stratified sampling and a hybrid of the two.  Importance sampling can
be selected by the application program, while the choice between
stratified sampling and the hybrid is done by~\program{VEGAS} based on
the dimension of the integration domain.  The
implementation~\cite{Lepage:1978:vegas,*Lepage:1980:vegas} very terse
and hard to understand.  Even popular
textbooks~\cite{Press/etal:1992:NumRecC,
*Press/etal:1992:NumRec77,*Press/etal:1996:NumRec90} reproduce the
code without further explanation.  The reimplementation that was
required for~\cite{Ohl:1998:VAMP}, aims to be more readable.

\begin{empcmds}
  vardef layout =
    setup_frame;
    numeric equ_div, adap_div, rx, ry, rxp, rxm, ryp, rym;
    equ_div = 3;  adap_div = 8;
    rx = 5.2; ry = 3.6;
    rxp = ceiling rx; rxm = floor rx;
    ryp = ceiling ry; rym = floor ry;
    numeric pi; pi = 180;
    vardef adap_fct_x (expr x) = (x + sind(2*x*pi)/8) enddef;
    vardef weight_x (expr x) = (1 + 2*sind(1*x*pi)**2) / 3 enddef;
    vardef adap_fct_y (expr x) = (x + sind(4*x*pi)/16) enddef;
    vardef weight_y (expr x) = (1 + 2*sind(2*x*pi)**2) / 3 enddef;
    vardef grid_pos (expr i, j) =
      (adap_fct_y(j/adap_div))[(adap_fct_x(i/adap_div))[ll,lr],
                               (adap_fct_x(i/adap_div))[ul,ur]]
    enddef;
    vardef grid_square (expr i, j) =
      grid_pos (i,j) -- grid_pos (i+1,j) -- grid_pos (i+1,j+1)
        -- grid_pos (i,j+1) -- cycle
    enddef;
  enddef;
  vardef decoration =
    fill (lr shifted (weight_y(0)*(weight_width,0))
             for y = .1 step .1 until 1.01:
               .. y[lr,ur] shifted (weight_y(y)*(weight_width,0))
             endfor
             -- ur -- lr -- cycle) shifted (weight_dist,0) withcolor 0.7white;
    fill (ul shifted (weight_x(0)*(0,weight_width))
             for x = .1 step .1 until 1.01:
               .. x[ul,ur] shifted (weight_x(x)*(0,weight_width))
             endfor
             -- ur -- ul -- cycle) shifted (0,weight_dist) withcolor 0.7white;
    picture px, py;
    px = btex $p_1(x_1)$ etex; py = btex $p_2(x_2)$ etex;
    label.top (image (unfill bbox px; draw px),
                .5[ul,ur] shifted (0,weight_dist));
    label.rt (image (unfill bbox py; draw py),
                .75[lr,ur] shifted (weight_dist,0));
    label.lrt (btex $0$ etex, ll);
    label.bot (btex $x_1$ etex, .5[ll,lr]);
    label.bot (btex $1$ etex, lr);
    label.ulft (btex $0$ etex, ll);
    label.lft (btex $x_2$ etex, .5[ll,ul]);
    label.lft (btex $1$ etex, ul);
  enddef;
\end{empcmds}
\begin{figure}
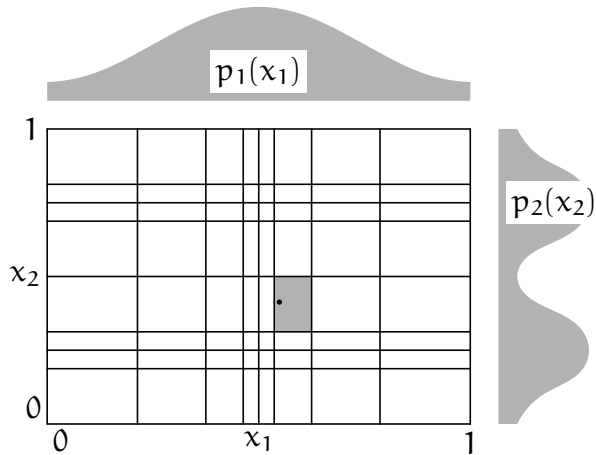

  \begin{center}
    \begin{emp}(75,58)
      layout;
      fill grid_square (rxm,rym) withcolor 0.7white;
      pickup pencircle scaled .7pt;
      for i = 0 upto adap_div:
        draw grid_pos(i,0) -- grid_pos(i,adap_div);
        draw grid_pos(0,i) -- grid_pos(adap_div,i);
      endfor
      pickup pencircle scaled 2pt;
      drawdot grid_pos(rx,ry);
      decoration;
    \end{emp}
  \end{center}
  \caption{\label{fig:VEGAS-importance}%
    \protect\program{VEGAS} grid structure for importance sampling}
\end{figure}
In pure importance sampling, \program{VEGAS} adjusts a grid like the
two-dimensional illustration in figure~\ref{fig:VEGAS-importance}.
The sampling proceeds by first selecting a hypercube at random and
subsequently selecting a point in this hypercube according to a
uniform distribution. This algorithm generates points distributed
according to~(\ref{eq:VEGAS-onedim}) and~(\ref{eq:VEGAS-ansatz}).

\begin{figure}
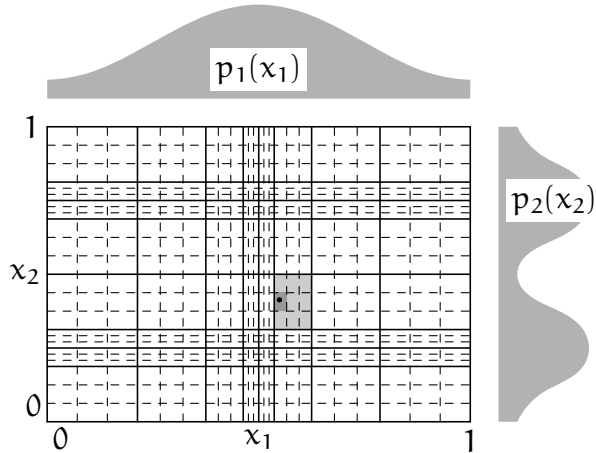

  \begin{center}
    \begin{emp}(75,58)
      layout;
      vardef grid_sub_pos (expr i, di, j, dj) =
        (dj/equ_div)[(di/equ_div)[grid_pos(i,j),grid_pos(i+1,j)],
                     (di/equ_div)[grid_pos(i,j+1),grid_pos(i+1,j+1)]]
      enddef;
      vardef grid_sub_square (expr i, di, j, dj) =
        grid_sub_pos (i,di,j,dj)
          -- grid_sub_pos (i,di+1,j,dj)
          -- grid_sub_pos (i,di+1,j,dj+1)
          -- grid_sub_pos (i,di,j,dj+1)
          -- cycle
      enddef;
      fill grid_square (rxm,rym) withcolor 0.8white;
      fill grid_sub_square (rxm,0,rym,1) withcolor 0.6white;
      pickup pencircle scaled .7pt;
      for i = 0 upto adap_div:
        draw grid_pos(i,0) -- grid_pos(i,adap_div);
        draw grid_pos(0,i) -- grid_pos(adap_div,i);
      endfor
      pickup pencircle scaled .5pt;
      for i = 0 upto (adap_div-1):
        for j = 1 upto (equ_div-1):
          draw grid_sub_pos(i,j,0,0)
                 -- grid_sub_pos(i,j,adap_div,0) dashed evenly;
          draw grid_sub_pos(0,0,i,j)
                 -- grid_sub_pos(adap_div,0,i,j) dashed evenly;
        endfor
      endfor
      pickup pencircle scaled 2pt;
      drawdot grid_pos(rx,ry);
      decoration;
    \end{emp}
  \end{center}
  \caption{\label{fig:VEGAS-stratified}%
    \protect\program{VEGAS} grid structure for genuinely stratified sampling.}
\end{figure}
In stratified sampling, faster convergence is attempted by distributing
sampling points more uniformly than random samples.
In~\program{VEGAS}, stratified sampling is implemented by dividing the
bins of the adaptive grid further as shown in
figure~\ref{fig:VEGAS-stratified}.  The number of divisions is chosen
such that the sum of two samples per bin approximates the desired
total number of samples.  Now the sampling proceeds by selecting two
random points from \emph{each} bin of the finer grid.  This procedure
shows better convergence for integration than importance sampling.

\begin{table}
  \caption{\label{tab:dimen}%
    To stratify or not to stratify: number of calls required for a
    grid of 25 bins in each direction, as a function of the dimension
    of the integration domain.}
  \begin{center}
    \begin{tabular}{c|c}
       $n_{\text{dim}}$
           & $N_{\text{calls}}^{\max}(n_g=25)$\\\hline
        2  & $1\cdot10^{3}$ \\
        3  & $3\cdot10^{4}$ \\
        4  & $8\cdot10^{5}$ \\
        5  & $2\cdot10^{7}$ \\
        6  & $5\cdot10^{8}$
    \end{tabular}
  \end{center}
\end{table}
As can be seen from table~\ref{tab:dimen}, the algorithm for
stratified sampling as described above can only be used in low
dimensions.  Already the eight-dimensional four-particle phase space
would require many billions of sampling points, which is usually not
possible. 

\begin{empcmds}
  numeric pi;
  pi = 180;
  vardef adap_fct_one (expr x) =
    (x + sind(2*x*pi)/8)
  enddef;
  vardef adap_fct_two (expr x) =
    (x + sind(4*x*pi)/16)
  enddef;
  vardef adap_fct (expr x) =
     adap_fct_two (x)
  enddef;
  vardef drawbar expr p =
    draw ((0,-.5)--(0,.5)) scaled 1mm shifted p
  enddef;
\end{empcmds}

\begin{empcmds}
  vardef pseudo (expr xlo, xhi, ylo, yhi, 
                      equ_lo, equ_hi, equ_div,
                      adap_lo, adap_hi, adap_div,
                      r, do_labels, do_arrow) =
    pair equ_grid.lo, equ_grid.hi, adap_grid[]lo, adap_grid[]hi;
    ypart (equ_grid.lo) = ypart (equ_grid.hi);
    ypart (adap_grid[1]lo) = ypart (adap_grid[1]hi);
    ypart (adap_grid[2]lo) = ypart (adap_grid[2]hi);
    xpart (equ_grid.lo) = xpart (adap_grid[1]lo) = xpart (adap_grid[2]lo);
    xpart (equ_grid.hi) = xpart (adap_grid[1]hi) = xpart (adap_grid[2]hi);
    equ_grid.hi = (xhi, yhi);
    adap_grid[1]lo = .5[equ_grid.lo,adap_grid[2]lo];
    adap_grid[2]lo = (xlo, ylo);
    numeric rp, rm;
    rp = ceiling r;
    rm = floor r;
    pickup pencircle scaled .5pt;
    for i = adap_lo upto adap_hi:
        draw (i/adap_div)[adap_grid[1]lo,adap_grid[1]hi]
               -- (adap_fct(i/adap_div))[adap_grid[2]lo,adap_grid[2]hi]
             withcolor 0.7white;
    endfor
    if do_arrow:
      fill (rm/adap_div)[adap_grid[1]lo,adap_grid[1]hi]
             -- (rp/adap_div)[adap_grid[1]lo,adap_grid[1]hi]
             -- (adap_fct(rp/adap_div))[adap_grid[2]lo,adap_grid[2]hi]
             -- (adap_fct(rm/adap_div))[adap_grid[2]lo,adap_grid[2]hi]
             -- cycle withcolor 0.7white;
    fi
    if do_labels:
      label.lft (btex \texttt{0} etex, equ_grid.lo);
      label.rt (btex \texttt{ng} etex, equ_grid.hi);
    fi
    draw (equ_lo/equ_div)[equ_grid.lo,equ_grid.hi]
          -- (equ_hi/equ_div)[equ_grid.lo,equ_grid.hi];
    for i = equ_lo upto equ_hi:
      drawbar (i/equ_div)[equ_grid.lo,equ_grid.hi];
    endfor
    if do_labels:
      label.lft (btex $\xi$, \texttt{i: 0} etex, adap_grid[1]lo);
      label.rt (btex \texttt{n} etex, adap_grid[1]hi);
      label.lft (btex \texttt{x: 0} etex, adap_grid[2]lo);
      label.rt (btex \texttt{1} etex, adap_grid[2]hi);
    fi
    draw (adap_lo/adap_div)[adap_grid[1]lo,adap_grid[1]hi]
          -- (adap_hi/adap_div)[adap_grid[1]lo,adap_grid[1]hi];
    draw (adap_fct(adap_lo/adap_div))[adap_grid[2]lo,adap_grid[2]hi]
          -- (adap_fct(adap_hi/adap_div))[adap_grid[2]lo,adap_grid[2]hi];
    for i = adap_lo upto adap_hi:
      drawbar (i/adap_div)[adap_grid[1]lo,adap_grid[1]hi];
      drawbar (adap_fct(i/adap_div))[adap_grid[2]lo,adap_grid[2]hi];
    endfor
    if do_arrow:
      pickup pencircle scaled 1pt;
      pair cell, ia, grid__;
      ia = (r/adap_div)[adap_grid[1]lo,adap_grid[1]hi];
      cell = ia shifted (equ_grid.hi - adap_grid[1]hi);
      grid__ = (adap_fct(r/adap_div))[adap_grid[2]lo,adap_grid[2]hi];
      if do_labels:
        label.top (btex \texttt{cell - r} etex, cell);
      fi
      drawarrow cell -- ia;
      drawarrow ia -- grid__;
      if do_labels:
        label.bot (btex \texttt{x} etex, grid__);
      fi
    fi
  enddef;
\end{empcmds}

\begin{figure}
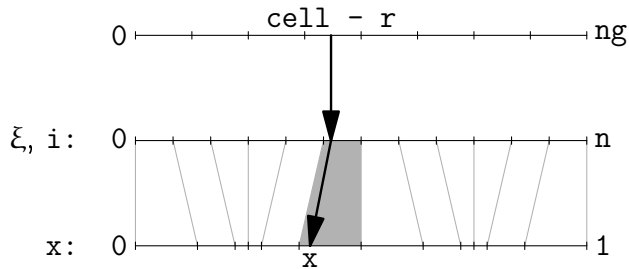

  \begin{center}
    \begin{emp}(120,40)
      pseudo (.3w, .8w, .1h, .8h, 0, 8, 8,  0, 12, 12, 5.2,   true, true);
    \end{emp}
  \end{center}
  \caption{\label{fig:VEGAS-pseudo-stratified}%
    \protect\program{VEGAS} grid structure for pseudo stratified sampling.}
\end{figure}
Nevertheless, the stratified sampling algorithm performs empirically
better than the importance sampling algorithm and it is worthwhile to
pursue the hybrid approach illustrated in
figure~\ref{fig:VEGAS-pseudo-stratified}.  Here the stratification
grid is no longer a subdivision of the adaptive grid.  Instead, the
points selected in the stratification grid are mapped linearly onto
the adaptive grid and further nonlinearly onto the integration
domain.  The adaptive grid is updated to minimize~$\Delta_1(f,g)$.
This hybrid
is, strictly speaking, theoretically not justified because importance
sampling assumes a \emph{random} distribution.  However, the ``more
uniform'' distribution appears to perform better in practical
applications and the hybrid approach is very successful.

\subsection{VAMP}
\label{sec:vamp}

\subsubsection{Introduction}

As mentioned before for the examples in
figure~\ref{fig:transformation-suffices}, the property of
factorization depends on the coordinate system.  Consider, for
example, the functions
\begin{subequations}
\label{eq:f1f2}
\begin{align}
  f_1(x_1,x_2) & = \frac{1}{(x_1-a_1)^2 + b_1^2} \\
  f_2(x_1,x_2) & = \frac{1}{\left(\sqrt{x_1^2+x_2^2}-a_2\right)^2 + b_2^2}
\end{align}
\end{subequations}
on~$M=(-1,1)\otimes(-1,1)$ with the
measure~$\textrm{d}\mu=\textrm{d}x_1\wedge\textrm{d}x_2$. Obviously,
$f_1$ is factorizable in Cartesian coordinates, while~$f_2$ is
factorizable in polar coordinates.  \program{VEGAS} will sample either function
efficiently for arbitrary~$b_{1,2}$ in suitable coordinate systems,
but there is no coordinate system in which \program{VEGAS} can sample the
sum~$f_1+f_2$ efficiently for small~$b_{1,2}$.

There is however a generalization~\cite{Ohl:1999:vamp} of the \program{VEGAS}
algorithm from factorizable distributions to sums of factorizable
distributions, where each term may be factorizable in a
\emph{different} coordinate system.  This larger class includes most
of the integrands appearing in particle physics and empirical studies
have shown a dramatic increase of accuracy for typical integrals.

\subsubsection{Maps}
\label{sec:maps}

The problem of estimating~$I(f)$ can be divided naturally into two
parts: parametrization of~$M$ and sampling of the function~$f$.  While
the estimate will not depend on the parametrization, the error will.

In general, we need an atlas with more that one chart~$\phi$ to cover
a manifold~$M$.  We will ignore this technical complication in the
following, because, for the purpose of integration, we can always
decompose~$M$ such that each piece is covered by a single chart.
Moreover, a single chart suffices in most cases of practical interest,
since we are at liberty to remove sets of measure zero from~$M$.  For
example, after removing a single point, the unit sphere can be covered
by a single chart.

Nevertheless, even if we are not concerned with the global properties
of~$M$ that require the use of more than one chart, the language of
differential geometry will allow us to use our geometrical intuition.
Instead of pasting together locally flat pieces, we will paste
together \emph{factorizable} pieces, which can be overlapping, because
integration is an additive operation.

For actual computations, it is convenient to use the same domain for
the charts of all manifolds.  The obvious choice for $n$-dimensional
manifolds is the open $n$-dimensional unit hypercube
\begin{equation}
  U = (0,1)^{\otimes n}\,.
\end{equation}
Sometimes, it will be instructive to view the chart as a
composition~$\phi=\psi\circ\chi$ with an irregularly
shaped~$P\in\mathbf{R}^n$ as an intermediate step
\begin{equation}
  \begin{fmfcd}(40,20)
    \fmfbottom{P,R}
    \fmftop{U}
    \fmfcdset{U}
    \fmfcdset{P}
    \fmfcdset[\mathbf{R}]{R}
    \fmfcdset{M}
    \fmfcdisomorph{U,M}{\phi}
    \fmfcdisomorph[right]{U,P}{\chi}
    \fmfcdisomorph{P,M}{\psi}
    \fmfcdmorph[right]{P,R}{f\circ\psi}
    \fmfcdmorph{M,R}{f}
    \fmfcdmorph{U,R}{f\circ\phi}
  \end{fmfcd}
\end{equation}
(in all commutative diagrams, solid arrows are reserved for
bijections and dotted arrows are used for other morphisms).
The integral~(\ref{eq:I(f)}) can now be written
\begin{equation}
  I(f) = \int_0^1\!\textrm{d}^nx\,
    \left|\frac{\partial\phi}{\partial x}\right| f(\phi(x))
\end{equation}
and it remains to sample~$|\partial\phi/\partial x|\cdot(f\circ\phi)$
on~$U$.  Below, it will be crucial that there is always more than one
way to map~$U$ onto~$M$
\begin{equation}
\label{eq:pi}
  \begin{fmfcd}(40,20)
    \fmfleft{U',U}
    \fmfright{P',P}
    \fmfcdset{U}
    \fmfcdset{U'}
    \fmfcdset{P}
    \fmfcdset{P'}
    \fmfcdset{M}
    \fmfcdisomorph{U,M}{\phi}
    \fmfcdisomorph{U',M}{\phi'}
    \fmfcdisomorph[right]{P,M}{\psi}
    \fmfcdisomorph[right]{P',M}{\psi'}
    \fmfcdisomorph[right]{U,U'}{\pi_U}
    \fmfcdisomorph{P,P'}{\pi_P}
    \fmfcdisomorph{U,P}{\chi}
    \fmfcdisomorph{U',P'}{\chi'}
  \end{fmfcd}
\end{equation}
and that we are free to select the map most suitable for our purposes.

The ideal choice for~$\phi$ would be a solution of the partial
differential equation $|\partial\phi/\partial x| = 1/(f\circ\phi)$,
but this is equivalent to an analytical evaluation of~$I(f)$ and is
impossible for the cases under consideration.  A more realistic goal
is to find a~$\phi$ such that $|\partial\phi/\partial
x|\cdot(f\circ\phi)$ has factorizable singularities and is therefore
sampled well by \program{VEGAS}.  This is still a non-trivial problem,
however.  For example, consider again the phase space integration for
gluon radiation $e^+e^-\to q\bar qg$.  From the Feynman diagrams in
figure~\ref{fig:3jets} we had seen that the squared matrix element has
singularities in both variables~$s_{1/2}=(q_{1/2}+k)^2$ that can not
be factorized.  On the other hand, it is straightforward to find
parametrizations that factorize the dependency on~$s_1$ or~$s_2$
\emph{separately}.

Returning to the general case, consider~$N_c$ different
maps~$\phi_i:U\to M$ and probability densities~$g_i:U\to [0,\infty)$.
Then the function
\begin{equation}
\label{eq:g(p)}
  g = \sum_{i=1}^{N_c} \alpha_i
     (g_i\circ\phi_i^{-1}) \left|\frac{\partial\phi_i^{-1}}{\partial p}\right|
\end{equation}
is a probability density~$g:M\to [0,\infty)$
\begin{equation}
  \int_M\! \textrm{d}\mu(p)\, g(p) = 1\,,
\end{equation}
as long as the~$g_i$ are probability densities themselves and
the~$\alpha_i$ are properly normalized
\begin{equation}
\label{eq:alpha}
 \int_0^1\!g_i(x)\textrm{d}^nx = 1\,,\;\;\;
 \sum_{i=1}^{N_c} \alpha_i = 1\,,\;\;\;
   0 \le \alpha_i \le 1 \,.
\end{equation}
{}From the definition~(\ref{eq:g(p)}), we have obviously
\begin{equation}
\label{eq:I(f)MC}
  I(f) = \sum_{i=1}^{N_c} \alpha_i
      \int_M\! g_i(\phi_i^{-1}(p))
          \left|\frac{\partial\phi_i^{-1}}{\partial p}\right|
          \textrm{d}\mu(p)\,
        \frac{f(p)}{g(p)}
\end{equation}
and, after pulling back from~$M$ to $N_c$~copies of~$U$
\begin{equation}
  I(f) = \sum_{i=1}^{N_c} \alpha_i
      \int_0^1\!g_i(x)\textrm{d}^nx\,
          \frac{f(\phi_i(x))}{g(\phi_i(x))}\,,
\end{equation}
we find a new estimator of the integral~$I(f)$
\begin{equation}
\label{eq:E(f)MC}
  E(f) = \sum_{i=1}^{N_c} \alpha_i
    \left\langle \frac{f\circ\phi_i}{g\circ\phi_i} \right\rangle_{g_i}\,.
\end{equation}
If the~$g_i$ in~(\ref{eq:I(f)MC}) and~(\ref{eq:E(f)MC}) are
factorized, they can be optimized using the classic
\program{VEGAS} algorithm~\cite{Lepage:1978:vegas,*Lepage:1980:vegas}
unchanged.  However, since we have to sample with a separate adaptive
grid for each channel, a new implementation~\cite{Ohl:1998:VAMP} is
required for technical reasons.

Using the $N_c^2$ maps~$\pi_{ij}=\phi_j^{-1}\circ\phi_i:U\to U$
introduced in~(\ref{eq:pi}), we can write
the~$g\circ\phi_i:U\to[0,\infty)$ in~(\ref{eq:E(f)MC}) as
\begin{equation}
\label{eq:gophi_i}
  g\circ\phi_i
     = \left|\frac{\partial\phi_i}{\partial x}\right|^{-1}
       \left( \alpha_i g_i + 
       \sum_{\substack{j=1\\j\not=i}}^{N_c} \alpha_j (g_j\circ\pi_{ij})
          \left|\frac{\partial\pi_{ij}}{\partial x}\right| \right)\,.
\end{equation}
{}From a geometrical perspective, the maps~$\pi_{ij}$ are just the
coordinate transformations from the coordinate systems in which the
other singularities factorize into the coordinate system in which the
current singularity factorizes.

Even if all~$g_i$ factorize, the~$g\circ\phi_i$ will not necessarily
factorize, provided the~$\phi_i$ are chosen appropriately.  The
maps~$\pi_{ij}$ are purely geometric objects, independent of the
physical model under consideration.  They are the coordinate
transformation from a coordinate system~$\Xi_i$ in which the $i$th
singularity factorizes to a coordinate system~$\Xi_j$ in which the
$j$th singularity factorizes.

The construction of the~$\pi_{ij}$ requires some mathematical input,
but it is required only once and the procedure allows economical
studies of the dependence of the cross section on kinematical cuts and
on external parameters through~$f$, because \texttt{VEGAS} can
optimize the~$g_i$ for each parameter set and set of cuts without
further human intervention.

Note that the integral in~(\ref{eq:I(f)MC}) does not change, when we
use~$\phi_i:U\to M_i\supseteq M$, if we extent~$f$ from~$M$ to~$M_i$
by the definition~$f(M_i\setminus M)=0$.
This is useful, for instance, when we want to
cover~$(-1,1)\otimes(-1,1)$ by both Cartesian and polar coordinates.
This causes, however, a problem with the~$\pi_{12}$
in~(\ref{eq:gophi_i}).  In the diagram
\begin{equation}
  \begin{fmfcd}(75,15)
    \fmfbottom{d1,U1,d2,U2,d3}
    \fmftop{P1,M1,M,M2,P2}
    \fmfcdset[U]{U1}
    \fmfcdset[U]{U2}
    \fmfcdset[P_1]{P1}
    \fmfcdset[M_1]{M1}
    \fmfcdset{M}
    \fmfcdset[M_2]{M2}
    \fmfcdset[P_2]{P2}
    \fmfcdisomorph[right]{U1,M1}{\phi_1}
    \fmfcdisomorph{U1,P1}{\chi_1}
    \fmfcdisomorph{P1,M1}{\psi_1}
    \fmfcdisomorph{U2,M2}{\phi_2}
    \fmfcdisomorph[right]{U2,P2}{\chi_2}
    \fmfcdisomorph[right]{P2,M2}{\psi_2}
    \fmfcdmorph{U1,U2}{\pi_{12}}
    \fmfcdmorph[right]{M,M1}{\iota_1}
    \fmfcdmorph{M,M2}{\iota_2}
  \end{fmfcd}
\end{equation}
the injections~$\iota_{1,2}$ are not onto and since~$\pi_{12}$ is
not necessarily a bijection anymore, the
Jacobian~$\left|\partial\pi_{ij}/\partial x\right|$ may be
ill-defined.  But since~$f(M_i\setminus M)=0$, we only need 
the unique bijections~$\phi'_{1,2}$ and~$\pi'_{12}$ that make the
diagram
\begin{equation}
  \begin{fmfcd}(90,15)
    \fmfbottom{d1,U1,U1',U2',U2,d3}
    \fmftop{P1,M1,M1',M2',M2,P2}
    \fmfcdset[U]{U1}
    \fmfcdset[U_1]{U1'}
    \fmfcdset[U_2]{U2'}
    \fmfcdset[U]{U2}
    \fmfcdset[P_1]{P1}
    \fmfcdset[M_1]{M1}
    \fmfcdset[M]{M1'}
    \fmfcdset[M]{M2'}
    \fmfcdset[M_2]{M2}
    \fmfcdset[P_2]{P2}
    \fmfcdisomorph{U1',M1'}{\phi'_1}
    \fmfcdisomorph[right]{U1,M1}{\phi_1}
    \fmfcdisomorph{U1,P1}{\chi_1}
    \fmfcdisomorph{P1,M1}{\psi_1}
    \fmfcdisomorph[right]{U2',M2'}{\phi'_2}
    \fmfcdisomorph{U2,M2}{\phi_2}
    \fmfcdisomorph[right]{U2,P2}{\chi_2}
    \fmfcdisomorph[right]{P2,M2}{\psi_2}
    \fmfcdmorph{U1',U1}{\iota^U_1}
    \fmfcdisomorph{U1',U2'}{\pi'_{12}}
    \fmfcdmorph[right]{U2',U2}{\iota^U_2}
    \fmfcdmorph[right]{M1',M1}{\iota_1}
    \fmfcdeq{M1',M2'}
    \fmfcdmorph{M2',M2}{\iota_2}
  \end{fmfcd}
\end{equation}
commute.

\subsubsection{Multi Channel Sampling}
\label{sec:MC}

Up to now, we have not specified the~$\alpha_i$, they are only subject
to the conditions~(\ref{eq:alpha}).  Intuitively, we expect the best
results when the~$\alpha_i$ are proportional to the contribution of their
corresponding singularities to the integral.  The option of tuning
the~$\alpha_i$ manually is not attractive if the optimal values depend
on varying external parameters.  Instead, we use a numerical
procedure~\cite{Kleiss/Pittau:1994:multichannel} for tuning
the~$\alpha_i$.

We want to minimize the variance~(\ref{eq:V(f,g)}) with respect to
the~$\alpha_i$. This is equivalent to minimizing
\begin{equation}
\label{eq:W(alpha)}
  W(f,\alpha) = \int_M\! g(p) \textrm{d}\mu(p)\,
       \left(\frac{f(p)}{g(p)}\right)^2
\end{equation}
with respect to~$\alpha$ with the subsidiary
condition~$\sum_i\alpha_i=1$.  After adding a Lagrange multiplier, the
stationary points of the variation are given by the solutions to the
equations
\begin{equation}
\label{eq:PDG(W)}
  \forall i: W_i(f,\alpha) = W(f,\alpha)
\end{equation}
where
\begin{equation}
  W_i(f,\alpha)
    = -\frac{\partial}{\partial\alpha_i} W(f,\alpha)
    = \int_0^1\!g_i(x)\textrm{d}^nx\,
          \left(\frac{f(\phi_i(x))}{g(\phi_i(x))}\right)^2
\end{equation}
and
\begin{equation}
   W(f,\alpha) = \sum_{i=1}^{N_c} \alpha_i W_i(f,\alpha)\,.
\end{equation}
It can easily be shown~\cite{Kleiss/Pittau:1994:multichannel} that the
stationary points~(\ref{eq:PDG(W)}) correspond to local minima.
If we use
\begin{equation}
   N_i = \alpha_i N
\end{equation}
to distribute~$N$ sampling points among the channels,
the~$W_i(f,\alpha)$ are just the contributions from channel~$i$ to the
total variance.  Thus we recover the familiar result from
stratified sampling, that the overall variance is minimized by
spreading the variance evenly among channels.

The~$W_i(f,\alpha)$ can be estimated with very little extra effort
while sampling~$I(f)$ (see~(\ref{eq:E(f)MC}))
\begin{equation}
\label{eq:Vi(alpha)}
  V_i(f,\alpha) =
    \left\langle \left(\frac{f\circ\phi_i}{g\circ\phi_i}\right)^2
       \right\rangle_{g_i}\,.
\end{equation}
Note that the factor of~$g_i/g$ in the corresponding formula
in~\cite{Kleiss/Pittau:1994:multichannel} is absent
from~(\ref{eq:Vi(alpha)}), because we are already sampling with the
weight~$g_i$ in each channel separately.

The equations~(\ref{eq:PDG(W)}) are non linear and can not be solved
directly.  However,  the solutions of~(\ref{eq:PDG(W)}) are a fixed
point of the prescription 
\begin{equation}
\label{eq:update}
  \alpha_i \mapsto \alpha_i'
      = \frac{\alpha_i \left(V_i(f,\alpha)\right)^\beta}
             {\sum_i\alpha_i \left(V_i(f,\alpha)\right)^\beta},
  \;\;\;(\beta>0)
\end{equation}
for updating the weights~$\alpha_i$.  There is no guarantee that this
fixed point will be reached from a particular starting value, such
as~$\alpha_i=1/N_c$, through successive applications
of~(\ref{eq:update}).  Nevertheless, it is clear
that~(\ref{eq:update}) will concentrate on the channels with large
contributions to the variance, as suggested by stratified
sampling. Furthermore, empirical studies show that~(\ref{eq:update})
is successful in practical applications.
The value~$\beta=1/2$ has been proposed
in~\cite{Kleiss/Pittau:1994:multichannel}, but it can be beneficial in
some cases to use smaller values like~$\beta=1/4$ to dampen
statistical fluctuations.

\subsubsection{Performance}
\label{sec:performance}

Both the implementation and the practical use of the multi channel
algorithm are more involved than the
application of the original \program{VEGAS} algorithm.  Therefore it is
necessary to investigate whether the additional effort pays off in
terms of better performance.

A Fortran implementation of this algorithm,
\program{VAMP}~\cite{Ohl:1998:VAMP}, has been used for empirical
studies.  This implementation features other improvements over
``\program{VEGAS} Classic'', most notably system independent and
portable support for parallel processing (see
section~\ref{sec:parallelization}) and support for unweighted
event generation.

There are two main sources of additional computational costs: at each
sampling point the function~$g\circ\phi_i$ has be evaluated, which
requires the computation of the~$N_c-1$ maps~$\pi_{ij}$ together with
their Jacobians and of the~$N_c-1$ probability distributions~$g_i$ of
the other \program{VEGAS} grids (see (\ref{eq:gophi_i})).

The retrieval of the current~$g_i$'s requires a bisection search in
each dimension, i.e.~a total of~$O((N_c-1)\cdot n_{\text{dim}}\cdot
\log_2 (n_{\text{div}}))$ executions of the inner loop of the search.
For simple integrands, this can indeed be a few times more costly than
the evaluation of the integrand itself.

The computation of the~$\pi_{ij}$ can be costly as well.  However,
unlike the~$g_i$, this computation can usually be tuned manually.
This can be worth the effort if many estimations of similar integrals
are to be performed.  Empirically, straightforward implementations of
the~$\pi_{ij}$ add costs of the same order as the evaluation of
the~$g_i$.

Finally, additional iterations are needed for adapting the
weights~$\alpha_i$ of the multi channel algorithm described
in~(\ref{sec:MC}).  Their cost is negligible, however, because
they are usually performed with far fewer sampling points than the
final iterations.

Even in cases in which the evaluation of~$g_i$ increases computation
costs by a whole order of magnitude, any reduction of the error by
more than a factor of~4 will make the multi channel algorithm
economical.  In fact, it is easy to construct examples in which the
error will be reduced by more than two orders of magnitude.  The
function
\begin{multline}
\label{eq:f(x)}
  f(x) = \frac{b}{144\atan(1/2b)}
      \biggl( \frac{3\pi\Theta(r_3<1)}{r_3^2((r_3-1/2)^2+b^2)}
             + \frac{2\pi\Theta(r_2<1,|x_3|<1)}{r_2((r_2-1/2)^2+b^2)} \\
             + \frac{\Theta(-1<x_1,x_2,x_3<1)}{x_1^2+b^2} \biggl)\,,
\end{multline}
with~$r_2=\sqrt{x_1^2+x_2^2}$ and~$r_3=\sqrt{x_1^2+x_2^2+x_3^2}$, is
constructed such that it can easily be normalized
\begin{equation}
  \int_{-1}^{1}\!\textrm{d}^3x\,f(x) = 1
\end{equation}
and allows a check of the result.  The three terms factorize in
spherical, cylindrical and Cartesian coordinates, respectively,
suggesting a three channel approach.  After five steps of weight
optimization consisting four iterations of $10^5$ samples, we have
performed three iterations of $10^6$ samples with the VAMP multi
channel algorithm.  Empirically, we found that we can perform four
iterations of $5\cdot10^5$ samples and three
iterations of $5\cdot10^6$ samples with the classic \program{VEGAS}
algorithm during the same time period.  Since the functional form of~$f$ is
almost as simple as the coordinate transformation, the fivefold
increase of computational cost is hardly surprising.

\begin{figure}
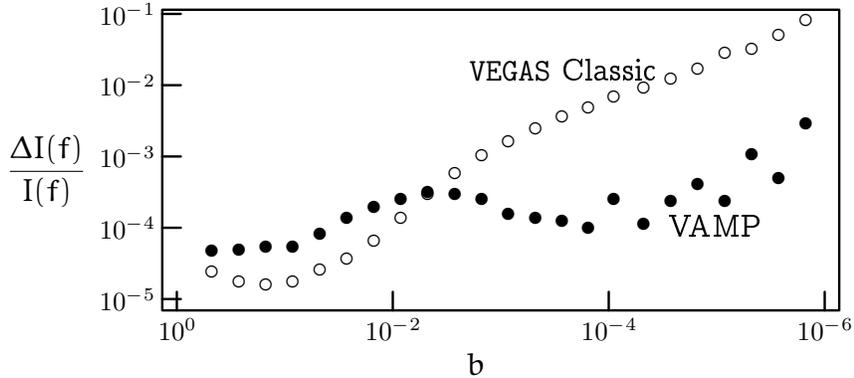

  \begin{center}
    \begin{empgraph}(90,40)
      vardef Formgen_ (expr q) = Formsci_ (q) enddef;
      pickup pencircle scaled 1pt;
      setcoords (-log, log);
      gdraw "vamp.data" plot btex $\bullet$ etex;
      glabel.lrt (btex VAMP etex, 17);
      gdraw "vegas.data" plot btex $\circ$ etex;
      glabel.ulft (btex \texttt{VEGAS} Classic etex, 17);
      autogrid (itick.bot, itick.lft);
      glabel.lft (btex $\displaystyle\frac{\Delta I(f)}{I(f)}$ etex, OUT);
      glabel.bot (btex $b$ etex, OUT);
    \end{empgraph}
  \end{center}
  \caption{\label{fig:bench}%
    Comparison of the sampling error for the integral of~$f$
    in~(\ref{eq:f(x)}) as a function of the width parameter~$b$ for
    the two algorithms at comparable computational costs.}
\end{figure}

In figure~\ref{fig:bench}, we compare the error estimates derived by
the classic \program{VEGAS} algorithm and by the three channel VAMP algorithm.
As one would expect, the multi channel algorithm does not offer any
substantial advantages for smooth functions (i.\,e.~$b>0.01$).
Instead, it is penalized by the higher computational costs.  On the
other hand, the accuracy of the classic \program{VEGAS} algorithm deteriorates
like a power with smaller values of~$b$.  At the same time, the
multi channel algorithm can adapt itself to the steeper functions,
leading to a much slower loss of precision.

The function~$f$ in~(\ref{eq:f(x)}) has been constructed as a showcase
for the multi channel algorithm, of course.  Nevertheless, more
complicated realistic examples from particle physics appear to gain
about an order of magnitude in accuracy.  Furthermore, the new
algorithm allows \emph{unweighted} event generation.  This is hardly
ever possible with the original \program{VEGAS} implementation, because the
remaining fluctuations typically reduce the average weight to very
small numbers.

A particularly attractive application is provided by automated tools
for the calculation of scattering cross sections.  While these tools
can currently calculate differential cross sections without manual
intervention, the phase space integrations still require hand tuning
of mappings for importance sampling for each parameter set.  The
present algorithm can overcome this problem, since it requires to
solve the geometrical problem of calculating the maps~$\pi_{ij}$
in~(\ref{eq:gophi_i}) for all possible invariants only \emph{once}.
The selection and optimization of the channels can then be performed
algorithmically.

\subsubsection{A Cheaper Alternative}

There is an alternative approach that avoids the evaluation of
the~$g_i$'s, sacrificing flexibility.  Fixing the~$g_i$ at unity, we
have for~$\tilde g:M\to [0,\infty)$
\begin{equation}
\label{eq:tildeg(p)}
  \tilde g = \sum_{i=1}^{N_c} \alpha_i
     \left|\frac{\partial\phi_i^{-1}}{\partial p}\right|
\end{equation}
and the integral becomes
\begin{equation}
  I(f) = \sum_{i=1}^{N_c} \alpha_i
      \int_M\! \left|\frac{\partial\phi_i^{-1}}{\partial p}\right|
          \textrm{d}\mu(p)\, \frac{f(p)}{\tilde g(p)}
       = \sum_{i=1}^{N_c} \alpha_i \int_0^1\!\textrm{d}^nx\,
          \frac{f(\phi_i(x))}{\tilde g(\phi_i(x))}\,.
\end{equation}
\program{VEGAS} can now be used to perform adaptive integrations of
\begin{equation}
  I_i(f) = \int_0^1\!\textrm{d}^nx\,
          \frac{f(\phi_i(x))}{\tilde g(\phi_i(x))}
\end{equation}
individually.  In some cases it is possible to construct a set
of~$\phi_i$ such that~$I_i(f)$ can estimated efficiently.
The optimization of the weights~$\alpha_i$ can again be effected by
the multi channel algorithm described in~(\ref{sec:MC}).

The disadvantage of this approach is that the optimal~$\phi_i$ will
depend sensitively on external parameters and the integration limits.
In the approach based on the~$g$ in~(\ref{eq:g(p)}), \program{VEGAS} can take
care of the integration limits automatically.

\section{Parallelization}
\label{sec:parallelization}

The computing needs of the~\ac{LC} can only be satisfied economically
if parallel computing on commodity hardware can be realized.
Traditionally, parallelization is utilized in~\ac{HEP} by running
independent Monte Carlo or analysis jobs in parallel and combining
their results statistically.  Nevertheless,
more fine grained parallelism is desirable in Monte Carlo jobs that
use adaptive algorithms\footnote{Already at LEP2, there are problems
with the parallelization of unweighted event generation if the maximum
weights have to be found in each parallel job
separately~\protect\cite{Moenig:1999:parallel}.}.

The problem of the parallelization of adaptive Monte Carlo integration
algorithms, in particular \program{VEGAS}, has gained some attention
recently~\cite{Krecker:1997:Parallel-Vegas,Veseli:1998:Parallel-Vegas}.
These implementations start from the classic implementation of
\program{VEGAS} and add synchronization barriers, either mutexes for
threads accessing shared memory or explicit message passing.  These
approaches result in compact code and achieve high performance, but
their low level nature obscures the mathematical structure.  Since
parallel computing is notoriously difficult, it is desirable to have a
well defined mathematical model that frees physicists from having to
waste too much effort dealing with technical computing details.
Therefore, we suggest a \emph{mathematical} model of parallelism for
adaptive Monte Carlo integration that is independent both of a
concrete paradigm for parallelism and of the programming language used
for an implementation.  We decompose the algorithm and prove that
certain parts can be executed in \emph{any} order without changing the
result.  As a corollary, we know that they can be executed in
parallel.

The algorithms presented below have been implemented successfully in
the library \program{VAMP}~\cite{Ohl:1998:VAMP}, along with the multi
channel algorithm described in section~\ref{sec:vamp}.

\subsection{Vegas}

As described in section~\ref{sec:vegas}, \program{VEGAS} uses two
different grids: an adaptive grid~$G^A$, which is used to adapt 
the distribution of the sampling points and a stratification
grid~$G^S$ for stratified sampling.  The latter is static and depends
only on the number of dimensions and on the number of sampling points.
Both grids factorize into \emph{divisions}~$d_{A,S}^i$
\begin{subequations}
\begin{align}
  G^A &= d^A_1 \otimes d^A_2 \otimes \cdots \otimes d^A_n \\
  G^S &= d^S_1 \otimes d^S_2 \otimes \cdots \otimes d^S_n \,.
\end{align}
\end{subequations}
The divisions come in three kinds
\begin{subequations}
\begin{align}
\label{eq:importance}
  d^S_i &= \emptyset   &&\text{(importance sampling)} \\
\label{eq:stratified}
  d^A_i &= d^S_i/m     &&\text{(stratified sampling)} \\
\label{eq:pseudo}
  d^A_i &\not= d^S_i/m &&\text{(pseudo-stratified sampling)}\,.
\end{align}
\end{subequations}
In the classic implementation of
\program{VEGAS}~\cite{Lepage:1978:vegas,*Lepage:1980:vegas}, all
divisions are of the same type.  In a more general
implementation~\cite{Ohl:1998:VAMP}, this is not required and it can be
useful to use stratification only in a few dimensions.

Two-dimensional grids for the cases~(\ref{eq:importance})
and~(\ref{eq:stratified}) have been illustrated in
figures~\ref{fig:VEGAS-importance} and~\ref{fig:VEGAS-stratified}.  In
case~(\ref{eq:importance}), there is no stratification grid and the
points are picked at random in the whole region according to~$G_A$.
In case~(\ref{eq:stratified}), the adaptive grid~$G_A$ is a regular
subgrid of the stratification grid~$G_S$ and an equal number of points
are picked at random in each cell of~$G_S$.  Since~$d^A_i = d^S_i/m$,
the points will be distributed according to~$G_A$ as well.  A
one-dimensional illustration of the most complicated
case~(\ref{eq:pseudo}) has been shown in
figure~\ref{fig:VEGAS-pseudo-stratified}.

\subsection{Formalization of Adaptive Sampling}
\label{sec:adaptive-sampling}

In order to discuss the problems with parallelizing adaptive
integration algorithms and to present a solution, it helps to introduce
some mathematical notation.  A sampling~$S$ is a map from the
space~$\pi$ of point sets and the space~$F$ of functions to the real
(or complex) numbers
\begin{equation}
\begin{aligned}
  S: \pi \times F & \to \mathbf{R} \\
     (p,f)        & \mapsto I = S(p,f)\,.
\end{aligned}
\end{equation}
For our purposes, we have to be more specific about the nature of the
point set~$p\in\pi$.  In general, this point set will be characterized by a 
sequence of pseudo random numbers~$\rho\in R$ and by one or more
grids~$G\in\Gamma$ used for importance or stratified sampling.  A
simple sampling
\begin{equation}
\label{eq:S0}
\begin{aligned}
  S_0: R \times \Gamma \times A \times F \times\mathbf{R}\times\mathbf{R}
         & \to R \times \Gamma \times A \times F \times\mathbf{R}\times\mathbf{R}\\
       (\rho, G, a, f, \mu_1, \mu_2) & \mapsto
            (\rho', G, a', f, \mu_1', \mu_2')
                = S_0 (\rho, G, a, f, \mu_1, \mu_2)
\end{aligned}
\end{equation}
estimates the $n$-th moments $\mu_n'\in\mathbf{R}$ of the
function~$f\in F$.  The integral and its standard deviation can be
derived easily from the moments
\begin{subequations}
\begin{align}
  I        &= \mu_1 \\
  \sigma^2 &= \frac{1}{N-1} \left(\mu_2 - \mu_1^2\right)\,,
\end{align}
\end{subequations}
while the latter are more convenient for the following discussion.
In addition, $S_0$ collects auxiliary information to be used in the
grid refinement, denoted by~$a\in A$.
The unchanged arguments~$G$ and~$f$ have been added to the result
of~$S_0$ in~(\ref{eq:S0}), so that~$S_0$ has identical domain and
codomain and 
can therefore be iterated.  Previous estimates~$\mu_n$ may be used
in the estimation of~$\mu_n'$, but a particular~$S_0$ is free to
ignore them as well.  Using a little notational freedom, we
augment~$\mathbf{R}$ and~$A$ with a special value~$\bot$, which will
always be discarded by~$S_0$.

In an adaptive integration algorithm, there is also a refinement
operation~$r:\Gamma\times A \to\Gamma$ that improves the grid based on
the auxiliary information~$a\in A$.  $r$ can be extended naturally
to the codomain of~$S_0$
\begin{equation}
\begin{aligned}
  r: R \times \Gamma \times A \times F \times\mathbf{R}\times\mathbf{R}
       & \to R \times \Gamma \times A \times F \times\mathbf{R}\times\mathbf{R}\\
     (\rho, G, a, f, \mu_1, \mu_2) & \mapsto
        (\rho, G', a, f, \mu_1, \mu_2) = r (\rho, G, a, f, \mu_1, \mu_2)\,,
\end{aligned}
\end{equation}
so that~$S=rS_0$ is well defined and we can specify $n$-step adaptive
sampling as
\begin{equation}
\label{eq:Sn}
  S_n = S_0 (rS_0)^n
\end{equation}
Since, in a typical application, only the estimate of the integral and
the standard deviation are used, a projection can be applied to the
result of~$S_n$: 
\begin{equation}
\label{eq:P}
\begin{aligned}
  P: R \times \Gamma \times A \times F \times\mathbf{R}\times\mathbf{R}
       & \to \mathbf{R}\times\mathbf{R}\\
       (\rho, G, a, f, \mu_1, \mu_2) & \mapsto (I,\sigma)\,.
\end{aligned}
\end{equation}
Then
\begin{equation}
  (I,\sigma) = P S_0 (rS_0)^n (\rho, G_0, \bot, f, \bot, \bot)
\end{equation}
and a good refinement prescription~$r$, such as \program{VEGAS}, will minimize
the~$\sigma$.

For parallelization, it is crucial to find a division of~$S_n$ or any
part of it into \emph{independent} pieces that can be evaluated in
parallel.  In order to be effective, $r$ has to be applied to
\emph{all} of~$a$ and therefore a synchronization of~$G$ before and
after~$r$ is appropriate.  Furthermore, $r$ usually uses only a tiny
fraction of the overall CPU time and it makes little sense to invest a lot of
effort into parallelizing it.  On the other hand, $S_0$ can be
parallelized naturally, because all operations are linear, including
he computation of~$a$.  We only have to make sure that the cost of
communicating the results of~$S_0$ and~$r$ back and forth during the
computation of~$S_n$ do not offset any performance gain from parallel
processing.

When we construct a decomposition of~$S_0$ and proof that it does not
change the results, i.e.
\begin{equation}
  S_0 = \iota S_0 \phi
\end{equation}
or
\begin{equation}
  \begin{CD}
    \bigoplus_{i=1}^N G_i @>{\bigoplus_{i=1}^N S_0}>> \bigoplus_{i=1}^N G_i \\
    @A{\phi}AA                                        @V{\iota}VV           \\
    G                     @>S_0>>                     G
  \end{CD}\,,
\end{equation}
where~$\phi$ is a forking operation and~$\iota$ is a joining
operation, we are faced with the technical problem of a parallel
random number source~$\rho$.

As made explicit in~(\ref{eq:S0}), $S_0$ changes the state of the
random number generator~$\rho$. Demanding \emph{identical} results,
imposes therefore a strict ordering on the operations and defeats
parallelization.  In principle, it is possible to devise
implementations of~$S_0$ and~$\rho$ that circumvent this problem by
distributing subsequences of~$\rho$ in such a way among processes that
the results do not depend on the number of parallel processes.

However, a reordering of the random number sequence will only change
the result by an amount on the order of
the statistical sampling error, as long as the scale of the
allowed reorderings is \emph{bounded} and much smaller than the period
of the random number generator\footnote{Arbitrary reorderings on the
scale of the period of the pseudo random number generator have to be
forbidden, of course.}.  Below, we will
therefore use the notation $x\approx y$ for ``equal for an appropriate
finite reordering of the~$\rho$ used in calculating~$x$ and~$y$''.
For our purposes, the relation~$x\approx y$ is strong enough and
allows simple and efficient implementations.

\subsection{Multilinear Structure of the Sampling Algorithm}
\label{sec:multi-linear}

Since~$S_0$ is essentially a summation, it is natural to expect a
linear structure 
\begin{subequations}
\label{eq:S0-parallel}
\begin{equation}
  \bigoplus_i S_0(\rho_i, G_i, a_i, f, \mu_{1,i}, \mu_{2,i})
     \approx S_0 (\rho, G, a, f, \mu_1, \mu_2)
\end{equation}
where
\begin{align}
  \rho  &= \bigoplus_i \rho_i \\
  G     &= \bigoplus_i G_i \\
  a     &= \bigoplus_i a_i \\
  \mu_n &= \bigoplus_i \mu_{n,i}
\end{align}
\end{subequations}
for appropriate definitions of ``$\oplus$''. For the moments, we have
standard addition
\begin{equation}
  \mu_{n,1} \oplus \mu_{n,2} = \mu_{n,1} + \mu_{n,2}
\end{equation}
and since we only demand equality up to reordering, we only need that
the~$\rho_i$ are statistically independent.  This leaves us with~$G$
and~$a$ and we have to discuss importance sampling and stratified
sampling separately.

\subsubsection{Importance Sampling}
In the cases of naive Monte Carlo without grid optimization and of
importance sampling, the natural
decomposition of~$G$ is to take~$j$ copies of the same
grid~$G/j$ that are identical to~$G$, each with one $j$-th of the
total sampling points.  As long as the~$a$ are linear themselves, we
can add them up just like the moments
\begin{equation}
  a_1 \oplus a_2 = a_1 + a_2
\end{equation}
and we have found a decomposition~(\ref{eq:S0-parallel}).  In the
case of \program{VEGAS}, the~$a_i$ are sums of function values at the sampling
points.  Thus they are obviously linear and this approach is
applicable to \program{VEGAS} in the importance sampling mode.

\subsubsection{Stratified Sampling}
The situation is more complicated in the case of stratified sampling.
The first complication is that in pure stratified sampling there are
only two sampling points per cell.  Splitting the grid in two pieces
as above can provide only a very limited amount of parallelization.  The
second complication is that the~$a$ are no longer linear, since they
correspond to a sampling of the variance per cell and no longer of
the function values themselves.

However, as long as the samplings contribute to disjoint bins only, we
can still ``add'' the variances by combining bins.  The solution is
therefore to divide the grid into disjoint bins along the divisions of
the stratification grid and to assign a set of bins to each processor.

Finer decompositions will incur higher communications costs and other
resource utilization.  An implementation based on~PVM is described
in~\cite{Veseli:1998:Parallel-Vegas}, which minimizes the overhead by
running identical copies of the grid~$G$ on each processor.  Since
most of the time is usually spent in function evaluations, it makes
sense to run a full~$S_0$ on each processor, skipping function
evaluations everywhere but in the region assigned to the processor.
This is a neat trick, which is unfortunately tied to the computational
model of message passing systems such as~PVM and~MPI.  More
general paradigms can not be supported since the separation of the
state for the processors is not explicit (it is implicit in the
separated address space of the PVM or MPI processes).

However, it is possible to implement~(\ref{eq:S0-parallel}) directly
in an efficient manner.  This is based on the observation that the
grid~$G$ used by \program{VEGAS} is factorized into divisions~$D^j$ for each
dimension
\begin{equation}
\label{eq:factorize}
  G = \bigotimes_{j=1}^{n_{\text{dim}}} D^j
\end{equation}
and that decompositions of the~$D^j$ induce decompositions of~$G$
\begin{multline}
\label{eq:decomp}
  G_1 \oplus G_2
    = \left(
        \bigotimes_{j=1}^{i-1} D^j
          \otimes D^i_1 \otimes \bigotimes_{i=j+1}^{n_{\text{dim}}} D^j
      \right)
      \oplus
      \left(
        \bigotimes_{j=1}^{i-1} D^j
          \otimes D^i_2 \otimes \bigotimes_{i=j+1}^{n_{\text{dim}}} D^j
      \right) \\
    = \bigotimes_{j=1}^{i-1} D^j
        \otimes \left( D^i_1 \oplus D^i_2 \right)
        \otimes \bigotimes_{j=i+1}^{n_{\text{dim}}} D^j\,.
\end{multline}
We can translate~(\ref{eq:decomp}) directly to code that performs the
decomposition~$D^i = D^i_1 \oplus D^i_2$ discussed below and simply
duplicates the other divisions~$D^{j\not=i}$.  A decomposition along
multiple dimensions is implemented by a recursive application
of~(\ref{eq:decomp}).

In \program{VEGAS}, the auxiliary information~$a$ inherits a factorization
similar to the factorization~(\ref{eq:factorize}) of the grid
\begin{equation}
\label{eq:factorize'}
  a = (d^1,\ldots,d^{n_{\text{dim}}})
\end{equation}
but not a multilinear structure.  Instead, \emph{as long as the
decomposition respects the stratification grid}, we find the in place
of~(\ref{eq:decomp})
\begin{equation}
\label{eq:decomp'}
  a_1 \oplus a_2
    = (d^1_1 + d^1_2,\ldots, d^i_1 \oplus d^i_2, \ldots,
       d^{n_{\text{dim}}}_1 + d^{n_{\text{dim}}}_2)
\end{equation}
with ``$+$'' denoting the standard addition of the bin contents and
``$\oplus$'' denoting the aggregation of disjoint bins.  If the
decomposition of the division would break up cells of the
stratification grid, the equation~(\ref{eq:decomp'}) would be
incorrect, because, as discussed above, the variance is not linear.

Now it remains to find a decomposition
\begin{equation}
  D^i = D^i_1 \oplus D^i_2
\end{equation}
for both the pure stratification mode and the pseudo stratification
mode of \program{VEGAS} (see figures~\ref{fig:VEGAS-importance}
and~\ref{fig:VEGAS-stratified}).  In the pure stratification mode, the
stratification grid is strictly finer than the adaptive grid and we
can decompose along either of them immediately.  Technically, a
decomposition along the coarser of the two is straightforward.  Since
a typical adaptive grid already has more than 25~bins, a decomposition along
the stratification grid offers no advantages for medium scale
parallelization ($O(10)$ to~$O(100)$ processors) and the decomposition
along the adaptive grid has been implemented.  This scheme is
particularly convenient, because the sampling
algorithm~$S_0$ can be applied \emph{unchanged} to the individual
grids resulting from the decomposition.

\begin{figure}
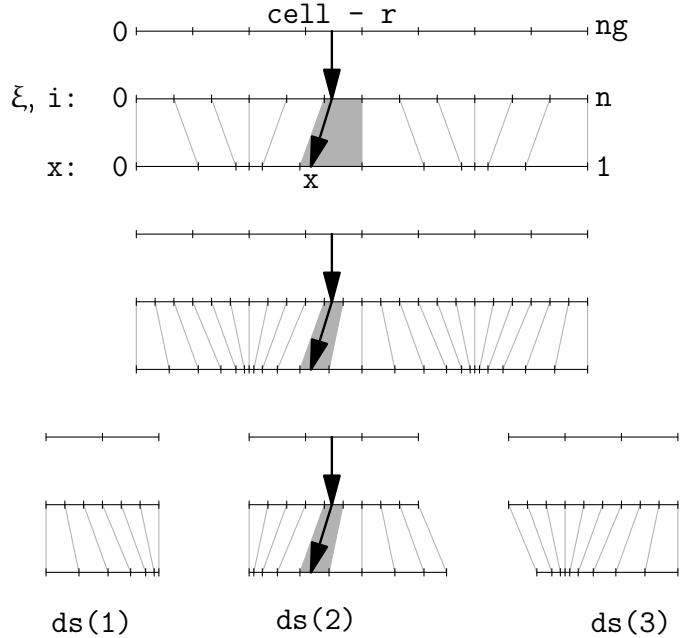

  \begin{center}
    \begin{emp}(120,90)
       pseudo (.3w, .8w, .7h, .9h, 0, 8, 8,  0, 12, 12, 5.2,   true, true);
       pseudo (.3w, .8w, .4h, .6h, 0, 8, 8,  0, 24, 24, 5.2*2, false, true);
       pseudo (.2w, .7w, .1h, .3h, 0, 2, 8,  0,  6, 24, 5.2*2, false, false);
       pseudo (.3w, .8w, .1h, .3h, 2, 5, 8,  6, 15, 24, 5.2*2, false, true);
       pseudo (.4w, .9w, .1h, .3h, 5, 8, 8, 15, 24, 24, 5.2*2, false, false);
       label.urt (btex \texttt{ds(1)} etex, (.2w, 0));
       label.top (btex \texttt{ds(2)} etex, (.5w, 0));
       label.ulft (btex \texttt{ds(3)} etex, (.9w, 0));
    \end{emp}
  \end{center}
  \caption{\label{fig:pseudo-fork}%
    Forking one dimension~\texttt{d} of a grid into three parts
    \texttt{ds(1)}, \texttt{ds(2)}, and~\texttt{ds(3)}.  The picture
    illustrates the most complex case of pseudo stratified sampling
    (see figure~\protect\ref{fig:VEGAS-pseudo-stratified}).}
\end{figure}

For pseudo stratified sampling (see figure~\ref{fig:VEGAS-pseudo-stratified}), the
situation is more complicated, because the adaptive and the
stratification grid do not share bin boundaries.  Since \program{VEGAS} does
\emph{not} use the variance in this mode, it would be theoretically
possible to decompose along the adaptive grid and to mimic the
incomplete bins of the stratification grid in the sampling algorithm.
However, this would cause technical complications, destroying the
universality of~$S_0$.  Instead, the adaptive grid is subdivided in
a first step in
\begin{equation}
  \mathop{\textrm{lcm}}
     \left( \frac{\mathop{\textrm{lcm}}(n_f,n_g)}{n_f}, n_x \right)
\end{equation}
bins\footnote{The coarsest grid covering the division of~$n_g$ bins
into~$n_f$ forks has $n_g / \mathop{\textrm{gcd}}(n_f,n_g) =
\mathop{\textrm{lcm}}(n_f,n_g) / n_f$ bins per fork.}. such that the
adaptive grid is strictly finer than the stratification grid.  This
procedure is shown in figure~\ref{fig:pseudo-fork}.
\begin{figure}
  \begin{equation*}
    \begin{CD}
      \parbox{35mm}{\includegraphics{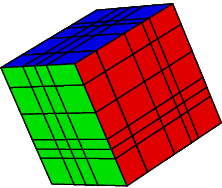}}
         @>{\text{fork along 3}}>>
            \parbox{35mm}{\includegraphics{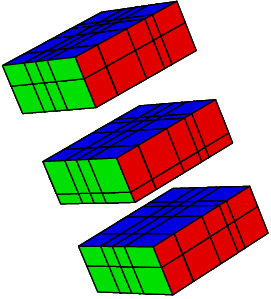}}\\
      @. 
        @V{\text{fork along 2}}VV           \\
      \parbox{35mm}{\includegraphics{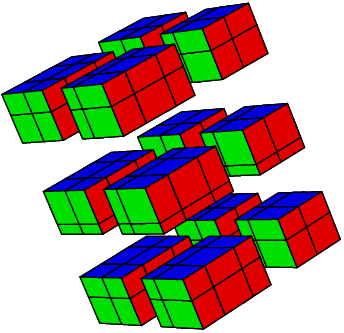}}
         @<{\text{fork along 1}}<<
            \parbox{35mm}{\includegraphics{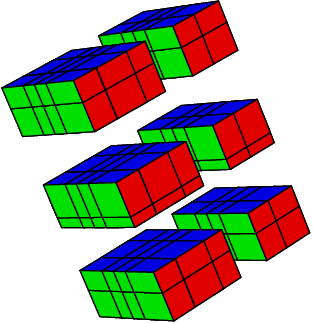}}
    \end{CD}
  \end{equation*}
  \caption{\label{fig:fork3}%
    Recursively fork a $6\times5\times4$-grid for distribution among
    $12 = 3\times2\times2$ processors.}
\end{figure}
The code for forking and duplicating a single dimension suffices as
building block for general recursion~\cite{Ohl:1998:VAMP}, as
illustrated in figure~\ref{fig:fork3}.

\subsection{Random Numbers}

The implementation~\cite{Ohl:1998:VAMP} of this procedure takes
advantage of the ability of Knuth's new preferred
generator~\cite{Knuth:1998:TAOCP2} to generate \emph{provable}
statistically independent subsequences.  However, since the state of
the random number generator is explicit in all procedure calls, other
means of obtaining subsequences can be implemented in a trivial
wrapper.

The results of the parallel example will depend on the number of
processors, because this effects the subsequences being used.  Of
course, the variation will be compatible with the statistical error.
It must be stressed that the results are deterministic for a given
number of processors and a given set of random number generator seeds.
Since parallel computing environments allow to fix the number of
processors, debugging of exceptional conditions remains therefore possible.

\section{Dedicated Phase Space Algorithms}
\label{sec:rambo}

No description of Monte Carlo algorithms for phase space integration
would be complete without mentioning the beautiful \textit{RAMBO}
algorithm~\cite{Kleiss/Stirling/Ellis:1986:RAMBO} for massless
$n$-particle phase space. RAMBO proceeds by generating massless
momentum vectors isotropically, without regard to energy or momentum
conservation. In a second step, these vectors are boosted with a
Lorentz transformation to their common rest frame to enforce momentum
conservation and the lengths of massless vectors can be
rescaled by a common factor to ensure energy conservation without
spoiling momentum conservation.  It can be
shown~\cite{Kleiss/Stirling/Ellis:1986:RAMBO} that this transformation
has a unit Jacobian.  Therefore the phase space distribution can be
generated without a rejection step.  The only obvious additional cost is
that a $n$-particle final state uses $4n$~random numbers for
$3n-4$~degrees of freedom.

Unfortunately, the transformation from the $4n$~random numbers to the
$3n-4$~independent variables has a substantial hidden cost: the
relation of the integration variables to the kinematical variables in
which matrix elements have singularities is lost and adaptive Monte
Carlo algorithms are almost useless.  For this reason, RAMBO is mainly
useful for smooth background distributions.
There is a modification (called \textit{MAMBO}) for light massive
particles~\cite{Kleiss/Stirling:1992:MAMBO}, but the Jacobians are now
no longer unity and the resulting event weights make the algorithm
not very efficient.

\chapter{Linear Colliders}
\label{sec:lc}

\begin{cutequote*}{Ingeborg Bachmann}{B\"ohmen liegt am Meer}
  Wie B\"ohmen sie bestand,
  das einen sch\"onen Tags zum Meer begnadigt wurde\\
  und jetzt am Wasser liegt.
\end{cutequote*}

The marquee measurement in the area of gauge boson physics at a linear
collider is without a doubt the precision determination of their
self-couplings, known as~\ac{TGC}.  As has been stressed in
chapter~\ref{sec:sm/eft}, the measurements a
LEP2~\cite{LEPEWWG:1999:global} (see
figure~\ref{fig:tgc-vancouver98}), have been a very important
confirmation, but can not have been surprising, because a consistent
description of the low energy data leaves little room for deviations
from the \ac{SM}~predictions at a level that would be detectable at
LEP2~\cite{Gounaris/etal:1996:LEP2-TGC}.

As will be shown in section~\ref{sec:tgc}, this is qualitatively
different at a~\ac{LC}, because higher energy and luminosity allow to
probe the couplings at a level where deviations due to different
realizations of~\ac{EWSB} are possible.

\section{Triple Gauge Couplings}
\label{sec:tgc}

The most popular parametrization of the most general Lorentz invariant
$W^+W^-Z^0$- and $W^+W^-\gamma$-vertices has been proposed by Hagiwara
et~al.~\cite{Hagiwara/etal:1987:TGC}
\begin{multline}
\label{eq:hagiwara}
  L_{WWV} / g_{WWV} = \\
     i g_1^V   \left( W_{\mu\nu}^\dagger W^\mu V^\nu
                       - W_\mu^\dagger V_\nu W^{\mu\nu} \right)
     + i \kappa_V W_\mu^\dagger W_\nu V^{\mu \nu }
     + i \frac{\lambda_V}{M_W^2}
           W_{\lambda\mu}^\dagger W^\mu_{\hphantom{\mu}\nu} V^{\nu\lambda} \\
     - g_4^V W_\mu^\dagger W_\nu
            \left( \partial^\mu V^\nu + \partial^\nu V^\mu \right)
     + g_5^V \epsilon^{\mu\nu\lambda\rho}
            \left( W_\mu^\dagger \partial_\lambda W_\nu -
                   \partial_\lambda W_\mu^\dagger W_\nu \right) V_\rho \\
     + i \tilde \kappa_V W_\mu^\dagger W_\nu \tilde V^{\mu\nu}
     + i \frac{\tilde\lambda_V}{M_W^2} W_{\lambda\mu}^\dagger
             W^\mu_{\hphantom{\mu}\nu} \tilde V^{\nu\lambda}\,,
\end{multline}
where~$V$ is either~$Z^0$ or~$\gamma$, while $g_{WW\gamma}=-e$
and~$g_{WWZ}=-e\cot\theta_w$ are the \ac{SM} couplings.  All
field-strengths tensors in~(\ref{eq:hagiwara}) are 
abelian $V_{\mu\nu}=\partial_\mu V_\nu-\partial_\nu V_\mu$ with the
tilde denoting the dual tensor. At tree level in the~\ac{SM}, the
parameters in~(\ref{eq:hagiwara}) take the values 
\begin{subequations}
\begin{align}
  \Delta\kappa_{\gamma,Z} = \kappa_{\gamma,Z} - 1 & = 0\\
  \Delta g_1^{\gamma,Z} = g_1^{\gamma,Z} - 1 &= 0 \\
  \lambda_{\gamma,Z} & = 0 \\
  \tilde \lambda_{\gamma,Z} = \tilde \kappa_{\gamma,Z} =
     g_4^{\gamma,Z} = g_5^{\gamma,Z} &= 0\,.
\end{align}
\end{subequations}
Electromagnetic~$\textrm{U}(1)_Q$ gauge invariance
dictates~$g_{1,5}^\gamma=0$ at zero momentum transfer.
The parametrization~(\ref{eq:hagiwara}) is calculationally convenient,
because the parameters translate directly to coefficients of the same
size in the Feynman rules.  Therefore, the matrix elements used in
most Monte Carlo event generators are expressed in terms
of~(\ref{eq:hagiwara}).

Another advantage of the parameters in~(\ref{eq:hagiwara}) is that
they translate immediately, without unnaturally small or large
coefficients, to model independent physical observables for the
electromagnetic properties of charged vector bosons, such as the
magnetic dipole
\begin{subequations}
\begin{align}
  \mu_W &= \frac{e}{2m_W}
     \Bigl(g_1^\gamma + \kappa_\gamma + \lambda_\gamma\Bigr)\,, \\
\intertext{electric quadrupole}
  Q^e_W &= -\frac{e}{m_W^2}
     \Bigl(\kappa_\gamma - \lambda_\gamma\Bigr)\,, \\
\intertext{electric dipole}
  d_W   &= \frac{e}{2m_W}
     \Bigl(\tilde\kappa_\gamma + \tilde\lambda_\gamma\Bigr)\,, \\
\intertext{and magnetic quadrupole}
  Q^m_W &= -\frac{e}{m_W^2}
     \Bigl(\tilde\kappa_\gamma - \tilde\lambda_\gamma\Bigr)
\end{align}
\end{subequations}
moments of
$W^\pm$~bosons~\cite{Durand/etal:1962:moments,*Kim/Tsai:1973:moments}.

On the other hand, (\ref{eq:hagiwara})~does not offer any indication
to the probable size of the individual contributions.  Completely
model independent fits of all 14~parameters simultaneously are not
possible, even at a high luminosity~\ac{LC}. Furthermore, the
parameters in~(\ref{eq:hagiwara}) are not necessarily constant and
could be polynomials in the invariant masses or even general form
factors.  Obviously, $g_5^V$, $\tilde\kappa_V$, and~$\tilde\lambda_V$
are $CP$-violating couplings and their study could be postponed.

A more systematic approach starts from a
basis~\cite{Buchmueller/Wyler:1986:operators} for the operators that
can appear in
an~$\textrm{SU}(2)_L\otimes\textrm{U}(1)_Y/\textrm{U}(1)_Q$ effective
Lagrangian and classifies them according to their order in the
momentum expansion.  For non-linear realizations, \ac{NDA} provides
the necessary hierarchy.  A systematic study can then start from the
lowest dimensional operators and can incorporate symmetries like the
custodial~$\textrm{SU}(2)_c$, which is strongly suggested by the low
energy electroweak precision data.

\begin{figure}
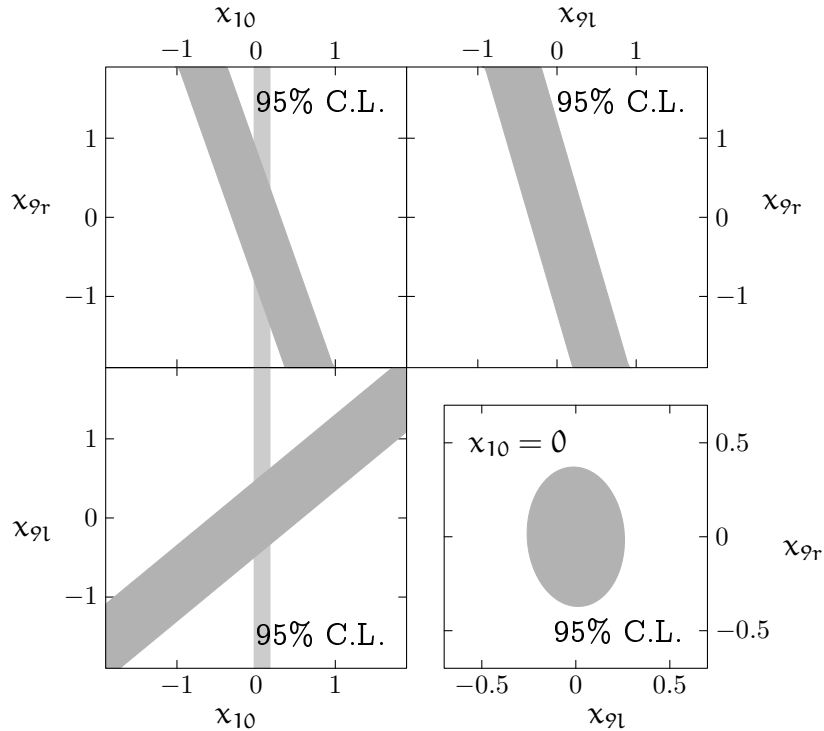

  \begin{center}
    \begin{empcmds}
      Gtemplate.itick := origin--(3bp,0);
      Gtemplate.otick := (-3bp,0)--origin;
      Gmarks := 3;
    \end{empcmds}
    \begin{empcmds}
      width = 40mm;
      height = 40mm;
      dwidth = 5mm;
      dheight = dwidth;
    \end{empcmds}
    \begin{empcmds}
      rotlab = 0;
      xrange = 1.9;
      yrange = 1.9;
      xxrange = 0.7;
      yyrange = 0.7;
    \end{empcmds}
    \begin{empcmds}
      vardef oblique (expr xa, xb) =
        (xa,yrange)--(xa,-yrange)--(xb,-yrange)--(xb,yrange)--cycle
      enddef;
    \end{empcmds}
    \begin{empcmds}
      vardef yoblique (expr xa, xb) =
        numeric xax, xbx;
        xax = max (xa,-xxrange);
        xbx = min (xb,xxrange);
        (xax,yyrange-0.2)--(xax,-yyrange-0.2)--(xbx,-yyrange-0.2)--(xbx,yyrange-0.2)--cycle
      enddef;
    \end{empcmds}
    \begin{emp}(110,90)
      pickup pencircle scaled 0.5pt;
      draw begingraph (width, height);
        setrange (-xrange, -yrange, xrange, yrange);
        Autoform := "
        Autoform := "";   autogrid (itick.bot,itick.rt);
        path zy; gdata ("liz_zy", $, augment.zy ($1, $2);)
        gfill oblique (-0.025,0.175) withcolor 0.8white;
        gfill zy--cycle withcolor 0.7white;
        glabel.llft (btex 95\%{} C.L. etex, (1.7,1.7));
        glabel.lft (btex $x_{9r}$ etex rotated rotlab, OUT);
        glabel.top (btex $x_{10}$ etex, OUT);
      endgraph shifted (0, height);
      draw begingraph (width, height);
        setrange (-xrange, -yrange, xrange, yrange);
        Autoform := "
        Autoform := "";   autogrid (itick.bot,itick.lft);
        path xy; gdata ("liz_xy", $, augment.xy ($1, $2);)
        gfill xy--cycle withcolor 0.7white;
        glabel.llft (btex 95\%{} C.L. etex, (1.7,1.7));
        glabel.top (btex $x_{9l}$ etex, OUT);
        glabel.rt (btex $x_{9r}$ etex rotated -rotlab, OUT);
      endgraph shifted (width, height);
      draw begingraph (width, height);
        setrange (-xrange, -yrange, xrange, yrange);
        Autoform := "
        Autoform := "";   autogrid (itick.top,itick.rt);
        path zx; gdata ("liz_zx", $, augment.zx ($1, $2);)
        gfill oblique (-0.025,0.175) withcolor 0.8white;
        gfill zx--cycle withcolor 0.7white;
        glabel.ulft (btex 95\%{} C.L. etex, (1.7,-1.7));
        glabel.bot (btex $x_{10}$ etex, OUT);
        glabel.lft (btex $x_{9l}$ etex rotated rotlab, OUT);
      endgraph;
      draw begingraph (width-dwidth, height-dheight);
        setrange (-xxrange, -yyrange, xxrange, yyrange);
        Autoform := "
        Autoform := "";   autogrid (itick.top,itick.lft);
        path zx'; gdata ("liz_xy0", $, augment.zx' ($1, $2);)
        gfill zx'--cycle withcolor 0.7white;
        glabel.ulft (btex 95\%{} C.L. etex, (0.6,-0.6));
        glabel.lrt (btex $x_{10} = 0$ etex, (-0.6,0.6));
        glabel.bot (btex $x_{9l}$ etex, OUT);
        glabel.rt (btex $x_{9r}$ etex rotated -rotlab, OUT);
      endgraph shifted (width+dwidth, 0);
    \end{emp}
  \end{center}
  \caption{\label{fig:x9/10}%
    Sensitivity on the parameters of a non-linearly
    realized \protect\acs{EWSB}~sector
    with~$\int\mathcal{L}=\unit[200]{{fb}^{-1}}$
    and~$\sqrt s = \unit[800]{GeV}$~\protect\cite{Nippe/Ohl:1996:TGV}.
    The vertical bars indicate the constraint on $x_{10}$ derived from
    LEP1 analyses~\protect\cite{Erler/Langacker:1995:x10}. The blind
    direction $(2\sin^2\theta_w, -2\cos^2\theta_w,
    \cos^2\theta_w-\sin^2\theta_w)$ does not contribute to the triple
    gauge couplings.}
\end{figure}
\begin{figure}
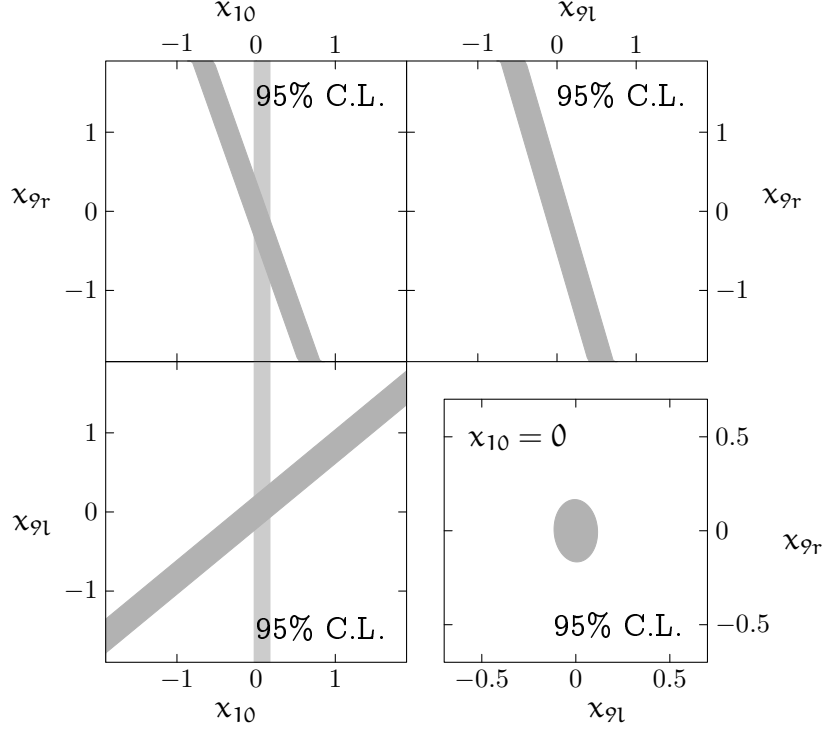

  \begin{center}
    \begin{emp}(110,90)
      pickup pencircle scaled 0.5pt;
      draw begingraph (width, height);
        setrange (-xrange, -yrange, xrange, yrange);
        Autoform := "
        Autoform := "";   autogrid (itick.bot,itick.rt);
        path zy; gdata ("liz_zy_ab", $, augment.zy ($1, $2);)
        gfill oblique (-0.025,0.175) withcolor 0.8white;
        gfill zy--cycle withcolor 0.7white;
        glabel.llft (btex 95\%{} C.L. etex, (1.7,1.7));
        glabel.lft (btex $x_{9r}$ etex rotated rotlab, OUT);
        glabel.top (btex $x_{10}$ etex, OUT);
      endgraph shifted (0, height);
      draw begingraph (width, height);
        setrange (-xrange, -yrange, xrange, yrange);
        Autoform := "
        Autoform := "";   autogrid (itick.bot,itick.lft);
        path xy; gdata ("liz_xy_ab", $, augment.xy ($1, $2);)
        gfill xy--cycle withcolor 0.7white;
        glabel.llft (btex 95\%{} C.L. etex, (1.7,1.7));
        glabel.top (btex $x_{9l}$ etex, OUT);
        glabel.rt (btex $x_{9r}$ etex rotated -rotlab, OUT);
      endgraph shifted (width, height);
      draw begingraph (width, height);
        setrange (-xrange, -yrange, xrange, yrange);
        Autoform := "
        Autoform := "";   autogrid (itick.top,itick.rt);
        path zx; gdata ("liz_zx_ab", $, augment.zx ($1, $2);)
        gfill oblique (-0.025,0.175) withcolor 0.8white;
        gfill zx--cycle withcolor 0.7white;
        glabel.ulft (btex 95\%{} C.L. etex, (1.7,-1.7));
        glabel.bot (btex $x_{10}$ etex, OUT);
        glabel.lft (btex $x_{9l}$ etex rotated rotlab, OUT);
      endgraph;
      draw begingraph (width-dwidth, height-dheight);
        setrange (-xxrange, -yyrange, xxrange, yyrange);
        Autoform := "
        Autoform := "";   autogrid (itick.top,itick.lft);
        path zx'; gdata ("liz_xy0_ab", $, augment.zx' ($1, $2);)
        gfill zx'--cycle withcolor 0.7white;
        glabel.ulft (btex 95\%{} C.L. etex, (0.6,-0.6));
        glabel.lrt (btex $x_{10} = 0$ etex, (-0.6,0.6));
        glabel.bot (btex $x_{9l}$ etex, OUT);
        glabel.rt (btex $x_{9r}$ etex rotated -rotlab, OUT);
      endgraph shifted (width+dwidth, 0);
    \end{emp}
  \end{center}
  \caption{\label{fig:x9/10_ab}%
    The sensitivity ellipses of figure~\protect\ref{fig:x9/10}
    rescaled to the new high luminosity TESLA
    with~$\int\mathcal{L}=\unit[1]{{ab}^{-1}}$.}
\end{figure}
The operators of dimension four in the momentum expansion contributing
to the~\ac{TGC}
are~\cite{Falk/etal:1991:anomalous}
\begin{multline}
\label{eq:liz}
  L = - i\frac{x_{9l}}{16\pi^2} \mathop{\mathrm{tr}}
          \Bigl( g W^{\mu\nu} D_\mu U^\dagger D_\nu U \Bigr) \\
      - i\frac{x_{9r}}{16\pi^2} \mathop{\mathrm{tr}}
          \Bigl( g' B^{\mu\nu} D_\mu U^\dagger D_\nu U \Bigr)
      + \frac{x_{10}}{16\pi^2} \mathop{\mathrm{tr}}
          \Bigl( U^\dagger g' B_{\mu\nu} U g W^{\mu\nu} \Bigr)\,.
\end{multline}
The contributions of the interaction~(\ref{eq:liz}) to the~\ac{TGC}
can be expressed in terms of the
general parametrization~(\ref{eq:hagiwara}) by comparing
coefficients\footnote{The erroneous non-singular transformation
formula in~\cite{Falk/etal:1991:anomalous} is corrected for the
special case $x_{9l}=x_{9r}$ in~\cite{Holdom:1991:anomalous} and for
the general case in~\cite{Appelquist/Wu:1993:anomalous}.} of the
trilinear terms
\begin{subequations}
\label{eq:liz2hagi}
\begin{align}
  g_1^Z &= 1 - \frac{e^2}{2\sin^2\theta_w\cos^2\theta_w}
               \frac{x_{9l}}{16\pi^2}               
             + \frac{e^2}{\cos^2\theta_w(\cos^2\theta_w-\sin^2\theta_w)}
               \frac{x_{10}}{16\pi^2} \\
  \kappa_Z &= 1 - \frac{e^2}{2\sin^2\theta_w}\frac{x_{9l}}{16\pi^2}
                + \frac{e^2}{2\cos^2\theta_w}\frac{x_{9r}}{16\pi^2}
                + \frac{2e^2}{\cos^2\theta_w\sin^2\theta_w}
                  \frac{x_{10}}{16\pi^2}\\
  \kappa_\gamma &= - \frac{e^2}{2\sin^2\theta_w}\frac{x_{9l}}{16\pi^2}
                   - \frac{e^2}{2\sin^2\theta_w}\frac{x_{9r}}{16\pi^2}
                   - \frac{e^2}{\sin^2\theta_w}\frac{x_{10}}{16\pi^2}\,.
\end{align}
\end{subequations}
Unfortunately, the transformation~(\ref{eq:liz2hagi}) is singular and
the blind direction
\begin{equation}
\label{eq:blind}
  \begin{pmatrix} x_{9l} \\ x_{9r} \\ x_{10} \end{pmatrix}_{\text{blind}}
    \propto
  \begin{pmatrix}
     2\sin^2\theta_w \\
     -2\cos^2\theta_w \\
     \cos^2\theta_w-\sin^2\theta_w
  \end{pmatrix}
\end{equation}
in the parameter space can not be observed in measurements of
the~\ac{TGC} alone.  However, LEP1 analyses
(e.\,g.~\cite{Erler/Langacker:1995:x10}) provide independent
constraints on~$x_{10}$, which can lift the degeneracy.

The coefficients in~(\ref{eq:liz}) have been normalized such
that their natural size according to~\ac{NDA} (see
sections~\ref{sec:msm} and~\ref{sec:NDA}) is of order one.
Consequently, the all important question from a physics perspective is:
\emph{Are experiments at a~\ac{LC} sensitive to values for $x_{9l}$,
$x_{9r}$, $x_{10}$ of order one?}

Detector effects appear to be small and preliminary studies can
therefore work on the generator level with simple geometrical
acceptance cuts~\cite{Nippe:1996:generator_level,
*Burgard/Petzold:1999:generator_level}.  Much more important than
detector effects are the distortions of the distributions from hard,
almost collinear initial state photon radiation, when the photon is
lost in the beam-pipe.  Since these effects cause a migration of
events in the phase space, simple acceptance corrections do not
suffice and more complicated unfolding procedures are required.

In one study~\cite{Nippe/Ohl:1996:TGV} (see
also~\cite{Accomando/etal:1998:TESLAReview}) the unfolding has been
performed by generating a large sample of Monte Carlo events (ten
times the size of the expected data sample at TESLA) with
\program{WOPPER}~\cite{Anlauf/etal:1994:wopper1.0,
*Anlauf/etal:1994:wopper1.1,*Anlauf/etal:1996:wopper1.5} and summing
over all radiative and non-radiative events with the same experimental
signature.  The events were subsequently reweighted with
anomalous~\ac{TGC} using fast matrix elements
of~\program{WPHACT}~\cite{Accomando/Ballestrero:1997:WPHACT} to obtain
the likelihood function for the~\ac{TGC}.  Using a quadratic
approximation of the likelihood function for the ``old'' high
luminosity option at TESLA with
$\int\mathcal{L}=\unit[200]{{fb}^{-1}}$ and $\sqrt s =
\unit[800]{GeV}$, the sensitivities in figure~\ref{fig:x9/10} have
been derived.  A more careful analysis will have to take
non-linearities in the likelihood function into account\footnote{The
non-linearities had been neglected because the unfolding would have
exceeded the limited computing resources in a non-linear fit.}, but
they can not change the conclusion that \emph{the \ac{LC} is
definitely sensitive to values for $x_{9l}$, $x_{9r}$, $x_{10}$ of
order one,} if constraints on~$x_{10}$ are taken into account.

Increasing the integrated luminosity according to the ``new'' TESLA
high luminosity option by a factor of five to
$\int\mathcal{L}=\unit[1]{{ab}^{-1}}$, as shown in
figure~\ref{fig:x9/10_ab}, shows that such a machine would allow
precision tests of the~\ac{TGC}, at the level of $10\%$~of reasonable
deviations from the~\ac{SM}, caused by different realizations of the
\ac{EWSB}~sector.


\section{Single $W^\pm$ Production}
\label{sec:singleW}

At energies that are high compared to the masses of all participating
particles, all cross sections with a finite limit for massless
particles have to fall off as dictated by dimensional
analysis
\begin{subequations}
\begin{equation}
\label{eq:1/s}
  \sigma(s) \propto \frac{1}{s}\,.
\end{equation}
Cross sections that have contributions from the exchange of light
particles in the $t$-channel do not necessarily have a finite limit
and can diverge in the forward direction if the mass of the exchanged
particle is send to zero.  Such cross sections can scale like
\begin{equation}
\label{eq:log(s)}
  \sigma(s) \propto \frac{1}{\mu^2} \ln\left(\frac{s}{\mu^2}\right)\,,
\end{equation}
\end{subequations}
where $\mu^2$~denotes a typical kinematical scale set by a minimum
momentum transfer.

\begin{figure}
  \begin{center}
    \hfil\\
    \begin{fmfgraph*}(40,30)
      \fmfleft{ep,em}
      \fmfright{nubar,u,dbar,em'}
      \fmflabel{$e^+$}{ep}
      \fmflabel{$e^-$}{em}
      \fmflabel{$\bar\nu$}{nubar}
      \fmflabel{$e^-$}{em'}
      \fmflabel{$u$}{u}
      \fmflabel{$\bar d$}{dbar}
      \fmf{fermion,tension=2}{em,vem}
      \fmf{fermion}{vem,em'}
      \fmf{fermion}{nubar,vep}
      \fmf{fermion,tension=2}{vep,ep}
      \fmf{photon,label=$\gamma$,label.side=left}{vem,tgc}
      \fmf{dbl_plain_arrow,label=$W^+$,label.side=left}{vep,tgc,vud}
      \fmf{fermion,tension=0.5}{dbar,vud,u}
      \fmfdot{vem,vep,vud,tgc}
    \end{fmfgraph*}\\
    \hfil
  \end{center}
  \caption{\label{fig:singleW}%
    ``Single-$W$'' contribution to~$e^+e^-\to e^-\bar\nu u\bar d$.}
\end{figure}
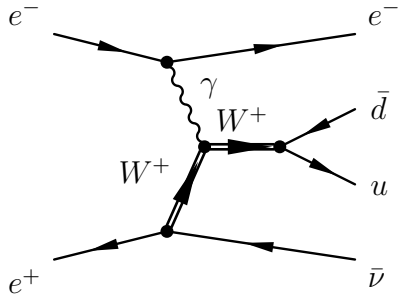
For the $t$-channel neutrino exchange in $W^\pm$~pair production, the
forward singularity is cut off kinematically by the $W^\pm$-mass and
the cross section is still dominated by~(\ref{eq:1/s}) in the
$\unit{TeV}$-region.  The situation is different for ``single-$W$''
production as shown in figure~\ref{fig:singleW}, because the electron
or positron that goes undetected in the forward direction does not
provide a kinematical cut-off.  The observable cross section is still
small at LEP2, but it will be larger at
a~\ac{LC} than that of $W^\pm$~pair production.

A unique physics opportunity of single-$W$ production in the context
of~\ac{TGC} is that, unlike $W^\pm$~pair production, the diagram in
figure~\ref{fig:singleW} is sensitive to the $W^+W^-\gamma$-couplings
alone, because the $Z^0$-exchange is not singular.  Therefore it can
be used to disentangle $W^+W^-Z^0$- and $W^+W^-\gamma$-couplings
without studying $\gamma\gamma$-collisions in a dedicated collider.

For LEP2, $W^\pm$~pair production has been studied
extensively~\cite{Beenakker/Berends/etal:1996:LEP2-WW} and many
reliable Monte Carlo event generators have been
prepared~\cite{Bardin/Kleiss/etal:1996:LEP2-WWMC}.  A more recent
dedicated comparison of theoretical predictions for single-$W$
production has revealed that the situation is less favorable for
single-$W$ production.  At \ac{LC} energies, unacceptable variations
of the predictions for the total cross section on the order of up to
five percent have been
reported~\cite{Ballestrero:1999:singleW,Boos:1999:singleW}.  It
appears that these discrepancies can be traced back to incomplete
implementations of effects from the finite electron mass.

The lesson to be learned is that a simple translation of the electron
mass into a lower cut-off on the electron scattering angle will not 
suffice at a~\ac{LC} and more complete calculations in the forward 
region are required, if single-$W$ production is to be used for
precision physics, like the disentangling of $W^+W^-Z^0$- and
$W^+W^-\gamma$-couplings.

The classification of the gauge invariant classes of Feynman diagrams
for single-$W$ production (see figure \ref{fig:cc20}) allows to
improve the handling of radiative corrections.  The single-$W$ and
$W^\pm$~pair production contributions to the same four fermion final
state can be folded with a separate radiator function, without spoiling
gauge invariance.  Since the two classes have very different
characteristic scales and the interferences are small, this improves
the predictions considerably.

\section{Beamstrahlung}
\label{sec:circe}

Another issue that differentiates a~\ac{LC} from LEP2 is the
non-trivial beam energy spectrum caused by \emph{beamstrahlung.}
Unlike initial state radiation, beamstrahlung is not characterized by
the scale of the hard interaction, but by the collective interaction
of the crossing bunches, which must be much denser than the bunches in
storage rings to obtain sufficient luminosity.  Therefore,
beamstrahlung is a more complex physical phenomenon, but it is
independent of the hard interaction under study and can be
parametrized in a single universal beam spectrum for each collider
design at each beam energy.

The comparison of analytical
approximations~\cite{Chen/Yokoya:1988:Beamstrahlung,*Chen:1992:Beamstrahlung}
with the results of microscopic
simulations~\cite{Chen/etal:1995:CAIN,Schulte:1996:Guinea_Pig} reveals
that the former are not adequate.  A successful pragmatical solution
is to parametrize the results of the microscopic simulations with a
simple factorized \emph{ansatz}~\cite{Ohl:1997:Circe}
\begin{equation}
  D_{p_1p_2} (x_1,x_2,s) = d_{p_1} (x_1)  d_{p_2} (x_2)
\end{equation}
where~$x_{e^\pm,\gamma}=E_{e^\pm,\gamma}/E_{\text{beam}}$ and
\begin{subequations}
\label{eq:beta}
\begin{align}
  d_{e^\pm} (x) &= a_0 \delta(1-x) + a_1 x^{a_2} (1-x)^{a_3} \\
  d_\gamma (x) &= a_4 x^{a_5} (1-x)^{a_6}\,.
\end{align}
\end{subequations}
For~$a_{0,1,4}\ge0$ and~$a_{2,3,5,6}\ge-1$, the ansatz~(\ref{eq:beta})
corresponds to a positive and integrable distribution can reproduce
the soft singularities from multiple emission at~$x_{e^\pm}\to1$
and~$x_\gamma\to0$ for~$a_{3,5}<0$.

The ansatz~(\ref{eq:beta}) is able to fit the results of microscopic
simulations well~\cite{Ohl:1997:Circe}.  Fits for the latest collider
parameter sets are available as distributions and as efficient random
number generator in the library~\Kirke/~\cite{Ohl:circe}.

\section{Quadruple~Gauge~Couplings and $W^\pm$~Scattering}
\label{sec:qgc}

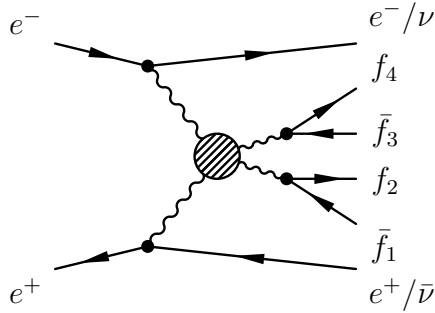
\begin{figure}
  \begin{center}
    \begin{fmfgraph*}(40,30)
      \fmfleft{ep,em}
      \fmfright{nub,f1,f2,f3,f4,nu}
      \fmflabel{$e^-$}{em}
      \fmflabel{$e^+$}{ep}
      \fmflabel{$e^+/\bar\nu$}{nub}
      \fmflabel{$e^-/\nu$}{nu}
      \fmflabel{$\bar f_1$}{f1}
      \fmflabel{$f_2$}{f2}
      \fmflabel{$\bar f_3$}{f3}
      \fmflabel{$f_4$}{f4}
      \fmf{fermion,t=3}{em,wm}
      \fmf{fermion}{wm,nu}
      \fmf{fermion}{nub,wp}
      \fmf{fermion,t=3}{wp,ep}
      \fmf{boson}{wp,v,wpd}
      \fmf{boson}{wm,v,wmd}
      \fmf{fermion,t=0.5}{f1,wmd,f2}
      \fmf{fermion,t=0.5}{f3,wpd,f4}
      \fmfdot{wm,wp,wmd,wpd}
      \fmfblob{(.15w)}{v}
    \end{fmfgraph*}
  \end{center}
  \caption{\label{fig:VV->VV}%
    Vector boson scattering subprocess in six-fermion production.}
\end{figure}

While LEP2 is, for gauge boson physics, essentially a four fermion
production machine, a~\ac{LC} can also produce final states with six
or more fermions, which have contributions from quadruple gauge boson
couplings.  In addition to the coupling of an $s$-channel
$Z^0$~or~$\gamma$ to three vector bosons, there are the kinematical
configurations corresponding to the fusion of almost on-shell vector
bosons, as shown in figure~\ref{fig:VV->VV}.  This part of the phase
space is particularly attractive for studying the interactions of
strongly interacting vector bosons.

Realistic estimates for the physics reach in this channl have already
been obtained 
using an improved equivalent particle approximation for the
``production'' of the vector bosons in the intermediate
state~\cite{Boos/etal:1998:WWWW} (see
also~\cite{Accomando/etal:1998:TESLAReview}).  Interesting physics can
be probed, but a high luminosity option is required.

The construction of a complete unweighted event generator
for~$e^+e^-\to6f$, including general gauge boson couplings, is still a
challenge.  Complete calculations for selected channels
exist~\cite{Accomando/etal:1997:6f,Gangemi/etal:1998:6f,
*Gangemi/etal:1999:6f}, which are expected to be
reliable in the regions of the phase space corresponding to top pair
production and Higgs production.  More work is required for a proper
treatment of the part of the phase space that corresponds to the
scattering of almost on-shell $W^\pm$'s as in figure~\ref{fig:VV->VV}.
Currently, work is underway that uses the Monte Carlo methods
described in section~\ref{sec:vamp} to bridge this gap.

\chapter{Conclusions}
\label{sec:conclusion}

\begin{cutequote}{Albert Michelson}
    While it is never safe to say that the future of Physical
    Science has no marvels even more astonishing than those of the
    past, it seems probable that most of the grand underlying
    principles have been firmly established and that further advances
    are to be sought chiefly in the rigorous applications of these
    principles to all the phenomena which come under our notice.  It
    is here that the science of measurement shows its
    importance---where quantitative results are more to be desired than
    qualitative work.  An eminent physicist has remarked that the
    future truths of Physical Science are to be looked for in the sixth
    place of decimals.
\end{cutequote}

There is ample model independent evidence from low energy data
that weak interactions exchange vector and axial-vector
quantum numbers, coupling to charged and neutral currents.

The attempt to incorporate these interactions in a
renormalizable~\ac{QFT} in which perturbation theory is well defined
without any short distance cut-off has singled out spontaneously
broken gauge theories in which the weak interactions are described by
the exchange of massive gauge bosons.   The same conclusion is reached
by attempting to enforce high energy tree-level unitarity of
scattering amplitudes. 

As discussed in chapter~\ref{sec:sm/eft} (with a little help from
appendix~\ref{sec:RG}), gauge theories retain their special status,
even if the electroweak~\ac{SM} is interpreted as an~\ac{EFT}, which
may have to be replaced by a more comprehensive theory at a energy scale
beyond the physics reach of current colliders.  Instead of
guaranteeing the consistency of formal\footnote{I.\,e.~without regard
to the convergence of the perturbative series.} perturbation theory to
arbitrarily high energies, the improved high energy behavior is used
for organizing the expansion of the effective interaction in a
more systematic way than would be possible with non-gauge vector
bosons.  Selecting the gauge theory among all equivalent~\acs{EFT}s
that describe the same physics (on-shell matrix elements) does not
sacrifice any generality.

Chapter~\ref{sec:me} described how to systematically organize
tree-level calculations for many particle final states in such a way
that non-abelian gauge invariance can be maintained by partial
sums.  This is important in state-of-the-art calculations, that
would exhaust available human and computing resources if they had to
be done in a single step.  In chapter~\ref{sec:ps}, methods for
improving numerical convergence and parallelization of computations
have been presented, that can be used for obtaining high precision
theoretical predictions for event rates in channels with many
particles in the final state, as discussed briefly in
chapter~\ref{sec:lc}.

Currently, a \emph{pessimistic} scenario can not be excluded, in which
a fundamental Higgs particle will be found, together with a few
supersymmetry partners of \ac{SM}~particles, compatible with
the~\ac{MSSM}.  While these discoveries would be a tremendous
achievement of experimental physics and the culmination of decades of
theoretical work, it would probably put the answers to the open
questions in the physics of flavor out of our reach.  However, we must
not forget which eminent \emph{``future truths of Physical Science''}
were found when Michelson had a look at \emph{``the sixth place of
decimals.''}  Rumors about the completion of physics are almost
always greatly exaggerated \ldots

\end{fmffile}
\begin{fmffile}{\jobname 2}
\fmfset{arrow_ang}{10}
\fmfset{curly_len}{2mm}
\fmfset{wiggly_len}{3mm}
\appendix
\chapter*{Appendix}
\chapter{The Renormalization Group}
\label{sec:RG}

\begin{cutequote*}{Thomas Carlyle}{The~French Revolution%
  \protect\footnote{Translation and quote from \textsc{Thomas Mann}:
    \textit{Betrachtungen eines Unpolitischen}.}}
  Wisse, da\ss{} dies Universum das ist, was es zu sein vorgibt: ein
  Unendliches.  Versuche nie im Vertrauen auf deine logische
  Verdauungskraft, es zu verschlingen; sei vielmehr dankbar, wenn du
  durch geschicktes Einrammen dieses oder jenes festen Pfeilers in das
  Chaos verhinderst, da\ss{} es dich verschlinge.
\end{cutequote*}

Until a few year back, textbooks on elementary particle physics
and~\ac{QFT} have hardly mentioned the method of~\ac{EFT} and have
usually given treatments of the~\ac{RG} that are have been more
formalistic than lucid.  Consequently, these subjects have had more of
an oral tradition than canonical scriptures.  Fortunately this has
changed now with the advent of two excellent textbooks offering both a
pedagogical~\cite{Peskin/Schroeder:QFT:Text} and a
systematical~\cite{Weinberg:QFTv1:Text,Weinberg:QFTv2:Text}
exposition of~\ac{QFT}.

These lectures have been given to an audience of graduate students
working in experiments and on theoretical projects.  Therefore they
had to live up to the combined challenge of presenting theorists new
angles on a subject they had already studied, while at the same time
offering concrete applications to experimentalists, without burying
them in formalism.

The intuitive reasoning in these lectures avoids the horrendous
combinatorial complications of perturbative renormalization theory.
Nevertheless, mathematical proofs of the intuitive observations still
need the technical results of traditional perturbative renormalization
theory.

These lectures deliberately use a strictly perturbative language in
order to appeal to the audience's intuition that is trained on Feynman
diagrams.   Many results remain valid outside of~\ac{PT},
but no attempt is made to mention this systematically.

\section{Introduction}
\label{sec:introduction}

\begin{cutequote*}{J.~W.~v.~Goethe}{Faust~I}
Ein Teil von jener Kraft,\\
Die stets das B\"ose will und stets das Gute schafft.
\end{cutequote*}

\emph{What is the size of leptons and quarks?} According to the most
recent report of the \ac{PDG}~\cite{PDG:1996}, the compositeness scale
is larger than~\unit[1.5]{TeV} for all of them, but even the best
limits fall short of~\unit[5]{TeV}.  In the \ac{SM}, we assume all
fermions to be strictly pointlike.  As you all know, the~\ac{SM} is
very successful.  Yet, we can't pin down the size of its constituents
below~$\unit[0.2]{TeV^{-1}}$.  \emph{How is this possible?}

The easy way out is to argue that our accelerators are too
small to probe any deeper. This is \emph{too} easy, however, because
we claim to control the \ac{SM} at the quantum (i.\,e.\ one-loop)
level.  The description of a typical LEP1 experiment~$e^+e^-\to q\bar
q$ involves loop diagrams like the one in
figure~\ref{fig:LEP1-loop}. And in this diagram, the loop momentum~$k$
is \emph{not} bounded anywhere and there is a part of the integration
region where it exceeds the energy scale~$\Lambda\approx1/\Delta x$.

\begin{figure}
  \begin{center}
    \begin{fmfgraph*}(40,30)
      \fmfstraight
      \fmfleft{p,e} \fmfright{q,qbar}
      \fmf{fermion}{e,v1,p}
      \fmf{fermion}{q,v3,v2,v4,qbar}
      \fmf{photon}{v1,v2}
      \fmffreeze
      \fmf{gluon,right=.3,label=$k$}{v3,v4}
      \fmfdotn{v}{4}
    \end{fmfgraph*}
  \end{center}
  \caption{\label{fig:LEP1-loop}%
    One-loop contribution to~$e^+e^-\to q\bar q$.}
\end{figure}
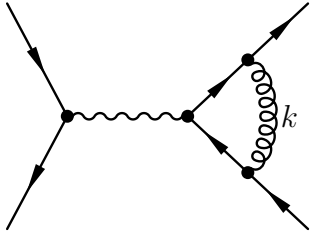

At this point we can rephrase the question: \emph{Why is the
compositeness scale irrelevant to physics at LEP1 energies?}  The
answer will be that the compositeness interactions
\begin{equation}
\label{eq:contact}
  \Delta L
    = \frac{1}{2\Lambda^2}\, \bar\psi\Gamma_\mu\psi\, \bar\psi\Gamma^\mu\psi
\end{equation}
are indeed \emph{irrelevant} and the term \emph{``irrelevant''} will
be given a precise technical definition below.  This might be a bit
surprising, because most of us have learned in graduate school that
interactions like~(\ref{eq:contact}) are \emph{verboten} in loop
calculations, because they make the theory non-renormalizable.

A closely related question is: \emph{why don't we have to understand
M-theory to do phenomenological particle physics?}  According to the
theoretical party line, we expect a lot of new particles (grand
unification gauge and Higgs bosons, higher string and membrane modes,
etc.) somewhere above~$\unit[10^{16}]{GeV}$.  Why do they have so
little impact on our bread and butter phenomenology that involves loop
diagrams?

\subsection{History}
\label{sec:history}

\ac{QFT} was introduced shortly after \ac{QM} in the 1920s.  But
while~\ac{QM} was put on a solid mathematical basis in the 1930s, it
took much longer to make sense out of~\ac{QFT}.

\begin{figure}
  \begin{center}
    \begin{fmfgraph}(30,10)
      \fmfstraight
      \fmfleft{l,dl} \fmfright{r,dr}
      \fmf{fermion}{l,vl}
      \fmf{fermion,tension=0.5}{vl,vr}
      \fmf{photon,tension=0,left}{vl,vr}
      \fmf{fermion}{vr,r}
      \fmfdot{vl,vr}
    \end{fmfgraph}
  \end{center}
  \caption{\label{fig:selfenergy}%
    Electron self energy}
\end{figure}
Soon (in the early 1930s), it was discovered that the second order
of~\ac{PT} gave non-sensical results in \ac{QED}.  For 
example, the electron self energy (see figure~\ref{fig:selfenergy},
but the calculators in the 1930s had to work without the benefits of
Feynman diagrams, of course) came out infinite.  Even though the lowest
order predictions were very successful, the inability to calculate
corrections was unsettling, to put it mildly.  By the late 1940s, a
workable scheme was established for absorbing the infinities in
renormalized couplings constants and masses.  Nevertheless, this
process of \emph{renormalization} was considered a nuisance and it was
expected that a complete theory would not need it.

Even though the renormalization procedure apparently worked in
practical applications, it took another twenty years of hard
theoretical work until a mathematical proof that the procedure works
in all orders of~\ac{PT} was established in the 1960s.

Today, the formalism of perturbative \ac{QFT} is well established and
the electroweak precision measurements at LEP1 leave no doubt that its
results are phenomenologically sound.  Nevertheless, it should be
remembered that no mathematical proof for the existence of an
interacting \ac{QFT} in four space-time dimensions has been given that
works outside of the realms of \ac{PT}.

\subsection{Renormalizability}
\label{sec:renormalizability}

\begin{figure}
  \begin{center}
    \begin{fmfgraph*}(30,15)
      \fmfstraight
      \fmfleft{l} \fmfright{r}
      \fmf{photon,label=$p$}{l,vl}
      \fmf{fermion,tension=0.3,left,label=$k$}{vl,vr}
      \fmf{fermion,tension=0.3,left,label=$k-p$}{vr,vl}
      \fmf{photon,label=$p$}{vr,r}
      \fmfdot{vl,vr}
    \end{fmfgraph*}
  \end{center}
  \caption{\label{fig:vacpol}%
     Vacuum polarization}
\end{figure}
Let us look at a simple example to get a feel for the physics of
renormalization.  For the vacuum polarization diagram of \ac{QED}
(see figure~\ref{fig:vacpol})
\begin{align}
  i\Pi_{\mu\nu}(p) & = -i(p^2 g_{\mu\nu} - p_\mu p_\nu)\Pi(p^2)\\
  \Pi(p^2) &= \frac{\alpha}{3\pi}
              \int_{-p^2}^{-\infty}\frac{\mathrm{d}(-k^2)}{(-k^2)}
              + \text{finite}
\end{align}
we find a logarithmically divergent integral over~$k$.  Obviously, the
divergence comes from the region~$-k^2\to\infty$, where the
$ee\gamma$-vertex has not been tested experimentally.  A brute force
remedy is to cut-off the integral at an arbitrary, but very large,
scale~$\Lambda$
\begin{equation}
\label{eq:vacpol-unsubtracted}
  \tilde\Pi(p^2)
    = \frac{\alpha}{3\pi}
         \int_{-p^2}^{\Lambda^2}\frac{\mathrm{d}(-k^2)}{(-k^2)}
        + \text{finite}
    = \frac{\alpha}{3\pi} \ln\frac{\Lambda^2}{-p^2} + \text{finite}
\end{equation}
This leaves us with a logarithmic dependence on the unphysical
parameter~$\Lambda$ and is not acceptable in this form.
We can however impose a renormalization condition
\begin{equation}
  \left.\tilde\Pi(p^2)\right|_{-p^2=\mu^2} = 0
\end{equation}
that trades the dependence on the \emph{unphysical} cut-off~$\Lambda$
for the dependence on the \emph{physical} renormalization point~$\mu$
at which we define the charge. 
Even though the result is now finite
\begin{equation}
\label{eq:vacpol-subtracted}
  \Pi^R(p^2) = - \frac{\alpha}{3\pi} \ln\frac{-p^2}{\mu^2}
\end{equation}
we are faced with a potentially large logarithm~$\ln(p^2/\mu^2)$, that
reaches~$25$ for LEP2 energies~($\sqrt s\approx\unit[190]{GeV}$) with
a charge renormalized at low energies~($\mu\approx m_e$).  The small
coupling constant of \ac{QED} helps to keep this logarithm relatively
harmless, but since LEP1 has introduced the era of electroweak
precision physics, higher orders in~\ac{PT} are required
for reliable quantitative predictions.  We will see below, that there
are cases, where the effective coupling is much larger and a
resummation of the perturbation series is required.

It is far from obvious that the subtraction procedure, that I have
sketched so crudely, works to all orders in~\ac{PT}.  It
is not even obvious that all required subtractions correspond to a
renormalization of coupling constants and masses.  While I will not go
into detail, it will be helpful later to establish some nomenclature
at this point.

\begin{figure}
  \begin{center}
    \begin{fmfgraph*}(30,15)
      \fmfstraight
      \fmfleft{l} \fmfright{r}
      \fmf{photon,label=$p$}{l,ct,r}
      \fmfv{label=$\displaystyle-\frac{\alpha}{3\pi}\ln\frac{\Lambda^2}{\mu^2}$,
            label.angle=90,decor.shape=hexagon, decor.size=10pt, decor.filled=1}{ct}
    \end{fmfgraph*}
  \end{center}
  \caption{\label{fig:vacpol-counterterm}%
     Vacuum polarization counterterm}
\end{figure}
The subtraction leading from~(\ref{eq:vacpol-unsubtracted})
to~(\ref{eq:vacpol-subtracted}) is obviously equivalent to adding the
new \emph{counterterm} vertex depicted in
figure~\ref{fig:vacpol-counterterm} to the theory.  Thinking in terms
of counterterms instead of subtractions has the advantage that the
generalization to higher orders is more intuitive.

In order to get a feeling for the required counterterms, let us
consider the free field actions for scalars~$\phi$, spin-1/2
fermions~$\psi$, and vectors~$A_\mu$
\begin{subequations}
\label{eq:free-actions}
\begin{align}
  L_0^\phi &= \int\!\mathrm{d}^4x\,
      \left(\frac{1}{2} \frac{\partial\phi(x)}{\partial x_\mu}
                        \frac{\partial\phi(x)}{\partial x^\mu}
              - \frac{m_\phi^2}{2} \phi^2(x) \right) \\
  L_0^\psi &= \int\!\mathrm{d}^4x\,
       \left( \bar\psi(x) i\gamma_\mu \frac{\partial}{\partial x_\mu} \psi(x)
              - m_\psi \bar\psi(x)\psi(x) \right) \\
  L_0^A &= \int\!\mathrm{d}^4x\,
       \frac{-1}{4} F_{\mu\nu}(x) F^{\mu\nu}(x)\,,\;\;\;
      F_{\mu\nu}(x) = \frac{\partial A_\nu(x)}{\partial x^\mu}
                      - \frac{\partial A_\mu(x)}{\partial x^\nu}\,.
\end{align}
\end{subequations}
We are using units with~$\hbar=c=1$ and therefore the actions must be
dimensionless.  If we assign dimension~$+1$ to mass
\begin{equation}
\label{eq:dim(m)}
  \mathop{\text{dim}} (m) = 1\,,
\end{equation}
then length has dimension~$-1$ in these units
\begin{subequations}
\begin{align}
  \mathop{\text{dim}} \left(\mathrm{d}^4x\right) &= -4\\
\label{eq:dim(p)}
  \mathop{\text{dim}} \left(\frac{\partial}{\partial x_\mu}\right) &= 1\,.
\end{align}
\end{subequations}
Now we can read off the dimensions of the fields
from~(\ref{eq:free-actions}) immediately
\begin{subequations}
\label{eq:dim(phi)}
\begin{align}
  \mathop{\text{dim}} \left(\phi(x)\right) &= 1\\
  \mathop{\text{dim}} \left(\psi(x)\right) &= \frac{3}{2}\\
\label{eq:dim(A)}
  \mathop{\text{dim}} \left(A_\mu(x)\right) &= 1 \,.
\end{align}
\end{subequations}
These dimensions are reflected in the high energy behavior of the
propagators, which is~$1/p^{2d-4}$
\begin{subequations}
\label{eq:free-propagators}
\begin{align}
  \int\!\mathrm{d}^4x\,e^{ipx}
     \Braket{0|\mathrm{T}\phi(x)\phi(0)|0}
       &= \frac{i}{p^2-m_\phi^2+i\epsilon} \\
  \int\!\mathrm{d}^4x\,e^{ipx}
     \Braket{0|\mathrm{T}\psi(x)\bar\psi(0)|0}
       &= i\frac{\fmslash{p}+m_\psi}{p^2-m_\psi^2+i\epsilon} \\
\label{eq:free-A-propagators}
  \int\!\mathrm{d}^4x\,e^{ipx}
     \Braket{0|\mathrm{T}A_\mu(x)A_\mu(0)|0}
       &= \frac{-ig_{\mu\nu}}{p^2+i\epsilon} \,.
\end{align}
\end{subequations}

The power counting for loop integrals in Feynman diagrams can be
related directly to the dimensions of the interaction operators.  Let
me illustrate this by a some examples\footnote{A systematic
exposition too all orders in~\ac{PT} is well beyond the
scope of these lectures, but can be found in any \ac{QFT} text.}.  I
will give a more comprehensive account from the point of view of the
\ac{RG} in later sections.

Consider a one-loop integral with two~$\phi^6$ insertions.  Since the
loop integral
\begin{equation}
  \int\!\frac{\mathrm{d}^4k}{(2\pi)^4}\,\frac{1}{k^2(p-k)^2}
\end{equation}
is logarithmically divergent, we need a~$\phi^8$ counterterm
\begin{equation}
  \parbox{31\unitlength}{%
    \begin{fmfgraph}(30,15)
      \fmfleftn{l}{4}
      \fmfrightn{r}{4}
      \fmf{plain}{l1,vl}
      \fmf{plain}{l2,vl}
      \fmf{plain}{l3,vl}
      \fmf{plain}{l4,vl}
      \fmf{plain}{r1,vr}
      \fmf{plain}{r2,vr}
      \fmf{plain}{r3,vr}
      \fmf{plain}{r4,vr}
      \fmf{plain,left=0.5}{vl,vr,vl}
      \fmfdot{vr,vl}
    \end{fmfgraph}}
  \;\;\;+\;\;\;
  \parbox{16\unitlength}{%
    \begin{fmfgraph}(15,15)
      \fmfleftn{l}{4}
      \fmfrightn{r}{4}
      \begin{fmffor}{i}{1}{1}{4}
        \fmf{plain}{l[i],v,r[i]}
      \end{fmffor}
      \fmfdot{v}
    \end{fmfgraph}}\;\;\;=\text{finite}\,.
\end{equation}
Similarly, a~$\phi^4$ and a~$\phi^6$ operator need a~$\phi^6$
counterterm 
\begin{equation}
  \parbox{31\unitlength}{%
    \begin{fmfgraph}(30,15)
      \fmfleftn{l}{4}
      \fmfrightn{r}{4}
      \fmf{plain}{l1,vl}
      \fmf{phantom}{l2,vl}
      \fmf{phantom}{l3,vl}
      \fmf{plain}{l4,vl}
      \fmf{plain}{r1,vr}
      \fmf{plain}{r2,vr}
      \fmf{plain}{r3,vr}
      \fmf{plain}{r4,vr}
      \fmf{plain,left=0.5}{vl,vr,vl}
      \fmfdot{vr,vl}
    \end{fmfgraph}}
  \;\;\;+\;\;\;
  \parbox{16\unitlength}{%
    \begin{fmfgraph}(15,15)
      \fmfleftn{l}{4}
      \fmfrightn{r}{4}
      \fmf{plain}{l1,v,r1}
      \fmf{phantom}{l2,v}\fmf{plain}{v,r2}
      \fmf{phantom}{l3,v}\fmf{plain}{v,r3}
      \fmf{plain}{l4,v,r4}
      \fmfdot{v}
    \end{fmfgraph}}\;\;\;=\text{finite}
\end{equation}
and two~$\phi^4$ operators need a~$\phi^4$ counterterm 
\begin{equation}
  \parbox{31\unitlength}{%
    \begin{fmfgraph}(30,15)
      \fmfleftn{l}{4}
      \fmfrightn{r}{4}
      \fmf{plain}{l1,vl}
      \fmf{phantom}{l2,vl}
      \fmf{phantom}{l3,vl}
      \fmf{plain}{l4,vl}
      \fmf{plain}{r1,vr}
      \fmf{phantom}{r2,vr}
      \fmf{phantom}{r3,vr}
      \fmf{plain}{r4,vr}
      \fmf{plain,left=0.5}{vl,vr,vl}
      \fmfdot{vr,vl}
    \end{fmfgraph}}
  \;\;\;+\;\;\;
  \parbox{16\unitlength}{%
    \begin{fmfgraph}(15,15)
      \fmfleftn{l}{4}
      \fmfrightn{r}{4}
      \fmf{plain}{l1,v,r1}
      \fmf{phantom}{l2,v,r2}
      \fmf{phantom}{l3,v,r3}
      \fmf{plain}{l4,v,r4}
      \fmfdot{v}
    \end{fmfgraph}}\;\;\;=\text{finite}\,.
\end{equation}
This suggests a pattern, which can indeed be verified by a more
detailed analysis.  If we have multiple insertions of operators with
dimension greater than four, then counterterms of higher and higher
dimension are needed. On the other hand, operators of dimension less
than or equal to four require counterterms of dimension less than or
equal to four.

This pattern can already be seen from dimensional arguments.  If an
operator has dimension greater than four, it must come with a coupling
constant of negative dimension, e.\,g.
\begin{equation}
  \frac{1}{\Lambda^2} \frac{1}{6!} \phi^6(x)\,,\;\;\;
        \mathop{\text{dim}}(\Lambda) = 1\,.
\end{equation}
If the loop integral does not depend on the scale in this coupling
constant, then the product of two such operators will have a coupling
constant of an even more negative dimension. The product will
therefore be of even higher dimension, e.\,g.
\begin{equation}
  \frac{1}{\Lambda^2} \frac{1}{6!} \phi^6(x)
  \frac{1}{\Lambda^2} \frac{1}{6!} \phi^6(y)
    \to \frac{1}{16\pi^2} \frac{1}{\Lambda^4} \frac{1}{8!} \phi^8(x)\,.
\end{equation}

Since all building blocks have a positive dimension
(see~(\ref{eq:dim(p)}) and~(\ref{eq:dim(phi)})), there are only a
finite number of operators of a given dimension. Therefore,
interactions of dimension less then four can be renormalized by a
\emph{finite} number of counterterms\footnote{While this is rather
intuitive from the one-loop examples, the mathematical proof is
extremely complicated in the general case.  It took forty years of
efforts~\protect\cite{Bogoliubov/Parasiuk:1957:BPHZ,*Hepp:1966:BPHZ,
*Zimmermann:1969:BPHZ} of some of the brightest theoretical physicists
of their time to 
disentangle the multi-loop calculations when the loops are not simply
nested, but are \emph{overlapping}.}.  As long as a \emph{finite}
number of counterterms suffices, the theory retains its predictive
power.  As a result, interactions of dimension four are called
\emph{renormalizable} and interactions of dimension less than four are
called \emph{super-renormalizable}.  There are additional
complications with the preservation of symmetries, which do not affect
the general argument, however.

It was then very reasonable to expect that nature is described by a
renormalizable \ac{QFT}, such as \ac{QED}, \ac{QCD} or the \ac{SM}.
These theories have been very successful and as a result, all
other---\emph{non-renormalizable}---theories fell from grace and
became second class citizens.

However, two very important questioned were apparently never posed,
let alone answered:
\begin{enumerate}
  \item \emph{Why is nature described by a renormalizable \ac{QFT}?}
    I will show later that the \ac{RG} will provide a satisfactory
    answer to this question. 
  \item \emph{Why should our low energy \ac{QFT} be correct to
    arbitrarily high energies?}   The contribution of string theory to
    particle physics that this question is no longer taboo.
\end{enumerate}

\subsection{The Renormalization Group}
\label{sec:rg-intro}

The great Cinderella story of theoretical particle physics since~1970
has been the rise of renormalization from a nuisance \textit{(sweeping
things under the rug)} to a powerful tool.  The turning points of this
epic are
\begin{epochs}{1999}
  \item[1971] Wilson's \ac{RG}~\cite{Wilson:1971,*Wilson:1971b}
    for critical phenomena (i.\,e.~phase transitions) and strong interactions
    proved that renormalization could help physics insight, instead of
    hindering it.
  \item[1972] The proof of renormalizability of spontaneously broken
    gauge theories~\cite{tHooft:1971:1,*tHooft:1971:2,
      *tHooft/Veltman:1972:Combinatorics,tHooft/Veltman:1972:DimReg}
    established a viable candidate for a \ac{QFT} of weak interactions
    (now known as \ac{SM}).
  \item[1973] Asymptotic
    freedom~\cite{Gross/Wilczek:AsymptoticFreedom:Orig,
    *Politzer:AsymptoticFreedom:Orig} paved the way for~\ac{QCD}, the
    \ac{QFT} of strong interactions.
  \item[1979] Weinberg's \ac{EFT} paper~\cite{Weinberg:1979:EFT}
    summarized and formalized the folklore of systematic expansions in
    the energy.
  \item[1990s] Even the non-renormalizable theories have been
    reinstated as first class citizens under the name of \ac{EFT}.
\end{epochs}
By the time of Weinberg's paper, the physics and the formalism was
well understood by the experts, but it remained for some time part of
the oral tradition that was covered insufficiently by textbooks.

\section{The Renormalization Group}
\label{sec:rg}

\begin{cutequote*}{Antonio Machado}{Proverbios~y~cantares, VI}
caminante, no hay camino \\
se hace camino al andar.
\end{cutequote*}
  
We evaluate the predictions of \ac{QFT} by doing integrals: loop
integrals from Feynman diagrams in~\ac{PT} and Feynman
path integrals beyond~\ac{PT}.  The \ac{RG} has a very
intuitive interpretation~\cite{Polchinski:1984:RG} (see also
\cite{Polchinski:1992:EFT}) in the evaluation of Feynman's path integral 
\begin{equation}
\label{eq:pathintegral}
  \int\!\mathcal{D}\phi\,e^{iS(\phi)}
\end{equation}
and I encourage everyone to consult the reference.  But I can not
assume a previous exposure to~(\ref{eq:pathintegral}) and a detailed
discussion is beyond the scope of these lectures.  Therefore I will
concentrate on Feynman diagrams and ~\ac{PT}.

\subsection{Cracking Integrals}
\label{sec:integration}

There are (at least) three common ways to do integrals in physics:
either from the anti-derivative or from the theorem of residues or
from a Riemann- (or Cauchy- or Lebesgue- or \ldots) construction.  The
first method is familiar to all of you from high school days, but it
is frowned upon in certain circles because the second method enables
physicists to do ``harder'' integrals and the third method allows
mathematicians to integrate more (i.\,e.~``wilder'') functions.  As
one might guess from this introduction, I will now sing the praise of
the first method.

The problem of calculating the integral
\begin{equation}
\label{eq:simple-integral}
  I(f;a,b) = \int_a^b\!\mathrm{d}x\,f(x)
\end{equation}
is equivalent to solving the first order \ac{ODE}
\begin{equation}
\label{eq:simple-ode}
  \frac{\mathrm{d}}{\mathrm{d}x}F(x) = f(x),\quad F(b) = C\,,
\end{equation}
since the integral is given by~$I(f;a,b) = F(a) - C$.  This relation
is familiar, because it is used in simple cases in the opposite
direction to obtain solutions to~(\ref{eq:simple-ode})
from~(\ref{eq:simple-integral}).  In fact, we will use it frequently
below.

Solving the \ac{ODE}~(\ref{eq:simple-ode}) might appear more difficult
than doing the integral~(\ref{eq:simple-integral}) and this is true in
the general case. However, I will try to convince you that, for many
applications in particle physics, the \ac{ODE} can give us very reliable
approximate results, where the value of the corresponding
integral is hard (or even impossible) to estimate.  In addition, the
\ac{ODE} can give us more physical insight than brute force
estimates of integrals.  I will also argue that there are cases, where
the integral~(\ref{eq:simple-integral}) does not exist, while the
\ac{ODE}~(\ref{eq:simple-ode}) can still give us the physically
correct answer.

\subsection{Wilson's Insight}
\label{sec:wilson}

\begin{figure}
  \begin{center}
    \begin{fmfgraph}(30,15)
      \fmfstraight
      \fmfleft{l,dl} \fmfright{r,dr}
      \fmf{fermion}{l,vl}
      \fmf{fermion,tension=0.5}{vl,vr}
      \fmf{phantom,tag=1,tension=0,left}{vl,vr}
      \fmf{fermion}{vr,r}
      \fmffreeze
      \fmfipath{p[]}
      \fmfiset{p0}{vpath1(__vl,__vr)}
      \fmfiset{p1}{subpath (0,.35length(p0)) of p0}
      \fmfiset{p2}{subpath (.65length(p0),length(p0)) of p0}
      \fmfforce{point infinity of p1}{v1}
      \fmfforce{point 0 of p2}{v2}
      \fmf{fermion,left}{v1,v2,v1}
      \fmf{fermion,left}{v1,v2,v1}
      \fmfi{photon}{p1}
      \fmfi{photon}{p2}
      \fmfdot{vl,vr,v1,v2}
    \end{fmfgraph}
  \end{center}
  \caption{\label{fig:nested}%
    Nested divergencies}
\end{figure}
We have seen heuristically in section~\ref{sec:introduction} that the
divergencies and large logarithms in \ac{PT} originate from
logarithmic integrals in loop momenta.  For \emph{nested
divergencies} (see figure~\ref{fig:nested}), it is intuitively clear
that the proper procedure is to proceed from the inside out and to
subtract the subdivergencies first.  Then only the overall divergency
from the outermost loop remains.

\begin{figure}
  \begin{center}
    \begin{fmfgraph}(30,10)
      \fmfstraight
      \fmfleft{l} \fmfright{r}
      \fmf{fermion}{l,v1}
      \fmf{fermion,tension=0.7}{v1,v2,v3,v4}
      \fmf{photon,tension=0,left}{v1,v3}
      \fmf{photon,tension=0,right}{v2,v4}
      \fmf{fermion}{v4,r}
      \fmfdotn{v}{4}
    \end{fmfgraph}
  \end{center}
  \caption{\label{fig:overlapping}%
    Overlapping divergencies}
\end{figure}
More intricate is the treatment of \emph{overlapping divergencies}
(see figure~\ref{fig:overlapping}), where each subdivergency has to be
subtracted without double counting.  The proof that the subtraction
algorithm is mathematically sound to all orders in \ac{PT} is an
extremely complicated combinatorial
exercise~\cite{Bogoliubov/Parasiuk:1957:BPHZ,*Hepp:1966:BPHZ,
*Zimmermann:1969:BPHZ}.

Nevertheless, it can be given a lucid physical interpretation, if we
change our point of view.  Why should we insist of doing the loop
integrals first and subtract divergencies later?  As we have seen
in~(\ref{eq:vacpol-unsubtracted}) and~(\ref{eq:vacpol-subtracted}), the 
leading logarithmic dependence on the renormalization scale can be
recovered from the dependence on the cut-off by dimensional analysis.
Wouldn't it be easier to calculate the dependence on the cut-off
directly?

As we shall see now, this heuristic argument can be made precise and
results in a powerful calculational tool, the \ac{RG}.  The essence of
the method lies in deriving a differential equation, the \ac{RGE},
that can be
integrated to give the desired result.  It will turn out that the
calculation never involves any infinite numbers and in most cases, the
coefficients of the \ac{PT} will never be large.

\subsection{The Sliding Cut-Off}
\label{sec:polchinski}

For simplicity, I will now concentrate on massless $\phi^4$-theory.
This is the theory of a spinless, uncharged, massless particle, that
interacts through a local $\phi^4$ vertex.  The arguments will go
through for more realistic \ac{QFT}s, but the technical complications
arising from gauge invariance would obscure the point that I'm trying
to make.  I will deal later with the effects from massive particles.

Let us assume for the moment that the theory has been defined with a
cut-off in momentum space, i.\,e.~a propagator
\begin{equation}
   iD_\Lambda(k^2) = \frac{i}{k^2+i\epsilon} \Theta(|k|\le\Lambda)\,.
\end{equation}
The cut-off $|k|\le\Lambda$ needs some explanation, because
it appears not to be Lorentz invariant.  In fact,
$\Theta(|k|\le\Lambda)$ is just a symbolic notation.  A simple minded
cut-off like~$|k^2|\le\Lambda^2$ is invariant, but it is not
restrictive enough, because~$k$ can grow along light-like directions
($k^2=0$) without being cut off.

\begin{figure}
  \begin{center}
    \begin{fmfgraph*}(70,40)
      \fmfi{plain}{(.5w,0)--(.5w,h)}
      \fmfi{plain}{(0,.5h)--(w,.5h)}
      \fmfi{dashes}{(0,.40h){right}..(.5w,.5h)..{right}(w,.6h)}
      \fmfi{dots}{(.45w,0){up}..(.5w,.5h)..{up}(.55w,h)}
      \fmfiv{d.shape=cross,d.size=2mm}{(.2w,.5h)}
      \fmfiv{d.shape=cross,d.size=2mm}{(.8w,.5h)}
      \fmfiv{l=$\mathop{\text{Re}}k_0$,l.angle=-90}{(.95w,.5h)}
      \fmfiv{l=$\mathop{\text{Im}}k_0$,l.angle=180}{(.5w,.95h)}
    \end{fmfgraph*}
  \end{center}
  \caption{\label{fig:wick}%
    Wick rotation}
\end{figure}
A more appropriate definition can be found by using the ``Wick
rotation''.  As can be seen from figure~\ref{fig:wick}, in each loop
integration, the integration contour in the complex~$k_0$ plane can be
deformed from the dashed curve to the dotted curve without crossing
singularities.  Then we can perform the substitution~$(k^0,\vec k) =
(ik^0_E,\vec k_E)$ and the Minkowski ``length'' becomes a Euclidean
length $k^2 = (k^0)^2 - {\vec k}^2 = - (k_E^0)^2 - {\vec k}^2 = -
k_E^2$.  The Lorentz invariant cut-off~$k_E^2\le\Lambda^2$ is now
effective, because there are no light-like directions in Euclidean
space.  We shall interpret~$\Theta(|k|\le\Lambda)$ in this fashion.

We can now compare the theories defined with two different
cut-offs~$\Lambda$ and $\Lambda'<\Lambda$.  It turns out that we can
change the theory in such a way that the physics does \emph{not}
change when we lower the cut-off, as long as the external momenta
remain below the new cut-off.  To see how this works, let us introduce
a graphical notation
\begin{subequations}
\begin{align}
  iD_\Lambda(k^2) &=
    \parbox{21\unitlength}{%
      \begin{fmfgraph}(20,1)
         \fmfleft{l}
         \fmfright{r}
         \fmf{plain}{l,r}
      \end{fmfgraph}}\\
  iD_{\Lambda'}(k^2) &=
    \parbox{21\unitlength}{%
      \begin{fmfgraph}(20,1)
         \fmfleft{l}
         \fmfright{r}
         \fmf{dashes}{l,r}
      \end{fmfgraph}}\\
  iD_{\Lambda}(k^2) - iD_{\Lambda'}(k^2) &=
    \parbox{21\unitlength}{%
      \begin{fmfgraph}(20,1)
         \fmfleft{l}
         \fmfright{r}
         \fmf{dbl_plain}{l,r}
      \end{fmfgraph}}
\end{align}
\end{subequations}
for propagators in the full theory, the low energy theory and their
difference.  For the one-loop correction to the propagator we find
below the new cut-off
\begin{equation}
  \parbox{21\unitlength}{%
    \begin{fmfgraph}(20,5)
       \fmfstraight
       \fmfleft{l,dl}
       \fmfright{r,dr}
       \fmf{dashes}{l,v}
       \fmf{dashes}{v,r}
       \fmf{plain}{v,v}
       \fmfdot{v}
    \end{fmfgraph}}
  = \parbox{21\unitlength}{%
    \begin{fmfgraph}(20,5)
       \fmfstraight
       \fmfleft{l,dl}
       \fmfright{r,dr}
       \fmf{dashes}{l,v,v,r}
       \fmfdot{v}
    \end{fmfgraph}}
  + \parbox{21\unitlength}{%
    \begin{fmfgraph}(20,5)
       \fmfstraight
       \fmfleft{l,dl}
       \fmfright{r,dr}
       \fmf{dashes}{l,v}
       \fmf{dashes}{v,r}
       \fmf{dbl_plain}{v,v}
       \fmfdot{v}
    \end{fmfgraph}}\,.
\end{equation}
Therefore the physics will not change, if we introduce a new vertex
\begin{equation}
\label{eq:tadpole}
  \parbox{21\unitlength}{%
    \begin{fmfgraph}(20,5)
       \fmfstraight
       \fmfleft{l}
       \fmfright{r}
       \fmf{dashes}{l,v,r}
       \fmfblob{5mm}{v}
    \end{fmfgraph}}
  = \parbox{21\unitlength}{%
    \begin{fmfgraph}(20,5)
       \fmfstraight
       \fmfleft{l,dl}
       \fmfright{r,dr}
       \fmf{dashes}{l,v}
       \fmf{dashes}{v,r}
       \fmf{dbl_plain}{v,v}
       \fmfdot{v}
    \end{fmfgraph}}
\end{equation}
into the low energy theory.  When we calculate~(\ref{eq:tadpole})
explicitely, it will be useful later to calculate the more general
integral
\begin{equation}
\label{eq:finite-I(n,D)}
  I(n,D) = \int\frac{\mathrm{d}^Dk}{(2\pi)^D}
             \left(\frac{1}{k^2+i\epsilon}\right)^n
             \Theta(\Lambda'\le|k|\le\Lambda)\,,
\end{equation}
defined in~$D$ space-time dimensions. As discussed above, we will
perform a ``Wick rotation''
\begin{equation}
  I(n,D) = (-1)^n i \int\frac{\mathrm{d}^Dk_E}{(2\pi)^D}
             \left(\frac{1}{k_E^2}\right)^n
             \Theta({\Lambda'}^2\le k_E^2\le\Lambda^2)\,.
\end{equation}
The volume of the surface of the $D$-dimensional sphere is given by
\begin{equation}
  \int\mathrm{d}\Omega_D = \frac{2\pi^{D/2}}{\Gamma(D/2)}
\end{equation}
and therefore
\begin{equation}
  I(n,D) = \frac{(-1)^ni}{(4\pi)^{D/2}\Gamma(D/2)}
             \int_{{\Lambda'}^2}^{\Lambda^2}\!\mathrm{d}k_E^2
             \,(k_E^2)^{D/2-1-n}\,,
\end{equation}
which can be calculated explicitely
\begin{subequations}
\begin{align}
  I(D/2,D) &= \frac{(-1)^ni}{(4\pi)^{D/2}\Gamma(D/2)}
              \ln\frac{\Lambda^2}{{\Lambda'}^2} \\
  \left.I(n,D)\right|_{n\not=D/2}
           &= \frac{(-1)^ni}{(4\pi)^{D/2}\Gamma(D/2)} \frac{1}{D/2-n}
              \left(\Lambda^{D-2n}-{\Lambda'}^{D-2n}\right)\,.
\end{align}
\end{subequations}
In the case of the tadpole diagram~(\ref{eq:tadpole}) we find finally
\begin{equation}
\label{eq:tadpole-result}
  \int\frac{\mathrm{d}^4k}{(2\pi)^4} \frac{i}{k^2+i\epsilon}
             \Theta(\Lambda'\le|k|\le\Lambda)
    = iI(1,4)
    = \frac{1}{(4\pi)^2} \left(\Lambda^2-{\Lambda'}^2\right)
\end{equation}
for the new vertex.
In the four point function
\begin{equation}
  \parbox{21\unitlength}{%
    \begin{fmfgraph}(20,10)
       \fmfstraight
       \fmfleftn{l}{2}
       \fmfrightn{r}{2}
       \fmf{dashes}{l1,vl,l2}
       \fmf{dashes}{r1,vr,r2}
       \fmf{plain,tension=0.5,left=0.5}{vl,vr,vl}
       \fmfdot{vl,vr}
    \end{fmfgraph}}
  = \parbox{21\unitlength}{%
    \begin{fmfgraph}(20,10)
       \fmfstraight
       \fmfleftn{l}{2}
       \fmfrightn{r}{2}
       \fmf{dashes}{l1,vl,l2}
       \fmf{dashes}{r1,vr,r2}
       \fmf{dashes,tension=0.5,left=0.5}{vl,vr,vl}
       \fmfdot{vl,vr}
    \end{fmfgraph}}
  + \parbox{21\unitlength}{%
    \begin{fmfgraph}(20,10)
       \fmfstraight
       \fmfleftn{l}{2}
       \fmfrightn{r}{2}
       \fmf{dashes}{l1,vl,l2}
       \fmf{dashes}{r1,vr,r2}
       \fmf{dashes,tension=0.5,left=0.5}{vl,vr}
       \fmf{dbl_plain,tension=0.5,left=0.5}{vr,vl}
       \fmfdot{vl,vr}
    \end{fmfgraph}}
  + \parbox{21\unitlength}{%
    \begin{fmfgraph}(20,10)
       \fmfstraight
       \fmfleftn{l}{2}
       \fmfrightn{r}{2}
       \fmf{dashes}{l1,vl,l2}
       \fmf{dashes}{r1,vr,r2}
       \fmf{dashes,tension=0.5,left=0.5}{vr,vl}
       \fmf{dbl_plain,tension=0.5,left=0.5}{vl,vr}
       \fmfdot{vl,vr}
    \end{fmfgraph}}
  + \parbox{21\unitlength}{%
    \begin{fmfgraph}(20,10)
       \fmfstraight
       \fmfleftn{l}{2}
       \fmfrightn{r}{2}
       \fmf{dashes}{l1,vl,l2}
       \fmf{dashes}{r1,vr,r2}
       \fmf{dbl_plain,tension=0.5,left=0.5}{vl,vr,vl}
       \fmfdot{vl,vr}
    \end{fmfgraph}}\,,
\end{equation}
we need more vertices.  First a modified four particle vertex
\begin{equation}
\label{eq:fish}
  \parbox{21\unitlength}{%
    \begin{fmfgraph}(20,10)
       \fmfstraight
       \fmfleftn{l}{2}
       \fmfrightn{r}{2}
       \fmf{dashes}{l1,v,l2}
       \fmf{dashes}{r1,v,r2}
       \fmfblob{5mm}{v}
    \end{fmfgraph}}
  = \parbox{21\unitlength}{%
    \begin{fmfgraph}(20,10)
       \fmfstraight
       \fmfleftn{l}{2}
       \fmfrightn{r}{2}
       \fmf{dashes}{l1,vl,l2}
       \fmf{dashes}{r1,vr,r2}
       \fmf{dbl_plain,tension=0.5,left=0.5}{vl,vr,vl}
       \fmfdot{vl,vr}
    \end{fmfgraph}}\,.
\end{equation}
Under the assumption that the external
momenta satisfy~$|p|\ll\Lambda'<\Lambda$, we can do the loop integral
in~(\ref{eq:fish}) easily to leading order
\begin{multline}
\label{eq:finite-fish}
  \frac{1}{2} \int\frac{\mathrm{d}^4k}{(2\pi)^4}
      \frac{i\Theta(\Lambda'\le|k|\le\Lambda)}{k^2+i\epsilon}
      \frac{i\Theta(\Lambda'\le|p-k|\le\Lambda)}{(p-k)^2+i\epsilon} \\
  = -\frac{1}{2} \frac{i}{16\pi^2} \ln\frac{\Lambda^2}{{\Lambda'}^2}
      + O\left(\frac{p^2}{\Lambda^2}\right)\,.
\end{multline}
The exact expression is more complicated, due to the finite
integration limits, but we will not need its explicit form.  More
important is the observation that all integrals exist, since they are
over a compact domain in Euclidean space, far from any singularities.
Below, after motivating the physics, we shall see how to perform
calculations more efficiently.
Anyway, there are three such diagrams and the new vertex is
\begin{equation}
  -\frac{3}{2} \frac{1}{16\pi^2} \ln\frac{\Lambda^2}{{\Lambda'}^2}
   \frac{g^2}{4!} \phi^4(x)\,.
\end{equation}
There is also a new six particle vertex
\begin{equation}
  \parbox{21\unitlength}{%
    \begin{fmfgraph}(20,10)
       \fmfstraight
       \fmfleftn{l}{3}
       \fmfrightn{r}{3}
       \begin{fmffor}{i}{1}{1}{3}
         \fmf{dashes}{l[i],v}
         \fmf{dashes}{r[i],v}
       \end{fmffor}%
       \fmfblob{5mm}{v}
    \end{fmfgraph}}
  = \parbox{21\unitlength}{%
    \begin{fmfgraph}(20,10)
       \fmfstraight
       \fmfleftn{l}{3}
       \fmfrightn{r}{3}
       \begin{fmffor}{i}{1}{1}{3}
         \fmf{dashes}{l[i],vl}
         \fmf{dashes}{r[i],vr}
       \end{fmffor}%
       \fmf{dbl_plain,tension=3}{vl,vr}
       \fmfdot{vl,vr}
    \end{fmfgraph}}\,,
\end{equation}
that reads (again neglecting small terms~$O(p^2/\Lambda^2)$):
\begin{equation}
  \frac{1}{\Lambda^2} \frac{g^2}{6!} \phi^6(x)\,.
\end{equation}

The general procedure consists in a stepwise reduction of the
scale~$\Lambda'$
\begin{equation}
      \Lambda>\Lambda'>\Lambda''>\Lambda'''>\ldots
\end{equation}
where the Lagrangian is adjusted in each step such that the low energy
observables below the current cut off scale is unchanged.  If we allow
arbitrary powers of the field~$\phi$ and derivatives, the low energy
physics can indeed be reproduced fully, as shown above.

\subsection{Renormalization Group Flow}
\label{sec:rgflow}

The most general interaction can be expanded
\begin{equation}
  L(x) = \sum_i g_i O_i(x)
\end{equation}
in an infinite series of local operators
\begin{equation}
  O(x) = \left\{ \phi^2(x), \phi^4(x), (\partial\phi)^2(x),
     \phi^6(x), (\phi\partial\phi)^2(x), \ldots\right\}\,,
\end{equation}
where an infinite sum will not necessarily correspond to a local
interaction.  A discrete renormalization group transformation of the
scale dependent coupling constants
\begin{equation}
\label{eq:discrete-rg}
  g_i(\Lambda) \to g_i(\Lambda')
    = \sum_j\Gamma_{ij}(\Lambda',\Lambda)g_j(\Lambda)
\end{equation}
is defined by the requirement that the low energy physics remains
unchanged.

However, discrete transformations like~(\ref{eq:discrete-rg}) are
technically hard to handle.  It is much more transparent to consider
a continuous transformation with infinitesimal generators instead.
Then the~\ac{RGE} is a system of coupled non-linear ordinary
differential equations
\begin{equation}
  {\Lambda'}^2 \frac{\mathrm{d}g_i(\Lambda')}{\mathrm{d}{\Lambda'}^2}
    = \sum_j \gamma_{ij} g_j(\Lambda')\,.
\end{equation}

\begin{figure}
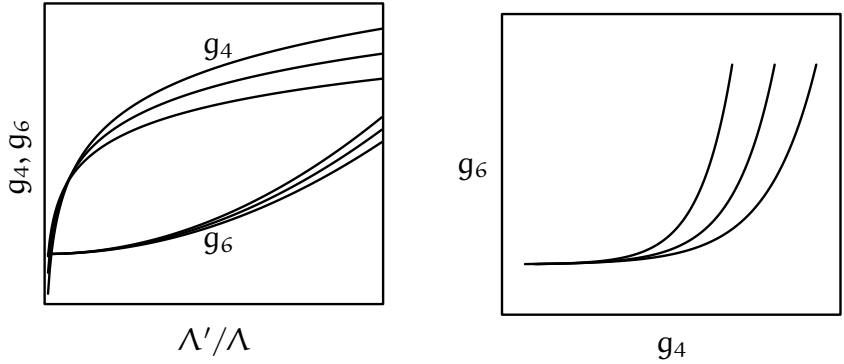

  \begin{center}
    \begin{empgraph}(45,40)
      pi := 3.14159; g := 10;
      vardef ln expr x = (mlog x) / 256 enddef;
      vardef gfour expr mu = g - 3/16/pi/pi*g*g*ln(1/mu) enddef;
      vardef gsix expr mu = 2 + 1/2*g*mu*mu enddef;
      pickup pencircle scaled 1pt;
      path p[], q[];
      for mu = 0.01 step 0.005 until 1:
        for gg = 9 step 1 until 11:
          g := gg;
          augment.p[gg] (mu, gfour mu);
          augment.q[gg] (mu, gsix mu);
        endfor
      endfor
      setcoords (linear, linear);
      setrange (0, 0, 1, 12);
      gdraw p[9]; gdraw p[10]; gdraw p[11]; glabel.top (btex $g_4$ etex, 100);
      gdraw q[11]; gdraw q[10]; gdraw q[9]; glabel.bot (btex $g_6$ etex, 100);
      otick.bot ("", 0) withcolor white;
      glabel.lft (btex $g_4,g_6$ etex rotated 90, OUT);
      glabel.bot (btex $\Lambda'/\Lambda$ etex, OUT);
    \end{empgraph}
    \qquad
    \begin{empgraph}(45,40)
      pi := 3.14159; g := 10;
      vardef ln expr x = (mlog x) / 256 enddef;
      vardef gfour expr mu = g - 3/16/pi/pi*g*g*ln(1/mu) enddef;
      vardef xisg expr gsix = sqrt (2*(gsix-2)/g) enddef;
      pickup pencircle scaled 1pt;
      path p[];
      for gsix = 2.02 step 0.02 until 10:
        for gg = 9 step 1 until 11:
          g := gg;
          augment.p[gg] (gfour (xisg gsix), gsix);
        endfor
      endfor
      setcoords (linear, linear);
      setrange (4, 0, 12, 12);
      gdraw p[9]; gdraw p[10]; gdraw p[11];
      otick.bot ("", 4) withcolor white;
      glabel.lft (btex $g_6$ etex, OUT);
      glabel.bot (btex $g_4$ etex, OUT);
    \end{empgraph}
  \end{center}
  \caption{\label{fig:g4g6}%
    The evolution of the coefficients~$g_4$ and~$g_6$ as a function of
    a cut-off scale~$\Lambda'$ (left hand side) and after elimination
    of the unphysical cut-off scale~$\Lambda'$ (right hand side).}
\end{figure}
As a concrete example, consider the evolution of the
coefficients~$g_4$ and~$g_6$
\begin{equation}
  L_{\text{int}}(x) =
     \frac{g_4}{4!} \phi^4(x) + \frac{g_6}{6!} \phi^6(x) + \ldots
\end{equation}
with a change of the cut-off scale~$\Lambda'$, as depicted on the left
hand side of figure~\ref{fig:g4g6}.  By construction, the low-energy
physics does not change along the trajectories in
figure~\ref{fig:g4g6}, as long as each coupling uses the
\emph{same}~$\Lambda'$.  The cut-off scale~$\Lambda'$ is therefore
just \emph{conventional} and determines what is considered to be part
of the lowest order Lagrangian and what is to be calculated from it in
higher orders.  Thus the scale~$\Lambda'$ is redundant and can be
eliminated from the physical parameter space.  This physics is
therefore not determined by a point in the parameter space, but by a
whole \emph{trajectory} instead.

Since~$\Lambda'$ is redundant, it is desirable to eliminate it
completely.  As long as~$g_4(\Lambda')$ can be inverted (which is
always possible in fixed orders of~\ac{PT}), the physics
on the left hand side of figure~\ref{fig:g4g6} can be recast as the
right hand side of the same figure, where the dimensionful
cut-off scale~$\Lambda'$ has be traded for the dimensionless coupling
parameter~$g_4$.  This is an example of the celebrated phenomenon of
\emph{dimensional
transmutation}~\cite{Coleman/Weinberg:1973:transmutation}, of which
the relation of the strong coupling~$\alpha_S$ and the~\ac{QCD}
scale~$\Lambda_{\text{QCD}}$ is the most famous example.

So far, we have ignored the dependence on the upper cut-off
scale~$\Lambda$, which was assumed to be fixed and much larger than any
physics scale.  After the introduction of the \ac{RG} trajectories,
the scale~$\Lambda$ obtains a straightforward interpretation as the
starting point of the trajectory.  In this interpretation, the low
energy physics will not depend on~$\Lambda$ if the~\ac{RGE} can be
solved for all~$\Lambda$.  However, the existence of global solutions
is not a trivial condition and there are several cases to consider.

\begin{figure}
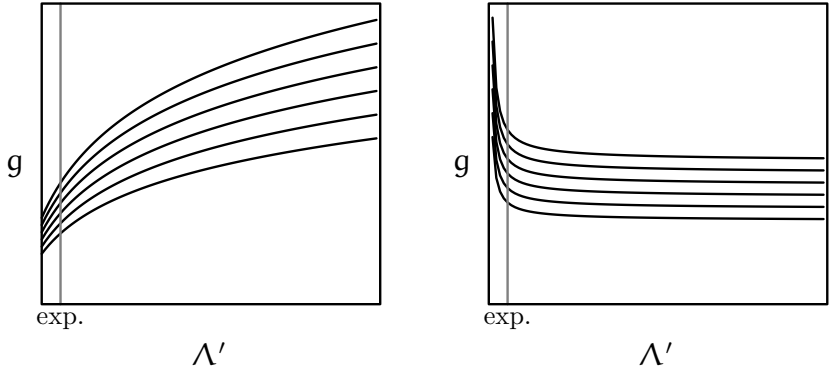

  \begin{center}
    \begin{empgraph}(45,40)
      vardef ln expr x = (mlog x) / 256 enddef;
      pickup pencircle scaled 1pt;
      setcoords (linear, linear);
      setrange (1, 0, 10, whatever);
      for x = 0.7 step 0.1 until 1.3:
        path p;
        y := x;
        for mu = 1 step 0.1 until 10:
          augment.p (mu, y + x * ln(mu));
        endfor
        gdraw p;
      endfor
      grid.bot ("exp.", 1.5) withcolor 0.5white;
      glabel.lft (btex $g$ etex, OUT);
      glabel.bot (btex $\Lambda'$ etex, OUT);
    \end{empgraph}
    \qquad
    \begin{empgraph}(45,40)
      vardef ln expr x = (mlog x) / 256 enddef;
      pickup pencircle scaled 1pt;
      setcoords (linear, linear);
      setrange (1, 0, 10, whatever);
      for x = 0.7 step 0.1 until 1.3:
        path p;
        y := x;
        for mu = 1.1 step 0.1 until 10:
          augment.p (mu, y + 0.1 * x / ln(mu));
        endfor
        gdraw p;
      endfor
      grid.bot ("exp.", 1.5) withcolor 0.5white;
      glabel.lft (btex $g$ etex, OUT);
      glabel.bot (btex $\Lambda'$ etex, OUT);
    \end{empgraph}
  \end{center}
  \caption{\label{fig:lambda2infinity}%
    All trajectories on the left hand side extend
    to~$\Lambda'\to\infty$ and a continuum limit exists.
    All trajectories on the right hand side even remain perturbative
    for~$\Lambda'\to\infty$.}
\end{figure}
In the scenario depicted on the left hand side of
figure~\ref{fig:lambda2infinity}, all trajectories extend
to~$\Lambda'\to\infty$.  Therefore, the starting point of the
evolution can be pushed back to~$\Lambda\to\infty$, which means that
the cut-off can be removed altogether and a continuum limit exists.
However, one must be careful that the trajectories might lead through
non perturbative regions with~$g\gg1$ and that a perturbative
determination of the evolution equation is unreliable in these regions.

Even more favorable is the scenario depicted on the right hand side of
figure~\ref{fig:lambda2infinity}, where all trajectories that start in
the low energy regime, where couplings are determined experimentally,
remain perturbative for all~$\Lambda'\to\infty$.  Perturbative
calculations are reliable everywhere in this scenario.  The most
favorable special case of this scenario is \emph{asymptotic
freedom},~\cite{Gross/Wilczek:AsymptoticFreedom:Orig,
*Politzer:AsymptoticFreedom:Orig}
in which the couplings vanish for~$\Lambda'\to\infty$ and low orders
of~\ac{PT} provide reliable predictions for high energy
experiments.

\begin{figure}
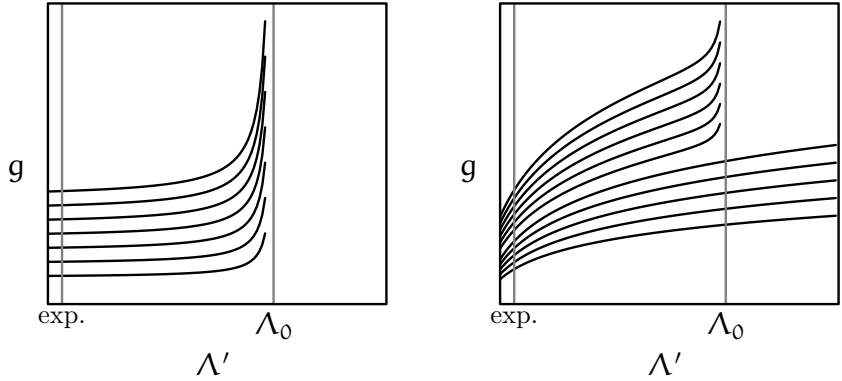

  \begin{center}
    \begin{empgraph}(45,40)
      vardef ln expr x = (mlog x) / 256 enddef;
      pickup pencircle scaled 1pt;
      setcoords (linear, linear);
      setrange (1, 0, 13, whatever);
      for x = 0.4 step 0.2 until 1.6:
        path p;
        y := 2x;
        for mu = 1 step 0.1 until 8.8:
          augment.p (mu, y + x / (9-mu));
        endfor
        gdraw p;
      endfor
      grid.bot ("exp.", 1.5) withcolor 0.5white;
      grid.bot (btex $\Lambda_0$ etex, 9) withcolor 0.5white;
      glabel.lft (btex $g$ etex, OUT);
      glabel.bot (btex $\Lambda'$ etex, OUT);
    \end{empgraph}
    \qquad
    \begin{empgraph}(45,40)
      vardef ln expr x = (mlog x) / 256 enddef;
      pickup pencircle scaled 1pt;
      setcoords (linear, linear);
      setrange (1, 0, 13, whatever);
      for x = 1.0 step 0.1 until 1.6:
        path p;
        y := x;
        for mu = 1 step 0.1 until 8.9:
          augment.p (mu, y + x * ln(mu) +  x / (9-mu));
        endfor
        gdraw p;
      endfor
      for x = 0.5 step 0.1 until 1.0:
        path p;
        y := x;
        for mu = 1 step 0.1 until 13:
          augment.p (mu, y + x * ln(mu));
        endfor
        gdraw p;
      endfor
      grid.bot ("exp.", 1.5) withcolor 0.5white;
      grid.bot (btex $\Lambda_0$ etex, 9) withcolor 0.5white;
      glabel.lft (btex $g$ etex, OUT);
      glabel.bot (btex $\Lambda'$ etex, OUT);
    \end{empgraph}
  \end{center}
  \caption{\label{fig:lambdanot2infinity}%
    The low energy theory on the left hand side can never be valid
    above~$\Lambda_0$. The low energy theory on the right hand side
    can be valid above~$\Lambda_0$, but only for a restricted part of
    the parameter space.}
\end{figure}
In stark contrast to the previous two scenarios, no trajectory on the
left hand side of figure~\ref{fig:lambdanot2infinity} can be extended
from the low energy domain beyond the scale~$\Lambda_0$.  Thus
the~\ac{QFT} that describes the present experiments can not be valid
above~$\Lambda_0$, unless non-perturbative effects change the
trajectories significantly close to~$\Lambda_0$.  Such a scenario is
not entirely bad news, because it predicts that something
``interesting'' will happen around~$\Lambda_0$: at least strong
interactions, maybe even ``new physics.''

An interesting combination of the scenarios in
figures~\ref{fig:lambda2infinity} and~\ref{fig:lambdanot2infinity} is
shown on the right hand side of figure~\ref{fig:lambdanot2infinity}:
some trajectories can be extended to $\Lambda'\to\infty$, while others
are confined to~$\Lambda'<\Lambda_0$.  Typically, there will be a
critical coupling~$g_0(\Lambda')$ and trajectories
with~$g(\Lambda')<g_0(\Lambda')$ can be extended, while trajectories
with~$g(\Lambda')>g_0(\Lambda')$ can not.  As a result, there will be
upper limits on the couplings at any given scale, if the theory is
intended to be defined at arbitrary high scales.  The most famous
example for this phenomenon is provided by the upper limits on the
Higgs self coupling resulting in an upper limit in the Higgs mass
$m_H^2 = g/2\cdot \langle\phi\rangle^2$ in
the~\ac{SM}~\cite{Cabibbo/etal:1979:Roman-Plots}.

\subsection{Relevant, Marginal Or Irrelevant?}
\label{sec:relevant}

While the graphical representation or the~\ac{RG} flow is very
suggestive for a few couplings, the untruncated problem has to deal
with \emph{infinitely many} couplings.  Therefore, it is not obvious
that the intuition gained from the low-dimensional examples can be
used in the infinite dimensional space of couplings.  This is
particularly problematic, because operators of higher and higher dimension
produce more and more divergent contributions to Feynman diagrams and
need even higher dimensional counterterms (see
section~\ref{sec:renormalizability}).

Fortunately, \emph{``the first shall be last and the last shall be
first''} and the higher dimensional operators turn out to be harmless
from the point of view of the~\ac{RG}.  If there was no interaction,
the couplings would be constant.  In this case, we can estimate 
matrix elements of the operators of dimension~$d$ by dimensional
analysis as (in four space-time dimensions)
\begin{equation}
\label{eq:dimensional}
  \left(\frac{|p|}{\Lambda}\right)^{d-4}
\end{equation}
where~$\Lambda$ is the high energy scale at which the ``bare''
couplings are defined and~$d$ is the power counting dimension of the
operator, counted according to (\ref{eq:dim(p)})
and~(\ref{eq:dim(phi)}).  In~(\ref{eq:dimensional}), we can identify
three different cases according to the value~$d$
\begin{itemize}
  \item $d<4$: these operators are termed \emph{relevant}, because
    their contributions are becoming more important at low energies,
    where experiments are performed.
  \item $d=4$: these operators are termed \emph{marginal}.
  \item $d>4$: these operators are termed \emph{irrelevant}, because
    their contributions are becoming less important at low energies,
    where experiments are performed.
\end{itemize}
Therefore, the initial conditions at high energies for higher
dimensional (``non renormalizable'') interactions play no important
r\^ole in low energy physics.

Unless there are very strong interactions, relevant and irrelevant
operators will retain their property, even if interactions are switched
on.  This can be seen from the evolution
\begin{subequations}
\begin{align}
   {\Lambda'}^2\frac{\mathrm{d}g_2(\Lambda')}{\mathrm{d}{\Lambda'}^2}
      &= {\Lambda'}^2 \frac{1}{16\pi^2} g_4(\Lambda') + \ldots \\
   {\Lambda'}^2\frac{\mathrm{d}g_4(\Lambda')}{\mathrm{d}{\Lambda'}^2}
      &= \frac{3}{2} \frac{1}{16\pi^2} g_4^2(\Lambda') + \ldots \\
   {\Lambda'}^2\frac{\mathrm{d}g_6(\Lambda')}{\mathrm{d}{\Lambda'}^2}
      &= \frac{1}{{\Lambda'}^2} g_4^2(\Lambda') + \ldots 
\end{align}
\end{subequations}
of the low dimensional operators:
\begin{itemize}
  \item $d=2$ (i.\,e.~$g_2$ or~$m^2$): the rate of change of this
    coupling is of dimension~2: $O({\Lambda'}^2)$.  Unless
    there is significant \emph{fine tuning} of the initial
    conditions, the value of the coupling will be of the order of the
    high energy starting point of the evolution:
    $g_2=O(\Lambda^2)\gg{\Lambda'}^2$.
  \item $d=6$ (i.\,e.~$g_6$): the rate of change of the coupling is of
    dimension~$-2$: $O({\Lambda'}^{-2})$.  Therefore the
    value of the initial condition at the high energy scale is indeed
    irrelevant with respect to the value induced by the evolution.
\end{itemize}
\begin{figure}
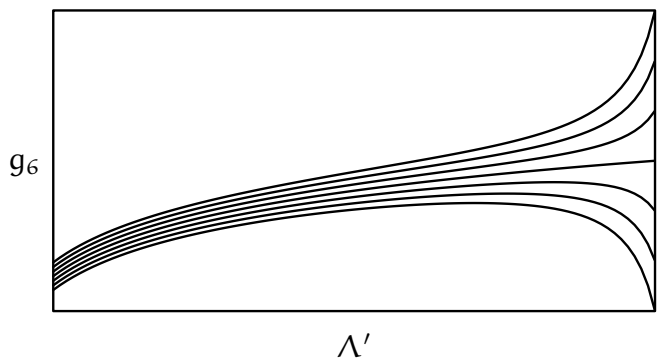

  \begin{center}
    \begin{empgraph}(80,40)
      pi := 3.14159; g := 10;
      vardef ln expr x = (mlog x) / 256 enddef;
      pickup pencircle scaled 1pt;
      setcoords (linear, linear);
      setrange (0.1, -6, 1, 6);
      for x = -0.6 step 0.2 until 0.6:
        path p;
        for mu = 0.1 step 0.01 until 1:
          augment.p (mu, 2*(ln mu) + x/(1.1-mu*mu));
        endfor
        gdraw p;
      endfor
      otick.bot ("", 1) withcolor white;
      glabel.lft (btex $g_6$ etex, OUT);
      glabel.bot (btex $\Lambda'$ etex, OUT);
    \end{empgraph}
  \end{center}
  \caption{\label{fig:infrared}%
    Contributions of irrelevant operators do \emph{not} necessarily
    vanish in the infrared, but the dependence of these contributions
    on their initial conditions at high energy vanishes in the
    infrared.}
\end{figure}
To avoid a possible misunderstanding induced by the technical term ``irrelevant
operator'', it is important to stress that the contributions of
irrelevant operators do \emph{not} necessarily vanish in the infrared.
Instead, the dependence of these contributions on their initial
conditions at high energy vanishes, because the \ac{RG}~trajectories
flow together in the infrared, as shown in figure~\ref{fig:infrared}.

Thus the initial conditions for the irrelevant operators have no effect
on the results of low energy experiments. In fact, the coefficients of
the irrelevant operators can be set to zero at the high scale without
changing the observable physics at lower energies.  Therefore, the
phenomena can be described by a renormalizable theory.

To avoid another possible misunderstanding, this observation does
not ``prove'' that only a renormalizable~\ac{QFT} can describe the
observed phenomena.  It only states that there is always are
renormalizable~\ac{QFT} that is indistinguishable from a
non-renormalizable theory at low energies.  Since renormalizable
theories are technically more convenient and depend on less
parameters, common sense and Occam's razor suggest to
prefer the renormalizable~\ac{QFT} with only relevant or marginal
interactions at the high scale over the others with the same infrared
behavior.

The technical advantage of renormalizable theories lies in the fact
that, for a given set of conserved quantum numbers, there is only a
finite number of relevant or marginal operators, which is even quite
small.  This simplifies calculations and increases the predictive
power of these theories because fewer free parameters have to be
fitted in experiments.

In fact, from the point of view of the~\ac{RG}, the relevant (``super
renormalizable'') operators are more problematic than the non
renormalizable ones.  The most prominent examples are masses of light
particles.  Unless there is a symmetry that forces the mass to be
small, most trajectories will lead to a value of the mass of the
order of the high energy cut off scale, which is incompatible with the
assumption that the particle is light.  The selection of a trajectory
that flows towards a light mass in the infrared requires unnatural
fine tuning in this case.

Fermion masses can be protected
by a chiral symmetry and need no fine tuning.  The same applies to the
Goldstone bosons of a spontaneous symmetry breaking.  Goldstone's
theorem protects their masses as well.  Other scalars need to be
related to fermions through a supersymmetry in order to be protected
against a naturally large mass.

The marginal operators are the most interesting from the theoretical
point of view.  Without interaction, their contributions do not change
with the scale.  Therefore even weak interactions can make the
marginal couplings grow strong or become weak asymptotically.  From
the~\ac{RGE} for~$g_4$
\begin{equation}
\label{eq:RGE-phi4}
   {\Lambda'}^2\frac{\mathrm{d}g_4(\Lambda')}{\mathrm{d}{\Lambda'}^2}
      = \beta (g_4(\Lambda'))
\end{equation}
it can be seen that the behavior of the coupling depends on the sign
of the $\beta$-function
\begin{itemize}
  \item $\beta>0$ (as in the scalar example considered above:
    $\beta=3g_4^2/(32\pi^2)$): the coupling grows strong in the
    ultraviolet and becomes weak in the infrared,
  \item $\beta<0$ (e..\,g.~\ac{QCD}, see section~\ref{sec:af}): the
    coupling grows strong in the infrared and becomes weak in the
    ultraviolet.
\end{itemize}
The solution of~(\ref{eq:RGE-phi4}) for our scalar example is
\begin{equation}
   g_4(\Lambda') =
      \frac{g_4(\Lambda)}{1 + \frac{3}{32\pi^2} g_4(\Lambda)
                            \ln\frac{\Lambda^2}{{\Lambda'}^2}}\,.
\end{equation}
Figure~\ref{fig:higher-orders} displays more contributions to the
$\beta$-function for~$g_4$, but all these contributions
are~$O(\Lambda^{-2})$ and can be ignored in leading order.

\begin{figure}
  \begin{center}
    \begin{fmfgraph}(20,15)
       \fmfstraight
       \fmfleftn{l}{2}
       \fmfrightn{r}{2}
       \begin{fmffor}{i}{1}{1}{2}
         \fmf{dashes}{l[i],v,r[i]}
       \end{fmffor}
       \fmf{dbl_plain,left=-90}{v,v}
       \fmfdot{v}
    \end{fmfgraph}\qquad\qquad
    \begin{fmfgraph}(20,15)
       \fmfstraight
       \fmfleftn{l}{2}
       \fmfrightn{r}{2}
       \begin{fmffor}{i}{1}{1}{2}
         \fmf{dashes}{l[i],v,r[i]}
       \end{fmffor}
       \fmf{dbl_plain,left=90}{v,v}
       \fmf{dbl_plain,left=-90}{v,v}
       \fmfdot{v}
    \end{fmfgraph}
  \end{center}
  \caption{\label{fig:higher-orders}%
    Suppressed contributions to the $\beta$-function for~$g_4$, which
    are~$O(\Lambda^{-2})$.}
\end{figure}
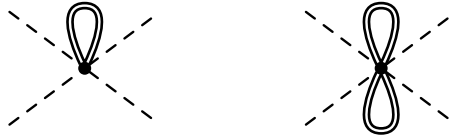

If the interactions are strong enough, relevant operators can
become marginal or even irrelevant and vice versa.  However, this is
impossible in the perturbative domain and therefore outside our
present scope.
The non perturbative calculation of path integrals via the solution of
a~\ac{RGE} is in principle possible, but technically very complicated.
The known solutions require a truncation of the~\ac{RGE} and the
systematic error introduced in this fashion is very hard to
quantify.

\subsection{Leading Logarithms}
\label{sec:lla}

In addition to providing an explanation for the fortunate circumstance
that nature seems to be described by a renormalizable~\ac{QFT}, the
study of the \ac{RG}~flows provides a more reliable calculational tool
than the direct perturbative evaluation of Feynman diagrams.

In theories with widely separated mass scales, \ac{PT}
suffers from large coefficients in the perturbative series that are
caused by logarithmic integrals from one mass scale to the next.  In
contract, the coefficients of the~\ac{RGE} never contain more than one
scale, which is determined by the slice of momentum space that is
being integrated out.  Thus these coefficients can not contain large
logarithms and the theory remains perturbative as long as the
couplings are weak.  Observable large logarithms are recovered only 
in the final stage of the calculation, the solution of the~\ac{RGE}.
Therefore, large unphysical logarithms that cancel in observables can
never appear in intermediate steps.

\subsection{Callan Symanzik Equations}
\label{sec:cse}

The calculation of integrals over a small slice of momentum space
like~(\ref{eq:finite-I(n,D)}) or~(\ref{eq:finite-fish}) is technically
demanding, because the finite integration limits make it impossible to
use many of the standard tricks of the trade, like the shifting of
integration variables.  In addition, fixed cut-offs manifestly violate
gauge invariance, which makes their practical application in gauge
theories~\cite{Warr:1988:RGE1,*Warr:1988:RGE2} very awkward. 

At this point, the relation of the \ac{RG}~flow to perturbative
renormalization comes in handy.  A change of the lower cut off
scale~$\Lambda'$ corresponds to a change of the Lagrangian that keeps
the physics unchanged.  In perturbative renormalization, the
renormalization scale~$\mu$ plays a similar r\^ole as a scale that
characterizes how much of the interaction is absorbed into the
Lagrangian in the form of local counter terms.  Therefore, it is tempting
to identify~$\Lambda'$ with~$\mu$.

The analogy is not complete, however. The \emph{hard} cut
off~$\Lambda'$ is replaced by counter terms that correspond to a
\emph{soft} cut off, which is switched on gradually.  This is the price
to pay for gauge invariance and technical convenience.  If we were to
sum \emph{all} orders of the perturbative series, both methods will
give the same result.  For a truncated series, however, differences of
higher order will remain.

Consider a general $n$-point function
\begin{equation}
  G^{(n)}(x_1,x_2,\ldots,x_n; g, \mu)
    = \Braket{0|\mathrm{T}\phi(x_1)\phi(x_2)\ldots\phi(x_n)|0}\,,
\end{equation}
which is renormalized at~$\mu$, i.\,e.~the invariants of the momenta
in the renormalization conditions are equal to~$\mu$.  A change
of~$\mu$ will not change the physics.  Instead, the theory is shifted
to another point on the \emph{same} trajectory.  Therefore, there is a
new value~$g'$ for the renormalized coupling, such that only the
unobservable normalization of the field operators is
changed~$\phi(x)\to1/\sqrt{Z}\phi(x)$.  Thus
\begin{equation}
  G^{(n)}(x_1,x_2,\ldots,x_n; g', \mu')
    = Z^{-n/2}(\mu,\mu') G^{(n)}(x_1,x_2,\ldots,x_n; g, \mu)\,.
\end{equation}
As above, it is more convenient to replace the discrete \ac{RG}
transformations by continuous transformations with infinitesimal
generators
\begin{equation}
  \mu' \frac{\mathrm{d}}{\mathrm{d}\mu'}
      \left( Z^{n/2}(\mu,\mu') G^{(n)}(x_1,\ldots,x_n; g', \mu')
         \right) = 0
\end{equation}
where $\mathrm{d}/\mathrm{d}\mu'$ is a \emph{total} derivative, that
includes the change of the coupling~$g$ as well.  Therefore,
the~\acf{CSE} reads
\begin{equation}
\label{eq:CSE}
  \left(\mu \frac{\partial}{\partial\mu}
        + \beta(g)\frac{\partial}{\partial g}
        + n\gamma(g) \right)
    G^{(n)}(x_1,\ldots; g, \mu) = 0\,,
\end{equation}
with
\begin{subequations}
\begin{align}
  \beta(g) &= \mu\frac{\mathrm{d}}{\mathrm{d}\mu} g(\mu) \\
  \gamma(g) &= \frac{1}{2Z(\mu_0,\mu)}
    \mu\frac{\mathrm{d}}{\mathrm{d}\mu} Z(\mu_0,\mu)\,.
\end{align}
\end{subequations}
The $\beta$-function is dimensionless.  In a massless theory, it can
therefore not depend on~$\mu$ for dimensional reasons.  In theories
with masses, there are always renormalization schemes that retain this
property~\cite{Weinberg:1973:RG}.  In other schemes,
$\beta$ can depend on ratios of~$\mu$ and masses.

However, it is \emph{not} obvious that~$\gamma$ does not depend
on~$\mu_0$ through the ratio $\mu/\mu_0$.  In renormalizable theories,
the limit~$\mu_0\to\infty$ exists and~$\gamma$ can not depend
on~$\mu_0$.  As we have explained in section~\ref{sec:relevant} above,
we can always chose a trajectory 
that corresponds to a renormalizable theory and drop the dependence
of~$\gamma$ on~$\mu_0$.

The actual calculation of the coefficient functions~$\beta$
and~$\gamma$ proceeds by calculating a sufficiently large set of
renormalized \greenfunction{}s and finding functions~$\beta$
and~$\gamma$ such that~(\ref{eq:CSE}) is satisfied.  The non-trivial
statement of the \ac{CSE} is that one universal function~$\beta$
for each coupling and one universal function $\gamma$ for each field
suffices to satisfy~(\ref{eq:CSE}) for \emph{all} \greenfunction{}s.

If there is more than one coupling and more than one field, the
generalization is obvious
\begin{equation}
  \left(\mu \frac{\partial}{\partial\mu}
        + \sum_{i=1}^k \beta_i(g)\frac{\partial}{\partial g_i}
        + \sum_{j=1}^m n_j\gamma_j(g) \right)
    G^{(n_1,\ldots,n_m)}(x_1,\ldots; g_1,\ldots, g_k, \mu) = 0\,.
\end{equation}
For example, the \ac{QED} vertex functions obeys the~\ac{CSE}
\begin{equation}
  \left(\mu \frac{\partial}{\partial\mu}
        + \beta(e)\frac{\partial}{\partial e}
        + 2\gamma_\psi(e) + \gamma_A(e) \right)
    \Braket{0|\mathrm{T}\psi(x_1)\bar\psi(x_2)A_\nu(x_3)|0} = 0\,.
\end{equation}

\subsection{Running Couplings and Anomalous Dimensions}
\label{sec:anodim}
As a concrete example for solving the~\ac{CSE}, consider a four-point
function
\begin{equation}
  G^{(4)} (P;g,\mu) = 
  \parbox{24\unitlength}{%
    \fmfframe(4,2)(4,3){%
      \begin{fmfgraph*}(15,10)
         \fmfstraight
         \fmfleftn{l}{2}
         \fmfrightn{r}{2}
         \fmflabel{$p_1$}{l1}
         \fmflabel{$p_2$}{l2}
         \fmflabel{$p_3$}{r2}
         \fmflabel{$p_4$}{r1}
         \begin{fmffor}{i}{1}{1}{2}
           \fmf{plain}{l[i],v,r[i]}
         \end{fmffor}
         \fmfblob{.3w}{v}
      \end{fmfgraph*}}}
     = \mathop{\text{F.T.}}
         \Braket{0|\mathrm{T}\phi(x_1)\phi(x_2)\phi(x_3)\phi(x_4)|0}
\end{equation}
in the deep Euclidean domain (this does not correspond to a physical
amplitude, but can be part of one):
\begin{subequations}
\begin{align}
  p_i^2 &= - P^2 \\
  p_ip_j &= 0\,.
\end{align}
\end{subequations}
Lowest order~\ac{PT} gives for the connected (but not
amputated) part
\begin{equation}
\label{eq:G4-perturbative}
  G^{(4)} (P;g,\mu)
    = \left(\frac{-i}{P^2}\right)^4 (-ig) + O(g^2)\,.
\end{equation}
Ignoring corrections from mass terms, dimensional analysis predicts
that to all orders in~\ac{PT}
\begin{equation}
  G^{(4)} (P;g,\mu) = \left(\frac{-i}{P^2}\right)^4 \hat G^{(4)} (P/\mu;g)\,,
\end{equation}
i.\,e.~that~$G^{(4)}$ is a homogenous function and solves the
differential equation
\begin{equation}
\label{eq:homogenous}
  \mu\frac{\partial}{\partial\mu} G^{(4)} (P;g,\mu)
   = - \left( 8 + P\frac{\partial}{\partial P} \right) G^{(4)} (P;g,\mu)\,.
\end{equation}
Plugging~(\ref{eq:homogenous}) into the \ac{CSE}~(\ref{eq:CSE}), we can
trade the dependence of the renormalization scale for the momentum
dependence and obtain the \acf{RGE} for the four point function
\begin{equation}
\label{eq:RGE}
  \left(P \frac{\partial}{\partial P}
        - \beta(g)\frac{\partial}{\partial g}
        + 8 - 4\gamma(g) \right)
    G^{(4)} (P;g,\mu) = 0\,.
\end{equation}
The \ac{RGE} can be solved with standard techniques for the
integration of a \ac{PDE}.  The introduction of a \emph{running
coupling constant}~$\bar g(P;g)$
\begin{subequations}
\begin{align}
  P \frac{\mathrm{d}}{\mathrm{d}P} \bar g(P;g)
            &= \beta(\bar g(P;g)) \\
   g(\mu;g) &= g
\end{align}
\end{subequations}
absorbs the differential operator
\begin{equation}
  \left(P \frac{\partial}{\partial P}
        - \beta(g)\frac{\partial}{\partial g} \right)
     \bar g(P;g) = 0
\end{equation}
and the general solution can be written as
\begin{equation}
\label{eq:RGE-solution}
  G^{(4)} (P;g,\mu) = \left(\frac{-i}{P^2}\right)^4 
    \mathcal{G}^{(4)} (\bar g(P;g))
     \cdot \exp\left(4\int\limits_g^{\bar g(P;g)}\!\mathrm{d}g'\,
           \frac{\gamma(g')}{\beta(g')}\right)\,,
\end{equation}
where~$\mathcal{G}^{(4)}$ is an \emph{arbitrary} function of one
variable, which is \emph{not} determined by the~\ac{RG}.  However,
\emph{matching} the solution~(\ref{eq:RGE-solution}) of the \ac{RGE}
to the perturbative result~(\ref{eq:G4-perturbative}) at~$P=\mu$
gives
\begin{equation}
 \mathcal{G}^{(4)} (\bar g) = -i\bar g + O({\bar g}^2)\,.
\end{equation}

In general, we can identify two elements in the solution of
the~\ac{RGE}.  One part of the solution is given by the perturbative
result at a given order with the coupling constant replaced by the
running coupling at the appropriate scale.  In addition, there is an
exponential factor modifying the scaling behavior of each field in
the $n$-point function under consideration.  Since the~$\gamma$ appear
in the numerator of the exponent, they are usually called
\emph{anomalous dimensions}.

{}From the $\beta$-function for~$\phi^4$
\begin{equation}
  \beta(g) = \frac{3}{16\pi^2} g^2 + O(g^3)
\end{equation}
follows the running coupling
\begin{equation}
  \bar g(P;g) = \frac{g}{1 - \frac{3}{16\pi^2} g \ln\frac{P}{\mu}}\,.
\end{equation}

The solution~(\ref{eq:RGE-solution}) has the nice feature that
\emph{all} large logarithms~$\ln(P^2/\mu^2)$ are collected in two
universal places, that are common to all \greenfunction{}s: the running
coupling~$\bar g(P;g)$ and the exponents of the
scale factors.  Thus the large logarithms have been resummed
successfully in universal quantities and~\ac{PT} is
reliable, as long as~$\bar g(P;g)$ is small.

Unfortunately, this procedure does not work as straightforwardly for
all \greenfunction{}s.  The \ac{RGE} is much harder to solve if not all
external momenta scale uniformly.  If there is more than one relevant
mass scale, the coefficient functions themselves can contain large
logarithms.  The classical example for large logarithms that can not
be obtained from the~\ac{RG} are the Sudakov double
logarithms~\cite{Sudakov:1956:logs} in exclusive scattering processes
\begin{equation}
  \parbox{27\unitlength}{%
    \fmfframe(3,3)(3,3){%
      \begin{fmfgraph*}(20,15)
         \fmfstraight
         \fmfleft{l}
         \fmfrightn{r}{2}
         \fmflabel{$Q^2$}{l}
         \fmflabel{$m^2$}{r1}
         \fmflabel{$m^2$}{r2}
         \fmf{photon}{l,v}
         \fmf{fermion,tension=0.5}{v1,v,v2}
         \fmf{plain}{r1,v1}
         \fmf{plain}{v2,r2}
         \fmffreeze
         \fmf{photon,right=0.3}{v1,v2}
         \fmfdot{v1,v,v2}
      \end{fmfgraph*}}}
     = -\frac{\alpha}{2\pi} \ln^2\left(\frac{Q^2}{m^2}\right)
\end{equation}

\subsection{Asymptotic Freedom}
\label{sec:af}

As discussed earlier, the behavior of the $\beta$-function in the
vicinity of~$g=0$ determines the properties of~\ac{PT}.
For a single coupling, the $\beta$-function must vanish for~$g=0$,
because without interactions, there is no mechanism that could induce
running of the coupling.

\begin{figure}
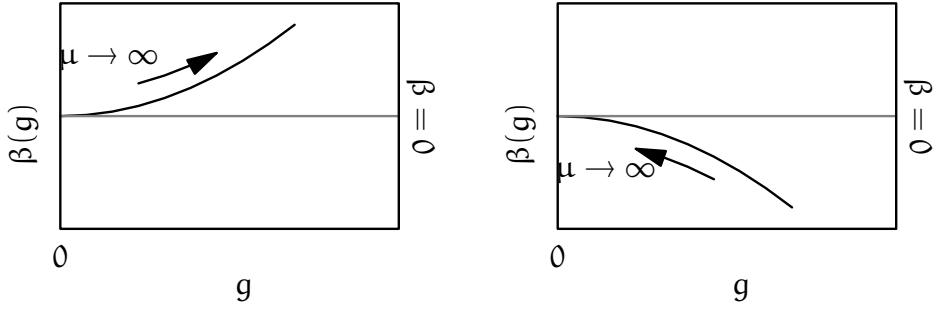

  \begin{center}
    \begin{empgraph}(45,30)
      pickup pencircle scaled 1pt;
      setcoords (linear, linear);
      setrange (0, -0.5, 1.3, 0.5);
      path p;
      for x = 0 step 0.1 until 1:
        augment.p (x, x*x/2);
      endfor
      gdraw p;
      gdrawarrow (subpath (3,6) of p shifted (0,0.1));
      glabel.ulft (btex $\mu\to\infty$ etex, 1);
      otick.bot (btex $0$ etex, 0) withcolor white;
      grid.rt (btex $\beta=0$ etex rotated -90, 0) withcolor 0.5white;
      glabel.lft (btex $\beta(g)$ etex rotated 90, OUT);
      glabel.bot (btex $g$ etex, OUT);
    \end{empgraph}
    \qquad
    \begin{empgraph}(45,30)
      pickup pencircle scaled 1pt;
      setcoords (linear, linear);
      setrange (0, -0.5, 1.3, 0.5);
      path p;
      for x = 0 step 0.1 until 1:
        augment.p (x, -x*x/2);
      endfor
      gdraw p;
      gdrawarrow (subpath (6,3) of p shifted (0,-0.1));
      glabel.llft (btex $\mu\to\infty$ etex, 2);
      otick.bot (btex $0$ etex, 0) withcolor white;
      grid.rt (btex $\beta=0$ etex rotated -90, 0) withcolor 0.5white;
      glabel.lft (btex $\beta(g)$ etex rotated 90, OUT);
      glabel.bot (btex $g$ etex, OUT);
    \end{empgraph}
  \end{center}
  \caption{\label{fig:IR/UV-free}%
    $\beta$-functions for infrared free (left) and ultraviolet free
    (right) theories.  The arrow denotes the direction of the
    evolution for a rising scale.}
\end{figure}
\begin{figure}
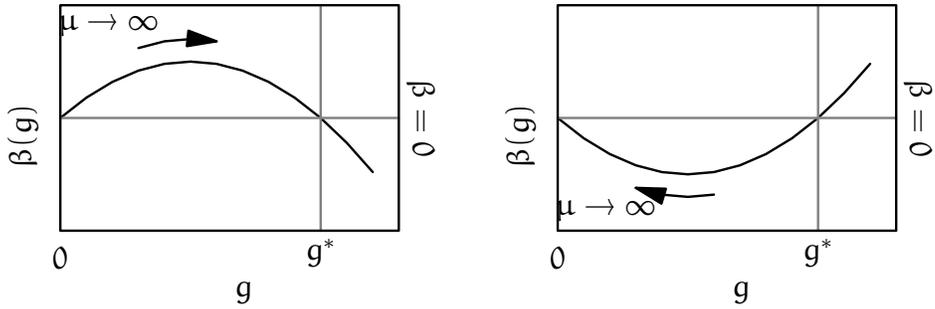

  \begin{center}
    \begin{empgraph}(45,30)
      pickup pencircle scaled 1pt;
      setcoords (linear, linear);
      setrange (0, -0.5, 1.3, 0.5);
      path p;
      for x = 0 step 0.1 until 1.3:
        augment.p (x, x*(1-x));
      endfor
      gdraw p;
      gdrawarrow (subpath (3,6) of p shifted (0,0.1));
      glabel.ulft (btex $\mu\to\infty$ etex, 1);
      otick.bot (btex $0$ etex, 0) withcolor white;
      grid.bot (btex $g^*$ etex, 1) withcolor 0.5white;
      grid.rt (btex $\beta=0$ etex rotated -90, 0) withcolor 0.5white;
      glabel.lft (btex $\beta(g)$ etex rotated 90, OUT);
      glabel.bot (btex $g$ etex, OUT);
    \end{empgraph}
    \qquad
    \begin{empgraph}(45,30)
      pickup pencircle scaled 1pt;
      setcoords (linear, linear);
      setrange (0, -0.5, 1.3, 0.5);
      path p;
      for x = 0 step 0.1 until 1.3:
        augment.p (x, -x*(1-x));
      endfor
      gdraw p;
      gdrawarrow (subpath (6,3) of p shifted (0,-0.1));
      glabel.llft (btex $\mu\to\infty$ etex, 2);
      otick.bot (btex $0$ etex, 0) withcolor white;
      grid.bot (btex $g^*$ etex, 1) withcolor 0.5white;
      grid.rt (btex $\beta=0$ etex rotated -90, 0) withcolor 0.5white;
      glabel.lft (btex $\beta(g)$ etex rotated 90, OUT);
      glabel.bot (btex $g$ etex, OUT);
    \end{empgraph}
  \end{center}
  \caption{\label{fig:UV/IR-fp}%
    $\beta$-functions with ultraviolet (left) and infrared (right)
    stable fixed points. The arrow denotes the direction of the
    evolution for a rising scale.}
\end{figure}
Figures~\ref{fig:IR/UV-free} and~\ref{fig:UV/IR-fp} display the four
possible different scenarios.  On the left hand side of
figure~\ref{fig:IR/UV-free}, we have $\beta>0$ everywhere and the
running coupling grows in the ultraviolet without limits.  However,
\ac{PT} is reliable in the infrared.  

On the right hand side of figure~\ref{fig:IR/UV-free}, we have
$\beta<0$ everywhere and the running coupling grows in the infrared
without limits. \ac{PT} is reliable in the ultraviolet,
but problematic in the infrared.  The theory of strong
interactions, \ac{QCD}, has this
property~\cite{Gross/Wilczek:AsymptoticFreedom:Orig,
*Politzer:AsymptoticFreedom:Orig}, called \emph{asymptotic freedom}
and it is known that non abelian gauge theories are the only class of
theories in four space-time dimensions that can exhibit asymptotic freedom.

In the left hand side of figure~\ref{fig:UV/IR-fp}, the coupling grows
in the ultraviolet, but the $\beta$-function makes a turn and the
coupling becomes constant after running into an \emph{ultraviolet
stable fixed point} at~$g^*$, because~$\beta(g^*)=0$.  The opposite
phenomenon appears in the right hand side of
figure~\ref{fig:UV/IR-fp}, where~$g^*$ is an \emph{infrared stable
fixed point}.

\subsection{Dimensional Regularization}

As mentioned above, the \ac{RG} functions~$\beta(g)$ and~$\gamma(g)$
are universal.  They are determined by calculating~$\mu\partial
G^{(n)}/\partial\mu$ for a sufficient set of~$G^{(n)}$ and solving
the~\ac{CSE} for~$\beta(g)$ and~$\gamma(g)$.

A typical integral appearing in these equations is
\begin{equation}
  \hat I_{n,m}(D,M^2)
     = \mu^{D-4}\int\!\frac{\mathrm{d}^Dk}{(2\pi)^D}
        \frac{\left(k^2\right)^n}{\left(k^2-M^2+i\epsilon\right)^m}\,,
\end{equation}
where the factor~$\mu^{4-D}$ has been added in order to make the
mass dimension of~$\hat I$ independent of the dimension of space time.
As long as~$m-n-D/2$ and~$D/2+n$ are not equal to negative whole
numbers, the integral can be evaluated
\begin{multline}
  \hat I_{n,m}(D,M^2) = \\
     \frac{(-1)^{n+m}i}{16\pi^2}
          \left(M^2\right)^{2+n-m}
      \cdot \left(\frac{M^2}{16\pi^2\mu^2}\right)^{D/2-2}
            \frac{\Gamma(m-n-D/2)\Gamma(D/2+n)}{\Gamma(m)\Gamma(D/2)}
\end{multline}
and the logarithmic divergency of~$\hat I_{0,2}(4,M^2)$ shows up as
the pole of the Euler $\Gamma$-function.  This divergency is
regularized by moving away from~$D=4$
\begin{equation}
  \hat I_{0,2}(4-2\epsilon,M^2)
     = \frac{i}{16\pi^2}
          \left(\frac{M^2}{16\pi^2\mu^2}\right)^{-\epsilon}
            \Gamma(\epsilon)\,.
\end{equation}
$\Gamma(\epsilon)$ can be expanded in a Laurent series, that starts
with a single pole
\begin{equation}
  \hat I_{0,2}(4-2\epsilon,M^2)
     = \frac{i}{16\pi^2} \left( \frac{1}{\epsilon}
          + \ln\frac{\mu^2}{M^2} + 2 \ln4\pi - \gamma_E \right)\,.
\end{equation}
In a massless theory, the dependence on~$\mu$ must be identical to the
dependence on the renormalization scale and the expansion of the
$\Gamma$-function shows that it suffices to determine the residues of
the poles in~$\epsilon=2-D/2$.  Even in massive theories, there are
renormalization schemes that retain this convenient
property~\cite{Weinberg:1973:RG}.

This procedure of \emph{dimensional
regularization}~\cite{tHooft/Veltman:1972:DimReg} and \emph{minimal
subtraction} of the poles in~$1/(D-4)$ is particularly useful in
gauge theories, because gauge invariance is not violated by the
regularization and no additional counter terms are needed in order to
maintain gauge invariance.  Since gauge invariance softens
divergencies, dimensional regularization avoids the overestimation of
radiative corrections induced by a naive application of regularization
procedures that violate gauge invariance.

\subsection{Gauge Theories}

In~\ac{QED}, we find for the lowest order \ac{RG}~functions
\begin{subequations}
\begin{align}
\label{eq:QED-vacpol}
  \beta(e) &= \frac{e^3}{12\pi^2} \\
  \gamma_\psi(e) &= \frac{e^2}{16\pi^2} \\
  \gamma_A(e) &= \frac{e^2}{12\pi^2}\,.
\end{align}
\end{subequations}
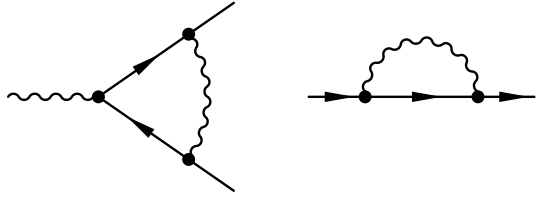
\begin{figure}
  \begin{center}
    \begin{fmfgraph}(30,25)
       \fmfstraight
       \fmfleft{l}
       \fmfrightn{r}{2}
       \fmf{photon}{l,v}
       \fmf{fermion,tension=0.5}{v1,v,v2}
       \fmf{plain}{r1,v1}
       \fmf{plain}{v2,r2}
       \fmffreeze
       \fmf{photon,right=0.3}{v1,v2}
       \fmfdot{v,v1,v2}
    \end{fmfgraph}\qquad
    \begin{fmfgraph}(30,25)
       \fmfstraight
       \fmfleft{l}
       \fmfright{r}
       \fmf{fermion}{l,v1}
       \fmf{fermion,tension=0.5}{v1,v2}
       \fmf{fermion}{v2,r}
       \fmffreeze
       \fmf{photon,left}{v1,v2}
       \fmfdot{v1,v2}
    \end{fmfgraph}
  \end{center}
  \caption{\label{fig:QED-WI}%
    The contributions of the vertex function and the self energy to the
    $\beta$-function in \protect\acs{QED} cancel.}
\end{figure}
There is a peculiarity of~\ac{QED}: the Ward identity enforces that
the contributions of the vertex function and the self energy (see
figure~\ref{fig:QED-WI}) to the $\beta$-function cancel to all orders
and the $\beta$-function can be calculated from the vacuum
polarization (figure~\ref{fig:vacpol}) alone.

{}From~(\ref{eq:QED-vacpol}), we see that $\beta_{\text{QED}} > 0$ and
that~\ac{QED} is not asymptotically free.  Higher order calculations
have not revealed an ultraviolet fixed point either.

A more complicated calculation in a non-abelian $\textrm{SU}(N_C)$ gauge
theory with $N_f$~quarks has the 
result~\cite{Gross/Wilczek:AsymptoticFreedom:Orig, 
*Politzer:AsymptoticFreedom:Orig}
\begin{equation}
\label{eq:beta-non-abelian}
  \beta(g) = \frac{e^3}{16\pi^2}
    \left( C_F \frac{N_f}{2} - \frac{11}{3} N_C \right)\,,
\end{equation}
where
\begin{equation}
  C_F = \frac{N_C^2-1}{2N_C}\,.
\end{equation}
In~\ac{QCD} (i.\,e.~$N_C=3$), this means
\begin{equation}
\label{eq:beta-QCD}
  \beta_{\text{QCD}}(g) = \frac{e^3}{16\pi^2}
    \left( \frac{2}{3}N_f - 11 \right)
\end{equation}
and the theory is indeed asymptotically free for~$N_f<33/2$.
As a result, the running coupling of~\ac{QCD} falls with rising energy:
\begin{equation}
   \alpha_s(Q^2) = \frac{4\pi}{\beta_0\ln(Q^2/\Lambda_{\text{QCD}}^2)}
\end{equation}
where $\unit[0.15]{GeV}<\Lambda_{\text{QCD}}<\unit[0.25]{GeV}$
and $\beta_0=11-2N_f/3$.

\begin{figure}
  \begin{center}
    \begin{fmfgraph}(25,10)
       \fmfstraight
       \fmfleft{l}
       \fmfright{r}
       \fmf{gluon}{l,v1}
       \fmf{gluon,tension=0.3,left}{v1,v2,v1}
       \fmf{gluon}{v2,r}
       \fmfdot{v1,v2}
    \end{fmfgraph} \qquad
    \begin{fmfgraph}(25,10)
       \fmfstraight
       \fmfleft{l}
       \fmfright{r}
       \fmf{gluon}{l,v1}
       \fmf{ghost,tension=0.3,left}{v1,v2,v1}
       \fmf{gluon}{v2,r}
       \fmfdot{v1,v2}
    \end{fmfgraph}
  \end{center}
  \caption{\label{fig:QCD-vacpol}%
    Non abelian contributions to the vacuum polarization: gluon and
    ghost loops.}
\end{figure}
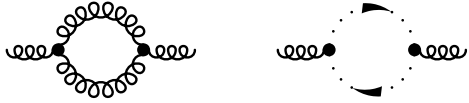
The most obvious difference to an abelian gauge theory like~\ac{QED} is the
existence of the genuinely non-abelian contributions to the vacuum
polarization displayed in figure~\ref{fig:QCD-vacpol}.  These diagrams
do indeed give a negative contribution to the $\beta$-function.

Intuitively, the negative contribution from the gluon and ghost loops
gives a fair description of the interplay of the positive quark
contribution and the negative non-abelian contributions.  Strictly
speaking, however this explanation is not correct, because the
cancellations of vertices and self energies familiar from~\ac{QED} are
not operative and more complicated diagrams also give contributions to
the $\beta$-function.


\section{Effective Field Theory}
\label{sec:eft2}

\begin{cutequote*}{Egon Friedell}{Kulturgeschichte Griechenlands}
Ein Bild erh\"alt ja gerade dadurch seine
Brauchbarkeit, da\ss{} es \emph{nicht} die Sache ist.
\end{cutequote*}

\subsection{Flavor Thresholds and Decoupling}
\label{sec:decoupling}

So far, all effects of particle masses have been neglected, in order
to highlight the simplicity of the~\ac{RG} approach.  This is often a
very good approximation.  For example at LEP2, fermion masses can be
neglected in four fermion production, unless one of the final state
particles is an electron and a $t$-channel singularity has to be
regularized.  The massless approximation is a good approximation,
because the LEP2 center of mass energy is far from all flavor
thresholds: $m_b\ll\sqrt{s_{\text{LEP2}}}<m_t$.  Top quarks can not be
produced and $m_b/M_W$~is a small number.

Close to a threshold, the~\ac{RG} loses some of its appeal, because
the value of the threshold introduces a second scale that enters the
calculation of the \ac{RG} functions and makes calculations more
complicated.  However, \acf{EFT}~\cite{Georgi:EFT:Review}
provides a systematic way of dealing with flavor thresholds in a
\ac{RG}~calculation.

Recall the $\beta$-function of a
$\textrm{SU}(N_C)$~gauge theory~(\ref{eq:beta-non-abelian}).  The
quarks loops in the gluon vacuum polarization induce a dependence on
the number of quark flavors
\begin{equation}
\label{eq:beta-flavor}
  \parbox{31\unitlength}{%
    \begin{fmfgraph*}(30,15)
       \fmfstraight
       \fmfleft{l}
       \fmfright{r}
       \fmflabel{$p$}{l}
       \fmf{gluon}{l,v1}
       \fmf{fermion,tension=0.5,left}{v1,v2,v1}
       \fmf{gluon}{v2,r}
       \fmfdot{v1,v2}
    \end{fmfgraph*}}
  \propto
  \frac{e^3}{16\pi^2} C_F \frac{N_f}{2}
\end{equation}
and in a concrete calculation we have to answer the question for the
value of~$N_f$.  This question can not be answered by counting the
quarks in table~\ref{tab:SM} alone.  To see this, consider the typical
case~$m_b^2\ll -p^2\ll m_t$ and~$m_b^2\ll \mu^2\ll m_t$.  Then the
contribution of a top quark loop in~(\ref{eq:beta-flavor}) is
\begin{equation}
  \Pi^R(p^2) = O(\frac{p^2}{m_t^2},\frac{\mu^2}{m_t^2})
\end{equation}
and is very small below the threshold.

The purpose of the~\ac{RG} is to systematically resum the large
logarithms in the vacuum polarization and other diagrams, but we find that
there are no large logarithms below the threshold, as long as the
renormalization scale is chosen sensibly.  Therefore, the top flavor
must not be included in~(\ref{eq:beta-flavor}) below threshold
and~$N_f=5$ is the appropriate value between the bottom and top
thresholds.

The fact that a heavy particle does not contribute to \ac{RG}
functions below its threshold is non-trivial.  In the unrenormalized
version of~(\ref{eq:beta-flavor}), each flavor contributes equally to
the high energy tail of the loop integral, as long as the cut off is
significantly above the threshold of this flavor.  Only after
renormalization, it can be seen that each flavor also contributes
equally to the counter term and that the net contribution of a heavy 
flavor vanishes, if the \greenfunction{}s are renormalized below the
threshold of this flavor.

Since the~\ac{RG} never deals with the loop integrals over all of
momentum space, the cancellations described in the previous paragraph
take place automatically.  The $\beta$-function describes the rate of
change of the running coupling and if only active flavors contribute
to this change, the cancellation of contributions from heavy flavors
and their counter terms is already manifest.  This argument, which has
been made heuristically using~(\ref{eq:beta-flavor}) for the
$\beta$-function, can be made rigorous in the \emph{decoupling
theorem}~\cite{Appelquist/Carazzone:1975:decoupling}; also for the
anomalous dimensions.

However, there is one circumstance under which the effects of a heavy
particle can be felt in the \ac{RG} flow \emph{below} its threshold.
This the case if the large mass violates a symmetry of the theory.
The most celebrated example of this effect is the breaking of the
$\textrm{SU}(2)_c$ custodial
symmetry~\cite{Sikivie/etal:1980:custodial} of the~\ac{SM} through the
heavy top mass and the 
corresponding contribution to the $\rho$-parameter.  In a perturbative
calculation with masses, there is a contribution to~$\rho$ that is
proportional to the quadratic mass splitting in the third generation
\begin{equation}
  \Pi_{W^+W^-}(q^2) =
    \parbox{31\unitlength}{%
      \begin{fmfgraph*}(30,20)
        \fmfleft{l}
        \fmfright{r}
        \fmf{boson}{l,vl}
        \fmf{boson}{vr,r}
        \fmf{fermion,label=$t$,left,tension=0.5}{vl,vr}
        \fmf{fermion,label=$b$,left,tension=0.5}{vr,vl}
        \fmfdot{vl,vr}
      \end{fmfgraph*}}
   \propto \frac{\alpha}{\pi}
     \frac{m_t^2-m_b^2}{M_W^2} + \cdots
\end{equation}
If every~$\textrm{SU}(2)_L$ doublet was degenerate, the decoupling
theorem would apply and each heavy doublet would decouple from the low
energy phenomena.  The resulting low energy physics could be described
by an~\ac{EFT} built from $\textrm{SU}(2)_c$-invariant interactions.
If, as in the third generation of the~\ac{SM}, there is a strong
lifting of the degeneracy, there will be an effect on the low energy
couplings that manifests itself as a $\rho$-parameter that is
different from unity.  Then the low energy \ac{EFT} below the top
threshold has to contain $\textrm{SU}(2)_c$-violating operators.
Below the bottom threshold, there are no new
$\textrm{SU}(2)_c$~violating contributions from the fermion sector,
but the effects accumulated between~$m_t$ and~$m_b$ remain.  If there
would be no reason to expect a custodial symmetry or~$\rho=1$, the low
energy \ac{EFT} would contain $\textrm{SU}(2)_c$-violating operators
from the outset, of course.

The general result of the decoupling theorem is that, unless
symmetries are violated, heavy particles can be removed from a theory,
if only phenomena below the energy scale given by the particle's mass
are to be calculated.  The only residual effect of the heavy particle
remains in a renormalization of coupling constants and fields.

\subsection{Integrating Out or Matching and Running}
\label{sec:integrating-out}

\begin{figure}
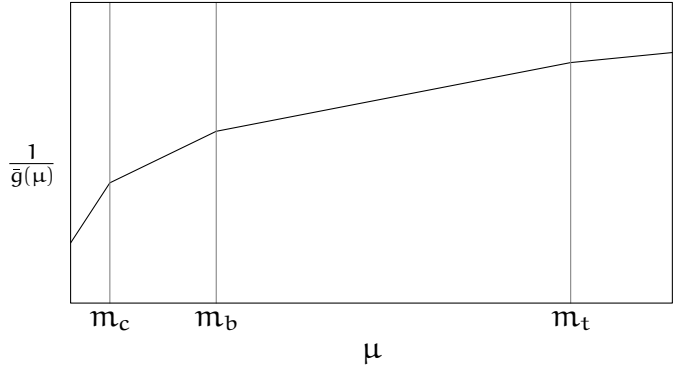

  \begin{center}
    \begin{empgraph}(80,40)
      path p;
      augment.p (1.0, 1/2);
      augment.p (1.5, 1/1);
      augment.p (4.5, 1/0.7);
      augment.p (175, 1/0.5);
      augment.p (500, 1/0.48);
      setcoords (log, linear);
      setrange (1, 0, 500, 2.5);
      gdraw p;
      grid.bot (btex $m_c$ etex, 1.5) withcolor 0.5white;
      grid.bot (btex $m_b$ etex, 4.5) withcolor 0.5white;
      grid.bot (btex $m_t$ etex, 175) withcolor 0.5white;
      glabel.lft (btex $\frac{1}{\bar g(\mu)}$ etex, OUT);
      glabel.bot (btex $\mu$ etex, OUT);
    \end{empgraph}
  \end{center}
  \caption{\label{fig:running-w/thresholds}%
    Running couplings with thresholds.  In an asymptotically free
    gauge theory, the running is slowed down at each new flavor
    threshold.}
\end{figure}
The decoupling of heavy flavors discussed in the previous section
provides a systematic calculational procedure for theories with more
than one mass scale.  For example, in order the calculate the strong
coupling at low energies from an experimental measurement at
a~$\unit[1]{TeV}$ \ac{LC}, one proceeds as follows:
\begin{enumerate}
  \item start with the measured value~$g_0$ at~$\mu_0=\unit[1]{TeV}\gg
     m_t$: $\bar g(\mu_0)=g_0$
  \item solve the massless~\ac{RGE} $\bar g(\mu)$ with the
     $\beta$-function for six active flavors from~$\mu=\mu_0$ down
     to~$\mu=m_t$
  \item start a second evolution step at~$\mu=m_t$, using the result
     of the first step as initial value
  \item solve the massless~\ac{RGE} $\bar g(\mu)$ with the
     $\beta$-function for five active flavors from $\mu=m_t$ down
     to~$\mu=m_b$.
  \item start a third evolution step at~$\mu=m_b$, \ldots
  \item and so on
\end{enumerate}
This procedure is known \emph{matching and running} and is illustrated
in figure~\ref{fig:running-w/thresholds}.  The matching at each
threshold is continuous, but not differentiable, because the rate of
change (i.\,e.~the $\beta$-function) changes discontinuously at each
threshold.

The approximation of the flavor thresholds by step functions is not
optimal, of course.  However, for many observables, the detailed
features of the threshold are much less important than the large
logarithms between thresholds and the matching and running procedure
provides a systematic way of resumming these logarithms.  In any case,
the calculations can be improved systematically by including higher
orders (NLO, NNLO) in the loop expansion, as well as power
corrections.

The most famous example (already mentioned in section~\ref{sec:sm?})
for a threshold effect is given by the running coupling of the three
\ac{SM} gauge groups.  The experimentally determined couplings do
\emph{not} meet~\cite{Amaldi/deBoer/Fuerstenau:1991:Unification} in
one point as would be a necessary condition of a~\ac{GUT}.  However,
matching the evolution of the couplings in the~\ac{SM} to the
evolution of the couplings in the~\ac{MSSM} at the threshold provided
by the supersymmetry breaking scale~$\mu\approx\unit[1]{TeV}$ causes
the three couplings to
meet~\cite{Amaldi/deBoer/Fuerstenau:1991:Unification}.  This may be an
accident, but it is the most suggestive phenomenological evidence for
supersymmetry so far.

\subsection{Important Irrelevant Interactions}
\label{sec:important-irrelevant}

We have argued in section~\ref{sec:relevant}, that operators of
dimension greater than four are irrelevant and that the influence of
their initial value at a high matching scale on the physics at much
lower energies can be ignored.

This argument can not be applied in two important cases.  If the
irrelevant interaction violates a symmetry that is respected by all
relevant and marginal operators, the effects of the irrelevant
interactions can be observable, even if they are small in absolute
terms.  A different problem arises if the hierarchy of scales is not
large enough and the lever arm of the~\ac{RG} is therefore
not large enough to
suppress irrelevant operators when running from one threshold to the
next. 

A typical example for the latter case is provided by the thresholds
for the top quark and the $W^\pm$ bosons.  The power
corrections~$M_W/m_t\approx0.45$ can not be neglected relative to the
logarithmic terms~$|\ln(M_W/m_t)|\approx0.8$.  The solution in cases
like $B^0\bar {B^0}$-mixing is to decouple the top quark and $W$ boson
\emph{simultaneously}
\begin{equation}
  \parbox{27\unitlength}{%
    \fmfframe(3,5)(3,5){%
      \begin{fmfgraph*}(20,20)
        \fmflabel{$b$}{b}
        \fmflabel{$d$}{d}
        \fmflabel{$\bar b$}{bbar}
        \fmflabel{$\bar d$}{dbar}
        \fmfleft{b,dbar}
        \fmfright{d,bbar}
        \fmf{fermion}{b,b'}
        \fmf{fermion}{d',d}
        \fmf{fermion}{bbar,bbar'}
        \fmf{fermion}{dbar',dbar}
        \fmf{fermion,label=$t$,label.side=right,tension=0.5}{b',d'}
        \fmf{fermion,label=$\bar t$,label.side=right,tension=0.5}{bbar',dbar'}
        \fmf{boson,label=$W$,label.side=left,tension=0.5}{b',dbar'}
        \fmf{boson,label=$W$,label.side=left,tension=0.5}{bbar',d'}
        \fmfdot{b',bbar',d',dbar'}
      \end{fmfgraph*}}} +
  \parbox{27\unitlength}{%
    \fmfframe(3,5)(3,5){%
      \begin{fmfgraph*}(20,20)
        \fmflabel{$b$}{b}
        \fmflabel{$d$}{d}
        \fmflabel{$\bar b$}{bbar}
        \fmflabel{$\bar d$}{dbar}
        \fmfleft{b,dbar}
        \fmfright{d,bbar}
        \fmf{fermion}{b,b'}
        \fmf{fermion}{d',d}
        \fmf{fermion}{bbar,bbar'}
        \fmf{fermion}{dbar',dbar}
        \fmf{fermion,label=$t$,label.side=left,tension=0.5}{b',dbar'}
        \fmf{fermion,label=$t$,label.side=left,tension=0.5}{bbar',d'}
        \fmf{boson,label=$W$,label.side=right,tension=0.5}{b',d'}
        \fmf{boson,label=$W$,label.side=right,tension=0.5}{bbar',dbar'}
        \fmfdot{b',bbar',d',dbar'}
      \end{fmfgraph*}}} =
  \Phi\left(\frac{M_W^2}{m_t^2}\right)\cdot
  \parbox{27\unitlength}{%
    \fmfframe(3,5)(3,5){%
      \begin{fmfgraph*}(20,20)
        \fmflabel{$b$}{b}
        \fmflabel{$d$}{d}
        \fmflabel{$\bar b$}{bbar}
        \fmflabel{$\bar d$}{dbar}
        \fmfleft{b,dbar}
        \fmfright{d,bbar}
        \fmf{fermion}{b,v,d}
        \fmf{fermion}{bbar,v,dbar}
        \fmfdot{v}
      \end{fmfgraph*}}}\,,
\end{equation}
where the function~$\Phi$ is calculated from the massive
loop~\cite{Inami/Lim:1981:function}.  Since
the scales in the loop are rather close, no large logarithms can
develop and the loop expansion for~$\Phi$ is well behaved.

The $|\Delta B|=2$ operator discussed in the previous paragraph is
also an example for first case, but the more common example is
provided by tree level charged current interactions.  Below~$M_W$, the
exchange of a~$W$ can be approximated by a local interaction
\begin{equation}
\label{eq:matching-W}
  \parbox{21\unitlength}{%
    \begin{fmfgraph*}(20,15)
      \fmfstraight
      \fmfleftn{l}{2}
      \fmfrightn{r}{2}
      \fmf{fermion}{l1,vl,l2}
      \fmf{fermion}{r2,vr,r1}
      \fmf{boson,label=$W$}{vr,vl}
      \fmfdot{vr,vl}
    \end{fmfgraph*}} =
  \parbox{21\unitlength}{%
    \begin{fmfgraph}(20,15)
      \fmfstraight
      \fmfleftn{l}{2}
      \fmfrightn{r}{2}
      \fmf{fermion}{l1,v,l2}
      \fmf{fermion}{r2,v,r1}
      \fmfdot{v}
    \end{fmfgraph}} + O\left(\frac{p^2}{M_W^2}\right)
\end{equation}
of dimension six
\begin{equation}
\label{eq:Fermi'}
  L_F = \frac{G_F}{\sqrt2}\, \bar\psi(1-\gamma_5)\gamma_\mu\psi\,
                             \bar\psi(1-\gamma_5)\gamma^\mu\psi
\end{equation}
which is irrelevant.  However, the Fermi interaction $L_F$ can not be
ignored, because it includes flavor changing terms that are absent in
all relevant or marginal operators provided by the electromagnetic and
strong interaction.

The~\ac{SM} $W^\pm$-exchange is matched to the local Fermi
interaction~(\ref{eq:Fermi'}) at~$\mu=M_W$ like~(\ref{eq:matching-W}).
Below the $W$-mass, all calculations are performed using the local
Fermi interaction.  The irrelevant operator~(\ref{eq:Fermi'}) is not
renormalizable in the traditional perturbative sense, because it
requires counter terms of higher dimensions.  In the~\ac{EFT}, this is
no problem, however, because these counter terms are all determined
by matching conditions at~$M_W$.  Consequently, there is no loss of
predictive power.

The radiative corrections can be summed below~$M_W$ by solving
a~\ac{RGE} that includes the anomalous dimension of the irrelevant
operator that is derived from diagrams like
\begin{center}
  \parbox{31\unitlength}{%
    \begin{fmfgraph}(30,20)
      \fmfstraight
      \fmfleftn{l}{2}
      \fmfrightn{r}{2}
      \fmf{phantom}{l1,v,l2}
      \fmf{phantom}{r1,v,r2}
      \fmfdot{v}
      \fmffreeze
      \fmf{fermion}{l1,v,l2}
      \fmf{fermion}{r1,v1,v,v2,r2}
      \fmf{gluon,right=0.3,tension=0}{v1,v2}
      \fmfdot{v1,v2}
    \end{fmfgraph}} \;\;\;\; + \ldots
\end{center}

\subsection{Mass Independent Renormalization}
\label{sec:mass-independent}

A useful intuitive picture for the matching and running procedure is
that more and more physics from higher scales is \emph{integrated out}
as the renormalization scale is lowered.  The physics is not lost in
this process, but is condensed from Feynman diagrams into the short
distance interaction
vertices.  One can imagine this as if all momentum integrations would
be carried out from above down to the renormalization scale and that
the remaining low energy momentum integrations would be cut off at the
renormalization scale.

This intuitive picture is, however, not entirely accurate.  As mentioned
above, a fixed momentum space cut off is technically inconvenient,
because of complicated integrals.  But more important is that the
fixed cut off is physically
dangerous (see e.\,g.~\cite{Burgess/London:1993:effective}),
because gauge symmetries are violated by cut-offs.
Therefore the dividing line between the short distance physics, which is
absorbed in effective local interactions, and the long distance physics
is gauge dependent and thus 
unphysical.   Typically, the unphysical short distance coefficients will be
spuriously large in magnitude, but their contributions will cancel in
physical observables against unphysical long distance contributions.  In order
to exploit the cancellation of unphysical contributions as soon as
possible, it is preferable to extend all loop integrals over the whole
momentum space and to replace the hard cut offs by counter terms.  The
price to pay is small, because the resulting divergencies can be
handled with traditional perturbative renormalization theory.
Calculations in mass independent scheme like minimal subtraction are
convenient and do respect gauge invariance.  The physically important
threshold effects are handled by matching.

\subsection{Naive Dimensional Analysis}
\label{sec:NDA}

If the theory at the high energy scale is known, \ac{EFT}~is but a
very powerful calculational tool. It allows to use the most
appropriate degrees of freedom at each scale by matching different
theories at thresholds.  Furthermore, large radiative corrections can
be summed systematically through the renormalization group.

In such a scenario, the short distance coefficients are calculated by
matching and running from the top to the bottom.  The opposite case is
when low energy data is fitted to an arbitrary effective Lagrangian
and this Lagrangian is evolved from bottom to top until it becomes
inconsistent, as shown in figure~\ref{fig:lambdanot2infinity}.  In
this case, reliable estimates of the scale of ``new physics'' can be made,
but the ``new physics'' above the threshold can rarely be determined
from the matching conditions at the threshold.

Nevertheless, we can use~\ac{EFT} to estimate the likely size of
contributions from ``new physics'' to observables at future
experiments using \acf{NDA}~\cite{Weinberg:1979:EFT,
Georgi/Manohar:1984:NDA,*Georgi/Randall:1986:NDA,*Georgi:1993:NDA,
Georgi:1984:BlueBook}.  We know from general \ac{RG}~arguments, that the
relevant operators will have larger coefficients at low energy than
the irrelevant operators of the same symmetry.  That this statement
remains true after renormalization was first pointed out
in~\cite{Weinberg:1979:EFT}.  Unfortunately these arguments alone give
no indication of the absolute magnitude of the coefficients.

{}From~(\ref{eq:tadpole-result}) and~(\ref{eq:finite-fish}) we see that
typical loop integrals that contain a single scale can be estimated by
a factor of
\begin{equation}
\label{eq:1/16pi2'}
  \frac{1}{16\pi^2} \approx 6.3 \cdot 10^{-3}\,,
\end{equation}
which is multiplied by a power of this scale as dictated by simple
dimensional analysis. A systematic analysis~\cite{Weinberg:1979:EFT,
Georgi/Manohar:1984:NDA,*Georgi/Randall:1986:NDA,*Georgi:1993:NDA}
reveals that the small number~(\ref{eq:1/16pi2'}) creates a hierarchy
that can be used classify terms.

One possible form\footnote{There are many equivalent
forms~\protect\cite{Coleman/Wess/Zumino:1969:CCWZ,
*Callan/Coleman/Wess/Zumino:1969:CCWZ}, but
\emph{all} involve the field~$\phi(x)$ only in the
combination~$\phi(x)/f$.} of the lowest order effective
Lagrangian in the momentum expansion~\cite{Weinberg:1979:EFT} for a
non-linearly realized broken abelian\footnote{In the general
non-abelian case, the formulae are more complicated, because the
factor group manifold~$G/H$ has to be
parametrized~\protect\cite{Coleman/Wess/Zumino:1969:CCWZ,
*Callan/Coleman/Wess/Zumino:1969:CCWZ}. These geometrical complications
have no impact on the power counting argument.} symmetry is
\begin{subequations}
\begin{equation}
  L = f^2 \tr\left((D_\mu U)^2 D^\mu U\right)
\end{equation}
where
\begin{equation}
  U(x) = e^{i\phi(x)/f}
\end{equation}
\end{subequations}
and~$f$ is a scale introduced by the symmetry breaking.  It
corresponds to the vacuum expectation value of a Higgs field in a
linear realization of the symmetry breaking.

Loops renormalize the operators of a given order as well as higher
orders in the momentum expansion.  For example, a loop with two lowest
order terms 
\begin{equation}
  \frac{1}{f^2}
    \left(\phi(x)\overleftrightarrow{\partial_\mu}\phi(x)\right)
    \left(\phi(x)\overleftrightarrow{\partial^\mu}\phi(x)\right)
\end{equation}
renormalized at a scale~$\mu^2$
\begin{equation}
\label{eq:NDA-fish}
  \parbox{31\unitlength}{%
    \begin{fmfgraph}(30,15)
      \fmfleftn{l}{2}
      \fmfrightn{r}{2}
      \fmf{plain}{l1,vl}
      \fmf{plain}{l2,vl}
      \fmf{plain}{r1,vr}
      \fmf{plain}{r2,vr}
      \fmf{plain,tension=0.5,left=0.5}{vl,vr,vl}
      \fmfdot{vr,vl}
    \end{fmfgraph}} = 
  \frac{1}{16\pi^2} \frac{P^2}{f^2} \ln\left(\frac{\mu^2}{P^2}\right)\cdot
  \parbox{16\unitlength}{%
    \begin{fmfgraph}(15,15)
      \fmfleftn{l}{2}
      \fmfrightn{r}{2}
      \fmf{plain}{l1,v}
      \fmf{plain}{l2,v}
      \fmf{plain}{r1,v}
      \fmf{plain}{r2,v}
      \fmfdot{v}
    \end{fmfgraph}} + \ldots \,,
\end{equation}
where~$P^2$ is an invariant built from the external momenta,
requires a counterterm from the next order of the momentum expansion
\begin{equation}
\label{eq:NDA-counterterm}
  \frac{1}{\Lambda^2} \frac{1}{f^2}
    \partial_\nu\phi(x)\partial_\mu\phi(x)
    \partial^\nu\phi(x)\partial^\mu\phi(x)\,,
\end{equation}
where~$\Lambda$ is the scale characterizing the momentum expansion.  A
priori, there is \emph{no} relation between the scales~$f$
and~$\Lambda$ for symmetry breaking and momentum expansion,
respectively.  However, unless there is some unnatural fine-tuning of
parameters at work, the rate of change of the couplings in the
\ac{RG}~running from~(\ref{eq:NDA-fish}) \emph{should} be of the same order
as the coefficients of the counterterm~(\ref{eq:NDA-counterterm}).
Equating coefficients gives
\begin{equation}
  \frac{1}{16\pi^2} \frac{1}{f^2}
  \approx \frac{1}{\Lambda^2}
\end{equation}
or the celebrated formula
\begin{equation}
\label{eq:Lambda4pif}
  \Lambda \approx 4\pi f\,.
\end{equation}
{}From another perspective, the \ac{RG} will evolve
towards~(\ref{eq:Lambda4pif}) in the infrared, unless a symmetry that
has not been made manifest in the effective Lagrangian forces the
\ac{RG} flow in a particular direction.  A physical threshold
between~$f$ and~$\Lambda$ would require matching and provides the only
way around~(\ref{eq:Lambda4pif}) in an~\ac{EFT}.  Such a physical
threshold will however be observable independently.

The heuristic argument for the
lowest order in the momentum expansion for scalar fields can be
extended systematically to all orders in the loop and momentum
expansions and to fermions and vector
fields~\cite{Georgi/Manohar:1984:NDA,*Georgi/Randall:1986:NDA,
*Georgi:1993:NDA}.  The result is that a natural interaction involving
scalars~$\phi$, fermions~$\psi$, derivatives~$\partial$ and gauge
bosons~$A$ has the size 
\begin{equation}
\label{eq:NDA}
  L^{(N_{\phi},N_{\psi},N_A,N_{\partial})} =
    f^2 \Lambda^2
    \left(\frac{\phi}{f}\right)^{N_{\phi}}
    \left(\frac{\psi}{f\sqrt{\Lambda}}\right)^{N_{\psi}}
    \left(\frac{eA}{\Lambda}\right)^{N_A}
    \left(\frac{\partial}{\Lambda}\right)^{N_{\partial}}
\end{equation}
for any quartet~$(N_{\phi},N_{\psi},N_A,N_{\partial})$ of natural
numbers.  The power counting rules assume that the vector boson~$A$ is
indeed a gauge boson, such that the two parts of the covariant
derivative~$\partial-ieA$ scale identically.  The rules would be much
more complicated for non-gauge bosons and their longitudinal
polarization states.

The important result of~\ac{NDA} is that the natural
strengths~(\ref{eq:NDA}) of interactions can be estimated, even in a
strongly interacting~\ac{EFT}.  The assumption that couplings have
their natural strengths is nothing more than an assumption, of course.
Nevertheless, the predictions from~\ac{NDA} work for low energy strong
interactions and~\ac{NDA} can be used as a guide while searching for
good places for hunting ``new physics.''  Higher dimensional operators
are most likely suppressed by a factor of~$1/(16\pi^2)$ and proposals
for experiments should take this into account.

\bibliography{hbl}

  \acrodef{1PI}{One Particle Irreducible}
  \acrodef{ABBJ}{Adler-Bell-Bardeen-Jackiw}
  \acrodef{BPHZ}{Bolgoliubov-Parasiuk-Hepp-Zimmermann}
  \acrodef{BRS}{Becchi-Rouet-Stora}
  \acrodef{CSE}{Callan-Symanzik Equation}
  \acrodef{DIS}{Deep Inelastic Scattering}
  \acrodef{DR}{Dimensional Regularization}
  \acrodef{EFT}{Effective Field Theory}
  \acrodef{EPP}{Elementary Particle Physics}
  \acrodef{EWSB}{Electro Weak Symmetry Breaking}
  \acrodef{GLAP}{Gribov-Lipatov-Altarelli-Parisi}
  \acrodef{GUT}{Grand Unified Theory}
  \acrodef{HEP}{High Energy Physics}
  \acrodef{LC}{Linear Collider}
  \acrodef{LHC}{Large Hadron Collider}
  \acrodef{LLA}{Leading Logarithmic Approximation}
  \acrodef{MS}{Minimal Subtraction}
  \acrodef{MSSM}{Minimal Supersymmetric Standard Model}
  \acrodef{NDA}{Naive Dimensional Analysis}
  \acrodef{ODE}{Ordinary Differential Equation}
  \acrodef{OPE}{Operator Product Expansion}
  \acrodef{PDE}{Partial Differential Equation}
  \acrodef{PDG}{Particle Data Group}
  \acrodef{PT}{Perturbation Theory}
  \acrodef{QAP}{Quantum Action Principle}
  \acrodef{QCD}{Quantum Chromodynamics}
  \acrodef{QED}{Quantum Electrodynamics}
  \acrodef{QFT}{Quantum Field Theory}
  \acrodef{QM}{Quantum Mechanics}
  \acrodef{RGE}{Renormalization Group Equation}
  \acrodef{RG}{Renormalization Group}
  \acrodef{SM}{Standard Model}
  \acrodef{SSB}{Spontaneous Symmetry Breaking}
  \acrodef{TEPP}{Theoretical Elementary Particle Physics}
  \acrodef{TGC}{Triple Gauge Couplings}
  \acrodef{pQCD}{perturbative \ac{QCD}}

\end{fmffile}
\end{empfile}
\end{document}